\documentclass[preprint]{ptephy_v1}
\preprintnumber{XXXX-XXXX} 
\usepackage{amsmath}
\usepackage{amssymb}
\usepackage{graphicx}
\usepackage{bm}
\usepackage{color}
\usepackage{hyperref}
\usepackage{graphicx}
\usepackage{upgreek}
\usepackage{scalerel}
\usepackage{multirow}
\usepackage{mathtools}

\usepackage{physics} 
\usepackage{mathdots}

\begin{document}

\title{Theory of Majorana zero modes in unconventional superconductors}


\author{Yukio Tanaka}
\affil{Department of Applied Physics, Nagoya University, Nagoya 464--8603, Japan 
}
\affil{Research Center for Crystalline Materials Engineering, Nagoya University, Nagoya 464-8603, Japan}

\author{Shun Tamura}
\affil{Institute for Theoretical Physics and Astrophysics, University of W\"{u}rzburg, D-97074 W\"{u}rzburg, Germany}

\author{Jorge Cayao}
\affil{Department of Physics and Astronomy, Uppsala University, Box 516, S-751 20 Uppsala, Sweden}



\begin{abstract}%
Majorana fermions are spin-1/2 neutral particles that are their own antiparticles and were initially predicted by Ettore Majorana in particle physics but their observation still remains elusive. 
The concept of Majorana  fermions has been borrowed into condensed matter physics, where, unlike particle physics,   Majorana fermions emerge as zero-energy quasiparticles that can be engineered by combining electrons and holes and have therefore been coined Majorana zero modes.  
In this review, we provide a pedagogical explanation of  the basic properties of Majorana zero modes in unconventional superconductors and their consequences in experimental observables, putting a special emphasis on the initial theoretical discoveries.  In particular, we first show that  Majorana zero modes are self-conjugated and emerge as a special type of zero energy surface Andreev bound states at the boundary of unconventional superconductors.  We then explore Majorana zero modes in one-dimensional spin-polarized $p$-wave superconductors,  where we address the formation of topological superconductivity and the physical realization in superconductor-semiconductor hybrids. 
In this part we highlight that Majorana quasiparticles appear as zero-energy edge states, exhibiting charge neutrality, spin-polarized, and spatial nonlocality as unique properties that can be already seen from their energies and wavefunctions. Next, we discuss analytically obtained Green's functions of $p$-wave superconductors and demonstrate that the emergence of Majorana zero modes is always accompanied by the formation of odd-frequency spin-triplet pairing as a unique result of the self-conjugate nature of Majorana zero modes. We finally address the signatures of Majorana zero modes in tunneling spectroscopy, including the anomalous proximity effect, and the phase-biased Josephson effect.
\end{abstract}

\subjectindex{xxxx, xxx}

\maketitle

\section{Introduction}
\label{intro}
Majorana fermions were initially introduced by Ettore Majorana in 1937 as real solutions to the Dirac equation, later known as the Majorana equation, which describes spin-1/2 particles that are identical to their own antiparticles \cite{majorana}. Since their conception, Majorana fermions have been the subject of intense research in nuclear and particle physics but their detection still remains an open problem \cite{RevModPhys.80.481,neutrinoless_2012,giuliani2012neutrinoless,dell2016neutrinoless,dolinski2019neutrinoless}. Despite the challenges at high energies, the Majorana equation has been shown to naturally appear at low energies in the condensed matter realm, where its solutions, the Majorana fermions, appear instead as emergent quasiparticles composed of electrons and holes in systems with superconducting order. As a matter of fact, quasiparticles in superconductors are described by the Bogoliubov-de Gennes equation \cite{Svidzinskii82,vonsovsky1982superconductivity,schmidt1997physics,zagoskin,abrikosov2017fundamentals,Tinkham,de2018superconductivity}, where the inherent particle-hole symmetry makes it to acquire the same form as the Majorana equation \cite{PhysRevB.61.9690,SATO2003126,10.1119/1.3549729,PhysRevB.81.224515,imura2012majorana,PhysRevLett.112.070604}.  Majorana quasiparticles in condensed matter emerge as edge or boundary states in a special type of superconductor known as topological superconductor. In two dimensions (2D), the topological superconducting phase can be created in spinless superconductors with $p_{x}\pm ip_{y}$ order parameters or pair potentials, which then hosts Majorana quasiparticles propagating along the boundaries or at zero-energy  bound to a vortex core \cite{jackiw1981zero,volovik1999fermion,doi:10.1143/JPSJ.68.994,PhysRevB.44.9667,PhysRevB.61.10267,Yakovenko2001,PhysRevLett.86.268,volovik2003universe,PhysRevLett.101.160401, PhysRevB.79.094504,Jackiw_2012}.  In one dimension (1D), the topological superconducting phase appears in spinless (or spin-polarized) $p$-wave superconductors where Majorana quasiparticles form at the edge at zero energy \cite{kitaev}. At this point, we note that the 1D spinless $p$-wave superconductor is often referred to as the Kitaev model. Moreover, Majorana states in condensed matter are also called Majorana zero modes (MZMs) or Majorana bound states (MBSs) for obvious reasons in order to distinguish them from their high-energy counterparts.   
MZMs or MBSs are sometimes called Majorana fermions but this might be an abuse of the language given that these quasiparticles are not exactly the originally thought by Ettore Majorana in high energy physics.
In contrast to the high-energy Majorana fermions,  the low-energy MZMs exhibit yet another distinguishing property which is based on the fact that braiding MZMs implements non-Abelian unitary transformations, revealing their non-Abelian statistics \cite{MOORE1991362,PhysRevLett.86.268,SATO2003126,PhysRevB.73.014505,PhysRevLett.103.020401,PhysRevB.79.094504,awoga2023controlling} that can be used for topologically protected quantum computation \cite{RevModPhys.80.1083,sarma2015majorana,Lahtinen_2017,Beenakker_2020,aguado2020majorana}. MZMs are also special because a pair of them defines a nonlocal fermion, where information is stored nonlocally and is thus immune to local sources of decoherence. Therefore, MZMs are not only of fundamental relevance, characterizing topological superconductivity, but they also exhibit interesting properties useful for possible applications in future quantum technologies. 

The origin of MZMs as boundary states in topological superconductors also reflects a less discussed property, that they in fact represent a special type of zero-energy surface Andreev bound states (ZESABSs) in unconventional superconductors \cite{kashiwaya00,tanaka12}. In contrast to conventional superconductors, where the order parameter is isotropic ($s$-wave) in space and spin-singlet, the order parameter in an unconventional superconductor is often anisotropic in space, with a nodal structure on the Fermi surface \cite{RevModPhys.63.239}. The research on ZESABSs in unconventional superconductors dates back to the 80s in the context of superfluid $^{3}$He \cite{PhysRevB.23.5788,10.1143/PTP.76.1237} and later on the pairing symmetry of high $T_{\rm c}$ cuprates \cite{Hu94,kashiwaya00}; for a review on cuprates, see Ref.\,\cite{RevModPhys.72.969}. Interestingly, one of the authors (Y. T.)  has shown that ZESABSs produce a zero-bias conductance peak (ZBCP) in high $T_{\rm c}$ cuprates  \cite{TK95} and revealed that tunneling spectroscopy in unconventional superconductors is a phase-sensitive probe \cite{KT96}. An outstanding feature of these studies is that the conductance formula derived in Refs.\,\cite{TK95,KT96,kashiwaya00} reveals that the ZESABSs induce perfect Andreev reflection, which is the responsible effect giving rise to ZBCPs and pioneered the search of MZMs via conductance signatures in topological superconductors, see next paragraph. The existence of ZBCPs has been confirmed by many experiments of tunneling spectroscopy in high $T_{c}$ cuprates 
\cite{Experiment1,Experiment2,Experiment3,Experiment4,Experiment5,Iguchi2000,Experiment6,Experiment7}, which also allowed to establish the spin-singlet $d$-wave pairing symmetry in cuprates  \cite{kashiwaya00,Lofwander2001,Deutscher}.  Along these lines, ZBCPs have also been observed via tunnel spectroscopy in  CeCoIn$_{5}$ \cite{Rourke} and PuCoIn$_{5}$ \cite{Daghero2012}, whose pair potential also  has spin-singlet $d$-wave symmetry. Furthermore,  ZESABSs have been detected in   grain boundary Josephson junctions made of cuprates \cite{Testa},  with signals leading to an enhancement and also a non-monotonic temperature dependence in the supercurrents. Interestingly,   Josephson junctions formed by $d$-wave superconductors have also been shown to host non-sinusoidal current-phase curves as a unique phenomenon due to the unconventional superconducting pair symmetry \cite{PhysRevLett.89.207002,PhysRevLett.90.117002,Ilichev,PhysRevB.76.224523,RevModPhys.76.411}, see also Ref.\,\cite{Tafuri_2005}. 

The formation of ZESABSs in unconventional superconductors is a universal phenomenon that is not restricted to superconductors with a pair potential having a nodal structure and spin-singlet symmetry as discussed in the previous paragraph. In fact, ZESABSs have also been shown to appear in spin-triplet $p$-wave  and spin-triplet $f$-wave superconductors  \cite{YTK97,YTK98,Honerkamp1998,Kusakabe1999,Yakovenko2001,Tanaka2002,Tanuma2002b}, {where their identification was entirely due to the appearance of robust ZBCPs. In the simplest situation, it has been shown that tunneling into the edge of a 1D spin-polarized $p$-wave superconductor gives rise to a  ZBCP quantized to $2e^{2}/h$ due to perfect reflection induced by an MZM, signaling the topological superconducting nature of such a state. This conclusion has its origin in the conductance formula derived by Y. T. in Refs.\,\cite{TK95,KT96,kashiwaya00} in the context of $d$-wave superconductors}; see also \cite{PhysRevLett.98.237002, PhysRevLett.103.237001, PhysRevB.82.180516}.  These findings enabled the potential of MZMs, and in general of ZESABSs,  for distinguishing between spin-singlet and spin-triplet superconductors. Motivated by this fact, Y. T. later obtained a charge transport formula in diffusive normal metal-unconventional superconductor junctions \cite{Proximityd,Proximityd2}, which revealed the robust nature of ZBCPs and further helped to clarify the impact of ZESABSs. Furthermore,  a zero-energy peak has been shown to appear in the local density of states (LDOS) of the diffusive normal metal (DN) when attached to a spin-triplet superconductor \cite{Proximityp}, a process that has been attributed to the penetration of a ZESABS generated at the interface and accompanies the leakage of superconducting correlations into DN. This phenomenon, together with the ZBCP, in spin-triplet  superconductors was coined anomalous proximity effect   \cite{Proximityp,Proximityp2,Proximityp3} to distinguish from the conventional proximity effect by spin-singlet superconductors which instead is characterized by a zero-energy dip in the LDOS \cite{Golubov88,Belzig96}.   

Further insights on the physical origin of the anomalous proximity effect have been obtained by analyzing the symmetries of the proximity induced superconducting correlations or Cooper pairs into the DN region. In this respect, it has been identified that the formation of ZESABSs in spin-singlet $d$-wave   superconductors accompanies the emergence of odd-frequency spin-singlet $p$-wave pairs, which, however, cannot penetrate into the DN part \cite{odd1,odd3,odd3b}. In contrast,   in spin-triplet $p$-wave superconductors, the induced pair correlations acquire odd-frequency spin-triplet $s$-wave symmetry which is robust against impurity scattering and thus penetrate into  DN, whose accompanied zero-energy LDOS peak can be detected as a ZBCP in a tunneling experiment  
\cite{odd1,Tshape,tanaka12}. We note that the odd-frequency pair symmetry mentioned here corresponds to a symmetry where the pair amplitude (or Cooper pair wavefunction) is an odd function under the exchange of time coordinates of  two electrons forming the Cooper pair, see  Refs.\,\cite{Berezinskii,Efetov2,golubov2009odd,tanaka12,LinderBalatsky,triola2020role,Cayao_2020} for relevant reviews. Since the ZESABSs in 1D spin-polarized $p$-wave superconductors are MZMs, the formation of odd-frequency spin-triplet $s$-wave pairing corresponds to a strong signature of Majorana physics and topological superconductivity \cite{tanaka12,Cayao_2020}. This intriguing relationship has been demonstrated in junctions based on $p$-wave superconductors \cite{Proximityp,Takagi2020,Spectralbulk,PhysRevB.100.115433,PhysRevB.101.214507}, $d$-wave superconductors \cite{Tamura2019PRB}, topological insulators \cite{PhysRevB.86.075410,PhysRevB.86.144506,PhysRevB.87.220506,PhysRevB.92.205424,Lu_2015,PhysRevB.92.100507,PhysRevB.96.155426,PhysRevB.96.174509,bo2016,PhysRevB.97.075408,PhysRevLett.120.037701,PhysRevB.97.134523,PhysRevB.101.180512,PhysRevB.106.L100502}, Weyl semimetals \cite{PhysRevB.100.104511,parhizgar2020large},  semiconductors with Rashba spin-orbit coupling \cite{Asano2013,Ebisu16,PhysRevB.98.075425,Tamura2019PRB},  
superconductor junctions with quantum anomalous Hall insulator \cite{Nakai2021}, superconducting doped topological insulator \cite{Mizushima2023},  
Majorana arrays \cite{PhysRevB.95.184506,PhysRevB.101.094506},  Sachdev-Ye-Kitaev setups \cite{PhysRevB.99.024506}, and interacting MZMs \cite{PhysRevB.92.121404}. Moreover, it has been revealed that unconventional superconductors hosting ZESABSs, such as spin-singlet $d$-wave or spin-triplet $p$-wave,  can be classified as topological superconductors, with topological invariants that distinguish the absence or presence of the anomalous proximity effect \cite{STYY11}. At the moment, there is a relevant experimental report detecting  ZBCPs in   ${\mathrm{CoSi}}_{2}/{\mathrm{TiSi}}_{2}$ heterostructures \cite{Lin2021,Lin2023}, which could be an indicator of anomalous proximity effect and of odd-frequency spin-triplet $s$-wave superconducting pairing.  Thus, the anomalous proximity effect, via zero-energy LDOS peaks, ZBCPs, and odd-frequency spin-triplet $s$-wave pairing, can serve as a powerful signature of topological superconductivity and MZMs. 

The motivation to discover MZMs is, therefore, enormous, including the characterization of a new state of matter and also their potential for future quantum applications.  There is, however, a crucial problem which is based on that the intrinsic spin-polarized $p$-wave superconductors, needed for realizing topological superconductivity and MZMs, are rare in nature. Nonetheless, it has been shown that 1D spin-polarized $p$-wave superconductivity and MZMs can be engineered by combining conventional spin-singlet $s$-wave superconductivity, spin-orbit coupling, and a strong magnetic field \cite{PhysRevB.77.220501,PhysRevB.79.094504,PhysRevLett.103.020401,PhysRevLett.104.040502,PhysRevLett.105.077001,PhysRevLett.105.177002,Alicea:PRB10,PhysRevLett.108.147003}, which represents one of the most studied physical realizations of topological superconductivity; for the realization using topological insulators, see Refs.\,\cite{Fu:PRL08,Fu:PRB09,RevModPhys.82.3045,RevModPhys.83.1057}, while for using chains of magnetic atoms, see Refs.\, \cite{PhysRevB.84.195442,PhysRevLett.111.186805,PhysRevB.88.020407,PhysRevLett.111.147202,PhysRevB.88.155420,PhysRevB.88.180503,EbisuKasai2015,PhysRevB.95.184511}. The simplicity of the recipe has inspired an impressive number of theoretical studies addressing the main properties of MZMs and their experimental signatures, including quantized ZBCPs \cite{PhysRevLett.109.227006,PhysRevB.86.180503,PhysRevB.86.224511,PhysRevB.87.024515,PhysRevResearch.3.023221,cao2023recent,lutchyn2018majorana,zhang2019next,prada2019andreev,frolov2020topological,flensberg2021engineered},  non-sinusoidal current-phase curves \cite{San-Jose:11a,PhysRevLett.112.137001,PhysRevB.94.085409,PhysRevB.96.165415,PhysRevB.96.205425,Ilan_2014,cayao2018andreev,cayao2018finite,luethi2023majorana}, anomalous proximity effect and odd-frequency pairing \cite{Asano2013,Spectralbulk,Tamura2019PRB}. On the experimental side, ZBCPs have been mostly pursued \cite{Mourik:S12,Chang2015Hard, Higginbotham2015, deng2016majorana, Albrecht16, PhysRevLett.119.176805, PhysRevLett.119.136803, Gul2018,zhang2019next,zhang2021large} but the unambiguous detection of MZMs still remains challenging; few experiments have also explored the Josephson effect, which show interesting promising signatures \cite{Tiira2017,PhysRevLett.125.116803,PhysRevLett.126.036802,moehle2022controlling}. One of the main current problems on the conductance experiments is that these systems also host  ZBCPs at finite magnetic fields due to topologically trivial zero-energy states \cite{PhysRevB.86.100503, PhysRevB.86.180503,PhysRevB.91.024514,San-Jose2016, PhysRevB.96.075161, PhysRevB.96.195430, PhysRevB.97.155425, PhysRevB.98.245407, PhysRevLett.123.107703, PhysRevB.100.155429, PhysRevLett.123.217003, 10.21468/SciPostPhys.7.5.061, Avila2019, PhysRevResearch.2.013377, PhysRevLett.125.017701, PhysRevLett.125.116803, PhysRevB.102.245431, Yu2021, prada2019andreev, Pal2018, doi:10.1126/science.abf1513, PhysRevB.101.195303, PhysRevB.98.155314, PhysRevB.97.161401, PhysRevB.101.014512, PhysRevB.104.134507, PhysRevB.104.L020501, Marra_2022x, PhysRevB.105.035148, Schuray2020, PhysRevB.102.045111, Grabsch2020, PhysRevB.102.245403, PhysRevB.103.144502,chen2022topologically,PhysRevB.105.144509,PhysRevB.106.014522,PhysRevB.107.184519,PhysRevLett.130.207001}, thus making it difficult to detect  MZMs clearly. By now it is well understood that, while ZBCPs indeed represent a necessary condition for MZMs, it is not sufficient and our efforts need to go beyond local conductance signatures \cite{PhysRevB.96.085418,PhysRevB.97.161401,PhysRevB.98.085125,PhysRevLett.123.117001,PhysRevB.102.045111,PhysRevB.104.L020501,baldo2023zero,PhysRevB.105.035148,PhysRevB.109.045132,PhysRevLett.130.116202,sugeta2023enhanced,PhysRevB.108.205405,PhysRevB.109.L081405}. In this regard, recent advances have reported the fabrication of high-quality samples, where  
a bottom-up engineering of a minimal Kitaev chain is being pursued \cite{dvir2023realization,zatelli2023robust,bordin2023crossed} and  
the control over the emergent spin-triplet superconductivity seems to be feasible  
\cite{Wang_Nat2022,Bordoloi_Nat2022,Wang_Natcom2023,PhysRevX.13.031031}. It is therefore reasonable to think that the unambiguous identification of true MZMs is very likely to occur in the near future. 

The aim of this review is to provide a pedagogical discussion about MZMs from the viewpoint of ZESABSs in unconventional superconductors, placing special emphasis on the initial theoretical discoveries. We start by introducing the BdG equations in unconventional superconductors in Section \ref{section2} and therein by simple arguments we demonstrate the emergence of ZESABSs and MZMs and their relationship. In Section \ref{section3} we discuss the properties of MZMs in 1D $p$-wave superconductors, highlighting their models, topological properties, and a brief discussion on the physical realization in the semiconductor-superconductor system. In Section \ref{section4}, we investigate the emergent pair correlations in 1D $p$-wave superconductors by means of Green's function, something that we believe is crucial for understanding MZMs but often ignored. In Sections \ref{section5}, \ref{section6}, and \ref{section7} we discuss experimental signatures of MZMs in the charge conductance, anomalous proximity effect, and Josephson effect.
 
We also note that this review does not intend to cover all the exciting developments in the field of topological superconductivity and MZMs. There are brilliant reviews covering different aspects of these topics and we would like the readers to refer to those works. For a review on Majorana fermions in nuclear, particle, and solid state physics, we refer to Ref.\,\cite{RevModPhys.87.137}. For readers interested in physical implementations of  $p$-wave superconductivity and experimental signatures of MZMs, we refer to Refs.\,\cite{leijnse2012introduction,Alicea_2012,Beenakker_2013,RevModPhys.87.1037,Aguadoreview17,lutchyn2018majorana,zhang2019next,prada2019andreev,frolov2020topological,flensberg2021engineered,Laubscher2021,Marra_2022}. For details on MZMs in topological superconductors and their topological properties, we refer to Refs.\,\cite{Sato_2017,doi:10.7566/JPSJ.85.072001}. For readers interested in MZMs in topological insulators, magnetic chains, and helical liquids, we refer to Refs.\,\cite{RevModPhys.82.3045,RevModPhys.83.1057,Tkachov_2013}, Ref.\,\cite{PAWLAK20191}, and Ref.\,\cite{Hsu_2021}, respectively. For readers interested in modeling MZMs in semiconductor-superconductor structures, we refer to Ref.\,\cite{Stanescu_2013}. For details on the potential application of MZMs topological quantum computing, we refer the readers to Refs.\,\cite{RevModPhys.80.1083,hassler2014majorana,sarma2015majorana,Lahtinen_2017,aguado2020majorana,Beenakker_2020}. For an extensive discussion on the emergent superconducting pair symmetries in unconventional superconductors and the relationship between topology and MZMs in topological superconductors, we refer to  Refs.\,\cite{tanaka12,Cayao_2020}.

\section{Surface Andreev bound states and Majorana zero modes in unconventional superconductors}
\label{section2}
In this section, we elucidate the formation of 
surface Andreev bound states (SABSs) and their relation to MZMs in unconventional superconductors. For this purpose, we first briefly discuss the structure of the pair potential in unconventional superconductors, outline the BdG approach to describe quasiparticles in superconducting systems, and 
finally, show how SABSs and MZMs emerge in unconventional superconductors.

\subsection{Order parameter of unconventional superconductors}
\label{subsection11}
Superconductors are characterized by a pair potential or order parameter ${\bf \Delta}$, which can be understood as a macroscopic wavefunction of  Cooper pairs whose symmetries determine the type of superconductor. Generally speaking, the pair potential  depends on two spatial coordinates  and is  a matrix in spin space,
\begin{equation}
\label{Deltamatrix}
{\bf \Delta}\left(\bm{r},\bm{r}' \right)=
\begin{pmatrix}
\Delta_{\uparrow\uparrow}\left(\bm{r},\bm{r}'\right)
& \Delta_{\uparrow\downarrow}\left(\bm{r},\bm{r}' \right)
\\
\Delta_{\downarrow\uparrow}\left(\bm{r},\bm{r}' \right)
&\Delta_{\downarrow\downarrow}\left(\bm{r},\bm{r}' \right)
\end{pmatrix}
\end{equation}
where $\bm{r},\bm{r}'$ and $\uparrow,\downarrow$ denote spatial coordinates and spins, respectively.  
Although Eq.\,(\ref{Deltamatrix}) corresponds to the general definition of the pair potential, for pedagogical purposes, we first analyze it in a spatially uniform 
system which is standard in the literature. In this case, it is useful to define the 
center of mass and relative coordinates according to 
$\bm{R}=(\bm{r} + \bm{r}')/2$, $\bm{x}=\bm{r}-\bm{r}'$. Hence, 
${\bf \Delta}\left(\bm{r},\bm{r}' \right)=
{\bf \Delta}\left(\bm{R},\bm{x} \right)$ 
does not depend on $\bm{R}$. 
After Fourier transformation with respect to 
$\bm{x}$, 
we can define ${\bf \Delta}\left(\bm{k}\right)$ 
with
\begin{equation}
\label{Deltamatrixk}
{\bf \Delta}\left(\bm{k} \right)=
\begin{pmatrix}
\Delta_{\uparrow\uparrow}\left(\bm{k} \right)
& \Delta_{\uparrow\downarrow}\left(\bm{k} \right)
\\
\Delta_{\downarrow\uparrow}\left(\bm{k} \right)
&\Delta_{\downarrow\downarrow}\left(\bm{k} \right)
\end{pmatrix}
.
\end{equation}
On one hand, for spin-singlet pair potential, the following relations are satisfied,  
\begin{equation}
\label{Deltasingletkspace}
\Delta_{\uparrow\uparrow}\left({\bm k} \right)
=\Delta_{\downarrow\downarrow}\left({\bm k} \right)=0, \ 
 \Delta_{\uparrow\downarrow}\left({\bm k} \right)=-\Delta_{\downarrow\uparrow}\left({\bm k} \right), \ 
 \Delta_{\uparrow\downarrow}\left({\bm k} \right)=\Delta_{\uparrow\downarrow}\left(-{\bm k} \right). 
 \end{equation}
On the other hand, for spin-triplet pair potential,  
${\bf \Delta}\left(\bm{k} \right)$ is 
can be written as 
\begin{equation}
{\bf \Delta}\left({\bm k} \right)= 
[\mathbf{d}\left({\bm k} \right)\cdot\boldsymbol{\sigma}]
i\sigma_2\,, 
\label{dvectork}
\end{equation}
where $\mathbf{d}\left({\bm k} \right) = 
-\mathbf{d}\left(-{\bm k} \right)$ represents the vector containing three spin-triplet components,  $\sigma_{j}$ is the $j$-th Pauli matrix in spin space, and $\boldsymbol{\sigma}$ the vector of 
Pauli matrices. More explicitly, for pairing between opposite spins with spin projection  $S_{z}=0$,  the components of
${\bf \Delta}\left({\bm k} \right)$ exhibit the relations given by
\begin{equation}
\label{DeltatripletS0k}
\Delta_{\uparrow\uparrow}\left({\bm  k}\right)=\Delta_{\downarrow\downarrow}\left({\bm k} \right)=0 \quad {\rm and}\quad \Delta_{\uparrow\downarrow}\left(\bm{k} \right)=\Delta_{\downarrow\uparrow}\left(\bm{k} \right)\,.
\end{equation}
Similarly, for equal-spin pairing with  $S_{z}=\pm1$, we obtain
\begin{equation}
\label{DeltatripletS0ke}
\Delta_{\uparrow\downarrow}\left({\bm  k}\right)=\Delta_{\downarrow\uparrow}\left({\bm k} \right)=0 \quad {\rm and}\quad \Delta_{\uparrow\uparrow}\left(\bm{k} \right)= \pm \Delta_{\downarrow\downarrow}\left(\bm{k} \right)\,.
\end{equation}
From the above discussion, we conclude that the spin structure of the pair potential gives rise to spin-singlet and spin-triplet superconductors. Furthermore, the dependence on the momentum $\bm{k}$  gives rise to distinct spatial symmetries, 
sometimes referred to as parity. 
For instance, superconductors with a spin-singlet pair potential that is isotropic in space exhibit $s$-wave symmetry; these superconductors are also known as conventional superconductors \cite{PhysRev.108.1175}.  Moreover, it is also possible to find more exotic dependencies, such as spin-singlet $d$-wave superconductors and spin-triplet $p$-wave superconductors. 

\begin{figure}[t]
\begin{center}
\includegraphics[width=9.5cm,clip]{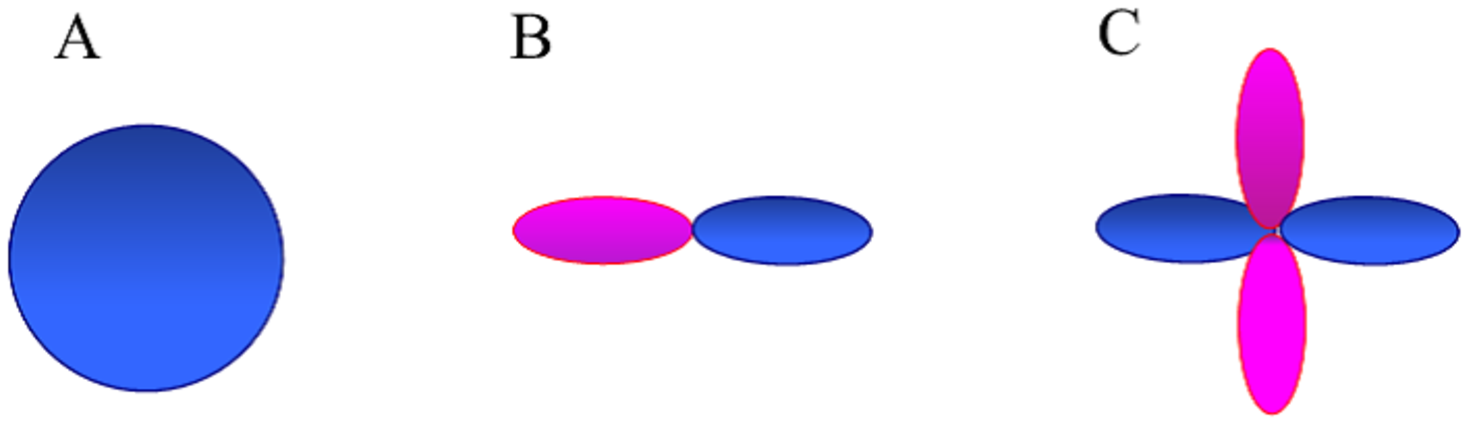}
\end{center}
\caption{Schematic representation of  pair potentials  with the following symmetries in momentum space:  A: $s$-wave, B: $p$-wave and C: 
$d$-wave.The blue and pink colors represent
opposite signs in momentum space.}
\label{Cooperpair}
\end{figure}

For a Cooper pair formed between two electrons near the Fermi surface, the pair potential $\Delta(\bm{k})$
discussed here acquires a simple form in momentum space. For spin-singlet $s$-wave, spin-triplet $p$-wave, and spin-singlet $d$-wave, respectively, they can be written as, 
\begin{equation}
\label{OPP}
\begin{split}
\Delta_{s}&=\Delta_{0}, \\
\Delta_{p}&=\Delta_{0}\hat{k}_{x} \\  
\Delta_{d}&=
\Delta_{0}\left(\hat{k}^{2}_{x}-\hat{k}^{2}_{y} \right), 
\end{split}
\end{equation}
where $(\hat{k}_{x},\hat{k}_{y})={\bm k}_{F}/k_{F}$ 
represent unit vector around the Fermi surface,
with ${\bm k}_{F}$ the Fermi wave vector and $k_{F}=|{\bm k}_{F}|$. 
These pair potentials are schematically shown 
in Fig.\,\ref{Cooperpair}. \par

In the case of systems that are not uniform in space, Eq.\, \ref{Deltamatrix} need to be described using two spatial coordinates. However, the spin structure for the distinct types of superconductors remains. In fact, for   spin-singlet superconductors, the components of the pair potential read,
\begin{equation}
\label{Deltasinglet}
\Delta_{\uparrow\uparrow}\left({\bm r},{\bm r}'\right)=\Delta_{\downarrow\downarrow}
\left({\bm r},{\bm r}'\right)=0 \quad {\rm and}\quad \Delta_{\uparrow\downarrow}\left({\bm r},{\bm r}'\right)=-\Delta_{\downarrow\uparrow}\left({\bm r},{\bm r}'\right)\,.
\end{equation}
Analogously, for spin-triplet superconductors with 
opposite-spin pairing with $S_{z}=0$, we obtain
\begin{equation}
\label{DeltatripletS0}
\Delta_{\uparrow\uparrow}\left({\bm r},\bm{r}'\right)=\Delta_{\downarrow\downarrow}\left({\bm r},{\bm r}'\right)=0 \quad {\rm and}\quad \Delta_{\uparrow\downarrow}\left(\bm{r},\bm {r}'\right)=\Delta_{\downarrow\uparrow}\left(\bm{r},\bm{r}' \right)\,,
\end{equation}
while  for equal spin-triplet superconductors with spin projections 
 $S_{z}=\pm1$, we find
\begin{equation}
\label{DeltatripletSpm1}
\Delta_{\uparrow\downarrow}\left({\bm r},\bm{r}'\right)=\Delta_{\downarrow\uparrow}\left({\bm r},{\bm r}'\right)=0 \quad {\rm and}\quad \Delta_{\uparrow\uparrow}\left({\bm r},{\bm r}' \right)=\pm\Delta_{\downarrow\downarrow}\left({\bm r},{\bm r}' \right)\,.
\end{equation} 
As for spatially uniform systems, the pair potential for non-uniform systems  can be compactly written as,
\begin{equation}
\Delta\left({\bm r},{\bm r}'\right)=
\left[\psi_s\left({\bm r},{\bm r}'\right)
+\mathbf{d}\left({\bm r},{\bm r}'\right)\cdot\boldsymbol{\sigma}
\right]i\sigma_2\,
\end{equation}
where $\psi_s$ describes the spin-singlet pair potential, while $\mathbf{d}$ represents the vector containing the three spin-triplet components.

Before closing this part, we note that the distinct types of pair potentials above can emerge intrinsically but they can also be engineered by combining conventional superconductors and other materials. While we do not specify the particular generating mechanism of superconductivity, we will try to point out the origin of a given type of superconductivity whenever necessary. We refer to Ref.\,\cite{RevModPhys.63.239} for a more detailed discussion on unconventional superconductors.

\subsection{Bogoliubov-de Gennes equations}
\label{subsection12}
To model the superconductors described in the previous subsection, it is common to employ the Bogoliubov-de Gennes equations, which are commonly written in Nambu space and given by \cite{PhysRevB.41.4017,kashiwaya00}

\begin{equation}
\label{generalizedBdG}
 \int d \bm{r}'
\mathcal{H}(\bm{r},\bm{r}')
\begin{pmatrix}
u_{\uparrow}\left(\bm{r}'\right) \\ 
u_{\downarrow}\left(\bm{r}' \right) \\ 
v_{\uparrow}\left(\bm{r}' \right) \\
v_{\downarrow}\left(\bm{r}' \right) 
\end{pmatrix}=
E
\begin{pmatrix}
u_{\uparrow}\left(\bm{r} \right) \\
u_{\downarrow}\left(\bm{r} \right) \\ 
v_{\uparrow}\left(\bm{r} \right) \\ 
v_{\downarrow}\left(\bm{r} \right)
\end{pmatrix}\,,
\end{equation}
where 
\begin{equation}
\mathcal{H}(\bm{r},\bm{r}')=
\begin{pmatrix}
\delta\left(\bm{r}-\bm{r}' \right)
h\left(\bm{r}' \right)
& 
0 &
\Delta_{\uparrow\uparrow}\left(\bm{r},\bm{r}'\right) &
\Delta_{\uparrow\downarrow}\left(\bm{r},\bm{r}'\right) \\
0 &
\delta\left(\bm{r}-\bm{r}'\right)h\left(\bm{r}'\right) & 
\Delta_{\downarrow\uparrow}\left(\bm{r},\bm{r}'\right) &
\Delta_{\downarrow\downarrow}\left(\bm{r},\bm{r}' \right) \\
-\Delta^{*}_{\uparrow\uparrow}\left(\bm{r},\bm{r}'\right) &
-\Delta^{*}_{\uparrow\downarrow}\left(\bm{r},\bm{r}' \right) &
-\delta\left(\bm{r}-\bm{r}'\right)h\left(\bm{r}' \right) & 
0 \\
-\Delta^{*}_{\downarrow\uparrow}\left(\bm{r},\bm{r}' \right) &
-\Delta^{*}_{\downarrow\downarrow}\left(\bm{r},\bm{r}' \right) &
0 &
-\delta\left(\bm{r}-\bm{r}'\right)h\left(\bm{r}' \right)
\end{pmatrix}
\end{equation}
is   the BdG Hamiltonian, with
\begin{equation}
\label{Hnormal}
h({\bm r})=-\frac{\hbar^{2}}{2m} \nabla^{2} + U({\bm r}) - \mu\,,
\end{equation}
describing the normal state part, while $\Delta_{\sigma\sigma'}({\bm r},{\bm r}')$ represents an anisotropic superconducting pair potential which is common in unconventional superconductors as seen in the previous section. 
Moreover, $\mu$ is the chemical potential measured from the band bottom, 
$U({\bm r})$ is the one-body potential, $m$ is the effective mass, while $u_{\sigma}$ and $v_{\sigma}$ the electron and hole components of the wavefunction, namely, $\Psi=(u_{\uparrow},u_{\downarrow},v_{\uparrow},v_{\downarrow})^{\rm T}$ written in the so-called Nambu space.  The system of equations offered by the matrix Eqs.\,(\ref{generalizedBdG}) is commonly referred to as Bogoliubov-de Gennes (BdG) equations.

Under certain circumstances, Eq.\,(\ref{generalizedBdG}) can be decomposed into two $2 \times 2$  matrices. For instance, for spin-singlet superconductors with a pair potential described by Eq.\,(\ref{Deltasinglet}), the two blocks of Eq.\,(\ref{generalizedBdG}) are given by
\begin{equation}
\int d \bm{r}'
\begin{pmatrix}
\delta\left(\bm{r}-\bm{r}'\right)h\left(\bm{r}'\right) & 
\Delta\left(\bm{r},\bm{r}'\right) \\
\Delta^{*}\left(\bm{r},\bm{r}'\right) &
-\delta\left(\bm{r}-\bm{r}'\right)h\left(\bm{r}'\right) 
\end{pmatrix}
\begin{pmatrix}
u_{\uparrow}\left(\bm{r}'\right) \\ 
v_{\downarrow}\left(\bm{r}'\right) 
\end{pmatrix}
=
E
\begin{pmatrix}
u_{\uparrow}\left(\bm{r}\right) \\
v_{\downarrow}\left(\bm{r} \right) 
\end{pmatrix}, 
\label{Bdg2*2singlet1}
\end{equation}
\begin{equation}
\int d \bm{r}'
\begin{pmatrix}
\delta \left(\bm{r}-\bm{r}' \right)h\left(\bm{r}' \right) & 
-\Delta\left(\bm{r},\bm{r}' \right) \\
-\Delta^{*}\left(\bm{r},\bm{r}'\right) &
-\delta\left(\bm{r}-\bm{r}' \right)h\left(\bm{r}' \right)
\end{pmatrix}
\begin{pmatrix}
u_{\downarrow}\left(\bm{r}'\right) \\ 
v_{\uparrow}\left(\bm{r}' \right) 
\end{pmatrix}
=
E
\begin{pmatrix}
u_{\downarrow}\left(\bm{r} \right) \\
v_{\uparrow}\left(\bm{r} \right) 
\end{pmatrix}\,. 
\label{Bdg2*2singlet2}
\end{equation}
It is thus straightforward to write the corresponding BdG equations for spin-singlet $s$-wave superconductors as,
\begin{equation}
\begin{pmatrix}
h\left(\bm{r} \right) & 
\Delta\left(\bm{r} \right) \\
\Delta^{*}\left(\bm{r} \right) &
-h\left(\bm{r} \right) 
\end{pmatrix}
\begin{pmatrix}
u_{\uparrow}\left(\bm{r}\right) \\ 
v_{\downarrow}\left(\bm{r} \right) 
\end{pmatrix}
=
E
\begin{pmatrix}
u_{\uparrow}\left(\bm{r}\right) \\
v_{\downarrow}\left(\bm{r} \right) 
\end{pmatrix}, 
\label{Bdg2*2singlet1s}
\end{equation}
\begin{equation}
\begin{pmatrix}
h\left(\bm{r} \right) & 
-\Delta\left(\bm{r} \right) \\
-\Delta^{*}\left(\bm{r} \right) &
-h\left(\bm{r} \right)
\end{pmatrix}
\begin{pmatrix}
u_{\downarrow}\left(\bm{r}\right) \\ 
v_{\uparrow}\left(\bm{r} \right) 
\end{pmatrix}
=
E
\begin{pmatrix}
u_{\downarrow}\left(\bm{r} \right) \\
v_{\uparrow}\left(\bm{r} \right) 
\end{pmatrix}\,, 
\label{Bdg2*2singlet2s}
\end{equation}
where we have used the fact that  $s$-wave pair potentials are local in space and satisfy $\Delta(\bm{r},\bm{r}')=\Delta(\bm{r})\delta({\bm r} - {\bm r}')$.
 
Similarly, for mixed-spin-triplet superconductors with a pair potential described by Eq.\,(\ref{DeltatripletS0}), the two blocks of Eq.\,(\ref{generalizedBdG}) are given by
\begin{equation}
\int d \bm{r}'
\begin{pmatrix}
\delta\left(\bm{r}-\bm{r}'\right)h(\bm{r}') & 
\Delta\left(\bm{r},\bm{r}'\right) \\
-\Delta^{*}\left(\bm{r},\bm{r}'\right) &
-\delta\left(\bm{r}-\bm{r}'\right)h(\bm{r}') 
\end{pmatrix}
\begin{pmatrix}
u_{\uparrow}\left(\bm{r}'\right) \\ 
v_{\downarrow}\left(\bm{r}'\right) 
\end{pmatrix}
=
E
\begin{pmatrix}
u_{\uparrow}\left(\bm{r}\right) \\
v_{\downarrow}\left(\bm{r}\right) 
\end{pmatrix}, 
\label{Bdg2*2triplet1}
\end{equation}
\begin{equation}
\int d \bm{r}'
\begin{pmatrix}
\delta\left(\bm{r}-\bm{r}'\right)h\left(\bm{r'}\right) & 
\Delta\left(\bm{r},\bm{r}'\right) \\
-\Delta^{*}\left(\bm{r},\bm{r}'\right) &
-\delta\left(\bm{r}-\bm{r}'\right) h(\bm{r}')
\end{pmatrix}
\begin{pmatrix}
u_{\downarrow}\left(\bm{r}'\right) \\ 
v_{\uparrow}\left(\bm{r}'\right) 
\end{pmatrix}
=
E
\begin{pmatrix}
u_{\downarrow}\left(\bm{r}\right) \\
v_{\uparrow}\left(\bm{r}\right) 
\end{pmatrix}. 
\label{Bdg2*2triplet2}
\end{equation}

Moreover, for equal-spin-triplet superconductors with an pair potential described by Eq.\,(\ref{DeltatripletSpm1}), the two blocks of Eq.\,(\ref{generalizedBdG}) can be written as
\begin{equation}
\int d \bm{r}'
\begin{pmatrix}
\delta\left(\bm{r}-\bm{r}'\right)h(\bm{r}') & 
\Delta\left(\bm{r},\bm{r}'\right) \\
-\Delta^{*}\left(\bm{r},\bm{r}'\right) &
-\delta\left(\bm{r}-\bm{r}'\right)h(\bm{r}') 
\end{pmatrix}
\begin{pmatrix}
u_{\uparrow}\left(\bm{r}'\right) \\ 
v_{\uparrow}\left(\bm{r}'\right) 
\end{pmatrix}
=
E
\begin{pmatrix}
u_{\uparrow}\left(\bm{r}\right) \\
v_{\uparrow}\left(\bm{r}\right) 
\end{pmatrix}, 
\label{Bdg2*2triplet3}
\end{equation}
\begin{equation}
\int d \bm{r}'
\begin{pmatrix}
\delta\left(\bm{r}-\bm{r}'\right)h\left(\bm{r'}\right) & 
\Delta\left(\bm{r},\bm{r}'\right) \\
-\Delta^{*}\left(\bm{r},\bm{r}'\right) &
-\delta\left(\bm{r}-\bm{r}'\right) h(\bm{r}')
\end{pmatrix}
\begin{pmatrix}
u_{\downarrow}\left(\bm{r}'\right) \\ 
v_{\downarrow}\left(\bm{r}'\right) 
\end{pmatrix}
=
E
\begin{pmatrix}
u_{\downarrow}\left(\bm{r}\right) \\
v_{\downarrow}\left(\bm{r}\right) 
\end{pmatrix}. 
\label{Bdg2*2triplet4}
\end{equation}

Then, to describe superconductors with a given symmetry in the pair potential and find their energies and wavefunctions, one needs to solve the equations written above which is not always a simple task. One evident difficulty is the second derivative appearing in the normal Hamiltonian $h$ given by Eq.(\ref{Hnormal}). Thus, to make it easier the discussion about the formation of SABSs, it is convenient to simplify the BdG equations.

\subsection{Andreev equations}
\label{subsection13}
To further simplify the BdG equations, here we discuss the case where they can be reduced to be a $2 \times 2$ matrix and apply the so-called Andreev approximation where the chemical potential is the largest energy scale. This approximation is sometimes referred to as quasiclassical approximation. 
Thus, to study spin-singlet, spin-triplet with $S_{z}=0$, and spin-triplet with $S_{z}=\pm1$, respectively, it is sufficient to consider Eqs.\,(\ref{Bdg2*2singlet1s}) or (\ref{Bdg2*2singlet2s}), Eqs.\,(\ref{Bdg2*2triplet1}) or (\ref{Bdg2*2triplet2}), 
and  Eqs.\,(\ref{Bdg2*2triplet3}) or (\ref{Bdg2*2triplet4}), see Refs.\,\cite{PhysRevB.41.4017,kashiwaya00}. Then, we start from 
\begin{equation}
\label{Bdgunvonventional}
\begin{split}
E\tilde{u}\left(\bm{r}\right) &= h\left(\bm{r}\right) \tilde{u}\left(\bm{r}\right)
+ \int d \bm{r^{\prime}}
\Delta\left(\bm{r},\bm{r^{\prime}}\right)
\tilde{v}\left(\bm{r^{\prime}}\right),
\\
E\tilde{v}\left(\bm{r}\right) &= -h\left(\bm{r}\right) \tilde{v}\left(\bm{r}\right)
\pm  \int d \bm{r^{\prime}}
\Delta^{*}\left(\bm{r},\bm{r^{\prime}}\right)
\tilde{u}\left(\bm{r^{\prime}}\right), 
\end{split}
\end{equation}
where 
$\tilde{u}\left(\bm{r} \right)=u_{\uparrow}\left(\bm{r} \right)$, 
$\tilde{v}\left(\bm{r} \right)=v_{\downarrow}\left(\bm{r} \right)$
for spin-singlet pairing and spin-triplet one with $S_{z}=0$, 
and
$\tilde{u}\left(\bm{r} \right)=u_{\uparrow}\left(\bm{r} \right)$, 
$\tilde{v}\left(\bm{r} \right)=v_{\downarrow}\left(\bm{r} \right)$
for equal spin-triplet paring with $S_{z}=\pm 1$. 
Here, $+(-)$ corresponds to 
spin-singlet (spin-triplet) superconductors. \par

In the quasiclassical description, it is common to consider  $\mu \gg|\Delta(\bm{r},\bm{r'})|$, which implies that we can consider
Cooper pair formed between two electrons on the Fermi surface and the 
pair potential is determined by the direction of the quasiparticle on the 
Fermi surface. The electron and hole wavefunction components $\tilde{u}(\bm{r})$ and $\tilde{v}(\bm{r})$
can be expressed by the 
product of rapidly oscillating term 
$e^{i \bm{k}_{F} \cdot \bm{r}}$
and slowly varying terms  
$u(\hat{\bm{k}},\bm{r})$ and 
$v(\hat{\bm{k}},\bm{r})$ 
as follows, 
\begin{equation}
\label{Psin}
\tilde{\Psi}\left(\bm{r}\right)
=
\left (
    \begin{array}{c}
        \tilde{u}\left(\bm{r}\right) \\
        \tilde{v}\left(\bm{r}\right)
    \end{array}
\right )
=
\left (
    \begin{array}{c}
        u(\hat{\bm{k}},\bm{r}) \\
        v(\hat{\bm{k}},\bm{r})
    \end{array}
\right ) e^{i \bm{k}_{F} \cdot \bm{r}}
=\hat{\Psi}(\hat{\bm{k}},\bm{r})
e^{ i{\bm{k}}_{F} \cdot \bm{r}}\,,
\end{equation}
where $\hat{\bm{k}}=\bm{k}_{F}/|\bm{k}_{F}|$ and $|\bm{k}_{F}|=\sqrt{2m\mu/\hbar^{2}}$. In the following, we ignore the second derivative of 
$\hat{\Psi}(\hat{\bm{k}},\bm{r})$. 
To find the integral in Eqs.\,(\ref{Bdgunvonventional}),  
we use the approximation that the 
size of the Cooper pair (effective range of inter-electron potential) 
is smaller than the length scale of the pair potential  
\cite{PhysRevB.41.4017}. 
Then,  for the first equation in Eqs,\,(\ref{Bdgunvonventional}), we obtain
\begin{equation}
\label{quasiclassicalappro1}
 \begin{split}
\int d \bm{r^{\prime}}
\Delta\left(\bm{r},\bm{r^{\prime}}\right)
\tilde{v}\left(\bm{r^{\prime}}\right)
&=
\int d \bm{r^{\prime}}
\Delta\left(\bm{x},\bm{R}\right)
\tilde{v}\left(\bm{r^{\prime}}\right)
\\
&=
e^{i \bm{k}_{F} \cdot \bm{r}}
\int d \bm{x}
\Delta\left(\bm{x},\bm{r}-\frac{{\bm x}}{2}\right)
v(\hat{\bm{k}},\bm{r}-\bm{x})
e^{- i \bm{k}_{F} \cdot \bm{x}}
\\
&\simeq
e^{ i \bm{k}_{F} \cdot \bm{r}}
\int d \bm{x}
\Delta(\bm{x},\bm{r})
v(\hat{\bm{k}},\bm{r})
e^{- i\bm{k}_{F} \cdot \bm{x}}
\\
&=
 e^{ i\bm{k}_{F} \cdot \bm{r}}
\Delta(\hat{\bm{k}},\bm{r})
v(\hat{\bm{k}},\bm{r}) 
= \Delta(\hat{\bm{k}},\bm{r})
\tilde{v}(\bm{r})\,,
 \end{split}
 \end{equation}
 while for the second equation in Eqs.\,(\ref{Bdgunvonventional}), we get 
 \begin{equation}
 \label{quasiclassicalappro2}
\begin{split} 
\int d \bm{r^{\prime}}
\Delta^{*}\left(\bm{r},\bm{r^{\prime}}\right)
\tilde{u}\left(\bm{r^{\prime}}\right)
&=
\int d \bm{r^{\prime}}
\Delta^{*}\left(\bm{x},\bm{R}\right)
\tilde{u}\left(\bm{r^{\prime}}\right)\\
&=
e^{i\bm{k}_{F} \cdot \bm{r}}
\int d \bm{x}
\Delta^{*}\left(\bm{x},\bm{r}-\frac{{\bm x}}{2}\right)
u\left(\hat{\bm{k}},\bm{r}-\bm{x}\right)
e^{- i\bm{k}_{F} \cdot \bm{x}}\\
&\simeq
e^{i \bm{k}_{F} \cdot \bm{r}}
\int d \bm{x}
\Delta^{*}\left(\bm{x},\bm{r}\right)
u\left(\hat{\bm{k}},\bm{r}\right)
e^{- i\bm{k}_{F} \cdot \bm{x}}\\
&=
e^{ i\bm{k}_{F} \cdot \bm{r}}
\left[\int d \bm{x}
\Delta\left(\bm{x},\bm{r}\right)
e^{ i\bm{k}_{F} \cdot \bm{x}} \right]^{*}
u\left(\hat{\bm{k}},\bm{r}\right)\\
&=
e^{i\bm{k}_{F} \cdot \bm{r}}
\Delta^{*}\left(-\hat{\bm{k}},\bm{r}\right)
u\left(\hat{\bm{k}},\bm{r}\right)=
\Delta^{*}\left(-\hat{\bm{k}},\bm{r}\right)
\tilde{u}\left(\bm{r}\right),
\end{split}
\end{equation}
 where ${\bm x}={\bm r}-{\bm r}'$, 
${\bf R}=({\bm r}+{\bm r}')/2$, and 
$\Delta(\hat{\bm{k}},\bm{r})$ satisfies
\begin{equation}
\Delta(\hat{\bm{k}},\bm{r})=
\int d \bm{x}
 e^{- i\bm{k}_{F} \cdot \bm{x}}
\Delta({\bm x},{\bm r})\,,
\end{equation}
which  represents  an effective pair 
potential with momentum 
$\hat{\bm{k}}$ at position ${\bm r}$. To arrive at the third equality in Eq.\,(\ref{quasiclassicalappro1}), we have   used $\Delta({\bm x},{\bm r}-{\bm x}/2)\approx \Delta({\bm x},{\bm r})-({\bm x}/2)\cdot \nabla \Delta({\bm x},{\bm r})\approx\Delta({\bm x},{\bm r})$,  a valid approximation when the size of the Cooper pair is much smaller than the coherence length of the 
pair potential.  Then,  Eqs.\,(\ref{Bdgunvonventional}) can be collected into a $2\times2$ matrix for $\Psi(\hat{\bm{k}},\bm{r})$ given by Eq.\,(\ref{Psin}), leading to an eigenvalue matrix equation given by
\begin{equation}
\label{Andreevequation}
\begin{pmatrix}
-i\hbar v_{F}{\hat{\bm{k}} \cdot \nabla} & \quad
\Delta(\hat{\bm{k}},\bm{r}) \\
\Delta^{*}(\hat{\bm{k}},\bm{r}) & \quad 
i\hbar v_{F}{\hat{\bm{k}} \cdot \nabla}
\end{pmatrix}\hat{\Psi}(\hat{\bm{k}},\bm{r})
=E\hat{\Psi}(\hat{\bm{k}},\bm{r})
\end{equation}
where $v_{F}$ is the Fermi velocity and we have used the relations   
$\Delta(-\hat{\bm{k}},\bm{r})=\Delta(\hat{\bm{k}},\bm{r})$ for spin-singlet superconductor, while  
$\Delta(-\hat{\bm{k}},\bm{r})=-\Delta(\hat{\bm{k}},\bm{r})$
for a spin-triplet superconductor.  Equation (\ref{Andreevequation}) represents the so-called Andreev equations, which are simpler than the BdG equations because they involve linear terms in the derivatives of the diagonals in contrast to the second derivatives of the BdG equations in Eqs.\,(\ref{generalizedBdG}).  As we will see below, this will be useful when finding the SABSs in unconventional superconductors.

\subsection{Surface Andreev bound state and Majorana bound states in unconventional superconductors}
\label{subsection2d}
\begin{figure}[b]
\begin{center}
\includegraphics[width=6.5cm,clip]{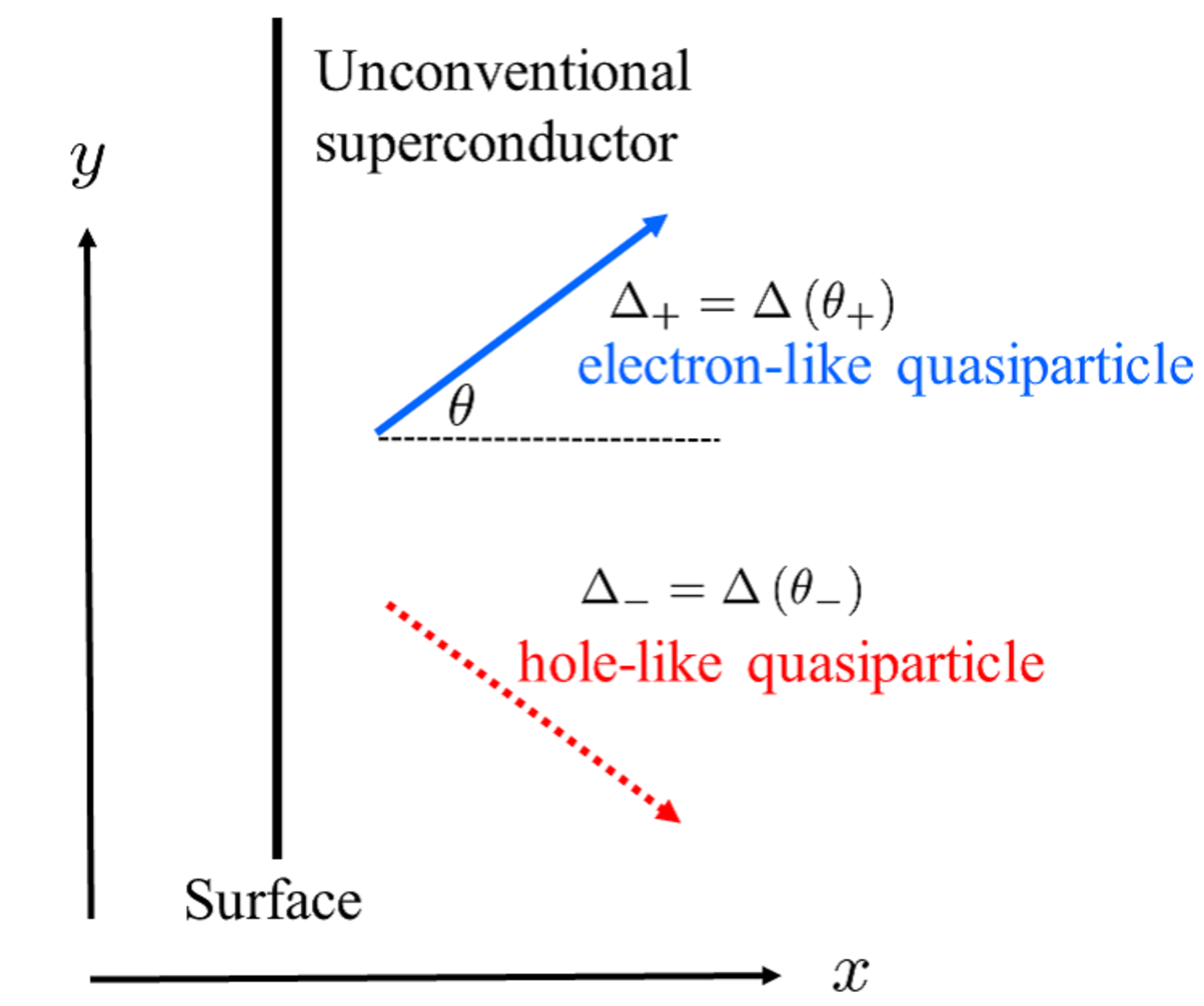}
\end{center}
\caption{A semi-infinite 2D NS junction along $x$ with its interface located at $x=0$ and an  unconventional superconductor for $x>0$. The surface of the junction is shown as a thick vertical line, while the electron-like and hole-like quasiparticles in the superconductor are depicted by solid blue and dotted red arrows and feel pair potentials parametrized by $\theta_{+}=\theta$ and $\theta_{-}=\pi-\theta$. The direction of the arrow denotes  
that of the group velocity of the quasiparticles.}
\label{figNSp}
\end{figure}

In this part, we employ the methodology presented in the previous two subsections and explore the emergence of SABSs in unconventional superconductors.  To characterize the formation of SABSs, we need to investigate the surface or boundary of unconventional superconductors using the BdG or Andreev equations given by Eqs.\,(\ref{generalizedBdG}) or (\ref{Andreevequation}).  For this purpose, we consider a two-dimensional unconventional superconductor junction with a flat surface located at $x=0$ and superconductor (S) defined for  $x>0$, see Fig.\,\ref{figNSp}. Thus, the perpendicular direction along $y$ is translationally invariant.  A quasiparticle in S thus exhibits a wavefunction that has the form of Eqs.\,(\ref{Psin}), which, taken into account that   momentum parallel to the interface is conserved, can be written as
\begin{equation}
\tilde{\Psi}(\bm{r})=\Psi(\hat{\bm{k}},x)\,e^{ik_{Fy}y}\,,
\end{equation}
where $\hat{\bm{k}}$ is related to the injection angle as 
\begin{equation}
\label{kFtheta}
e^{i\theta}\equiv \frac{k_{Fx}}{k_{F}}+i \frac{k_{Fy}}{k_{F}}
\end{equation}
where $k_{Fx}=k_{F}{\rm cos}\theta$, $k_{Fy}=k_{F}{\rm sin}\theta$, and  $k_{F}=|\bm{k}_{\rm F}|$. We note that Eq.\,(\ref{kFtheta}) implies that functions labeled by $\hat{\bm{k}}$ can be labeled by $\theta$. Thus, the electron and hole components of the wavefunction become dependent on $\theta$ and $x$, which means that $\Psi(\hat{\bm{r}},x)$ can be written as
\begin{equation}
\Psi(\theta,x)
=
\left(
\begin{array}{c}
u_{+}\left(\theta,x\right) \\
v_{+}\left(\theta,x\right)
\end{array}
\right) \exp\left(ik_{F}\cos \theta x\right)
+\left(
\begin{array}{c}
u_{-}\left(\theta,x\right) \\
v_{-}\left(\theta,x\right)
\end{array}
\right) \exp\left(-ik_{F}\cos \theta x\right)
\label{wavefunctiontotal}
\end{equation}
with   $-\pi/2 < \theta < \pi/2$. Then, within quasiclassical approximation, $\Psi(\theta,x)$ satisfies the Andreev equations given by Eqs.\,(\ref{Andreevequation}), which, for $x\neq0$,   read
\begin{equation}
\label{envelopefunction}
\begin{split}
Eu_{j}\left(\theta,x\right)&=
-\frac{i\hbar^{2}\sigma k_{F}\cos\theta}{m}\frac{d}{dx}
u_{j}\left(\theta,x\right) + \Delta(\theta_{j},x)\Theta\left(x\right)
v_{j}
\left(\theta,x\right), 
\\
Ev_{j}\left(\theta,x\right)&=
\frac{i\hbar^{2}\sigma k_{F}\cos\theta}{m}\frac{d}{dx}
v_{j}\left(\theta,x\right) + \Delta^{*}(\theta_{j},x)
\Theta(x)u_{j}\left(\theta,x\right)\,,
\end{split}
\end{equation}
where $j=+$ for $\sigma=1$ and $\theta_{j}=\theta$, while $j=-$ for $\sigma=-1$ and $\theta_{j}=\pi-\theta$. Also,   $\Delta(\theta_{j},x)$ represents the order parameter of the unconventional superconductor parametrized by the angle $\theta_{j}$; and $\Theta(x)$ is the Heaviside step function being it $\Theta(x)=1$ for $x>0$, in S. Now, we need to solve the above linear coupled differential equations. Before doing so, we note that, in general, the spatial dependence of pair potential should be determined selfconsistently, which is known to give reduced values of the pair potential near the interface. However, here we assume spatially  constant pair potential inside the superconductor  because it was previously shown that the SABS and also MZMs are not influenced  by the 
spatial depletion of the pair potential \cite{odd3,odd3b}. 

Then, the wavefunction $\Psi(\theta,x)$    can be written as 
\begin{equation}
\label{wavefunctionenvelop1}
\begin{split}
\begin{pmatrix}
u_{+}(\theta,x) \\
v_{+}(\theta,x)
\end{pmatrix} 
&=
c(E,\theta)
\begin{pmatrix}
u_{+} \\
v_{+}
\end{pmatrix} \exp\left(i\gamma_{+}x\right) \\
\begin{pmatrix}
u_{-}(\theta,x)\,, \\
v_{-}(\theta,x)
\end{pmatrix} 
&=
d(E,\theta)
\begin{pmatrix}
v_{-} \\
u_{-}
\end{pmatrix} \exp\left(i\gamma_{-}x\right) 
\end{split}
\end{equation}
where 
\begin{equation}
\label{coherenceuv}
\begin{split}
\displaystyle{
u_{\pm}}&=\displaystyle{\sqrt{\frac{1}{2}\left(1+\frac{\Omega_{\pm}}{E}\right)}}\,, \\
\displaystyle{ 
v_{+}}&=\displaystyle{\sqrt{\frac{1}{2}\left(1-\frac{\Omega_{+}}{E}\right)}\frac{\Delta^{*}\left(\theta_{+}\right)}
{|\Delta\left(\theta_{+}\right)|}}\\ 
\displaystyle{v_{-}}&=\displaystyle{\sqrt{\frac{1}{2}\left(1-\frac{\Omega_{-}}{E}\right)}\frac{\Delta\left(\theta_{-}\right)}
{|\Delta\left(\theta_{-}\right)|}
}\,,
\end{split}
\end{equation}
 and
\begin{equation}
\gamma_{0}
=\frac{|E| m}{\hbar^{2} k_{F}\cos \theta}\,, \ \ 
\gamma_{\pm}
=\frac{\Omega_{\pm}m}{\hbar^{2} k_{F}\cos \theta}\,, 
\label{gammapm}
\end{equation}
\begin{equation}
\Omega_{\pm}
\equiv \lim_{\delta \rightarrow 0} 
\sqrt{\left(E + i\delta \right)^{2} - |\Delta(\theta_{\pm})|^{2}}
= \left \{
\begin{array}{ll}
\sqrt{E^{2} - |\Delta(\theta_{\pm})|^{2}} & E  \geq |\Delta(\theta_{\pm})| \,,\\
i \sqrt{|\Delta(\theta_{\pm})|^{2}  - E^{2}} &  -|\Delta(\theta_{\pm})| \leq E \leq  |\Delta(\theta_{\pm})| \,,\\ 
-\sqrt{E^{2} - |\Delta(\theta_{\pm})|^{2}} & E  \leq -|\Delta(\theta_{\pm})|\,.
\end{array}
\right.
\label{Omegapm}
\end{equation}

Then, by introducing convenient parameters $\Gamma_{+}$ 
and $\Gamma_{-}$ given by 
\begin{equation}
\label{TKformulaGamma0}
\begin{split}
\Gamma_{+}&=\frac{v_{+}}{u_{+}}=\frac{\Delta^{*}(\theta_{+})}{E + \Omega_{+}}\,, \\ 
\Gamma_{-}&=\frac{v_{-}}{u_{-}}=\frac{\Delta(\theta_{-})}{E + \Omega_{-}}\,, 
\end{split}
\end{equation}
the wave function in S can be written as
\begin{equation}
\label{Psicd}
\Psi(\theta, x)
= c 
\begin{pmatrix}
1 \\ \Gamma_{+}
\end{pmatrix}
\exp( ik^{+} x)
+ d 
\begin{pmatrix}
\Gamma_{-} \\ 1 
\end{pmatrix}
\exp( -ik^{-} x)
\end{equation}
with $k^{\pm} = k_{Fx} \pm \gamma_{\pm}$. The wavefunction $\Psi(\theta, x)$ has two terms. The first and the second terms represent electron-like and hole-like quasiparticle wavefunctions.  Since ${\rm Im}{\gamma_{+}}>0$ and 
${\rm Im}{\gamma_{-}}<0$ are satisfied,  both terms do not have divergence for $x \rightarrow \infty$, as expected. For these reasons, a state at the boundary $x=0$ is called  SABS which is the focus of this part.

To fully characterize the wavefunction $\Psi(\theta, x)$ in S,  we need to find the coefficients $c$ and $d$ which can be found by imposing boundary conditions for the wavefunction at $x=0$ given by 
\begin{equation}
\Psi\left(\theta, x=0 \right)=0\,,
\end{equation}
which, when searching for a nontrivial solution, we find the condition of the existence of the SABS to be given by 
\begin{equation}
\label{SABSrelation}
1=\Gamma_{+}\Gamma_{-}\,,
\end{equation}
where $\Gamma_{\pm}$ are given by Eqs.\,(\ref{TKformulaGamma0}). Now, to obtain the energy of the SABS, we plug $\Gamma_{\pm}$ from Eqs.\,(\ref{TKformulaGamma0}), using Eqs.\,(\ref{coherenceuv}), into Eq.\,(\ref{SABSrelation}) and solve for $E$. When   $|\Delta_{+}|=|\Delta_{-}|$, where $\Delta_{\pm}=\Delta(\theta_{\pm})$, a simple expression is obtained for the SABSs, which are given by \cite{Tanaka2021}
\begin{equation}
\label{ABSUNSC}
E_{b}=\Delta_{0} \,{\rm cos}(\delta_{\phi}/2)\,{\rm sgn}[{\rm sin}(\delta_{\phi}/2)]\,, 
\end{equation}
where $\Delta_{0}$ is the maximum value of the pair potential,  $\delta_{\phi}=\phi_{-}-\phi_{+}$, and ${\rm exp}(i\phi_{\pm})=\Delta_{\pm}/|\Delta_{\pm}|$. Interestingly, when $\Delta_{+}=-\Delta_{-}$ is satisfied, i.e., when
\begin{equation}
\label{conditionDelta}
\Delta_{+}\Delta_{-}<0\,,
 \end{equation}
we obtain $\delta_{\phi}=\pi$, which, remarkably, gives rise to SABSs with zero energy  $E_{b}=0$. Thus, the condition given by Eq.\,(\ref{conditionDelta}) gives rise to zero-energy SABSs which we will denote as ZESABS \cite{kashiwaya00,Hu94}. The ZESABSs have attracted a lot of attention and have become one of the central topics in the context of high $T_{c}$ cuprates \cite{Hu94}. Furthermore, the ZESABSs have been observed by tunneling spectroscopy, giving rise to the establishment of the pairing symmetry in high $T_{c}$ cuprates to be  
spin-singlet $d$-wave. \cite{kashiwaya00,TK95}.  
In what follows we discuss ZESABSs for unconventional superconductors with specific profiles of their order parameters, describing spin-triplet $p$-wave and spin-singlet $d$-wave pair potentials.

\subsubsection{ZESABSs in $d$-wave superconductors}
\label{dwaveSABS}
\begin{figure}[b]
\begin{center}
\includegraphics[width=7.0cm,clip]{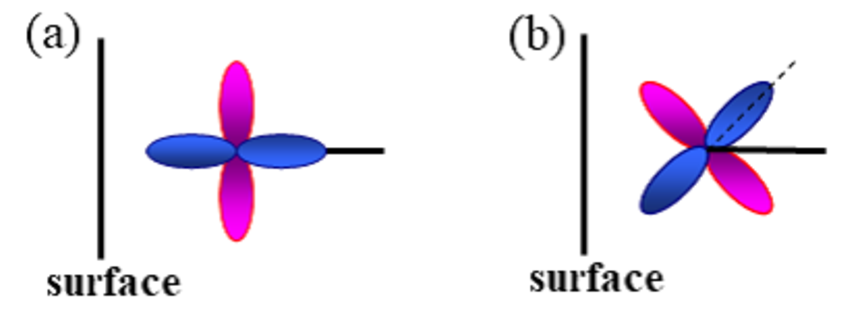}
\end{center}
\caption{Schematic figure showing  the surface of   a $d$-wave superconductor with misorientation angle $\alpha$: (a) $\alpha=0$, (b) $\alpha=\pi/4$. The blue and pink colors indicate values of opposite signs in the pair potential.}
\label{SABSdwave}
\end{figure}

 In the case of spin-singlet $d$-wave superconductors, we would like to remind that $d$-wave pairing can be realized in high $T_{c}$ cuprates. In this case, the $d$-wave pair potential can be expressed
   by the angle measured from the $a$-axis of the basal plane of cuprate since one of the directions of the lobe of $d$-wave pairing coincides with that of $a$-axis.  Then, we consider the junction with misorientation angle $\alpha$,  where $\alpha$ denotes the angle between the 
normal to the interface and $a$-axis of $d$-wave superconductor.  Thus, the pair potential $\Delta_{\pm}$  felt by quasiparticle with an injection angle $\theta$
is given by 
\begin{equation}
\Delta_{+}=\Delta(\theta_{+})=\Delta_{0}\cos\left(2\theta - 2\alpha \right),\quad \Delta_{-}=\Delta(\theta_{-})=\Delta_{0}\cos\left(2\theta + 2\alpha \right)\,.
\label{dwavemisorientation}
\end{equation}
The schematic of the surface of $\alpha=0$ and $\alpha=\pi/4$ is 
shown in Fig.\,\ref{SABSdwave}.  We now inspect when ZESABSs appear taking into account the condition discussed before $\Delta_{+}\Delta_{-} < 0$. i) For $|\alpha| \leq \pi/4$, a ZESABS is generated for 
$\pi/4 - |\alpha| < |\theta| < \pi/4 + |\alpha|$
\cite{Hu94,TK95,TK96a,KT96,kashiwaya00}.  
ii) For $\alpha=0$, ZESABS is absent for any $\theta$. 
iii) On the other hand, at $\alpha=\pm \pi/4$, a ZESABS appears for 
all $\theta$ except for the nodal direction $\theta=0$. 
Interestingly, the ZESABS here has a dispersionless flat band as a function of 
momentum $k_{Fy}=k_{F}\sin \theta$ \cite{Proximityp}.  This flat band ZESABS has a topological invariant defined for fixed $k_{Fy}$ \cite{STYY11}. 

\subsubsection{ZESABSs in spin-triplet $p$-wave superconductors}
\label{pwaveSABS}
In the case of spin-triplet   $p$-wave superconductors, we restrict our analysis first to 1D, where $\theta=0$. In this case, the pair potential is expressed by $\Delta_{+}=-\Delta_{-}=\Delta_{0}$, which is the condition for having a ZESABSs \cite{PhysRevB.23.5788,10.1143/PTP.76.1237,Kusakabe1999}. 
Hence, the  $p$-wave superconductor hosts a ZESABS at the boundary. Since these junctions are in 1D, the ZESABS  can be seen as an edge state. It is also important to remark that for spin-triplet $p_{x}$-wave pair potential with either   $S_{z}=0$ or $S_{z}=\pm1$ and where 
$\Delta(\theta_{\pm})$ is given by $\Delta(\theta_{\pm})=\pm \Delta_{0}
\cos \theta$, the ZESABS become flat as a function of $k_{y}$, as in the case discussed for $d$-wave superconductors in the previous subsection.
 
\subsubsection{Wavefunctions of the ZESABSs}
\label{sec: Majorana relation}
In this part, we analyze the wavefunctions of the ZESABSs found in the previous two subsections. 
We first focus on spin-singlet $d$-wave  and spin-triplet $p$-wave superconductors with $S_{z}=0$ and then analyze the case with $S_{z}=\pm1$. For the $d$-wave superconductor, we choose $\alpha=\pi/4$ in Eq.\,(\ref{dwavemisorientation}), which implies that the pair potential is then given by $\Delta(\theta_{+})=-\Delta(\theta_{-})=\Delta_{0}{\rm sin}(2\theta)$. This pair potential describes spin-singlet $d_{xy}$-wave superconductors.

{\bf 2D spin-singlet $d$-wave and spin-triplet  $p$-wave superconductors with $S_{z}=0$}:  In this case, for the $d$-wave superconductor, we choose $\alpha=\pi/4$ 
in Eq.\,(\ref{dwavemisorientation}) and the pair potential is then given by $\Delta(\theta_{+})=-\Delta(\theta_{-})=\Delta_{0}\sin 2\theta$. This order parameter describes spin-singlet $d_{xy}$-wave superconductors. For the spin-triplet superconductor, we consider $S_{z}=0$ with opposite spin pairing. Then, from Eq.\,(\ref{Psicd}), the wavefunctions of the ZESABSs at the boundary of  the spin-singlet $d$-wave and spin-triplet $p$-wave   are given \cite{STYY11}, respectively, by  
\begin{equation}
\label{dwaveZESwavefunction1n}
\begin{split}
\psi_{d_{xy}}^{(0)}(x)&=\begin{pmatrix}
u_{k_{Fy}\sigma}\left(x\right) \\
v_{k_{Fy}\bar{\sigma}}\left(x \right)
\end{pmatrix}
=\exp\left(i\frac{\pi}{4}{\rm sgn}(k_{Fy})\sigma_{s} \right)
\begin{pmatrix} 
1 \\
-i{\rm sgn}\left(k_{Fy} \right) \sigma_{s}
\end{pmatrix}e^{ik_{Fy} y - x/\xi_{d}}\sin \left(k_{Fx} x\right)\,,\\
\psi_{p}^{(0)}(x)&=\begin{pmatrix}
u_{k_{Fy}\sigma}\left(x \right) \\
v_{k_{Fy}\bar{\sigma}}\left(x \right)
\end{pmatrix}
=
\exp\left(i\frac{\pi}{4}\right)
\begin{pmatrix}
1\\
-i
\end{pmatrix}
e^{ik_{Fy} y-x/\xi_{p}}\sin\left(k_{Fx} x\right)\,,
\end{split}
\end{equation}
where $\sigma$($\bar{\sigma}$)
denotes the spin direction   with $\uparrow$($\downarrow$), 
$\downarrow$($\uparrow$) 
and $\uparrow$ corresponds to $\sigma_{s}=1$, 
and $\downarrow$ corresponds to $\sigma_{s}=-1$. The subscript $(0)$ in the wavefunction indicates that it corresponds to $S_{z}=0$ spin projection.  Moreover, we note that $\xi_{d}^{-1}= (2 \Delta_{0} \sin \theta)/(\hbar v_{F})$ represents the superconducting coherence length in the spin-singlet $d$-wave superconductor, while $\xi_{p}^{-1}= \Delta_{0}/(\hbar v_{F})$ the superconducting coherence length in the spin-triplet $p$-wave superconductor. 
At this point, we remember that the wavefunctions of the ZESABSs given by Eqs.\,(\ref{dwaveZESwavefunction1n}) are solutions of 
eq. (\ref{Bdgunvonventional}). 
Thus, their wavefunctions define the so-called BdG transformation which diagonalizes the BdG equations in terms of new quasiparticles also known as Bogoliubov quasiparticles or Bogolons.  Thus, it is   possible to define     the annihilation operator  of a Bogoliubov quasiparticle associated with the ZESABS as
 \begin{equation}
 \gamma^{\vphantom{\dagger}}_{k\sigma}(x)=
 \left[
 \psi_{j}^{(i)}(x)\right]^{T}\alpha_{k\sigma}
 \end{equation}
where $\psi_{j}^{(i)}$ is the wavefunction of the ZESABS given by Eqs.\,(\ref{dwaveZESwavefunction1n}) and $\alpha_{k\sigma}=(a_{k,\sigma},a^{\dagger}_{-k,\bar{\sigma}})^{\rm T}$ is the Nambu spinor.  Thus, plugging the expressions given by  Eqs.\,(\ref{dwaveZESwavefunction1n}),  the annihilation operator of the Bogoliubov quasiparticle  is written as
\begin{equation}
\gamma^{\vphantom{\dagger}}_{k_{Fy}\sigma}(x)=u^{\vphantom{\dagger}}_{k_{Fy} \sigma}(x)a^{\vphantom{\dagger}}_{k_{Fy} \sigma} 
+ v^{\vphantom{\dagger}}_{k_{Fy} \bar{\sigma}}(x) 
a_{-k_{Fy} \bar{\sigma}}^{\dagger}\,,
\end{equation}
where the coherence factors $u^{\vphantom{\dagger}}_{k_{Fy}}(x)$ 
and $v^{\vphantom{\dagger}}_{k_{Fy}}(x)$ are  the defined by the electron and hole entries of the wavefunction in Eq.\,(\ref{dwaveZESwavefunction1n}). It is now simple to identify that the coherence factors of the wavefunctions in Eq.\,(\ref{dwaveZESwavefunction1n}) satisfy  
$u^{\vphantom{*}}_{k_{Fy}\sigma}(x)=v_{-k_{Fy}\sigma}^{*}(x)$. As a result, the Bogoliubov operator satisfies the following expression
\begin{equation}
\gamma^{\vphantom{\dagger}}_{k_{Fy} \sigma}(x)
=\gamma^{\dagger}_{-k_{Fy}\bar{\sigma}}(x)
\label{Majoranadpwave}
\end{equation}
is satisfied.  This equation reveals an interesting  non-trivial relation 
between creation and annihilation operators between different spin sectors.  Even though Eq.\,(\ref{Majoranadpwave}) already reflects the self-conjugate nature pointed out for Majorana states in the introduction, it still does not represent quasiparticles with Majorana nature since Eq.\,(\ref{Majoranadpwave}) indicates the relation between different spin species.

{\bf 2D   spin-triplet  $p$-wave superconductors with $S_{z}=\pm1$}: We now turn our attention to 
2D  superconductors with   spin-triplet $p_{x}$-wave pair potentials having 
$S_{z}=\pm1$. In this case, the two electrons forming   Cooper pairs have the same spin direction. Having only one spin direction implies that Cooper pairs behave as spin-polarized or spinless.
The wave function of ZESABS is given by  
\begin{equation}
\label{pwaveZESwavefunction2n}
\psi_{p}^{(1)}(x)=
\begin{pmatrix}
u_{k_{Fy}\sigma}(x) \\
v_{k_{Fy}\sigma}(x)
\end{pmatrix}
=
\exp\left(i\frac{\pi}{4}\right)
\begin{pmatrix}
1\\
-i
\end{pmatrix}e^{ik_{Fy} y-x/\xi_{p}}\sin\left(k_{Fx} x\right) \,.
\end{equation}
where $k_{Fx}$ is determined from $k_{F}=\sqrt{k_{Fy}^{2}+k_{Fx}^{2}}$. As before, we now define a Bogoliubov annihilation operator as
\begin{equation}
\gamma^{\vphantom{\dagger}}_{k_{Fy}\sigma}
\left( x \right)
=u^{\vphantom{\dagger}}_{k_{Fy} \sigma}\left( x \right)
a^{\vphantom{\dagger}}_{k_{Fy} \sigma} 
+ v^{\vphantom{\dagger}}_{k_{Fy} \sigma}\left( x \right) 
a_{-k_{Fy} \sigma}^{\dagger}
\end{equation}
where the coherence factors obey $u^{\vphantom{*}}_{k_{Fy}\sigma}(x)=v_{-k_{Fy}\sigma}^{*}(x)$. Therefore, the Bogoliubov operator satisfies the following expression,
 \begin{equation}
 \label{Majoranawavefunction}
\gamma^{\vphantom{\dagger}}_{k_{Fy} \sigma}(x) =\gamma^{\dagger}_{-k_{Fy} \sigma}(x)\,,
\end{equation}
which reveals that the ZESABS in a spin-triplet $p$-wave superconductor with $S_{z}=\pm1$ exhibit Majorana character \cite{Ueno} and the quasiparticles described by Eq.\,(\ref{Majoranawavefunction}) can be referred to as MZMs or MBSs. 
Moreover, it is worth noting that, if the time reversal symmetry is satisfied,  MZMs appear as  Kramers pairs \cite{Ueno}. 
For a fully spin-polarized state with one of the  spin degrees of freedom  quenched, we can ignore the spin index $\sigma$ and write,
\begin{equation}
\label{MZMs}
\gamma^{\vphantom{\dagger}}_{k_{Fy}}(x)=\gamma^{\dagger}_{-k_{Fy}}(x)
\end{equation}
This situation corresponds to a single MZM appearing at the surface of a 
spin-polarized $p$-wave superconductor for fixed momentum $k_{Fy}$. The existence of a non-degenerate MZM at fixed $k_{Fy}$ has been extensively investigated. In particular, it has been shown that in 1D, where  $k_{Fy}=0$, the 1D spin-polarized (or spinless) $p$-wave superconductor realizes a topological phase that is characterized by the emergence of   MZMs that fulfill the relation given by Eq.\,(\ref{MZMs}). In the next section, we provide further details on this matter. 

\subsection{Summary}
In this section, we have first discussed the symmetries of the order parameters in unconventional superconductors and then outlined a brief introduction of the BdG formalism which is useful for describing quasiparticles in superconductors. Furthermore, we have employed these concepts and methods to demonstrate the formation of ZESABSs in spin-singlet $d$-wave and spin-triplet $p$-wave superconductors.  We highlighted that ZESABSs in spin-polarized $p$-wave superconductors fulfill the self-conjugate Majorana condition and thus become MZMs,   whose wavefunction is exponentially localized at the interface and decays towards the bulk of the superconductor following an oscillatory profile.

\section{Majorana zero modes in $p$-wave superconductors}
\label{section3}
In the previous section, we discussed that MZMs represent a special family of ZESABSs and emerge at the edge of a $p$-wave superconductor. In this part, we explore further the properties of such exotic zero-energy states, by investigating the continuum and tight-binding models of spin-polarized 1D $p$-wave superconductors. Furthermore, we also address their relation to topology and how they can be engineered in semiconductor-superconductor hybrids.

\subsection{Topological superconductivity and MZMs in $p$-wave superconductors}
In this part, we show that spin-polarized $p$-wave superconductors exhibit a topological phase that is characterized by the emergence of MZMs. We further show that this topological superconducting phase is characterized by a topological invariant. To address these points we explore the continuum and tight-binding descriptions of spin-polarized $p$-wave superconductors.

\subsubsection{Continuum model}
\begin{figure}[!t]
\centering
	\includegraphics[width=0.95\columnwidth]{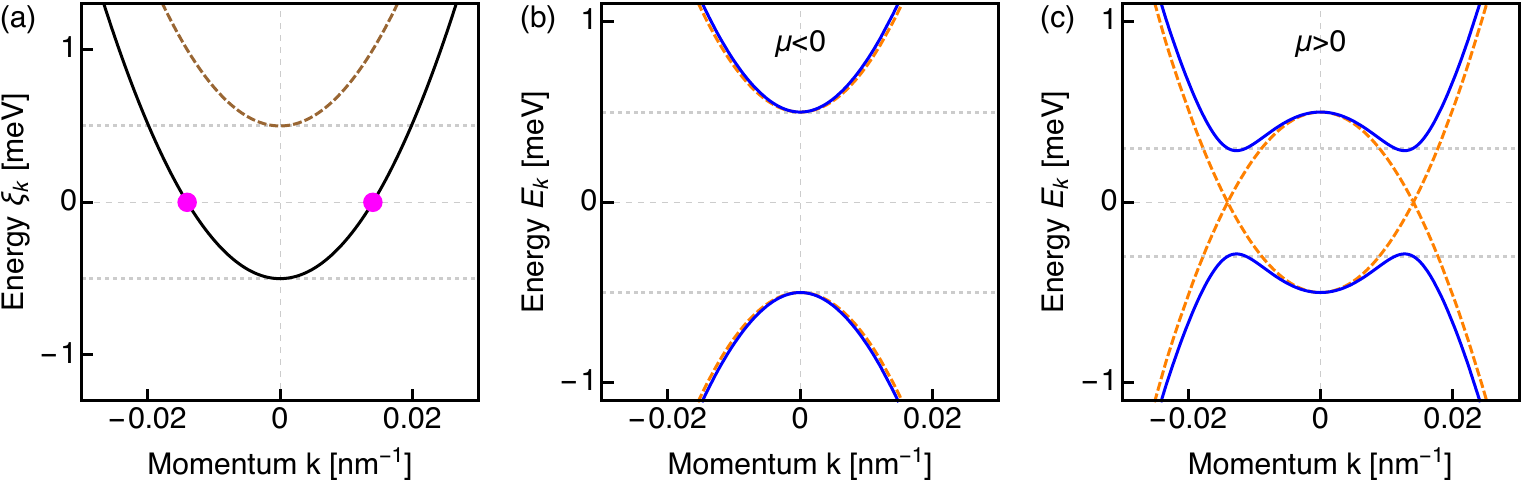}\\
 	\caption{(a) Energy as a function of momentum of the normal part of the $p$-wave superconductor for $\mu=0.5{\rm meV}>0$ (black curve) and $\mu=-0.5{\rm meV}<0$ (brown dashed curve).  The dotted gray horizontal line marks the bottom of the band given by $\mu>0$; for $\mu<0$ the band is shifted to higher energies.  For $\mu>0$ the system is metallic and exhibits an odd number of pairs of Fermi points denoted by magenta-filled circles. For  $\mu<0$, no states at the Fermi level are found and the system is then insulating. (b,c) Eigenvalues of the BdG Hamiltonian for a $p$-wave superconductor given by Eq.\,(\ref{Kitaevbulk}) as a function of momentum for $\mu=-0.5{\rm meV}<0$ (b) and $\mu=0.5{\rm meV}>0$ (c), where orange and blue curves correspond to $\Delta=0$  and $\Delta=0.3{\rm meV}\neq0$, respectively. A finite $\Delta$ opens a gap at the Fermi points.}
\label{FigKitaevC} 
\end{figure}

We start by exploring the continuum model of 1D spin-polarized  $p$-wave superconductors, which follows the discussion presented in Subsection \ref{subsection11} for unconventional superconductors. We note that  momentum is a good quantum number and the    BdG Hamiltonian   can be written as
\begin{equation}
\mathcal{H}=\frac{1}{2}\sum_{k}\psi_{k}^{\dagger}H(k)\psi_{k}\,,
\end{equation} 
where $\psi_{k}=(c_{k},c_{-k}^{\dagger})^{\rm T}$ is the Nambu spinor and
\begin{equation}
\label{Kitaevbulk}
H(k)=
\xi_{k} \tau_{z}+\Delta_{k}\tau_{x}\,,
\end{equation}
with $\xi_{k} =\hbar^{2}k^{2}/2m-\mu$, $\Delta_{k}=k\Delta$ is the momentum dependent $p$-wave order parameter, and $\mu$ the chemical potential. The eigenvalues of $H(k)$ are $E_{k,\pm}=\pm\sqrt{\xi_{k}^{2}+k^{2}|\Delta|^{2}}$, which form a gapped spectrum as long as $\mu\neq0$. Interestingly, at $\mu=0$ the spectrum becomes gapless at $k=0$, a critical point that separates two important regimes. For $\mu>0$, the system in the normal state ($\Delta=0$) has a Fermi surface and is thus metallic, while for $\mu<0$ the system with $\Delta=0$ is an insulator without a Fermi surface, see Fig.\,\ref{FigKitaevC}(a).  This is of particular relevance because it reveals that $H(k)$ exhibits two phases. The insulating phase ($\mu<0$) is gapped at $\Delta=0$ and is adiabatically connected to the spectrum of the gapped superconductor when $\mu<0$ and $\Delta\neq0$, see   Fig.\,\ref{FigKitaevC}(b). In contrast,   metallic phase $\mu>0$ is gapless at $\Delta=0$ but becomes gapped only at $\Delta\neq0$, see Fig.\,\ref{FigKitaevC}(c). Hence, the phases with $\mu>0$ and $\mu<0$ cannot be adiabatically connected without exhibiting a gap closing and reopening. Therefore, these two phases are topologically distinct. It has been shown that the phase with $\mu>0$ is topological, and we will explore more below.

\subsubsection{Tight-binding model}
The lattice version of the $p$-wave superconductor can be simply given by 
\cite{kitaev} 
\begin{equation}
\label{Kitaevlattice}
\hat{H}=-\mu_{\rm TB}\sum_{j=1}^{N}c_{j}^{\dagger}c_{j}+\sum_{j=1}^{N-1}\big[-t(c_{j}^{\dagger}c_{j+1}+c_{j+1}^{\dagger}c_{j})+\Delta_{\rm TB} c_{j}c_{j+1}+\Delta^{*}_{\rm TB}  c_{j+1}^{\dagger}c_{j}^{\dagger}\big]
\end{equation}
where $c_{j}$ ($c_{j}^{\dagger}$) destroys (creates)  a spinless fermionic state at site $j$, $\mu_{\rm TB}$   denotes the onsite energies in the tight-binding (TB) model, alike the chemical potential of the previous part, $t$ is the nearest-neighbor hopping, and $\Delta_{\rm TB}$ is the $p$-wave pair potential. We label the chemical potential and the pair potential with the subscript TB to distinguish it from their corresponding continuum counterparts. In this part we are interested in exploring the two phases identified in the previous subsection and also the impact when the system is of finite length. 

To show that this lattice model describes the physics discussed for the bulk Hamiltonian in Eq.\,(\ref{Kitaevbulk}), we consider Eq.\,(\ref{Kitaevlattice}) to form a closed loop with periodic boundary conditions such that $c_{1}=c_{N+1}$ and the second sum in Eq.\,(\ref{Kitaevlattice}) runs up to $N$. The closed loop Hamiltonian is then given by $\tilde{H}=\hat{H}-t[c_{N}^{\dagger}c_{N+1}+c_{N+1}^{\dagger}c_{N}]+\Delta_{\rm TB} c_{N} c_{N+1}+\Delta_{\rm TB}^{*} c_{N+1}^{\dagger} c_{N}^{\dagger}$. We can then perform a lattice Fourier transformation of $\tilde{H}$ and obtain 
\begin{equation}
\tilde{\mathcal{H}}=\frac{1}{2}\sum_{k}\psi_{k}^{\dagger}\tilde{H}(k)\psi_{k}+{\rm constant}\,,
\end{equation} 
where $\psi_{k}=(c_{k},c_{-k}^{\dagger})$ and 
\begin{equation}
\label{KitaevbulkTB}
\tilde{H}(k)=\tilde{\xi}_{k}\tau_{z}+\tilde{\Delta}_{k}\tau_{y}={\bm h}\cdot {\bm \tau}\,,
\end{equation}
with ${\bm h}=(0,\tilde{\Delta}_{k},\tilde{\xi}_{k})$ and ${\bm \tau}=(\tau_{x},\tau_{y},\tau_{z})$,  $\tilde{\xi}_{k}=-2t{\rm cos}(ka)-\mu_{\rm TB}$, $\tilde{\Delta}_{k}=-2\Delta_{\rm TB}{\rm sin}(ka)$, with $a$ being the lattice spacing. We can now inspect that Eq.\,(\ref{KitaevbulkTB}) indeed corresponds to the Hamiltonian describing the bulk Kitaev model given by Eq.\,(\ref{Kitaevlattice}). For this purpose, we obtain the spectrum of $\tilde{H}(k)$ which is   given by $E_{k,\pm}=\pm\sqrt{[2t{\rm cos}(ka)+\mu_{\rm TB}]^{2}+4|\Delta_{\rm TB}|^{2}{\rm sin}^{2}(ka)}$. Then, by expanding these eigenvalues for $k\sim 0$, it is clear that we recover the eigenvalues  and Hamiltonian of the continuum model after an appropriate redefinition of the TB parameters. By analogy to what we discussed in the bulk case, here we also obtain critical points when $\tilde{\xi}_{k}=0$ and $\tilde{\Delta}_{k}=0$. The pair potential vanishes when $k=0$ and $k=\pm \pi/a$. At these momenta, the normal dispersion vanishes when  $\mu_{\rm TB}=-2t$ and $\mu_{\rm TB}=+2t$, respectively.  These two lines $\mu_{\rm TB}=\pm 2t$ thus separate two gapped superconducting phases, $|\mu_{\rm TB}|<|2t|$ and $|\mu_{\rm TB}|>|2t|$, which are connected only by closing and reopening    the energy spectrum gap. Before we go further we note that   for $|\mu_{\rm TB}|<|2t|$, the system is metallic  with a Fermi surface and thus corresponds to the metallic phase discussed for the bulk Hamiltonian in previous section.

\begin{figure}[!t]
\centering
	\includegraphics[width=0.95\columnwidth]{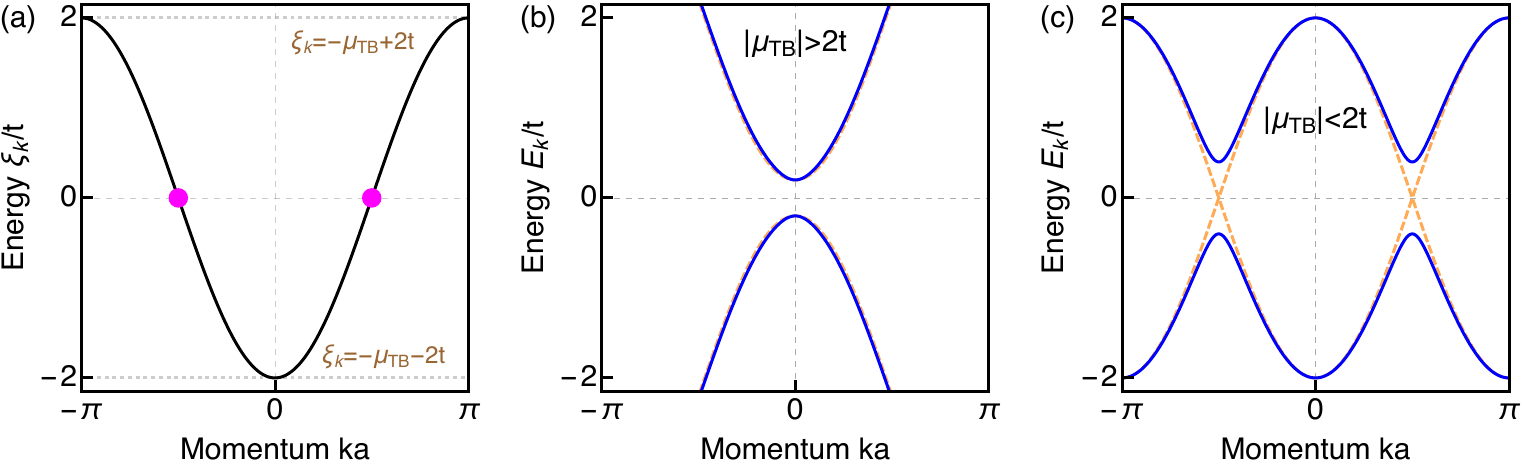}\\
 	\caption{(a) Energy as a function of momentum of the normal part of the $p$-wave superconductor within the TB description at $\mu_{\rm TB}=0$. The gray dotted horizontal lines mark indicate the points at which the energy gap closes when tuning the chemical potential $\mu_{\rm TB}$ appropriately. Inside the region between gray dotted lines, i.e., $|\mu_{\rm TB}|<2t$, the system is metallic and it exhibits an odd number of pairs of Fermi points denoted by magenta filled circles. Outside this region, however, the system is insulating. (b) Eigenvalues of the BdG Hamiltonian for a $p$-wave superconductor given by Eq.\,(\ref{KitaevbulkTB}) as a function of momentum for $\mu_{\rm TB}=2.2t>2t$, where orange and blue curves correspond to $\Delta_{\rm TB}=0$  and $\Delta_{\rm TB}=0.2{\rm meV}\neq0$, respectively. (c) Same as in (b) but with $\mu_{\rm TB}=0<2t$; a finite  $\Delta_{\rm TB}$ opens a gap  at $k=\pm\pi/2$.}
\label{FigKitaevTB} 
\end{figure}

Having understood the appearance of two phases which are connected via the closing and reopening of an energy gap, we now explore the tight-binding Hamiltonian given by Eq,\,(\ref{Kitaevlattice}) under open boundary conditions in these two phases. For this purpose, it is convenient to perform a transformation into a new set of operators given by
\begin{equation}
\label{MajoTrans}
c_{j}=\frac{1}{2}(\gamma_{j}^{A}+i\gamma_{j}^{B})\,,\quad c_{j}^{\dagger}=\frac{1}{2}(\gamma_{j}^{A}-i\gamma_{j}^{B}).
\end{equation}
where the operators $\gamma_{j}^{A(B)}$ are also known as Majorana operators. These Majorana operators satisfy
\begin{equation}
\label{conditionMajoranaOP}
\{\gamma_{i}^{\alpha},\gamma_{j}^{\beta}\}=2\delta_{ij}\delta_{\alpha\beta}\,,\quad \gamma_{j}^{\alpha}=\left[\gamma_{j}^{\alpha}\right]^{\dagger}\,, \quad \left[\gamma_{j}^{\alpha}\right]^{2}=1\,,
\end{equation}
and follow from the obtained Bogoliubov operators in Section \ref{section2}. Any fermionic operator that satisfies the  conditions given by Eqs.\,(\ref{conditionMajoranaOP}) is a Majorana operator. It is worth noting that the second expression in Eqs.\,(\ref{conditionMajoranaOP}) indeed expresses the essence borrowed from Majorana fermions: a particle created by the operator $\gamma^{\dagger}$ is identical to its antiparticle created by $\gamma$. We note that it is always possible to perform the Majorana transformation given by Eqs.\,(\ref{MajoTrans}); any fermion operator can be defined in terms of two new operators. However, it is worth pointing out that sometimes it only complicates the problem and does not lead to any novel physics. In our current case,  we will show that such decomposition leads to interesting physics.

\begin{figure}[t] 
   \includegraphics[width=0.7\columnwidth]{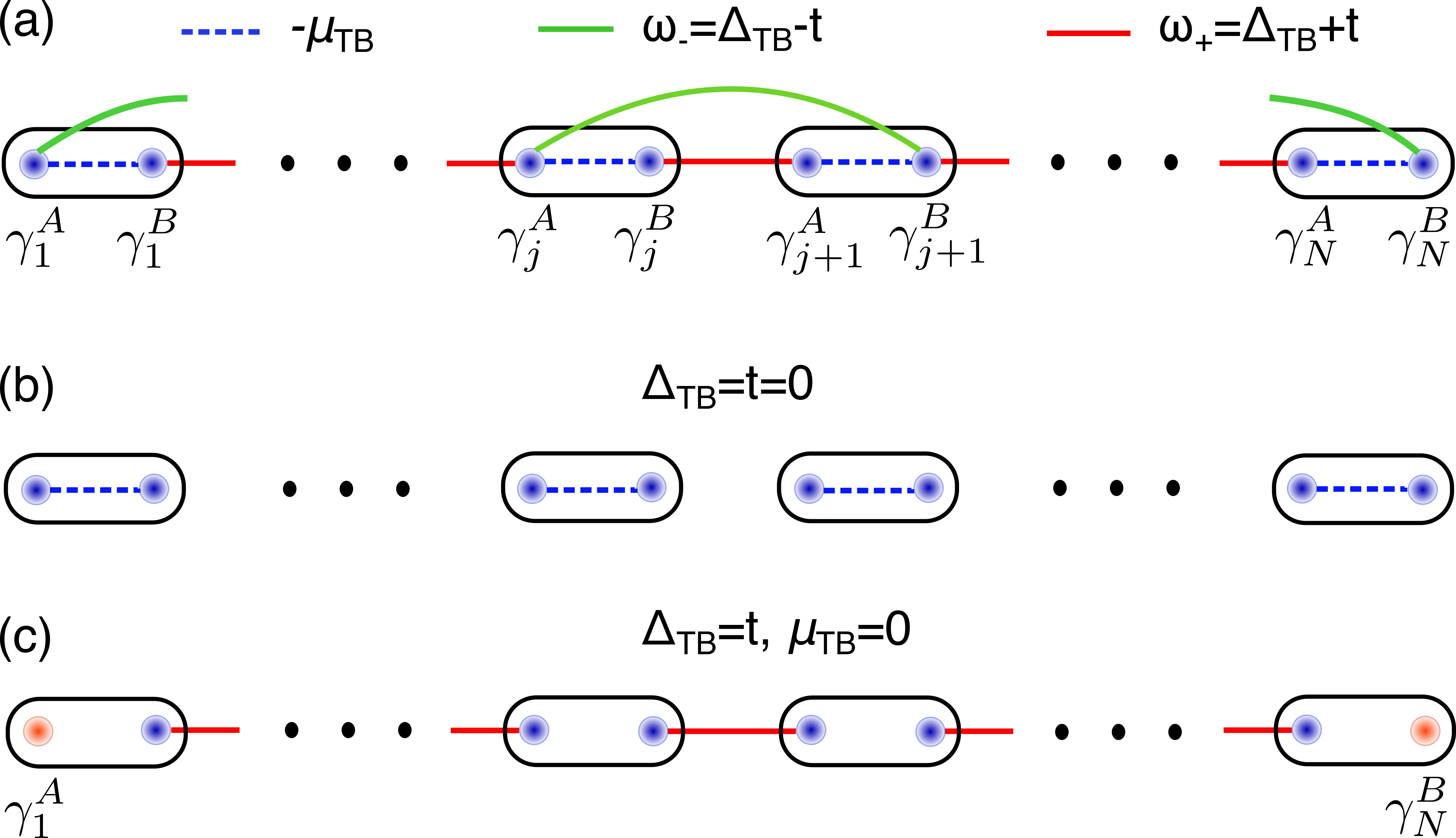} 
   \centering
     \caption{(a) The Kitaev chain with N sites in the fermionic and Majorana basis given by Eq.\,(\ref{kitaev0a}), depicted by boxes and blue gradient filled circles. In this case, $\mu_{\rm TB}$, $\omega_{-}$ and $\omega_{+}$ denote, respectively, the hopping amplitudes between Majorana operators of the same fermionic site $j$, between the first Majorana operator of site  $j$ with the second Majorana operator of site $j+1$, and between the second Majorana operator of site $j$ with the first Majorana operator of site $j+1$. (b) Representation of the Kitaev chain in the trivial phase, where Majorana operators of the same physical site are coupled to form a Dirac fermion as in Eq.\,(\ref{kitaev0b}). (c) Kitaev chain in the topological phase, where two unpaired Majorana operators appear at the end of the chain, see Eq.\,(\ref{kitaev0c}). Adapted from Ref.\,\cite{CayaoThesis}.
}
   \label{fig:kitaev2}
\end{figure}

Thus,   the Hamiltonian given by Eq,\,(\ref{Kitaevlattice}) in terms of these new operators reads  \begin{equation}
\label{kitaev0a}
H=-\frac{i\mu_{\rm TB}}{2}\sum_{j=1}^{N}\gamma^{A}_{j}\gamma^{B}_{j}+\frac{i}{2}\sum_{j=1}^{N-1}\Big[\omega_{+}\gamma^{B}_{j}\gamma^{A}_{j+1}+ \omega_{-}\gamma^{A}_{j}\gamma^{B}_{j+1}\Big]\,,
\end{equation}
where $\mu_{\rm TB}$, $\omega_{-}=\Delta_{\rm TB}-t$ and $\omega_{+}=\Delta_{\rm TB}+t$ represent, respectively, the hopping amplitudes between Majorana operators of the same fermionic site $j$, between the first Majorana operator of site $j$ with the second Majorana operator of site $j+1$, and between the second Majorana operator of site $j$ with the first Majorana operator of site $j+1$. In  Fig.\,\ref{fig:kitaev2}(a) we schematically show Eq.\,(\ref{kitaev0a}), where dashed blue, solid green, and solid red lines denote $\mu_{\rm TB}$, $\omega_{-}$ and $\omega_{+}$, respectively. Depending on the parameters $\mu_{\rm TB}$, $\Delta_{\rm TB}$ and $t$, the Hamiltonian  given by Eq\,(\ref{kitaev0a}) exhibits different interesting properties, which we discuss next. In particular, we illustrate the difference between the topological and trivial phases by looking at two special limits.

\textbf{Regime with $|\mu_{\rm TB}|>|2t|$:} In this case, we can consider $t=\Delta_{\rm TB}=0$ as an example. Thus, 
the Hamiltonian given by Eq.\,(\ref{kitaev0a}) reads
\begin{equation}
\label{kitaev0b}
H=-\frac{i\mu_{\rm TB}}{2}\sum_{j=1}^{N}\gamma^{A}_{j}\gamma^{B}_{j}\,.
\end{equation}
The sum in this equation tells us that Majorana operators from the same physical site are paired together to form a regular fermion, as depicted in   Fig.\,\ref{fig:kitaev2}(b). In this case,  the ground state is given by all states being empty ($\mu_{\rm TB}<0$) or occupied ($\mu_{\rm TB}>0$).  It is therefore clear that there is not nothing important happening in this regime, which is a consequence of the system being topologically trivial.

\textbf{Regime with ${|\mu_{\rm TB}|<|2t|}$:} To inspect this phase, we consider   $\mu_{\rm TB}=0$ and $\omega_{-}=0$. Then,  the Hamiltonian given by Eq.\,(\ref{kitaev0a}) acquires the following form,
\begin{equation}
\label{kitaev0c}
H=it\sum_{j=1}^{N-1}\gamma^{B}_{j}\gamma^{A}_{j+1}\,=\,i\Delta_{\rm TB}\Big[\gamma^{B}_{1}\gamma^{A}_{2}+\gamma^{B}_{2}\gamma^{A}_{3}+\cdots+\gamma^{B}_{N-1}\gamma^{A}_{N}\Big]\,,
\end{equation}
which clearly reveals that Majorana operators on adjacent lattice sites are coupled but, interestingly, 
operators $\gamma_{1}^{A}$ and $\gamma_{N}^{B}$ do not appear in the description, see Fig.\,\ref{fig:kitaev2}(c). Thus, the Majorana operators $\gamma_{1}^{A}$ and $\gamma_{N}^{B}$ represent Majorana modes emerging at zero energy and located at the ends of the chain. It is worth noting that a consequence of $\gamma_{1}^{A}$ and $\gamma_{N}^{B}$ not appearing in the Hamiltonian is that they do not couple to any Majorana operator and therefore commute with the Hamiltonian, namely,    $[H,\gamma_{1}^{A}]=[H,\gamma_{N}^{B}]=0$. This implies that Majorana states with $\gamma_{1}^{A}$ and $\gamma_{N}^{B}$ can be created without energy cost and, therefore, the ground state remains at the same energy.

Further insights are obtained by defining new fermionic operators for effectively writing Eq.\,(\ref{kitaev0c}) and other fermionic operators to describe the Majorana pair not present in 
Eq.\,(\ref{kitaev0c}). We start by  defining a new set of fermionic operators that involve only Majorana operators included in the Hamiltonian given by 
Eq.\,(\ref{kitaev0c}),
\begin{equation}
\begin{split}
d_{j}&=\frac{1}{2}\Big(\gamma_{j}^{B}+i\gamma_{j+1}^{A} \Big)\,,\quad d_{j}^{\dagger}=\frac{1}{2}\Big(\gamma_{j}^{B}-i\gamma_{j+1}^{A} \Big)\,,
\end{split}
\end{equation}
which leads to a new form of   Eq.\,(\ref{kitaev0c}) that  reads
\begin{equation}
\label{fermioMan}
H=2t\sum_{j=1}^{N-1}\Big(d_{j}^{\dagger}d_{j}-\frac{1}{2} \Big)\,.
\end{equation}
This Hamiltonian can be also written as
\begin{equation}
H=2t\sum_{j=1}^{N-1}d_{j}^{\dagger}d_{j}-t(N-1)\,,
\end{equation}
where the ground state energy is then given by $E_{0}=-t(N-1)$ for $t>0$.

In relation to   the two unpaired Majorana operators located at the ends of the chain, $\gamma_{1}^{A}$ and $\gamma_{N}^{B}$, they can be   fused into a fermion operator with the following transformation
\begin{equation}
\label{newopMaj}
f=\frac{1}{2}\Big(\gamma_{1}^{A}+i\gamma_{N}^{B}\Big)\,,\quad f^{\dagger}=\frac{1}{2}\Big(\gamma_{1}^{A}-i\gamma_{N}^{B}\Big)\,.
\end{equation} 
 Interestingly this fermion exhibits a very unusual property: it is nonlocal in space because it has contributions from both ends of the chain. Moreover, as already pointed out, creating such a nonlocal fermion requires zero energy because their composing Majorana operators are absent in Eq.\,(\ref{fermioMan}). At this point, we remember that  $[H,\gamma_{1}^{A}]=[H,\gamma_{N}^{B}]=0$, which implies that $Hf\ket{GS}=E_{0}f\ket{GS}$ and $Hf^{\dagger}\ket{GS}=E_{0}f^{\dagger}\ket{GS}$, where $\ket{GS}$ is the ground state with energy $E_{0}$. Thus, the ground states of $H$ can be spanned by $f$ and $f^{\dagger}$. This can be visualizing by writing the  occupation number operator   $n=f^{\dagger}f=(1+i\gamma_{1}^{A}\gamma_{N}^{B})/2$ and the fermion parity operator $P=(-1)^{n}=-i\gamma_{1}^{A}\gamma_{N}^{B}$. Then, $\bra{GS}P\ket{GS}=-\bra{GS}[2f^{\dagger}f-1]\ket{GS}=\mp1$ for occupied ($n=1$) or empty ($n=0$) ground state. Hence, we can use $n$ to label the ground state, which can be seen to have a  two-fold degeneracy arising from the two possible occupancies $n=0,1$. Thus, the two ground states are $\ket{0}$ and $\ket{1}$, which satisfy  $f\ket{0}=0$   and $\ket{1}=f^{\dagger}\ket{0}$, with $\braket{1}{0}=0$, which reveals that the ground state degeneracy of the Kitaev's model for $|2\mu_{\rm TB}|<2t$ is indeed two-fold. The parity state therefore defines a degenerate two-level quantum system, which is the definition of a quantum bit (qubit) and can be used to encode quantum information \cite{kitaev}. Furthermore, we remember that the number operator defining the parity corresponds to a spatially nonlocal fermion with zero energy. This implies that the parity state cannot be accessed with any local measurement on one of the bound states at the end of the chain, which shows that the information in such a qubit is stored non-locally and robust against local sources of decoherence \cite{kitaev}. See Refs.\,\cite{hassler2014majorana,aguado2020majorana,Beenakker_2020,Marra_2022} for a detailed discussion on how Majorana operators can be used for designing robust qubits.

We have therefore seen that the two phases $|\mu_{\rm TB}|<|2t|$ and $|\mu_{\rm TB}|>|2t|$ are different and can be distinguished because the former phase hosts MZMs at the edges.  We also note that these two phases are topologically distinct and can be distinguished by means of topological invariants. In the case of 1D topological superconductors, it has been shown that one can define a topological invariant given by $M=(-1)^{\nu}$, where $\nu$ represents the number of pairs of Fermi points in the Brillouin zone in the normal state ($\Delta=0$) spectrum of the Kitaev model under periodic boundary conditions Eq.\,(\ref{KitaevbulkTB}). In this formulation,    an odd (even) number of pairs of Fermi points signal the emergence of a topological  (trivial) phase. Interestingly, for $|\mu_{\rm TB}|<|2t|$, the number of pairs of Fermi points is one and therefore  $M=-1$, signaling that the phase with $|\mu_{\rm TB}|<|2t|$ is topological.  It is also important to note that, for $|\mu_{\rm TB}|<|2t|$, the system is metallic with a Fermi surface as it happens for the continuum model with $\mu>0$.   Before ending this part, we   note that MZMs in this case can be also characterized by a $Z_{2}$ topological index also known as Majorana number $M$, which can be calculated as \cite{kitaev}
\begin{equation}
M={\rm sgn}[{\rm Pf}A(k=0)]{\rm sgn}[{\rm Pf}A(k=\pi)]=\pm1\,,
\end{equation}
where $A$ represents the Hamiltonian written as an antisymmetric matrix given by 
\begin{equation}
A\left(k \right)= i U^{\dagger} H\left( k \right) U, \ \ 
U=
\frac{1}{\sqrt{2}}
\begin{pmatrix}
1 & -i \\
1 & i 
\end{pmatrix}
.
\end{equation}
 Here,
  $M=+1$ corresponds to topologically trivial states, while $M=-1$ characterizes the topologically nontrivial state. In Kitaev’s 1D chain, the Majorana number $M$ is given by
\begin{equation}
\label{Paffian}
M={\rm sgn}(\mu_{\rm TB}+2t){\rm sgn}(\mu_{\rm TB}-2t)
\end{equation}
with   ${\rm Pf}(A)^{2}={\rm det}(A)$. Therefore, Eq.\,(\ref{Paffian}) implies that the topological phase occurs for $2t>|\mu_{\rm TB}|$, given $\Delta_{\rm TB}\neq0$.  We can hence conclude that the phase where MZMs appear is a topological superconducting phase.

\subsection{Physical realization using semiconductor-superconductor hybrids}
\label{subsection3b}
As we have demonstrated in previous subsections, to realize topological superconductivity and MZMs, it is necessary to have a spin-triplet  $p$-wave superconductor. There are in principle two ways to have this exotic superconductor. One is to find materials with intrinsic spin-triplet  $p$-wave pair correlations, which, albeit intense efforts, such as in Refs.\,\cite{ishida1998spin,RevModPhys.74.235,RevModPhys.75.657,PhysRevLett.107.217001,Maeno2012,kallin2016chiral,matano2016spin,PhysRevX.7.011032,zhang2018observation,wang2018evidence,talantsev2019p,zhu2023intrinsic}, it is still represents an ongoing challenging. Another possibility is to engineer such unconventional superconductivity in the lab. In this regard, this has been predicted to be achieved by combining common ingredients, which include conventional spin-singlet $s$-wave superconductors,   spin-orbit coupling (SOC), and magnetic fields.  These ingredients have been exploited in semiconductors \cite{PhysRevB.77.220501,PhysRevB.79.094504,PhysRevLett.104.040502,PhysRevLett.105.077001,PhysRevLett.105.177002,Alicea:PRB10,PhysRevLett.108.147003}, topological insulators 
\cite{Fu:PRL08,Fu:PRB09,RevModPhys.82.3045,RevModPhys.83.1057,TanakaNagaosa2009}, and chains of magnetic atoms \cite{PhysRevB.84.195442,PhysRevLett.111.186805,PhysRevB.88.020407,PhysRevLett.111.147202,PhysRevB.88.155420,PhysRevB.88.180503}; see also Ref.\,\cite{PhysRevLett.103.020401}. In this part we discuss the realization of 1D topological superconductivity and MZMs by combining conventional spin-singlet $s$-wave superconductors, Rashba spin-orbit coupling, and an external magnetic field.  In particular, this part focuses on understanding the continuum model and its lattice representation.

\subsubsection{Continuum model}
We consider a single channel nanowire in one-dimension with SOC and Zeeman interaction, where the radius of the wire is small compared to the Fermi wavelength and there is a single 1D occupied mode. Thus, this 1D nanowire is modeled by  
\begin{equation}
\label{H0Rashba}
\mathcal{H}_{0}=\int dx\,\psi^{\dagger}(x)\,H_{0}\,\psi(x)\,,
\end{equation}
where  $\psi=(\psi_{\uparrow},\psi_{\downarrow})^{\rm T}$, with $\psi_{\sigma}$ represents the annihilation operator of an electron at position $x$ with spin $\sigma=\uparrow,\downarrow$, while $H_{0}$ is the Hamiltonian density that reads,
\begin{equation}
\label{H0Hamil}
H_{0}=\frac{p^{2}_{x}}{2m}-\mu-\frac{\alpha_{R}}{\hbar}\sigma_{y}p_{x}+B\sigma_{x}\,.
\end{equation}
 Here, the first term is the kinetic part with   $p_{x}=-i\hbar\partial_{x}$ being the momentum operator, $m$ is the effective electron's mass in the nanowire, and $\mu$ is the chemical potential which determines the filling of the nanowire and measured from the bottom of the band. The second term corresponds to the Rashba SOC Hamiltonian with the spin direction along the $y$-axis, where $\alpha_{R}$ represents the strength of Rashba SOC. The last term in Eq.\,(\ref{H0Hamil}) is the Zeeman Hamiltonian with $B=g\mu_{B}\mathcal{B}/2$ due to an external magnetic field $\mathcal{B}$ applied here along the nanowire $x$-axis, where $g$ is the wire's $g$-factor and $\mu_{B}$ the Bohr magneton. Note that the Zeeman field is perpendicular to the SOC axis, which, as we will see later, is crucial for achieving topological superconductivity.
 
\begin{figure} 
 \centering
   \includegraphics[width=0.95\columnwidth]{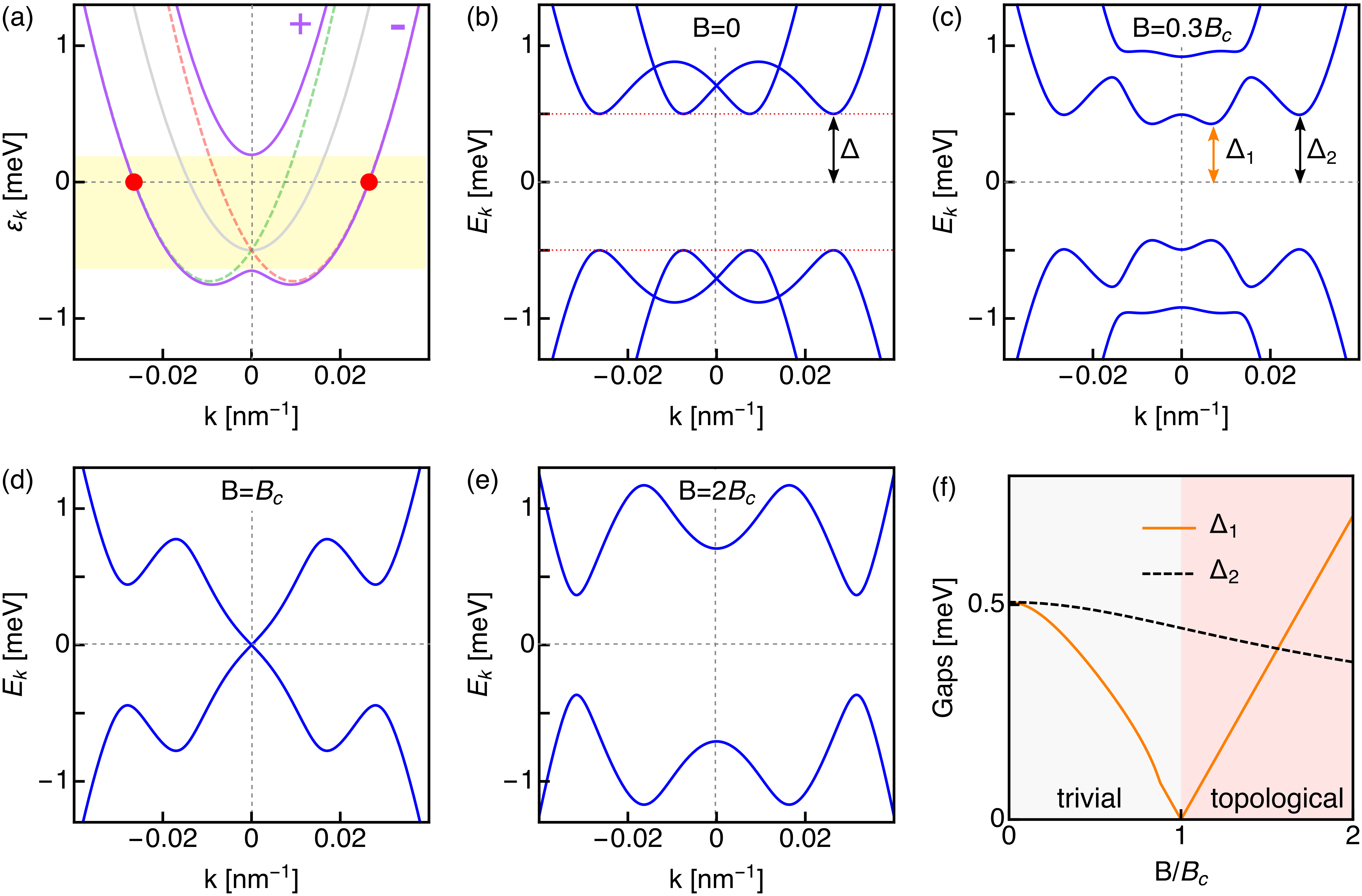} 
     \caption{(a) Energy bands  versus momentum $k$ in the normal state of a 1D Rashba semiconductor. Gray curve corresponds to spin degenerate bands for free electrons, where the chemical potential is measured from the bottom of the band.  The red and green dashed curves correspond to energy bands at finite Rashba SOC,  which split in spin but still remain degenerate at $k=0$. Magenta curves show the bands labeled by $\pm$ at finite Zeeman field when $B>\mu$, which opens a gap $k=0$ indicated by the yellow-shaded region. Inside this region, states at the Fermi level (red-filled circles) exhibit different spin projections and propagate in opposite directions,  giving rise to the concept of helicity.  (b-e) Evolution of the energy versus momentum in the superconducting state for different values of the Zeeman field defined in terms of the critical field $B_{c}=\sqrt{\mu^{2}+\Delta^{2}}$. (f) Zeeman dependence of the two energy gaps that appear in the energy bands of the Rashba superconductor at finite Zeeman field. The shaded gray (red) regions indicate their trivial (topological) origin.
}
   \label{RashbaSC}
\end{figure}

At this point, it is important to understand the ingredients of the Hamiltonian $H_{0}$ describing the Rashba nanowire given by  Eq.\,(\ref{H0Hamil}). For this purpose, we find the eigenvalues and eigenvectors of $H_{0}$. In terms of plane waves $\Psi_{k}(x)=\phi\,{\rm e}^{ikx}$,  we get
\begin{equation}
\label{H0vEnVec}
\varepsilon_{k,\pm}=\xi_{k}\pm\sqrt{B^{2}+\alpha_{R}^{2}k^{2}}\,,\quad \Psi_{k,\pm}(r)=\phi_{\pm}(k)\,\frac{1}{\sqrt{L}}{\rm e}^{ikx}\,,\quad  \phi_{\pm}(k)=\frac{1}{\sqrt{2}}\begin{pmatrix}
\pm\gamma_{k}\\
1
 \end{pmatrix}\,,
\end{equation}
where $\xi_{k}=\hbar^{2}k^{2}/(2m)-\mu$, and $\gamma_{k}=(i\alpha_{R} k+B)/\sqrt{B^{2}+\alpha_{R}^{2}k^{2}}$. To visualize the behaviour of the eigenvalues, in Fig.\,\ref{RashbaSC}(a) we present them as a function of momentum $k$ for distinct values of the parameters. In the absence of SOC and Zeeman energy, the energy bands are two superimposed parabolas that are degenerate in spin, see gray curve in Fig.\,\ref{RashbaSC}(a). A finite SOC, shifts the energy bands for each spin along the momentum axis such that they become centered at $\pm k_{soc}=\pm m\alpha_{R}/\hbar^{2}$ and by an overall energy $E_{soc}=m\alpha_{R}^{2}/2\hbar^{2}$, see red and green dashed curves in Fig.\,\ref{RashbaSC}(a). A finite Zeeman field, perpendicular to the SOC, lifts the degeneracy at $k=0$ and opens a gap of $2B$, as shown by magenta curves and yellow region in Fig.\,\ref{RashbaSC}(a). Interestingly, when $B>\mu$,   only the lowest band labeled by $-$ is partially occupied and the nanowire behaves as spin-polarized or spinless. In this regime, the system exhibits two Fermi points inside the gap corresponding to counter-propagating states with different spins, which implies that the spin projection is locked to the momentum. For this reason, this regime is also known as helical and the opened gap at $k=0$ is sometimes referred to as helical gap \cite{PhysRevB.91.024514,PhysRevB.90.235415,kammhuber2017conductance}. The behavior of these Rashba bands,  having an odd number of Fermi points, is similar to what is found when describing the normal bands of the Kitaev model, revealing the potential of the Rashba nanowire for topological superconductivity which is discussed next.  
 
To see the emergence of topological superconductivity, we now incorporate conventional spin-singlet $s$-wave superconductivity in the nanowire. In principle, the nanowire is coupled to a superconductor and then electrons in the nanowire feel an effective superconducting pair potential due to the so-called \emph{superconducting proximity effect} where superconducting correlations leak into the nanowire \cite{RevModPhys.36.225,Doh272}. Thus, the proximity effect involves tunneling between these systems, implicitly requiring a good interface between the nanowire and superconductor is necessary.  At a phenomenological level, we consider that conventional superconductivity in the nanowire is described by 
\begin{equation}
\label{hsc}
\mathcal{H}_{sc}=\int dx\Big[\mathbf{\Delta}\psi_{\uparrow}^{\dagger}(k)\psi_{\downarrow}^{\dagger}(-k)+\mathbf{\Delta}^{\dagger}\psi_{\downarrow}(-k)\psi_{\uparrow}(k)\Big]\,,
\end{equation}
where $\mathbf{\Delta}={\rm e}^{i\varphi}\Delta$ is the spin-singlet $s$-wave pair  potential and   $\varphi$ is the superconducting phase.  Even though $\mathbf{\Delta}$ is complex in general, in what follows we consider it to be real. Thus, taking into account the normal and superconducting parts described by Eqs.\,(\ref{H0Rashba}) and \,(\ref{hsc}),  the full system Hamiltonian is written as,
\begin{equation}
\label{fullHamiltonian}
\mathcal{H}=\mathcal{H}_{0}+\mathcal{H}_{sc}=\mathcal{H}_{kin}+\mathcal{H}_{SOC}+\mathcal{H}_{Z}+\mathcal{H}_{sc}\,.
\end{equation}
This Hamiltonian already provides a good start to analyze the impact of SOC, Zeeman field, and superconductivity. However, to see how a spinless $p$-wave superconductor can be engineered, it is instructive to write   the Hamiltonian given by Eq.\,(\ref{fullHamiltonian})  in the helical basis which diagonalizes $\mathcal{H}_{0}$ constructed from Eqs.\,(\ref{H0vEnVec}) as follows \cite{Alicea:PRB10,Alicea_2012},
 \begin{equation}
 \label{eq111}
 \psi(k)=\phi_{-}(k)\psi_{-}(k)+\phi_{+}(k)\psi_{+}(k)\,,
 \end{equation}
where $\psi_{\pm}$ are operators that annihilate states in the upper/lower bands labelled by $\pm$ 
at momentum $k$ with energy $\varepsilon_{k,\pm}$. Moreover,  $\phi_{\pm}$ correspond to the  normalized wave-functions   given by Eq.\,(\ref{H0vEnVec}). We further decompose Eq.\,(\ref{eq111})  into the two spinor components, which read
 \begin{equation}
 \label{basis1}
   \psi_{\uparrow}(k)=\frac{1}{\sqrt{2}}\left[-\gamma_{k}\psi_{-}(k)+\gamma_{k}\psi_{+}(k)\right]\,,\quad   \psi_{\downarrow}(k)=\frac{1}{\sqrt{2}}\left[\psi_{-}(k)+\psi_{+}(k)\right]\,.
 \end{equation}
 By introducing Eqs.\,(\ref{basis1}) into Eq.\,(\ref{fullHamiltonian}),   the full Hamiltonian in this new basis is given by
\begin{equation}
\label{fullHhelical}
\begin{split}
\mathcal{H}&=
\int \frac{dk}{2\pi}
\big[
\varepsilon_{k,+}\psi_{+}^{\dagger}(k)\psi_{+}(k)
+
\varepsilon_{k,-}\psi_{-}^{\dagger}(k)\psi_{-}(k)
\big]\\
&+
\bigg[\frac{\Delta_{--}(k)}{2}\psi_{-}^{\dagger}(k)\psi_{-}^{\dagger}(-k)
+\frac{\Delta_{++}(k)}{2}\psi_{+}^{\dagger}(k)\psi_{+}^{\dagger}(-k)\\
&+\Delta_{+-}(k)\psi_{+}^{\dagger}(k)\psi_{-}^{\dagger}(-k)+h.c
\bigg],
\end{split}
\end{equation}
where 
\begin{equation}
\label{pairings0}
\begin{split}
\Delta_{--}(k)&=\frac{i\alpha_{R} k \Delta}{\sqrt{B^{2}+\alpha_{R}^{2}k^{2}}}\,,\\
\Delta_{++}(k)&=\frac{-i\alpha_{R} k \Delta}{\sqrt{B^{2}+\alpha_{R}^{2}k^{2}}}\,,\\
 \Delta_{+-}(k)&=\frac{B \Delta}{\sqrt{B^{2}+\alpha_{R}^{2}k^{2}}}\,,
\end{split}
\end{equation}
represent the distinct pair potentials appearing in the nanowire due to the interplay of Rashba SOC, Zeeman interaction, and spin-singlet $s$-wave superconductivity. The first line in Eq.\,(\ref{fullHhelical}) represents the normal Rashba nanowire Hamiltonian, which, as expected, is diagonal in this helical basis. The second and third lines, respectively,     correspond to the superconducting part containing terms that describe pairing between states of the same  ($\Delta_{--,++}(k)$) and different ($\Delta_{+-}(k)$) bands.  The band index $\sigma=\pm$  plays the role of pseudospin (or spin broadly speaking) and therefore,  having pairing between states of the same band $\sigma$, can be seen to be similar to having spin-triplet pairing between electrons of the same spin, namely, spin-polarized pairing.  Analogously,  pairing between states of different $\sigma$ can be seen to be a sort of spin-singlet type of pairing in the band index. Another important property of the pair potentials in Eqs.\,(\ref{pairings0}) is that they exhibit an interesting dependence on momentum.  In fact,  the intraband pair potentials $\Delta_{--,++}(k)$ are odd functions in momentum $k$, which reflects their $p$-wave pair symmetry, similar to what we discussed in subsection \ref{subsection11}. In contrast, the interband $\Delta_{+-}$ is an even function in momentum and thus has  $s$-wave pair symmetry. We thus see that the Hamiltonian for a nanowire with Rashba SOC, Zeeman interaction, and conventional spin-singlet $s$-wave superconductivity is able to realize spin-polarized $p$-wave superconductivity as shown by Eq.\,(\ref{fullHhelical}), which, however, in general, coexists with spin-singlet $s$-wave superconductivity.

To further inspect the emergence of $p$-wave superconductivity, we explore the eigenvalues of Eq.\,(\ref{fullHhelical}). For this reason, we first write it in Nambu space as
 \begin{equation}
 \mathcal{H}=\frac{1}{2}\int \frac{dk}{2\pi}\Psi^{\dagger}(k)\,H_{BdG}\,\Psi(k)\,, \quad \Psi(k)=\begin{pmatrix}
\psi_{+}^{\dagger}(k),
\psi_{-}^{\dagger}(k),
\psi_{+}(-k),
\psi_{-}(-k)
\end{pmatrix}^{\dagger}
 \end{equation}
 where the BdG Hamiltonian is given by
 \begin{equation}
 H_{BdG}=
 \begin{pmatrix}
 \varepsilon_{k,+}&0&\Delta_{++}(k)&\Delta_{+-}(k)\\
 0&\varepsilon_{k,-}&-\Delta_{+-}(k)&\Delta_{--}(k)\\
 \Delta_{++}^{\dagger}(k)&-\Delta_{+-}^{\dagger}(k)&-\varepsilon_{-k,+}&0\\
 \Delta_{+-}^{\dagger}(k)&\Delta_{--}^{\dagger}(k)&0&-\varepsilon_{-k,-}
 \end{pmatrix}\,.
 \end{equation}
Then, by a simple diagonalization, we obtain the eigenvalues of $H_{BdG}$ which are given by
 \begin{equation}
 \label{SpectrumFull}
E_{\pm}^{2}(k)=|\Delta_{++}(k)|^{2}+\Delta_{+-}^{2}(k)+\frac{\varepsilon_{k,+}^{2}+\varepsilon_{k,-}^{2}}{2}
\pm|\varepsilon_{k,+}-\varepsilon_{k,-}|\sqrt{\Delta_{+-}^{2}(k)+\Big[\frac{\varepsilon_{k,+}+\varepsilon_{k,-}}{2}\Big]^{2}}\,,
\end{equation}
where $\varepsilon_{k,\pm}$ are given by Eq.\,(\ref{H0vEnVec}) and $\Delta_{\sigma\sigma'}$ by Eqs.\,(\ref{pairings0}).  In Figs.\,\ref{RashbaSC}(b-e) we show the evolution of the eigenvalues given by Eq.\,(\ref{SpectrumFull}) as a function momentum $k$ for different values of the Zeeman field $B$. At $B=0$, there are four bands  with a finite energy gap opened by $\Delta$ at the Fermi points $k_{F,\pm}$, see Fig.\,\ref{RashbaSC}(b). We remind that $k_{F,\pm}$ are found from $\varepsilon_{k,\pm}=0$ and given by $k_{F,\pm}=\sqrt{k_{F}^{2}+2k_{soc}^{2}\mp\sqrt{(k_{F}^{2}+2k_{soc}^{2})^{2}-k_{F}^{4}+k_{Z}^{4}}}$, with $k_{F}=\sqrt{2m\mu/\hbar^{2}}$, $k_{Z}=\sqrt{2mB/\hbar^{2}}$, and $k_{soc}=m\alpha_{R}/\hbar^{2}$. At finite Zeeman fields, the bands lift the spin degeneracy at high energies and also develop distinct energy gaps at low and high Fermi momenta denoted by $\Delta_{1(2)}$, see the orange and black double arrows in Fig.\,\ref{RashbaSC}(c). Since these distinct gaps are formed at the lower band   $E_{-}$ from Eq.\,(\ref{SpectrumFull}), they can be roughly defined as
\begin{equation}
\label{D1D2}
\Delta_{1}=E_{-}(k_{F,+})\,,\quad \Delta_{2}=E_{-}(k_{F,-})\,,
\end{equation}
It is thus evident that, because $k_{F,\pm}$ depend on the SOC, Zeeman field, and chemical potential, the gaps  $\Delta_{1(2)}$  will necessarily exhibit a  different behavior when varying such parameters.  Further increasing the Zeeman field the low momentum gap decreases such that the spectrum becomes gapless at $B=B_{c}$ in   Fig.\,\ref{RashbaSC}(d), where  
 \begin{equation}
\label{Bc}
B_{c}\equiv\sqrt{\Delta^{2}+\mu^{2}}
\end{equation}
defines a critical field. The high momentum gap $\Delta_{2}$, however, remains finite, with a value that can be roughly constant under strong SOC.  For $B>B_{c}$, the gap reopens such that for sufficiently large fields the lowest gap is $\Delta_{2}$ and the higher band is not anymore within the low-energy description, see  Fig.\,\ref{RashbaSC}(e). To further understand the gap closing and reopening, we plot $\Delta_{1,2}$ in Fig.\,\ref{RashbaSC}(f) at strong SOC,  which clearly shows the discussed behavior and reveals that $\Delta_{1}$ exhibits roughly a linear dependence on $B$. This can be further confirmed by evaluating the energy bands at $k=0$, which gives
\begin{equation}
\label{topotrans2}
 E_{-}(k=0)=|B-B_{c}|\,,
\end{equation}
and clearly shows the gap closing at $B=B_{c}$. Thus, the Rashba nanowire with Zeeman field and superconductivity hosts two distinct gapped phases which are connected only by a gap closing and reopening. The phase with  $B>B_{c}$ only contains the lowest band, where only pairing between states of the same band are possible and, therefore, behaves as it was effectively spin-polarized or spinless.

It was shown that the two phases, with $B>B_{c}$ and $B<B_{c}$, are topologically different, and $B=B_{c}$ defines a topological phase transition into a topological superconducting phase with MZMs \cite{PhysRevLett.105.177002,PhysRevLett.105.077001}. As in Kitaev's model discussed in the previous subsection, the topological invariant that distinguishes these phases is given by $M=(-1)^{\nu}$, where $\nu$ is the number of pairs of Fermi points in the normal dispersion, which is shown in Fig.\,\ref{RashbaSC}(a). Indeed,  the system hosts an odd number of pairs of Fermi points when $|\mu|<B$, which is fulfilled by   $B>B_{c}$. Thus, the regime with $B>B_{c}$ is expected to be topological. We anticipate that this topological phase contains MZMs, protected by the effective gap $\Delta_\mathrm{eff}=\mathrm{Min}(\Delta_1,\Delta_2)$, at the wire ends. Above a certain field $B_c^{(2)}$, the gap $\Delta_\mathrm{eff}$ saturates at $\Delta_2$ [Fig.\,\ref{RashbaSC}(f)], which is strongly dependent on the strength of SOC and approximated as \cite{PhysRevB.96.205425} $\Delta_{2}\approx 2\Delta E_{soc}/\sqrt{E_{soc}(2E_{soc}+\sqrt{B^{2}+4E_{soc}^{2}})}$. In this regime,  a single band is left in the description and pairing occurs between states of the same band.  In the following, we show that for high Zeeman fields $B\gg B_{c}$, the Hamiltonian $\mathcal{H}$ given by Eq.\,(\ref{fullHhelical})  describes the Kitaev's model.

\subsubsection{Strong magnetic fields and mimicking the Kitaev model}
To have further insight into the previous discussion, the system has to exhibit $p$-wave pairing symmetry according to Kitaev's model. Thus,
it is convenient to project the system Hamiltonian onto the lower band $-$.
This is allowed because for reaching the topological phase one needs a strong Zeeman field, then the upper band $+$, see Fig.\,\ref{RashbaSC}, can be removed from the low-energy problem. Therefore, we can write,
\begin{equation}
\label{fullHlowestband}
\begin{split}
\mathcal{H}&=
\int \frac{dk}{2\pi}\bigg[
\varepsilon_{k,-}\psi_{-}^{\dagger}(k)\psi_{-}(k)
+
\frac{\Delta_{--}(k)}{2}\psi_{-}^{\dagger}(k)\psi_{-}^{\dagger}(-k)
+\frac{\Delta_{--}^{\dagger}(k)}{2}\psi_{-}(-k)\psi_{-}(k)
\bigg],
\end{split}
\end{equation}
where the superconducting pairing potential, or order parameter, 
\begin{equation}
\Delta_{--}(k)=\frac{i\alpha_{R}k\Delta}{\sqrt{B^{2}+\alpha_{R}^{2}k^{2}}}
\end{equation}
 has $p$-wave symmetry.
 Now, we write previous Hamiltonian in the BdG form,
 \begin{equation}
 \mathcal{H}=\frac{1}{2}\int\frac{dk}{2\pi}\psi^{\dagger}(k) H_{BdG}\psi(k)\,,\quad 
 \psi(k)=\begin{pmatrix}
\psi_{-}(k)\\
\psi_{-}(-k)^{\dagger}
\end{pmatrix}\,,
 \end{equation}
 where
 \begin{equation}
 H_{BdG}=
 \begin{pmatrix}
\varepsilon_{k,-}&\Delta_{--}(k)\\
\Delta_{--}^{\dagger}(k)&-\varepsilon_{-k,-}\,
\end{pmatrix}\,,
 \end{equation}
 whose energy spectrum is given by
 \begin{equation}
 E_{k,-}=\pm\sqrt{\varepsilon_{k,-}^{2}+|\Delta_{--}(k)|^{2}}\,.
 \end{equation}
 The previouss equation is in essence the energy spectrum of a $p$-wave superconductor, thus being in concordance with Kitaev's model described in the previous section. Therefore, it is not a coincidence that the model given by the Hamiltonian Eq.\,(\ref{fullHamiltonian}) indeed describes Majorana-like physics when $B>B_{c}$. 
 
 From the point of view of topology, the infinite, Zeeman polarised, single-channel semiconducting nanowire in proximity to a conventional $s$-wave superconductor, described in this section, belongs to the so-called one-dimensional D class \cite{Altland:PRB97}, which has an invariant $\nu'$ that may be $\nu'=0$ (topologically trivial) or $\nu'=1$ (non-trivial).
This system undergoes a band topological transition from $\nu'=0$ to $\nu'=1$ when the Zeeman splitting $B$, perpendicular to the spin-orbit axis, exceeds a critical value $B_c=\sqrt{\mu_S^2+\Delta^2}$, where $\mu_S$ and $\Delta$ are the wire's Fermi energy and induced gap respectively.

 \begin{figure} 
   \centering
   \includegraphics[width=0.95\columnwidth]{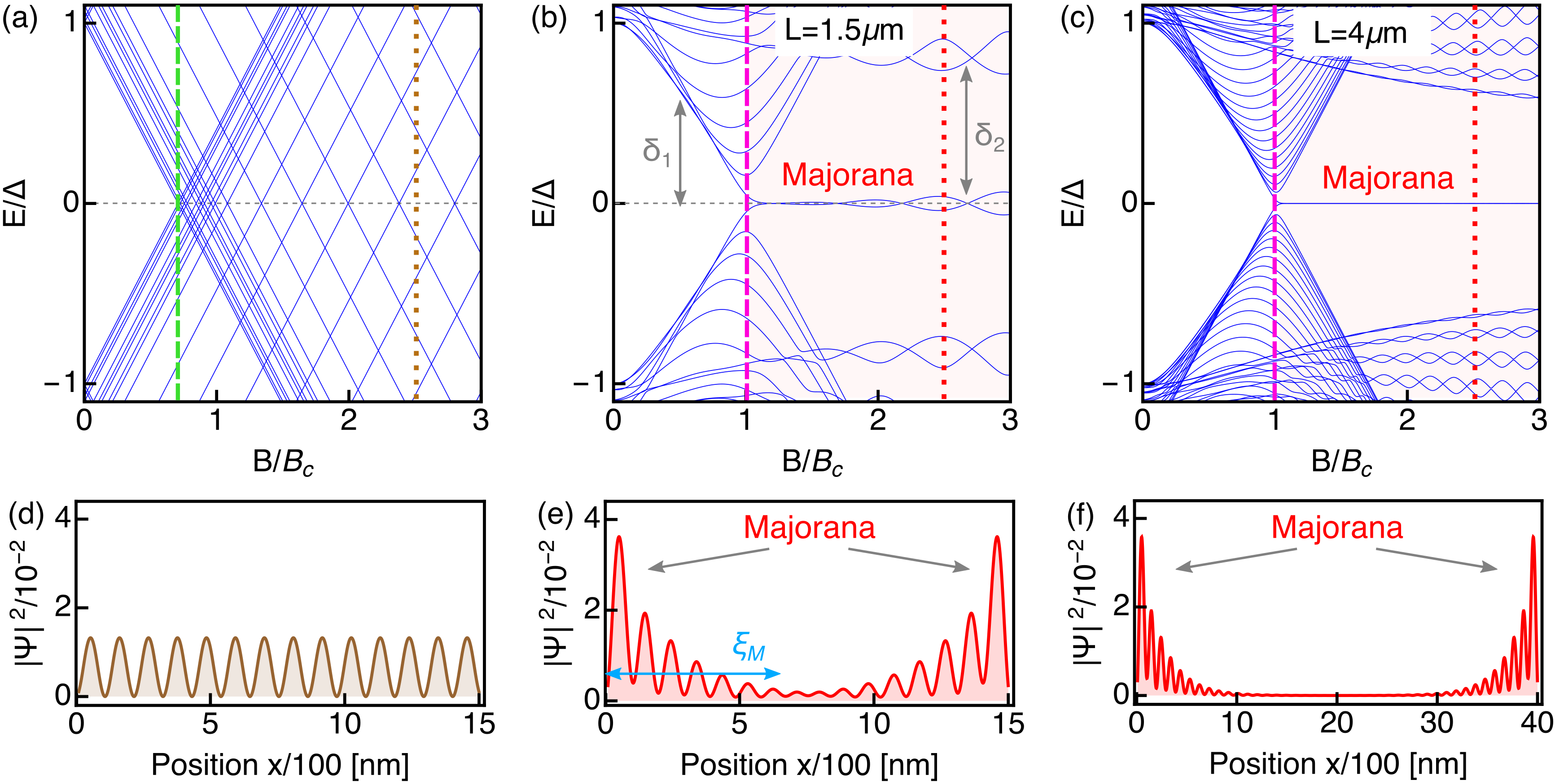} 
     \caption{Low-energy spectrum of a  nanowire with a finite length $L$ and spin-singlet $s$-wave superconductivity as a function of the Zeeman field.   (a) shows the spectrum at zero SOC at $L=1.5\mu$m, where the dashed green line mark $B=\Delta$. In (b,c) the spectrum at finite SOC is shown for two values of the length of the wire, where a magenta dashed line marks the topological phase transition $B=B_{c}$ into a topological phase with Majorana quasiparticles. Here, $\delta_{1(2)}$ denote the lowest gaps in the trivial (topological phases). In (d-f) the wavefunctions of the lowest state at high Zeeman fields in (d-f) is shown as a function of space. The values of the Zeeman fields is indicated by dotted vertical lines in    (a-c). Parameters: $\Delta=0.5$meV, $\mu=0.5$\,meV, $\alpha_{R}=40$meVnm.
}
   \label{RashbaSCNW}
\end{figure}

\subsubsection{Lattice model}
Now we inspect a finite superconducting nanowire with SOC, Zeeman field, and spin-singlet $s$-wave superconductivity. We thus discretize the Hamiltonian given by Eq.\,(\ref{fullHamiltonian}) into a spinful 1D tight-binding chain which is given by
\begin{equation}
\label{modelTBNW}
\begin{split}
H_\textrm{S}&=(2t-\mu)\sum_{\sigma n}c_{\sigma n}^\dagger c^{\phantom{\dagger}}_{\sigma n} 
-\mathop{\sum_{\langle n,n'\rangle}}_{\sigma}t\, c_{\sigma n'}^\dagger c^{\phantom{\dagger}}_{\sigma n} 
-i\mathop{\sum_{\langle n,n'\rangle}}_{\sigma,\sigma'}t^{\mathrm{SOC}\phantom{\dagger}}_{n'-n}c_{\sigma' n'}^\dagger \sigma^{y}_{\sigma'\sigma}c^{\phantom{\dagger}}_{\sigma n}\\
&+\sum_{\sigma,\sigma' n}B\,c_{\sigma' n}^\dagger \sigma^{x}_{\sigma'\sigma}c^{\phantom{\dagger}}_{\sigma n}+
\sum_{\sigma n}\Delta\, c_{\sigma n}^\dagger c^\dagger_{\bar\sigma n}+\mathrm{H.c}\,,
\end{split}
\end{equation} 
where $c_{\sigma n}$ annihilates a fermionic state with spin $\sigma$ at site $n$, $\mu$ is the chemical potential,  $t=\hbar^2/(2m a^2)$ is the hopping,   $m$ is the effective mass,   and $a$ is the lattice spacing. In Eq.\,(\ref{modelTBNW}), $\langle n,n'\rangle$ indicates hopping between nearest neighbor sites.  Moreover,   $\Delta$ represents the   induced spin-singlet $s$-wave pair potential, $t^\mathrm{SOC}_{\pm 1}=\pm \alpha_\mathrm{R}/(2a)$ is the the SOC hopping, where $\alpha_\mathrm{R}$ is the SOC strength that defines a SOC length as $\lambda_\mathrm{SOC}=\hbar^2/(m \alpha_\mathrm{R})$,  $B=g\mu_B \mathcal{B}/2$ is the Zeeman field due to an external magnetic field $B$, where $g$ is the g-factor, and $\mathcal{B}$ is the magnetic field along the wire. The length of the nanowire is given by $L=Na$, where $N$ is the number of lattice sites. 

Here we are interested in exploring the impact of the Zeeman field on the spectrum because we saw before that it induces a topological phase transition at $B=B_{c}$.  We thus perform a diagonalization of Eq.\,(\ref{modelTBNW}) and plot its spectrum as a function of the Zeeman field in Fig.\,\ref{RashbaSCNW}(a-c). In the absence of any SOC, a finite pair potential induces a gap in the low-energy spectrum at $B=0$, which does not host energy levels in agreement with the Anderson theorem \cite{Anderson-theorem}. In fact, induced superconductivity in the absence of SOC and Zeeman remains with spin-singlet symmetry as no mechanism is present to mix spin states. At finite Zeeman field but without SOC, the energy levels split in spin, where the up and down levels shift by $\pm B/2$, which then produces a reduction of the size of the induced superconducting gap. Thus,  different spin components cross zero energy at $B=\Delta$, signalling the closing of the induced superconducting gap, marked by the magenta dashed line. The closing of the superconducting gap can be understood to be a result of   Zeeman depairing  where spin-singlet Cooper pairs are destroyed. As $B$ acquires values larger than $\Delta$, there is a region of Zeeman fields roughly given by $\Delta<B<B_{c}$ where energy levels possess both spin components which, however, depends on having a finite $\mu$. For strong $B$, such as for $B>B_{c}$, the energy levels become spin-polarized and develop crossings at zero and finite energies.
 
 In the presence of SOC, the spectrum exhibits interesting differences, as we can observe in Fig.\,\ref{RashbaSCNW}(b,c). The first feature is that the gap closing now occurs at $B=B_{c}$, where $B_{c}$ is the critical field defined by Eq.\,(\ref{Bc}) which marks the topological phase transition separating the trivial and topological phases of the Rashba superconductor nanowire. This gap closing is more visible for longer wires, as expected the condition $B=B_{c}$ in Eq.\,(\ref{Bc}) corresponds to a bulk system.  Second, for $B>B_{c}$   the gap reopens and    
 the finite SOC moves all the Zeeman crossings to higher energies but, remarkably, leaves 
 only two energy levels that oscillate around zero energy. These zero-energy oscillatory level  are separated by an energy gap $\delta_{2}$ from the quasicontinuum, which is strongly dependent on the size of SOC. As the wire is made longer, the amplitude of the oscillations is reduced such that these levels acquire zero energy. For $L\rightarrow\infty$, $\delta_{1(2)}$ take the values of $\Delta_{1(2)}$ defined by Eqs.\,(\ref{D1D2}).
 The strong dependence on the length of the wire already suggests an interesting property related to their spatial behavior. Further insights on the spatial dependence of these intriguing zero-energy states is obtained from their wavefunctions $\Psi=(u_{\uparrow},u_{\downarrow},v_{\uparrow},v_{\downarrow})$, which is plotted in Fig.\,\ref{RashbaSCNW}(d-f) for a fixed strong  Zeeman field. While the energy level closest to zero energy without SOC exhibits homogeneous oscillations in space just as a particle in a box, the wavefunctions of the zero energy states at finite SOC develop an interesting profile. Their wavefunctions are localized at the edges and decay towards the center of the system in an exponentially oscillatory fashion, with a decay length $\xi_{\rm M}$ that is shorter for longer wires \cite{PhysRevB.86.180503,PhysRevB.86.220506,PhysRevB.87.024515,PhysRevB.96.205425}. It can be also verified that the charge density $|u_{\sigma}|^{2}-|v_{\sigma}|^{2}$ exhibits homogeneous oscillations for short wires and vanishes for long wires \cite{PhysRevB.91.045403}. These zero-energy energy levels define the emergence of  MZMs or  MBSs.
 
 \subsection{Summary}
 We have shown that 1D spin-polarized $p$-wave superconductors exhibit a topological phase where MZMs emerge as \emph{zero-energy}   edge states. Moreover, we have pointed out that MZMs also exhibit \emph{spatial nonlocality} due to their wavefunction localization at the edges of the wire. Finally, we have shown that 1D spin-polarized $p$-wave superconductivity can be realized by combining conventional spin-singlet $s$-wave superconductivity, SOC, and a Zeeman field perpendicular to the SOC axis.
 
\section{Green's functions of $p$-wave superconductors with MZMs}
\label{section4}
Having discussed the emergence of MZMs in 1D $p$-wave superconductors, here we would like to explore how they manifest on the local density of states (LDOS) and what is the nature of their superconducting pair correlations or simply pair amplitudes. To address these two questions, we calculate the Green's functions of 1D  $p$-wave superconductors, which permit us to directly obtain the LDOS and pair amplitudes. For completeness, here we first focus on the continuum model of 1D  $p$-wave superconductors and inspect the impact of having semi-infinite and finite length systems. Furthermore, for pedagogical purposes, we start by discussing how to obtain Green's functions by using scattering states.

\subsection{Brief discussion on obtaining Green's functions from scattering states}
\label{Subsection4a}
To obtain Green's functions, in this review, we follow the scattering formalism initially formulated by McMillan \cite{McMillan1968}. Within this framework, the retarded Green's functions can be simply obtained by considering a scattering region and appropriately combining incoming and outgoing waves as,
\begin{equation}
\label{McMillan}
G^{r}(x,x',E)=\begin{cases}
\alpha_{1}\Psi^{\left(+\right)}_{out}\left(x\right)\tilde{\Psi}^{t\left(+\right)}_{in}\left(x' \right) 
+ \alpha_{2}\Psi^{\left(+\right)}_{out}\left(x\right)\tilde{\Psi}^{t\left(-\right)}_{in}\left(x'\right) \\
\ + \alpha_{3}\Psi^{\left(-\right)}_{out}\left(x\right)\tilde{\Psi}^{t\left(+\right)}_{in}\left(x'\right) 
+ \alpha_{4}\Psi^{\left(-\right)}_{out}\left(x\right)\tilde{\Psi}^{t\left(-\right)}_{in}\left(x'\right)\,, 
\quad x'< x 
 \\
\beta_{1}\Psi^{\left(+\right)}_{in}\left( x \right)\tilde{\Psi}^{t\left(+\right)}_{out}\left( x' \right) 
+ \beta_{2}\Psi^{\left(-\right)}_{in}\left( x \right)\tilde{\Psi}^{t\left(+\right)}_{out}\left( x' \right) \\
\ + \beta_{3}\Psi^{\left(+\right)}_{in}\left( x \right)\tilde{\Psi}^{t\left(-\right)}_{out}\left( x' \right) 
+ \beta_{4}\Psi^{\left(-\right)}_{in}\left( x \right)\tilde{\Psi}^{t\left(-\right)}_{out}\left( x' \right)\,, 
 \quad
x<x'\,.
\end{cases}
\end{equation}
where the subscript $t$ denotes the transpose operation and  $\Psi^{(j)}_{out(in)}$  represents 
outgoing (incoming) waves for electron- and hole-like quasiparticles;
here the label  $j=+$ and $j=-$ denote the electron- and hole-like characters of the wavefunction. In Eqs.\,(\ref{McMillan}),   $\Psi^{(j)}_{m}$ and $\tilde{\Psi}^{(j)}_{m}$ represent solutions to the BdG equations defined by
\begin{equation}
\label{HamitonianPsi}
\begin{split}
\int \hat{H}\left(x,x' \right) \Psi\left(x' \right) dx' &= E 
\Psi\left(x \right)\,,\\
\int \hat{H}^{t}\left(x',x \right)\tilde{\Psi}\left(x' \right) dx' 
&= E \tilde{\Psi}\left(x \right)\,,
\end{split}
\end{equation}
where $\hat{H}(x,x')$ and $\hat{H}^{t}(x',x)$ are, respectively, BdG and transposed BdG Hamiltonians whose form depends on the respective problem at hand. 

To fully characterize $G^{r}(x,x',E)$, it is necessary to find the coefficients $\alpha_{i}$ and $\beta_{i}$, which is done by employing the equation of motion \cite{mahan2013many,zagoskin}
\begin{equation}
\label{EQOM}
[E-\hat{H}(x,x')]G^{r}(x,x',E)=\delta(x-x')\,.
\end{equation}
By integrating once (twice) around $x=x'$, this equation provides the necessary boundary conditions to find $\alpha_{i}$ and $\beta_{i}$. Once the retarded Green's function is fully characterized, we are in a position to obtain the LDOS and pair amplitudes. It is important to point out that in Nambu space the Green's function is a $2\times2$ matrix which can be written as
\begin{equation}
\label{GF}
G^{r}(x,x',E)=
\begin{pmatrix}
G_{0}^{r}(x,x',E)&F^{r}(x,x',E)\\
\bar{F}^{r}(x,x',E)&\bar{G}_{0}^{r}(x,x',E)
\end{pmatrix}\,,
\end{equation}
where the diagonal represents the normal electron-electron and hole-hole Green's functions, while the off-diagonal components correspond to the anomalous electron-hole and hole-electron Green's functions.  It is thus evident that the diagonal and off-diagonal parts of  Eq.\,(\ref{GF}) enable the calculation of the LDOS and superconducting pair amplitudes, respectively. In particular, the LDOS can be calculated as  
\begin{equation}
\label{LDOS}
\rho(E,x)=-\frac{1}{\pi}{\rm ImTr}[G_{0}(x,x,E)]\,,
\end{equation}
where the ${\rm Tr}$ is taken over the other degrees of freedom such as spin.

In relation to the pair amplitudes, if spin is an active degree of freedom, then $F^{r}(x,x',E)$ is a matrix and needs to be decomposed in order to know which type of superconducting pair symmetry is favored. Thus, we can define the spin components of the pair amplitudes as $f^{r}_{\sigma\sigma'}(x,x',E)=[F^{r}(x,x',E)]_{\sigma\sigma'}$, and, taking into account the Fermi-Dirac statistics, we find    the following condition
\begin{equation}
\label{antisymmetry}
f^{r}_{\sigma\sigma'}(x,x',E)=-f^{a}_{\sigma'\sigma}(x',x,-E)\,,
\end{equation}  
where we have used the advanced Green's function in passing to negative energies (or frequencies) $G^{a}(x,x',E)=[G^{r}(x',x,E)^{\dagger}]$. The condition given by Eq.\,(\ref{antisymmetry}) is extremely relevant because it determines all the possible emergent pair symmetries in the presence of the quantum numbers determined by spins, spatial coordinates and finite energy. Therefore, an allowed pair symmetry must be antisymmetric under the exchange of all the involved quantum numbers.

To see the possible pair symmetries in a system with active spin,  considering   Nambu basis   given by $(c_{\uparrow},c_{\downarrow},c^{\dagger}_{\uparrow},c_{\downarrow}^{\dagger})$, it is useful to write $F^{r}(x,x',E)$ as
\begin{equation}
F^{r}(x,x',E)=[f_{0}^{r}(x,x',E)\sigma_{0}+f_{j}^{r}(x,x',E)\sigma_{j}]i\sigma_{y}
\end{equation}
where $\sigma_{j}$ is the $j$-th Pauli matrix in spin space and repeated indices in the second term imply summation. Thus, $f_{0}^{r}$ and $f_{3}^{r}$ correspond to spin-singlet and mixed spin-triplet pairs with spin projection $S_{z}=0$ and written  as $f_{0(3)}^{r}=(f_{\uparrow\downarrow}^{r} \mp f_{\downarrow\uparrow}^{r})/2$. Similarly, $f_{1}^{r}=(f_{\downarrow\downarrow}^{r}- f_{\uparrow\uparrow}^{r})/2$ and $f_{2}^{r}=(f_{\downarrow\downarrow}^{r}+f_{\uparrow\uparrow}^{r})/2i$ give rise to $if_{2}^{r}\mp f_{1}^{r}$ which correspond to equal spin-triplet pairs with spin-projection $S_{z}=\pm1$. We thus see that pair amplitudes can have spin-singlet and spin-triplet symmetries. Moreover, to comply with the antisymmetry condition imposed by Eq.\,(\ref{antisymmetry}), each of the amplitudes $f_{i}^{r}$ can still be even or odd under the exchange of spatial coordinates and/or energy as long as the total exchange of quantum numbers makes $f_{i}^{r}$ antisymmetric. Thus, spin-singlet pairs are allowed to have the following symmetry: even-frequency, spin-singlet, even-parity (ESE) or odd-frequency, spin-singlet odd-parity (OSO) \cite{Balatsky,Balatsky2,Coleman} both pair symmetries consistent with the antisymmetry condition. For the spin-triplet pair amplitudes, they are allowed to have the following symmetries:   even-frequency, spin-triplet, odd-parity (ETO) or odd-frequency, spin-triplet even-parity (OTE) \cite{Berezinskii}. Therefore, when the pair amplitude depends on the frequency, spins, and spatial coordinates of the paired electrons,  there are four pair symmetries allowed by Fermi-Dirac statistics, see Refs.\,\cite{odd3,odd3b,LinderBalatsky,Cayao_2020,Tanaka2021}. We stress that, for spin-polarized $p$-wave superconductors, the equal-spin triplet component is relevant, implying the appearance of either ETO or OTE pair symmetries, which is the focus on next subsections.

\subsection{Bulk $p$-wave superconductor}
\label{Subsection4b}
Here we use the recipe discussed in the previous subsection and calculate the Green's function for a spin-polarized $p$-wave superconductor in the bulk.  Before going further, it is important to write down  the spatial dependence of the Hamiltonian  for a 1D spin-polarized $p$-wave superconductor, which, in general, is given by
\begin{equation}
\label{Hamiltonianmatrix}
\hat{H}\left(x,x'\right)=
\begin{pmatrix}
\delta\left(x-x' \right) 
\left(-\frac{\hbar^{2}}{2m} \frac{d^{2}}{dx'^{2}}-\mu \right)
& 
\Delta\left(x,x' \right)  \\
-\Delta^{*}\left(x,x' \right) & 
-\delta\left(x-x' \right)
\left(-\frac{\hbar^{2}}{2m} \frac{d^{2}}{dx'^{2}}-\mu \right)
\end{pmatrix}
\end{equation}

Now, we denote the electron and hole component of $\Psi(x)$ and $\tilde{\Psi}(x)$ as 
$u(x)$ and   $v(x)$, respectively. Then, by applying quasiclassical approximation  used in 
Eq.\,(\ref{quasiclassicalappro1}), we obtain 
\begin{equation}
\label{Deltabulk}
\begin{split}
\int \Delta(x,x') v(x')dx' &= \Delta(\hat{k},x)v(x)\,,, \\ 
\int \Delta^{*}(x,x') u(x') dx'&= \Delta^{*}(-\hat{k},x)u(x)\,,\\
\int \Delta(x',x) v(x') dx'    &= \Delta(-\hat{k},x)v(x), \\ 
\int \Delta^{*}(x',x) u(x') dx'&= \Delta^{*}(\hat{k},x)u(x)\,,
\end{split}
\end{equation}
where $\Delta(\hat{k},x)=\hat{k} \Delta_{0}$ is the $p$-wave pair potential which satisfies $\Delta(\hat{k},x)=-\Delta(-\hat{k},x)$, with  $\hat{k}={\rm sgn}(k)$ and  
$\Delta(k)=-\Delta(-k)$. By plugging Eqs.\,(\ref{Deltabulk}) into the BdG equations defined by Eq.\,(\ref{Hamiltonianmatrix}), and using Eqs.\,(\ref{HamitonianPsi}), we obtain the incoming and outgoing wavefunctions in the bulk. For $\mu\gg\Delta_{0}$ and $\mu\gg|E|$, they  read
\begin{equation}
\Psi^{+}_{out}\left(x \right)=\exp\left(ik^{+}x \right)
\begin{pmatrix}
1 \\
\Gamma
\end{pmatrix}, \ \ 
\Psi^{-}_{out}\left(x \right)=\exp\left(-ik^{-}x \right)
\begin{pmatrix}
-\Gamma \\
1
\end{pmatrix}, 
\end{equation}
\begin{equation}
\Psi^{+}_{in}(x)=\exp\left(-ik^{+}x \right)
\begin{pmatrix}
1 \\
-\Gamma
\end{pmatrix}, \ \ 
\Psi^{-}_{in}\left(x \right)=\exp\left(ik^{-}x \right)
\begin{pmatrix}
\Gamma \\
1
\end{pmatrix}, 
\end{equation}
\begin{equation}
\tilde{\Psi}^{+}_{out}\left(x \right)=\exp\left(ik^{+}x \right)
\begin{pmatrix}
1 \\
-\Gamma
\end{pmatrix}, \ \ 
\tilde{\Psi}^{-}_{out}\left(x \right)=\exp\left(-ik^{-}x \right)
\begin{pmatrix}
\Gamma \\
1
\end{pmatrix}, 
\end{equation}

\begin{equation}
\tilde{\Psi}^{+}_{in}\left(x \right)=\exp\left(-ik^{+}x \right)
\begin{pmatrix}
1 \\
\Gamma
\end{pmatrix}, \ \ 
\tilde{\Psi}^{-}_{in}\left(x \right)=\exp\left(ik^{-}x \right)
\begin{pmatrix}
-\Gamma \\
1
\end{pmatrix}\,,  
\end{equation}
where $\Gamma=\Delta_{0}/(E + \Omega)$, $\Omega$ is defined in Eq.\,(\ref{Omegapm}), and 
\begin{equation}
\label{momentapm}
k^{\pm}=\sqrt{\frac{2m}{\hbar^{2}}\left(\mu \pm \Omega\right)}\,. 
\end{equation}
 
Having found the incoming and outgoing wavefunctions, we can now calculate the Green's function for a bulk $p$-wave superconductor without any interface by using Eq.\,(\ref{McMillan}). Since there is no interface,  the process of both normal and Andreev reflections are absent, restricting the combination of incoming and outgoing waves in Eq.\,(\ref{McMillan}) to occur only between electron-like (and hole-like) quasiparticles, i.e., $\alpha_{2,3}=0$ and $\beta_{2,3}=0$. 
To find the rest of the coefficients ($\alpha_{1,4}$ and $\beta_{1,4}$), we use the boundary conditions defined by Eqs.\,
 (\ref{EQOM}), which here acquire the following forms
\begin{equation}
\label{boundaryConditions}
\begin{split}
\hat{G}^{R}\left(x+0,x,E\right)&=\hat{G}^{R}\left(x-0,x,E\right)\,,\\
\frac{\partial}{\partial x}\hat{G}^{R}\left(x,x',E \right)\Big|_{x=x'+0}
-\frac{\partial}{\partial x}\hat{G}^{R}\left(x,x',E \right)\Big|_{x=x'-0}&
=
\frac{2m}{\hbar^{2}}
\begin{pmatrix}
1 & 0 \\
0 & -1
\end{pmatrix}\,.
\end{split}
\end{equation}
This way we find $\alpha_{1,4}$ and $\beta_{1,4}$ and, after a simple manipulation considering $\mu\gg\Omega$ and $k^{\pm}=k_{F}\pm\gamma$, we obtain the Green's function in a bulk  spin-polarized $p$-wave superconductor given by,
\begin{equation}
\label{pwaveMcMillanbulk}
\begin{split}
\hat{G}^{R}\left(x,x',E \right)&=
\eta
\exp\left[i\gamma \mid x -x' \mid \right]
\biggl[ 
\frac{E}{\Omega}
\cos\left(k_{F} \mid x -x' \mid \right)
\hat{\tau}_{0}
+ i \sin\left(k_{F} \mid x -x' \mid \right)
\hat{\tau}_{3} \\
&+ i \frac{\Delta_{0}}{\Omega}
\sin\left[k_{F}\left(x-x' \right) \right]
\hat{\tau}_{1}
\biggr]
\end{split}
 \end{equation} 
where $\eta=m/(i \hbar^{2} k_{F})$, $\gamma=\Omega/(\hbar v_{F})$, $v_{F}=2\mu/(\hbar k_{F})$ is the Fermi velocity, $k_{F}=\sqrt{2m\mu/\hbar^{2}}$, and $\hat{\tau}_{i}$ represents the $i$-th Pauli matrix in Nambu space. It is now evident that, as discussed in Subsection \ref{Subsection4a}, the retarded Green's function obtained in Eq.\,(\ref{pwaveMcMillanbulk}) is a $2\times2$ matrix in Nambu space, with its diagonal and off-diagonal elements representing the normal and anomalous Green's functions. 
\begin{equation}
\begin{split}
G_{0}^{r}(x,x',E)&=\eta\exp\left[i\gamma \mid x -x' \mid \right]\left[\frac{E}{\Omega}
\cos\left(k_{F} \mid x -x' \mid \right)+i \sin\left(k_{F} \mid x -x' \mid \right)\right]\,,\\
F^{r}(x,x',E)&=\frac{i\eta\Delta_{0}}{\Omega} \exp\left[i\gamma \mid x -x' \mid \right] 
\sin\left[k_{F}\left(x-x' \right)\right]\,. 
\end{split}
\end{equation}
Thus, the LDOS obtained by using $G_{0}^{r}$ in Eq.\,(\ref{LDOS})  is given by 
$\rho(x,E)=-{\rm Im}[\eta E/(\pi \Omega)]$, as expected in a bulk 1D spin-polarized $p$-wave superconductor \cite{Spectralbulk}. In relation to the pair amplitude $F^{r}$, we first note that it corresponds to a spin-polarized pair potential, so its spin symmetry is triplet. Second,  $F^{r}$ is an odd function under the exchange of $x$ and $x'$, taking zero value locally in space at $x=x'$, a symmetry that reflects its $p$-wave  symmetry. Third, if we define $F^{a}(x,x',E)$ which is an advanced part of $F^{r}(x,x',E)$, $F^{r}(x,x',E)$ satisfies 
$F^{r}(x,x',-E)=F^{a}(x,x',E)$ pointing out  its even-frequency symmetry. 
In summary, the pair amplitude $F^{r}$ exhibits an even-frequency, spin-triplet, $p$-wave symmetry, an ETO pair symmetry class according to the antisymmetry condition given by Eq.\,(\ref{antisymmetry}) and consistent with the symmetry of the pair potential for spin-polarized $p$-wave superconductor considered here, see Eq.\,(\ref{Hamiltonianmatrix}).

\subsection{Semi-infinite $p$-wave superconductor}
\label{Subsection4c}
\begin{figure}[tb]
\begin{center}
\includegraphics[width=8.0cm]{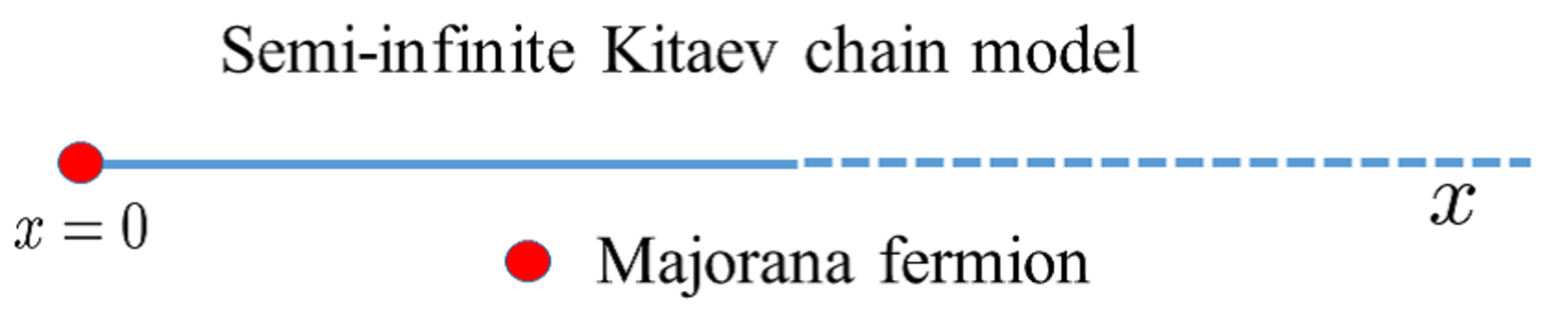}
\end{center}
\caption{Schematic representation of a semi-infinite 1D spin-polarized $p$-wave superconductor located for $x \geq 0$ and with a interface at $x=0$. In the topological phase, it hosts a single MZM at $x=0$. }
\label{Kitaevmodel1}
\end{figure}

In this part, we inspect the Green's function when having one interface in the 1D spin-polarized  $p$-wave superconductor discussed in the previous section. In particular, we consider a semi-infinite 1D spin-polarized  $p$-wave superconductor with the interface at $x=0$, as schematically shown in Fig.\,\ref{Kitaevmodel1}.

As discussed in Subsection \ref{Subsection4a}, to obtain the Green's function, we need to obtain the incoming and outgoing waves for $\hat{H}(x,x')$ and $\hat{H}^{t}(x,x')$, which in this case  are given by 
\begin{equation}
\label{EqsSemiinfinite}
\begin{split}
\Psi_{out}^{\left(1 \right)}\left(x \right)&=\exp\left(ik^{+}x \right)
\begin{pmatrix}
1 \\
\Gamma
\end{pmatrix}, \ \ 
\Psi_{out}^{\left(2\right)}\left(x \right)=\exp\left(-ik^{-}x \right)
\begin{pmatrix}
-\Gamma \\
1
\end{pmatrix}\\
\Psi_{in}^{\left(1 \right)}\left(x \right)&=
\begin{pmatrix}
\Gamma f_{1}\left(x \right) \\ i f_{2}\left(x \right)
\end{pmatrix}, \ \ 
\Psi_{in}^{(2)}\left(x \right)=
\begin{pmatrix}
i f_{3}\left(x \right) \\ \Gamma f_{1}\left(x \right)
\end{pmatrix}\\
\tilde{\Psi}_{out}^{\left(1\right)}\left(x\right)&=\exp\left(ik^{+}x\right)
\begin{pmatrix}
1 \\ -\Gamma
\end{pmatrix}, \ \ 
\tilde{\Psi}_{out}^{\left(2\right)}\left(x\right)=\exp\left(-ik^{-}x\right)
\begin{pmatrix}
\Gamma \\ 1
\end{pmatrix}\\
\tilde{\Psi}_{in}^{\left(1\right)}\left(x\right)&=
\begin{pmatrix}
-\Gamma f_{1}\left(x\right) \\ i f_{2}\left(x\right)
\end{pmatrix}, \ \ 
\tilde{\Psi}_{in}^{\left(2\right)}\left(x\right)=
\begin{pmatrix}
i f_{3}\left(x\right) \\ -\Gamma f_{1}\left(x\right)
\end{pmatrix}
\end{split}
\end{equation}
where $f_{1}\left(x \right)=\cos\left(k^{+}x \right)-\cos\left(k^{-}x \right)$,  
$f_{2}\left(x \right)=\Gamma^{2}\sin\left(k^{+}x \right)-\sin\left(k^{-}x \right)$, and $f_{3}\left(x\right)=\sin\left(k^{+}x\right)-\Gamma^{2}\sin\left(k^{-}x\right)$. Moreover, we emphasize that Eqs.\,(\ref{EqsSemiinfinite}) satisfy the boundary conditions at the interface $\Psi_{in}^{\left(1 \right)}\left(x=0 \right)=\Psi_{in}^{\left(2 \right)}\left(x=0 \right)=0$ and $\tilde{\Psi}_{in}^{\left(1\right)}\left(x=0 \right)=\tilde{\Psi}_{in}^{\left(2\right)}\left(x=0 \right)=0.$ Thus, by plugging Eqs.\,(\ref{EqsSemiinfinite}) into Eqs.\,(\ref{McMillan})
and using the boundary condition given by Eq.\,(\ref{boundaryConditions}), 
we obtain the retarded Green's function \cite{Spectralbulk}

\begin{equation}
\label{semi-infiniteKitaev}
\begin{split}
\hat{G}^{R}\left(x,x',E\right)&=\eta
\Biggl\{
\exp\left(i\gamma \mid x -x' \mid\right)\\
&\times 
\Bigg[
\frac{E}{\Omega}
\cos(k_{F}|x -x' |)
\hat{\tau}_{0}
 + i \sin(k_{F}|x -x'|)
\hat{\tau}_{3}
+ i \frac{\Delta_{0}}{\Omega}
\sin[k_{F}(x-x')]
\hat{\tau}_{1}
\Bigg] \\
&- \exp[i\gamma(x+x')]\cos[k_{F}(x+x')] \frac{E}{\Omega}
\hat{\tau}_{0}\\
&- i \exp[i\gamma(x+x')]\sin[k_{F}(x+x')] 
\hat{\tau}_{3} \\
&- i \exp[i\gamma(x+x')]\sin[k_{F}(x-x')] \frac{\Delta_{0}}{\Omega}
\hat{\tau}_{1} \\
&-2\exp[i\gamma(x+x')]\sin(k_{F}x)\sin(k_{F}x')\frac{\Delta_{0}}{E}
\begin{pmatrix}
\frac{\Delta_{0}}{\Omega} & 1 \\
-1 & \frac{\Delta_{0}}{\Omega} 
\end{pmatrix}
\Biggr\}.
\end{split}
\end{equation}
Despite its apparent complexity, this Green's function reveals interesting features due to the boundary effect. The first feature to note is that $\hat{G}^{R}$ here has two clear parts, one involving properties of the bulk and another involving properties of the boundary. In fact, the element in square brackets in the second line of Eq.\,(\ref{semi-infiniteKitaev}), and multiplied by the exponential coefficient, is exactly the bulk Green's function found in the previous subsection and given by Eq.\,(\ref{pwaveMcMillanbulk}). Thus, the rest of elements can be clearly interpreted to be an effect coming entirely from the boundary.

To further inspect the impact of the boundary, let's write down the normal and anomalous components of Eq.\,(\ref{semi-infiniteKitaev}), which read
\begin{equation}
\label{GFSEMIINFINITE}
\begin{split}
G_{0}^{r}(x,x',E)&=\eta\exp[i\gamma|x -x'|][\frac{E}{\Omega}
\cos(k_{F}|x -x'|)+i \sin(k_{F}|x -x'|)]\\
&- \frac{\eta E}{\Omega}\exp[i\gamma(x+x')]\cos[k_{F}(x+x')] 
-i\eta \exp[i\gamma(x+x')]\sin[k_{F}(x+x')]\\  
&- \frac{2\eta\Delta_{0}^{2}}{E\Omega}\exp[i\gamma(x+x')]\sin(k_{F}x)\sin(k_{F}x')\,,\\
F^{r}(x,x',E)&=\frac{i\eta\Delta_{0}}{\Omega} \exp[i\gamma|x -x'|] 
\sin[k_{F}(x-x' )]\\
&-\frac{i\eta\Delta_{0}}{\Omega} \exp[i\gamma(x+x')]\sin[k_{F}(x-x')]- \frac{2\eta\Delta_{0}}{E}\exp[i\gamma(x+x')]\sin(k_{F}x)\sin(k_{F}x')\,,
\end{split}
\end{equation}
where the first line of each expression for $G_{0}^{r}$ and $F^{r}$ represents the contributions coming from the bulk, while the remaining lines show the contribution due to the boundary. It is now useful to inspect the LDOS, which, using $G_{0}^{r}$ in Eq.\,(\ref{LDOS}), reads,
\begin{equation}
\label{LDOSKitaev}
\rho(x,E)=-\frac{1}{\pi}
{\rm Im}
\left\{
\eta \left[
\frac{E}{\Omega} 
-\exp(2i\gamma x)
[\frac{E}{\Omega}\cos(2k_{F}x) + i \sin(2k_{F}x) + \frac{2 \Delta_{0}^{2}}{\Omega E}\sin^{2}(k_{F}x)] 
\right]
\right\}.
\end{equation}  
This LDOS expression shows that, when a boundary is present as we have here, the total LDOS contains a bulk contribution (first term) and also a component that depends on space (second third, and fourth terms in square brackets). For subgap energies, $\Omega$ is imaginary as seen in Eqs.\,(\ref{Omegapm}), which implies that the exponential term in front of the square brackets exhibits an exponential and oscillatory decay for $x>0$ that, interestingly, strongly depends on $E$.  Of particular relevance is the behavior close to $E=0$, which can clearly induce a divergent profile 
around the edge component of the LDOS, see last term in Eq.\,(\ref{LDOSKitaev}). This property is a manifestation of the emergence of a MZM at the boundary. We note that the LDOS for a semi-infinite spin-singlet $s$-wave superconductor is given by \cite{Spectralbulk}  
\[
\rho_{s}=-\frac{1}{\pi}
{\rm Im} 
\left\{
\eta
\left[
\frac{E}{\Omega} - \exp(2i\gamma x)
\left[\frac{E}{\Omega}
\cos\left(2k_{F}x \right)
+i \sin \left(2k_{F}x \right)
\right]
\right]
\right\},
\]
which, although partially similar to Eq.\,(\ref{LDOSKitaev}),   does not contain a contribution that produces a zero energy peak (ZEP)
 at $E=0$.  Therefore, a ZEP in the LDOS can be seen to be a manifestation of MZMs at the edge of spin-polarized $p$-wave superconductors. 

When it comes to the pair amplitude $F^{r}$ in Eq.\,(\ref{GFSEMIINFINITE}), we also note two contributions arising from the bulk (first line) and boundary (second line). In subsection \ref{Subsection4a}, we identified that the symmetry of the bulk pair amplitude is ETO and therefore vanishes at $x=x'$. Interestingly, in Eqs.\,(\ref{GFSEMIINFINITE}) we see that $F^{r}$ retains contributions even when $x=x'$, which are given by
\begin{equation}
\label{OTEFsemiinfinite}
F^{r}(x,x,E)=- \frac{2\eta\Delta_{0}}{E}\exp\left(2ix\gamma\right)\sin^{2}\left(k_{F}x\right)\,,
\end{equation}
an expression that has strong similarities with the last term in the LDOS of Eq.\,(\ref{LDOSKitaev}).
This pair amplitude is local in space,  thus having  even parity ($s$-wave), and decays in an exponentially and oscillatory fashion as $x$ increases. Moreover,   it is spin-polarized, so it has spin-triplet symmetry. When it comes to the energy dependence (or frequency), it is evident that it is an odd function, with a divergent profile at $E=0$, revealing its odd-frequency symmetry; as imposed by Eq.\,(\ref{antisymmetry}), when $E\leftrightarrow-E$ we compare with the advanced counterpart of $F^{r}(x,x,E)$ in Eq.\,(\ref{OTEFsemiinfinite}). Thus,   the pair amplitude in Eq.\,(\ref{OTEFsemiinfinite}) has OTE symmetry, which is in agreement with the antisymmetry condition given by Eq.\,(\ref{antisymmetry}) discussed in Subsection \ref{Subsection4a}. For completeness, we also write   the pair amplitude in Eq.\,(\ref{OTEFsemiinfinite})  in Matsubara frequencies by performing  analytic continuation  as $E\rightarrow i\omega_{n}$ where $\omega_{n}=2\pi \kappa_{B} T(n + 1/2)$, which reads,
\begin{equation}
\label{OTEFsemiinfinitew}
F^{r}(x,x,\omega_{n})=- \frac{2\eta\Delta_{0}}{i\omega_{n}}\exp\left(-2x\gamma_{n}\right)\sin^{2}\left(k_{F}x\right)\,. 
\end{equation}
where $\gamma_{n}=\sqrt{\omega_{n}^{2} + \Delta_{0}^{2}}/\hbar v_{F}$. We thus see that Eq.\,(\ref{OTEFsemiinfinitew}) is an odd function under $\omega_{n}\leftrightarrow-\omega_{n}$.  The  OTE symmetry of $F^{r}(x,x,E)$, with intriguing dependences, reveals the nature of the emergent pair correlations in the presence of MZMs.   

\subsubsection{Quasiclassical Green's functions}
Further understanding of the relationship between  MZMs and OTE pair correlations is obtained by analyzing the quasiclassical Green's function \cite{Quasi}. For this purpose, it is useful to write the retarded Green's function found in Eq.\,(\ref{semi-infiniteKitaev}) as 
\begin{equation}
\begin{split}
\hat{G}^{R}\left(x,x',E \right)
&=
\hat{\mathcal{G}}^{++}\left(x,x'\right)\exp\left[ik_{F}\left(x -x' \right) \right]
+ \hat{\mathcal{G}}^{--}\left(x,x'\right)
\exp\left[-ik_{F}\left(x -x' \right) \right]\\
&+\hat{\mathcal{G}}^{+-}\left(x,x'\right)\exp\left[ik_{F}\left(x + x' \right) \right]
+\hat{\mathcal{G}}^{-+}\left(x,x'\right)\exp\left[-ik_{F}\left(x + x' \right) \right]. 
\end{split}
\end{equation}
where $\hat{\mathcal{G}}^{\alpha\beta}(x,x')
\equiv \hat{\mathcal{G}}^{\alpha\beta}(x,x',E)$, with $\alpha=\pm, \beta=\pm$, expresses the envelope of the Gor'kov Green's function given by
\begin{equation}
\label{semi-infiniteKitaev++--}
\begin{split}
\hat{\mathcal{G}}^{++}\left(x+0,x,E\right)&=\left(\frac{m}{2i\hbar v_{F}}\right)
\Biggl\{
\frac{E}{\Omega}
\hat{\tau}_{0}
+ 
\hat{\tau}_{3}
+  
\frac{\Delta_{0}}{\Omega}
\left[
1 - \exp\left(2i\gamma x \right)
\right]
\hat{\tau}_{1}
-
\frac{\Delta_{0}}{E}
\exp\left( 2i\gamma x \right)
\left[
\frac{\Delta_{0}}{\Omega}
\hat{\tau}_{0}
+ 
i\hat{\tau}_{2}
\right]
\Biggr\}\,,\\
\hat{\mathcal{G}}^{--}\left(x+0,x,E\right)&=\left(\frac{m}{2i\hbar v_{F}}\right)
\Biggl\{
\frac{E}{\Omega}
\hat{\tau}_{0}
- 
\hat{\tau}_{3}
-  
\frac{\Delta_{0}}{\Omega}
\left[
1 - \exp\left(2i\gamma x \right)
\right]
\hat{\tau}_{1}-
\frac{\Delta_{0}}{E}
\exp\left( 2i\gamma x \right)
\left[
\frac{\Delta_{0}}{\Omega}
\hat{\tau}_{0}
+ 
i\hat{\tau}_{2}
\right]
\Biggr\}
\end{split}
\end{equation}
where   $\hat{\tau}_{i}$ is the $i$-th Pauli matrix in Nambu space. 

Now, we use $\hat{\mathcal{G}}^{++}(x \pm 0,x,E)$ and $\hat{\mathcal{G}}^{--}(x \pm 0,x,E)$ and define a new Green's function, also referred to as quasiclassical Green's function \cite{Nagai1989}, as \begin{eqnarray}
i \hat{g}^{\alpha \alpha }\left(x, E \right) 
&\equiv& 
-i\left(\hat{\gamma}_{3}\right)^{\alpha \alpha } 
-2v_{F}\hat{\tau}_{3}
\hat{\mathcal{G}}^{\alpha \alpha }\left(x + 0,x, E \right)
\nonumber
\\
&=&
i\left(\hat{\gamma}_{3}\right)^{\alpha \alpha } 
-2v_{F} \hat{\tau}_{3}
\hat{\mathcal{G}}^{\alpha \alpha }\left(x-0,x, E \right)
\label{quasiclassicalNagai}
\end{eqnarray}
where $\hat{\gamma}_{3}^{\alpha\alpha}$ is obtained from 
$\hat{\gamma}_{3}^{++}=\hat{\tau}_{0}$ and $\hat{\gamma}_{3}^{--}=-\hat{\tau}_{0}$. Then, we obtain expressions for $\hat{g}^{++}(x, E)$ and $\hat{g}^{--}(x, E)$, given by
\begin{equation}
\label{Eilenberger}
\begin{split}
\hat{g}^{++}\left(x, E \right)&
= g_{+}\left(x,E\right)  \hat{\tau}_{3}
+ f_{1+}\left( x, E \right) \hat{\tau}_{1}
+ f_{2+}\left( x, E \right) \hat{\tau}_{2}\,,\\ 
\hat{g}^{--}\left(x, E \right)&
= g_{-}\left(x,E\right)  \hat{\tau}_{3}
+ f_{1-}\left( x, E \right) \hat{\tau}_{1}
+ f_{2-}\left( x, E \right) \hat{\tau}_{2}, 
\end{split}
\end{equation}
where 
\begin{equation}
\label{Eilenbergergx}
\begin{split}
g_{\pm}\left(x \right)
=g_{\pm}\left(x,E\right)
&=\left[ \frac{E}{\Omega} - 
\frac{\Delta_{0}^{2}}{E \Omega}
\exp\left(2i\gamma x \right)
\right]\,,\\
f_{1\pm}\left(x \right)
= f_{1\pm}\left(x,E\right)
&= - \frac{\Delta_{0}}{E}
\exp\left(2i\gamma x \right)\,,\\
f_{2\pm}\left( x \right)
=f_{2\pm}\left(x,E\right)
&=\pm i
\left[ 1 - 
\exp\left(2i\gamma x \right)
\right] 
\frac{\Delta_{0}}{\Omega}\,.
\end{split}
\end{equation}
Before further showing the properties of Eq.\,(\ref{Eilenbergergx}), we note that the quasiclassical Green's functions $\hat{g}^{\alpha\alpha}$ satisfy the normalization condition given by
\begin{equation}
\begin{split}
\left[\hat{g}^{++}\left(x,E \right) \right]^{2}&=\hat{\tau}_{0}\,,\\ 
\left[\hat{g}^{--} \left(x, E \right)\right]^{2}&=\hat{\tau}_{0}\,,
\end{split}
\end{equation}
which is specific to quasiclassical Green's functions and can be proved by using Eqs.\,(\ref{Eilenberger}) and (\ref{Eilenbergergx}), see Ref.\,\cite{Nagai1989}. Moreover, we note that $\hat{g}^{++}(x,E)$ and $\hat{g}^{--}(x,E)$ 
satisfy the following differential equation given by 
\begin{equation}
i v_{F} \partial _{x}
\hat{g}^{\alpha \alpha }\left(x, E \right) 
= -\alpha \left[ E\hat{\tau}_{3}
+\hat{\Delta}_{tr}, \hat{g}^{\alpha \alpha }
\left(x, E \right) \right]
\label{Eilenberger1d}
\end{equation}%
where $\hat{\Delta}_{tr}=i\Delta_{\alpha}\hat{\tau}_{2}$, with $\alpha=\pm$, $\Delta_{\pm}=\pm\Delta_{0}$. Eq.\,(\ref{Eilenberger1d}) is also known as Eilenberger equation, see Ref.\,\cite{Kopnin} for more details. 

\begin{figure}[tb]
\begin{center}
\includegraphics[width=15cm,clip]{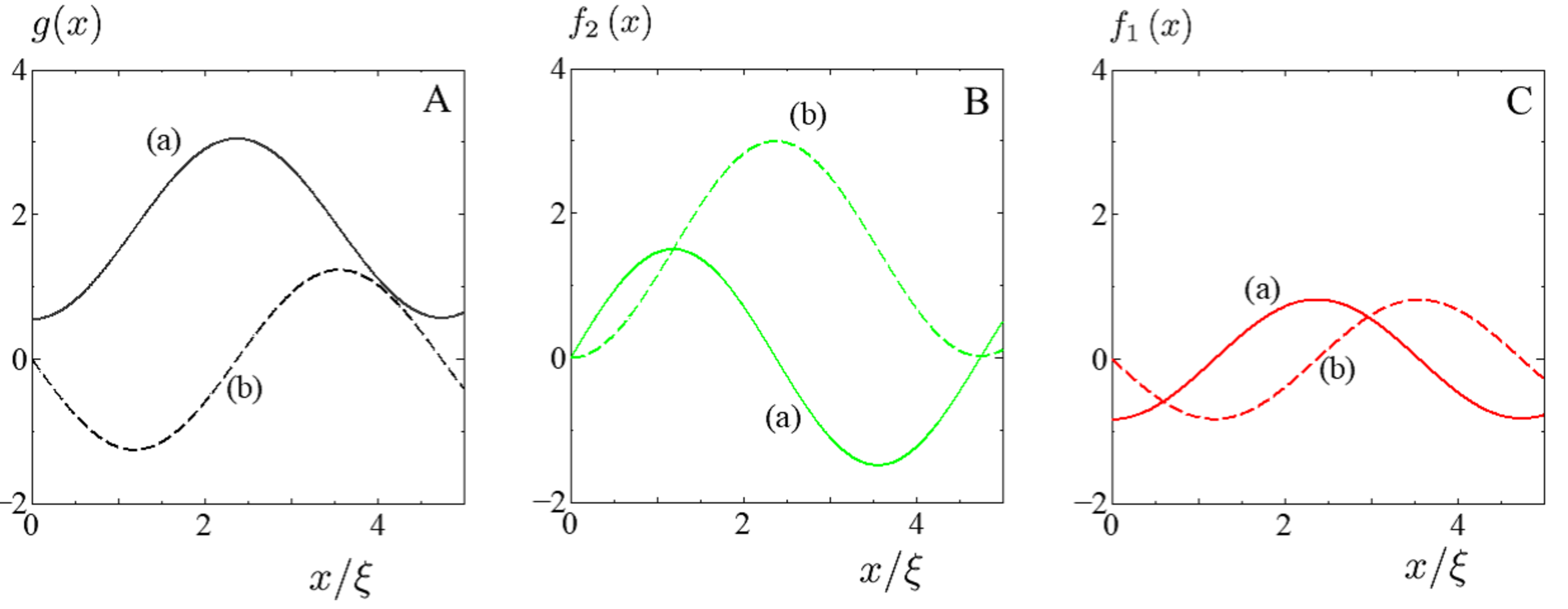}
\end{center}
\caption{
Spatial dependence of quasiclassical Green's functions at $E/\Delta_{0}=1.2$, where different panels correspond to  A: $g(x)$, B: $f_{2}(x)$, and C: $f_{1}(x)$. In each panel, the solid (a) and dashed (b) curves show the Re and Im parts, respectively. Parameters:   $\delta=0.001\Delta_{0}$,  $\xi=\hbar v_{F}/\Delta_{0}$. } 
\label{quasiclassical3}
\end{figure}

Now, we are in the position to discuss the properties of Eqs.\,(\ref{Eilenbergergx}) can be seen as the quasiclassical counterparts of the expressions found in Eqs.\,(\ref{GFSEMIINFINITE}) but with a small variation for the normal component. In fact, while $g_{\pm}(x)$  expresses the  normal Green's function with quasiparticle excitation, since $\hat{g}^{\alpha\alpha}(x,E)$ is defined multiplied by $i$ in Eq.\,(\ref{quasiclassicalNagai}),    the real part of $g_{\pm}(x)$  already represents the normalized LDOS, which by construction is normalized by its value in the normal state. Furthermore,  $f_{1\pm}(x)$ and $f_{2\pm}(x)$ represent the pair amplitudes which, as can be seen in Eqs.\,(\ref{Eilenbergergx}), correspond to odd-  and even-frequency symmetries, respectively \cite{odd3,odd3b}. 
As a note, we mention that, within the quasiclassical approach, the spatial coordinate $x$ is now the center of mass coordinate. Another feature to note that that $g_{\pm}(x)$ and $f_{2\pm}(x)$  exhibit contributions from bulk and boundary: the bulk term is independent of space, while the boundary term depends on space, in a similar manner as obtained for Eqs.\,(\ref{GFSEMIINFINITE}).
 
\begin{figure}[tb]
\begin{center}
\includegraphics[width=15cm,clip]{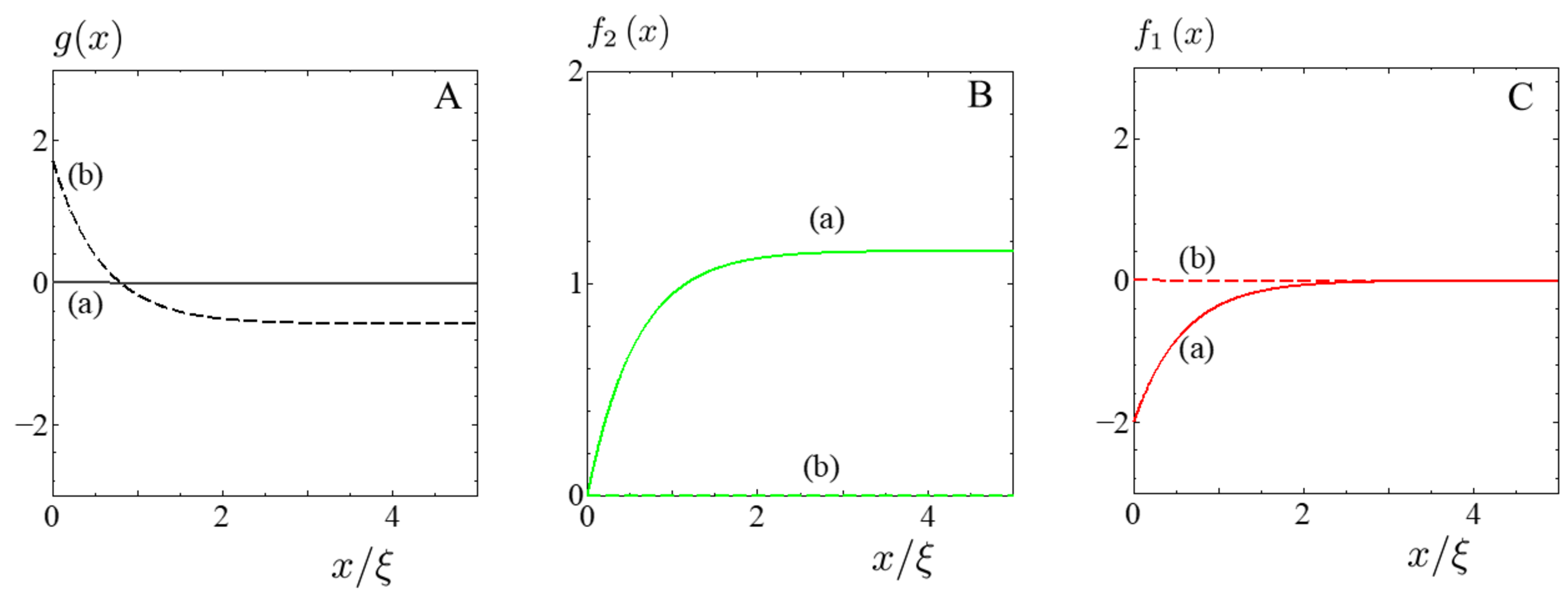}
\end{center}
\caption{Spatial dependence of quasiclassical Green's functions at $E/\Delta_{0}=0.5$, where different panels correspond to  A: $g(x)$, B: $f_{2}(x)$, and C: $f_{1}(x)$. In each panel, the solid (a) and dashed (b) curves show the Re and Im parts, respectively. Parameters:   $\delta=0.001\Delta_{0}$, 
$\xi=\hbar v_{F}/\Delta_{0}$. }
\label{quasiclassical2}
\end{figure}

\begin{figure}[tb]
\begin{center}
\includegraphics[width=15cm,clip]{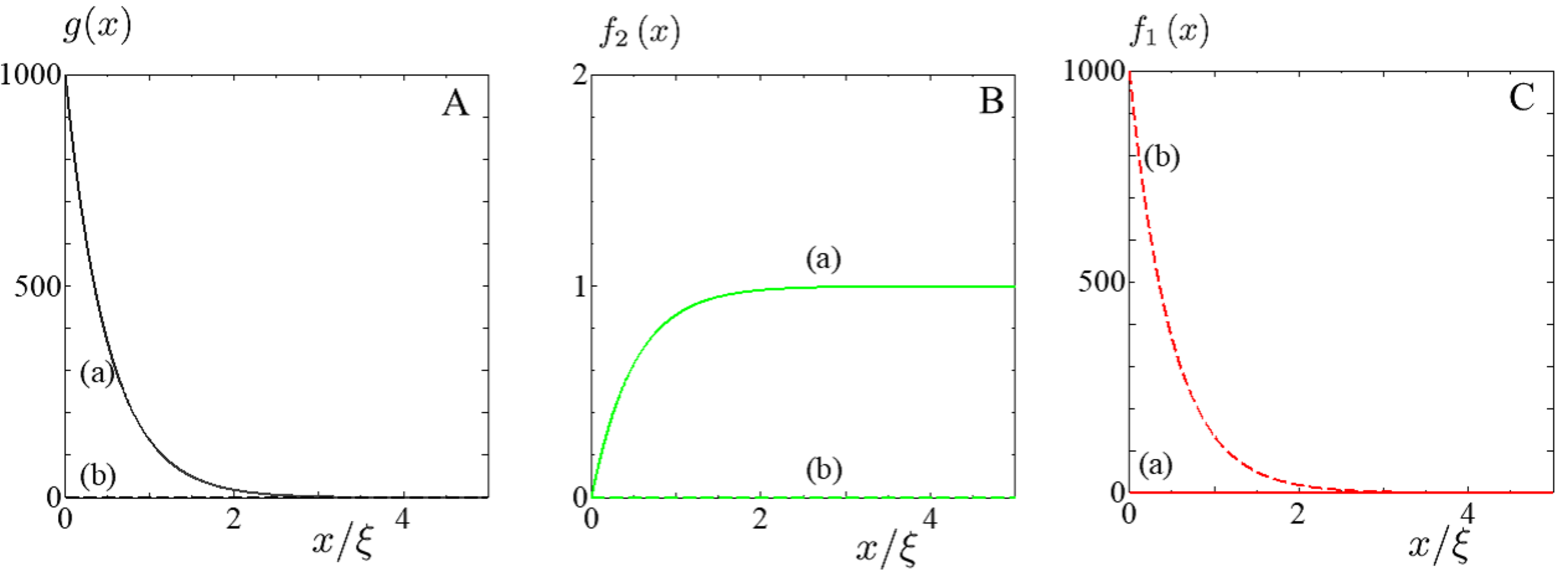}
\end{center}
\caption{
Spatial dependence of quasiclassical Green's functions at $E/\Delta_{0}=0$, where different panels correspond to  A: $g(x)$, B: $f_{2}(x)$, and C: $f_{1}(x)$. In each panel, the solid (a) and dashed (b) curves show the Re and Im parts, respectively. Parameters:   $\delta=0.001\Delta_{0}$, $\xi=\hbar v_{F}/\Delta_{0}$.}\label{quasiclassical1}
\end{figure}

In Figs.\ref{quasiclassical3}, \ref{quasiclassical2}, and 
\ref{quasiclassical1}, we show the spatial dependence of the quasiclassical Green's functions $g(x)=g_{+}(x)$, $f_{2}(x)=f_{2+}(x)$, and $f_{1}(x)=f_{1+}(x)$.  For numerical purposes,  we replace $E$ for $E + i\delta$ with infinitesimal imaginary number $\delta=0.001\Delta_{0}$.  When inspecting $g(x)$ and $f_{1,2}(x)$ at energies above the gap $E>\Delta_{0}$,  they develop an oscillatory behavior, as predicted by Eqs.\,(\ref{Eilenbergergx}). At $x=0$, $f_{2}(x)=0$, while  $g(x)$ and $f_{1}(x)$ exhibit finite values. For energies within the gap, $|E|<\Delta_{0}$, the factor $\gamma$ becomes imaginary according to Eq.\,(\ref{Omegapm}), inducing an exponential decay from the interface in the boundary contributions, with a decay length $\sim 2\xi$, see Fig.\,\ref{quasiclassical2}. For the chosen energy, the imaginary parts $g(x)$ and the real part of $f_{1,2}(x)$  become dominant, taking values of the same order. It is interesting to note that, the imaginary part of $g(x)$ decays towards the bulk of the system, a behavior that is also exhibited by the OTE pair amplitude $f_{2}(x)$. In contrast, the real part of the ETO amplitude increases as $x$ takes values deeper inside the superconductor, see Fig.\,\ref{quasiclassical2}. This is of course an expected behaviour because the pair potential of the bulk superconductor exhibits ETO pair symmetry.

\begin{figure}[h]
\begin{center}
\includegraphics[width=11cm,clip]{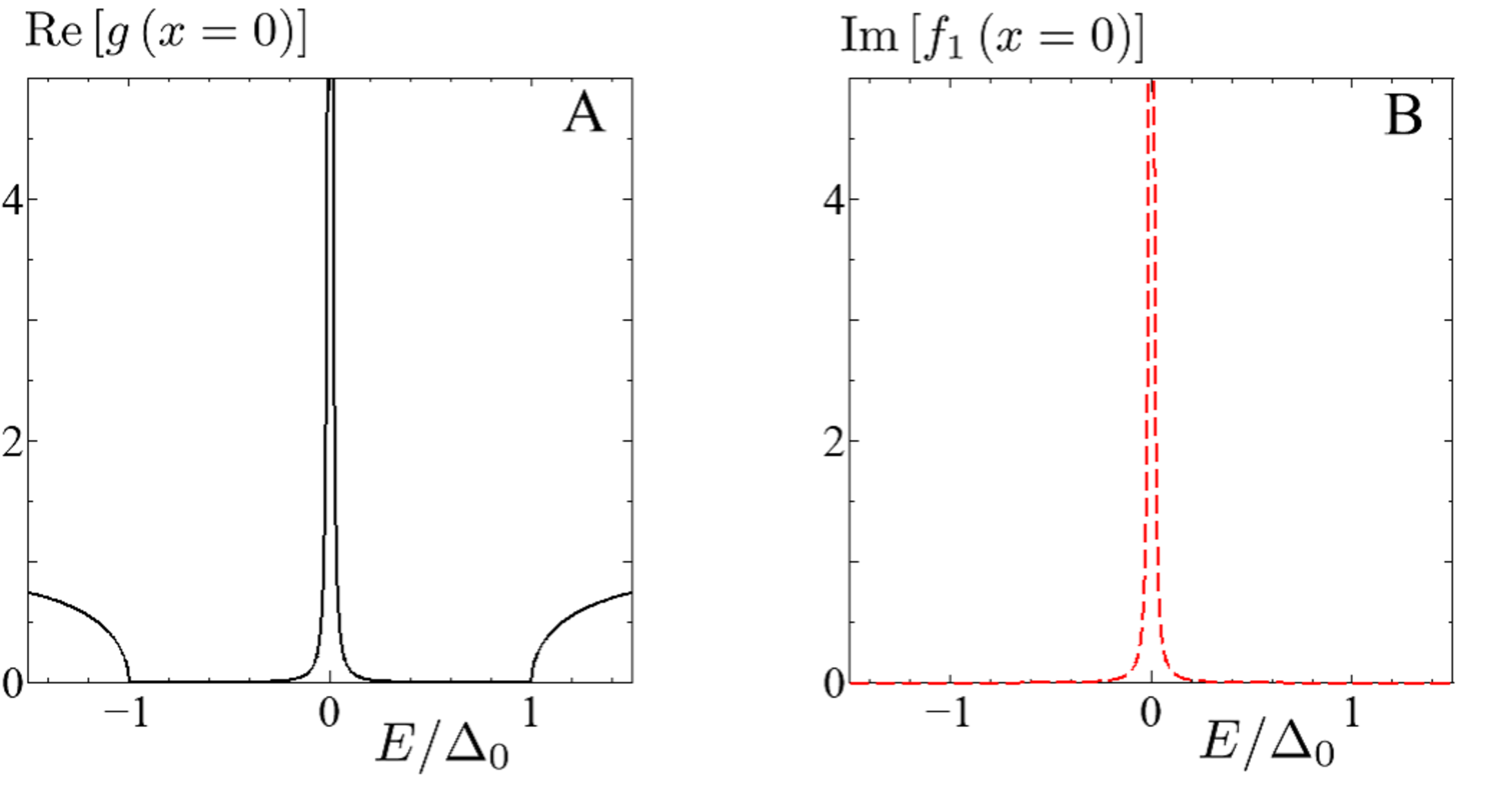}
\end{center}
\caption{Energy dependence of   LDOS  and OTE pair amplitude  at   $x=0$, where A: ${\rm Re}[g(x)]$ and B:${\rm Im}[f_{1}(x)]$. 
$\delta=0.001\Delta_{0}$.}
\label{LDOSandOdd1}
\end{figure}

\begin{figure}[h]
\begin{center}
\includegraphics[width=11cm,clip]{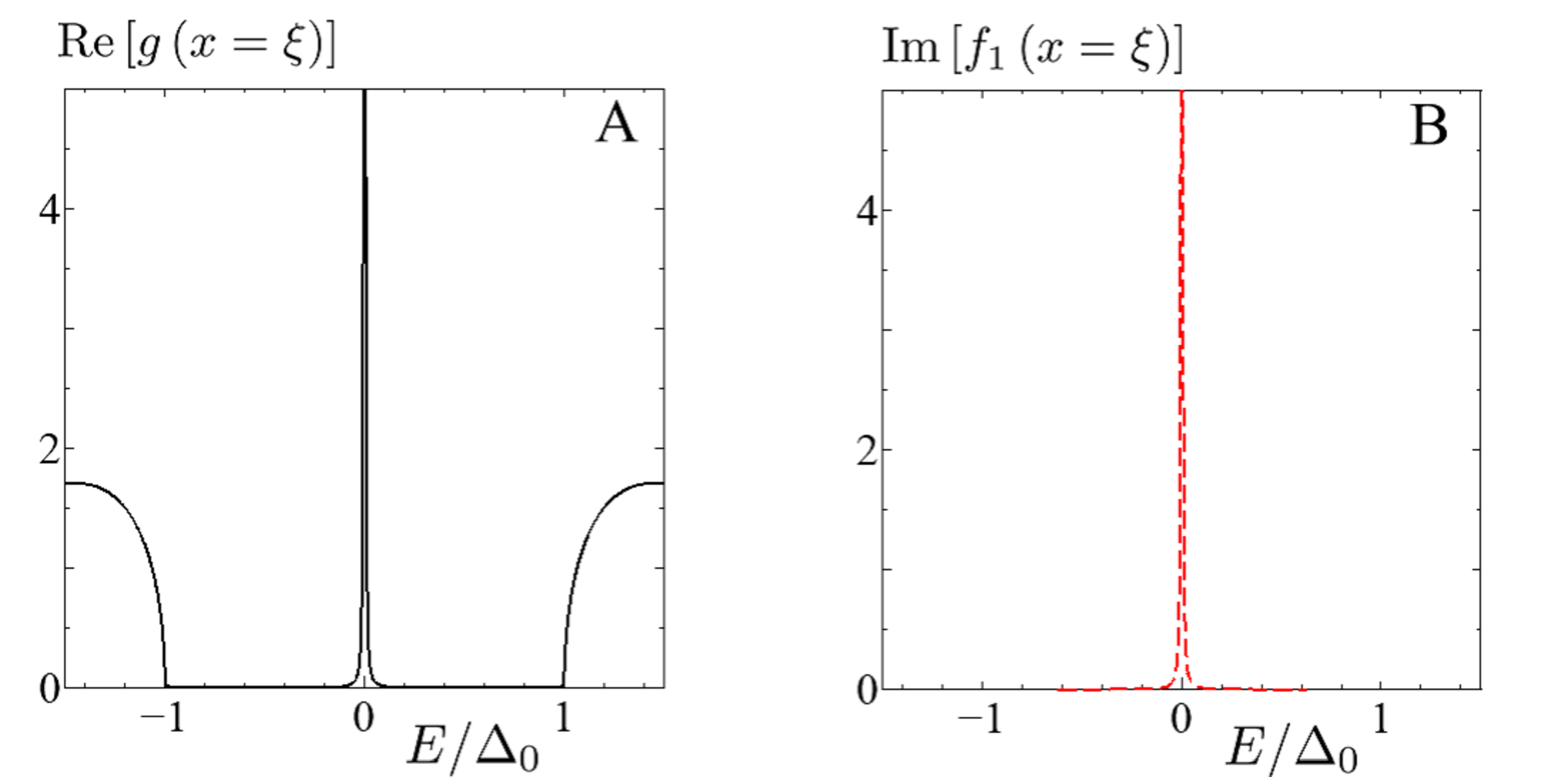}
\end{center}
\caption{Energy dependence of   LDOS  and OTE pair amplitude  at   $x=\xi$, where A: ${\rm Re}[g(x)]$ and B:${\rm Im}[f_{1}(x)]$. 
$\delta=0.001\Delta_{0}$.} 
\label{LDOSandOdd2}
\end{figure}

To close this part, we analyze the spatial dependence of the quasiclassical Green's functions at $E=0$, which is presented in Fig.\,\ref{quasiclassical1}. As we observe, the spatial dependence of the real part of $g(x)$ and the imaginary part of $f_{1}(x)$ coincide, developing a maximum at $x=0$ with huge values, followed by an exponential decay towards the bulk of the system. This peak at the boundary is caused by the presence of a MZM, since the spin-polarized $p$-wave superconductor we here consider is in the topological phase.  In consequence, this further demonstrates that the emergence of a MZM at the boundary is always accompanied by the generation of OTE pairing and that there is a one-to-one correspondence between them \cite{tanaka12,Asano2013}. This view can be further supported by   Figs. \ref{LDOSandOdd1} and  \ref{LDOSandOdd2} which show the energy dependence 
of the LDOS ${\rm Re}[g(x)]$   and the OTE pair amplitude ${\rm Im}[f_{1}(x)]$ at $x=0$. Since  
the real part of $g(x)$, namely, ${\rm Re}[g(x)]$
, represents the  LDOS normalized by its value in the normal state, there is a direct correspondence between ZESABS and  OTE pairing at very low energies. A huge peak forms at $E=0$   in both the LDOS and the OTE pair amplitude due to the emergence of an MZM at the boundary,  which demonstrates the strong relationship between LDOS and OTE at $E=0$.  It is worth noting that for finite subgap energies ( $|E|<\Delta_{0}$) the LDOS and OTE pairing have the same dependence, while only the LDOS captures the gap edges at $|E|=\Delta_{0}$. Nevertheless,  the intriguing signature of both OTE pairing and LDOS is their  ZEP due to an MZM at the boundary. As we enter into the superconductor, the height and width of the ZEP is suppressed as shown in    Fig. \ref{LDOSandOdd2} at   $x=\xi$. This behavior is indeed consistent with the Majorana origin of the ZEP, thus showing that both LDOS and OTE pair amplitude at the boundary of a topological superconductor can be used to detect MZMs and their exotic pair symmetry.

\subsection{Finite length $p$-wave superconductor}
\label{Subsection4d}
\begin{figure}
\begin{center}
\includegraphics[width=8.0cm]{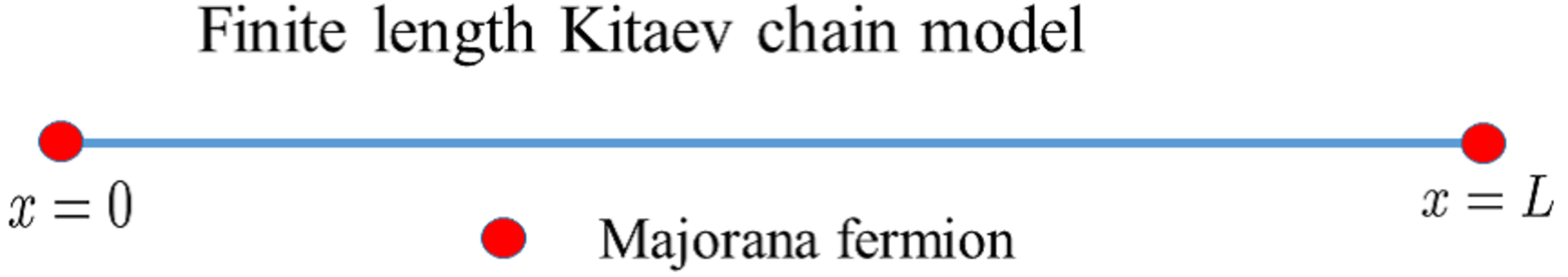}
\end{center}
\caption{Schematic  of the finite length Kitaev chain model 
for $0 \leq x \leq L$. 
Two Majorana bound states are located at $x=0$ and $x=L$. 
}
\label{Kitaevmodel2}
\end{figure}

In this part we explore the Green's function properties of a spin-polarized $p$-wave superconductor with finite length $L$, as schematically shown in Fig.\,\ref{Kitaevmodel2}. Since MZMs appear at the boundary in the topological phase, we expect that a MZM emerges at each end of the system as depicted by red filled circles in Fig.\,\ref{Kitaevmodel2}.
 
Before constructing the Green's function for a finite length 1D $p$-wave superconductor, it is instructive to first obtain the energy of the bound state, following the same spirit of Subsection \ref{subsection2d}. In this regard,  we recall that  the total wavefunction of the bound state of the 1D $p$-wave superconductor described by Eq.\,(\ref{Hamiltonianmatrix}), for $|E| <\Delta_{0}$   reads,
\begin{equation}
\label{wavefunctionfinitescp}
\Psi\left(x \right) =
A
\begin{pmatrix}
1 \\
\Gamma 
\end{pmatrix}
\exp\left(ik^{+}x \right)
+
B
\begin{pmatrix}
1 \\
-\Gamma
\end{pmatrix}
\exp\left( -ik^{+}x \right)
 +
C 
\begin{pmatrix}
\Gamma \\
1 
\end{pmatrix}
\exp\left( ik^{-}x \right)
+ D
\begin{pmatrix}
-\Gamma \\
1
\end{pmatrix}
\exp\left( -ik^{-}x \right)\,,
\end{equation}
which is composed of two electron-like (proportional to $A$ and $B$) and two hole-like (proportional to $C$ and $D$) wavefunctions at    momenta $\pm k^{\pm}$  given by Eq.\,(\ref{momentapm}). Now, we use hard-wall boundary conditions at both $x=0$ and $x=L$ and then write down a system of equations for the total wavefunction in Eq.\,(\ref{wavefunctionfinitescp}), namely, $\Psi(x=0)=0$ and  $\Psi(L=0)=0$. We thus obtain the following  system of equations for the coefficients,
\begin{equation}
\label{EQSfiniteSCp}
\begin{split}
A + B + \left(C - D \right)\Gamma&=0\,,\\ 
\left(A - B \right)\Gamma + C + D&=0\,,\\
A \exp\left( ik^{+}L \right) 
+ B \exp\left(-ik^{+}L \right) 
+\left[ C \exp \left(ik^{-}L \right) -D \exp\left(-ik^{-}L \right) 
\right]
\Gamma&=0\,,\\
\left[
A \exp\left(ik^{+}L \right)
- B \exp\left(-ik^{+} L \right) \right] \Gamma
+ C \exp\left(ik^{-}L\right) + D \exp\left(-ik^{-}L\right)&=0\,.
\end{split}
\end{equation}

A nontrivial solution of a bound state is then obtained by postulating that these four
equations should be satisfied for nonzero values of $A$,$B$,$C$,and $D$, in a similar way as we have done in Subsection \ref{subsection2d} when deriving the bound state energy at one boundary. Thus, in the present case, we obtain the bound state solution given by 
\begin{equation}
\label{boundstate}
E_{b}^{2} =\frac{2\Delta_{0}^{2}}{\cos \delta_{s} -\cos \delta_{f}}
\sin^{2}\left( \frac{\delta_{f}}{2} \right)\,,
\end{equation}
where  $\delta_{s}$ and $\delta_{f}$ are defined by    
\begin{equation}
\begin{split}
\exp\left[i \left(k^{+}-k^{-} \right)L \right]
&\equiv \exp\left( i\delta_{s} \right)\,,\\
\exp\left[i \left(k^{+} + k^{-} \right)L \right]
&\equiv \exp\left( i\delta_{f} \right)\,.
\end{split}
\end{equation}
The bound state expression given by  Eq.\,(\ref{boundstate}) can be further simplified for $\mu\gg \Delta_{0}$, where we can approximate $\cos \delta \sim {\rm cosh}(2\gamma_{B} L)$ and $\cos\delta_{1} \sim cos(2k_{F}L)$, with $\gamma_{B}={\sqrt{\Delta_{0}^{2}-E_{b}^{2}}}/{\hbar v_{F}}$. Then, the bound state energy from Eq.\,(\ref{boundstate})  reduces to
\begin{equation}
E^{2}_{b}
=\frac{2 \Delta_{0}^{2}}{{\rm cosh}(2\gamma_{B} L) - \cos(2k_{F}L)}
\sin^{2}(k_{F} L)\,. 
\label{boundstateapproximation}
\end{equation}
At this point, we are in a position to draw some conclusions about the bound state energy of a spin-polarized  $p$-wave superconductor of finite length $L$.  The first feature to notice is that there are two solutions, a positive and a negative, which, under general conditions of finite but short $L$, exhibit an oscillatory behavior with sizeable amplitudes and crossings at zero energy when $k_{F}L= n \pi$ with   $n\in\mathbb{Z}$  like in a Fabry-Perot interferometer. Furthermore, it can be shown that the amplitude of such oscillations increases as the chemical potential increases.  Interestingly, for very large $L$, the two bound state energies acquire vanishing values such that $E_{b}=0$.   These zero-energy bound states represent the two MZMs located at the ends $x=0$ and $x=L$ of the 1D spin-polarized $p$-wave superconductor.
 
Having understood that a finite topological superconductor hosts two MZMs, one at each end, whose energies depend on the system length, now we turn our attention to their impact on the Green's function. As discussed in subsection \ref{Subsection4a}, in order to construct Green's function, we must first obtain the outgoing and incoming wavefunctions associated to the BdG equations of $\hat{H}$ and $\hat{H}^{t}$, see Eqs.\,(\ref{HamitonianPsi}).  The outgoing and incoming wavefunctions associated to $\hat{H}$ are given by,
\begin{equation}
\label{finiteoutgoingincomingwave}
\begin{split}
\Psi_{out}^{(1)}\left( x \right)
&=
\begin{pmatrix}
1 \\ \Gamma 
\end{pmatrix} 
\exp\left[ ik^{+}\left( x - L \right) \right]
- a 
\begin{pmatrix}
\Gamma \\ 1 
\end{pmatrix} 
\exp\left[ ik^{-}\left( x - L \right) \right]
- b 
\begin{pmatrix}
1 \\ -\Gamma 
\end{pmatrix} 
\exp\left[ -ik^{+}\left( x - L \right) \right]\,,\\
\Psi_{out}^{(2)}\left( x \right)
&=
\begin{pmatrix}
-\Gamma \\ 1 
\end{pmatrix} 
\exp\left[ -ik^{-}\left( x - L \right) \right]
+ a 
\begin{pmatrix}
1 \\ -\Gamma 
\end{pmatrix} 
\exp\left[ -ik^{+}\left( x - L \right) \right]
- b 
\begin{pmatrix}
\Gamma \\ 1 
\end{pmatrix} 
\exp\left[ ik^{-}\left( x - L \right) \right]\,,\\
\Psi_{in}^{(1)}\left( x \right)
&=
\begin{pmatrix}
1 \\ -\Gamma 
\end{pmatrix} 
\exp\left( -ik^{+}x \right)
+ a 
\begin{pmatrix}
 -\Gamma \\ 1
\end{pmatrix} 
\exp\left( -ik^{-}x \right)
- b 
\begin{pmatrix}
1 \\ \Gamma 
\end{pmatrix} 
\exp\left( ik^{+} x \right)\,,\\
\Psi_{in}^{(2)}\left( x \right)
&=
\begin{pmatrix}
\Gamma \\ 1 
\end{pmatrix} 
\exp\left( ik^{-}x \right)
- a 
\begin{pmatrix}
1 \\ \Gamma  
\end{pmatrix} 
\exp\left( ik^{+}x \right)
- b 
\begin{pmatrix}
-\Gamma \\ 1 
\end{pmatrix} 
\exp\left( -ik^{-} x \right)\,,
 \end{split}
\end{equation}
which satisfy the boundary condition given by 
\[
\Psi_{out}^{(1,2)}\left( x=L \right)=0\,, \quad \Psi_{in}^{(1,2)}\left( x=0 \right)=0\,,
\]
while the coefficients $a$ and $b$ read
\begin{equation}
a = \frac{2\Gamma}{1 + \Gamma^{2}}, \ \ 
b= \frac{1 - \Gamma^{2}}{1 + \Gamma^{2}}\,.
\label{finiteincomingwave}
\end{equation}

For $\hat{H}^{t}$, the outgoing and incoming  wavefunctions are  given by 
\begin{equation}
\label{finiteoutgoingincomingwavetilde}
\begin{split}
\tilde{\Psi}_{out}^{(1)}\left( x \right)
&=
\begin{pmatrix}
1 \\ -\Gamma 
\end{pmatrix} 
\exp \left[ ik^{+}\left( x - L \right) \right]
+ a 
\begin{pmatrix}
-\Gamma \\ 1 
\end{pmatrix} 
\exp \left[ik^{-}\left( x - L \right) \right]
- b 
\begin{pmatrix}
1 \\ \Gamma 
\end{pmatrix} 
\exp \left[-ik^{+}\left( x - L \right) \right]\,,\\
\tilde{\Psi}_{out}^{(2)}\left( x \right)
&=
\begin{pmatrix}
\Gamma \\ 1 
\end{pmatrix} 
\exp \left[-ik^{-}\left( x - L \right) \right]
- a 
\begin{pmatrix}
1 \\ \Gamma 
\end{pmatrix} 
\exp \left[ -ik^{+}\left( x - L \right) \right]
- b 
\begin{pmatrix}
-\Gamma \\ 1 
\end{pmatrix} 
\exp \left[ik^{-}\left( x - L \right) \right]\,,\\
\tilde{\Psi}_{in}^{(1)}\left( x \right)
&=
\begin{pmatrix}
1 \\ \Gamma 
\end{pmatrix} 
\exp \left[-ik^{+}x  \right]
- a 
\begin{pmatrix}
\Gamma \\ 1 
\end{pmatrix} 
\exp \left[ -ik^{-}x \right]
- b 
\begin{pmatrix}
1 \\ -\Gamma 
\end{pmatrix} 
\exp \left[ ik^{+} x \right]\,,\\
\tilde{\Psi}_{in}^{(2)}\left( x \right)
&=
\begin{pmatrix}
-\Gamma \\ 1 
\end{pmatrix} 
\exp \left( ik^{-}x \right)
+ a 
\begin{pmatrix}
1 \\ -\Gamma 
\end{pmatrix} 
\exp \left( ik^{+}x \right)
- b 
\begin{pmatrix}
\Gamma \\ 1 
\end{pmatrix} 
\exp \left( -ik^{-} x \right)\,,
\end{split}
\end{equation}
which satisfy 
\[
\tilde{\Psi}_{out}^{(1,2)}\left( x=L \right)=0\,,\quad \tilde{\Psi}_{in}^{(1,2)}\left( x=0 \right)=0\,.
\]

\begin{figure}[tb]
\begin{center}
\includegraphics[width=0.8\columnwidth]{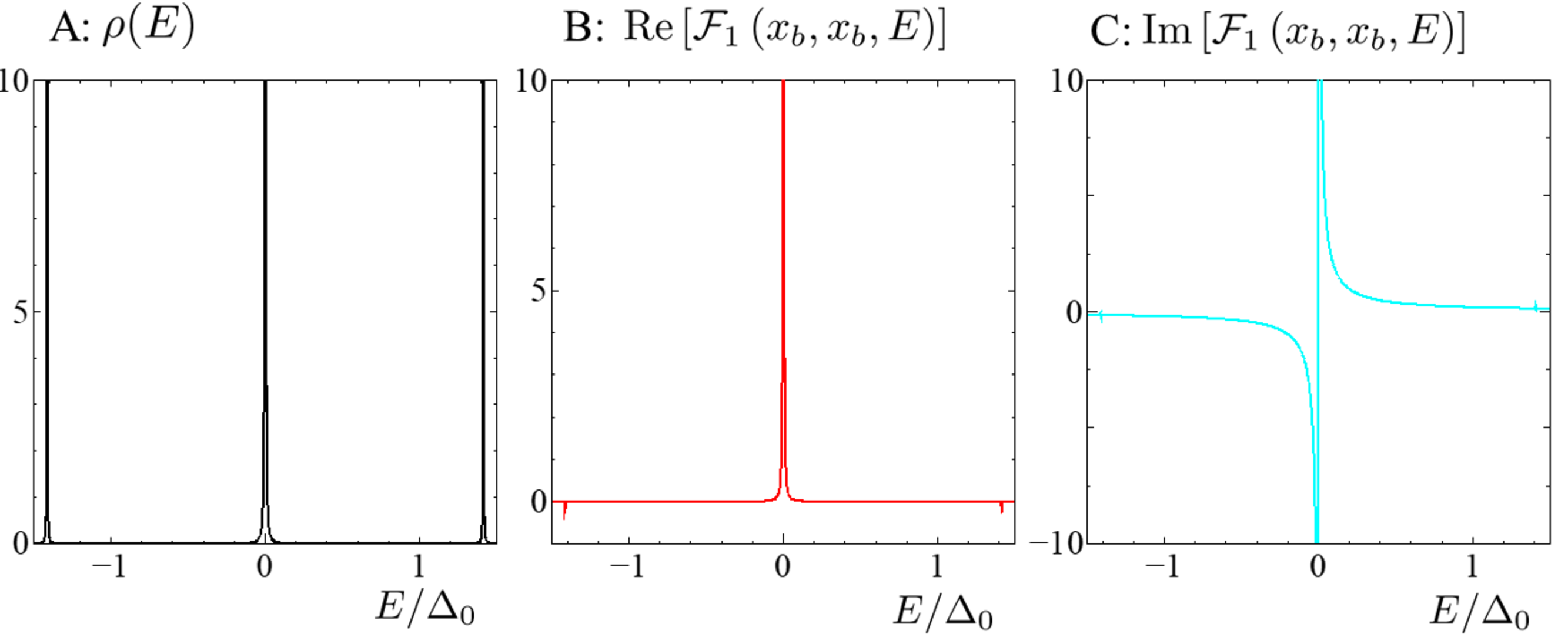}
\end{center}
\caption{Energy dependence of LDOS and local OTE pair amplitude in a finite 1D spin-polarized $p$-wave superconductor at $x_{b}=L/1113$ and  $Lk_{F}=1000\pi$: A: $\rho(E)=-{\rm Im}[\mathcal{G}(x_{b},x_{b},E)]$, B: ${\rm Re}[\mathcal{F}_{1}(x_{b},x_{b},E)]$
and C: ${\rm Im}[\mathcal{F}_{1}(x_{b},x_{b},E)]$. Parameters:
$\Delta_{0}=0.002E_{F}$, and  $\delta=0.001\Delta_{0}$.}
\label{FiniteKitaev4}
\end{figure}

\begin{figure}[tb]
\begin{center}
\includegraphics[width=0.8\columnwidth]{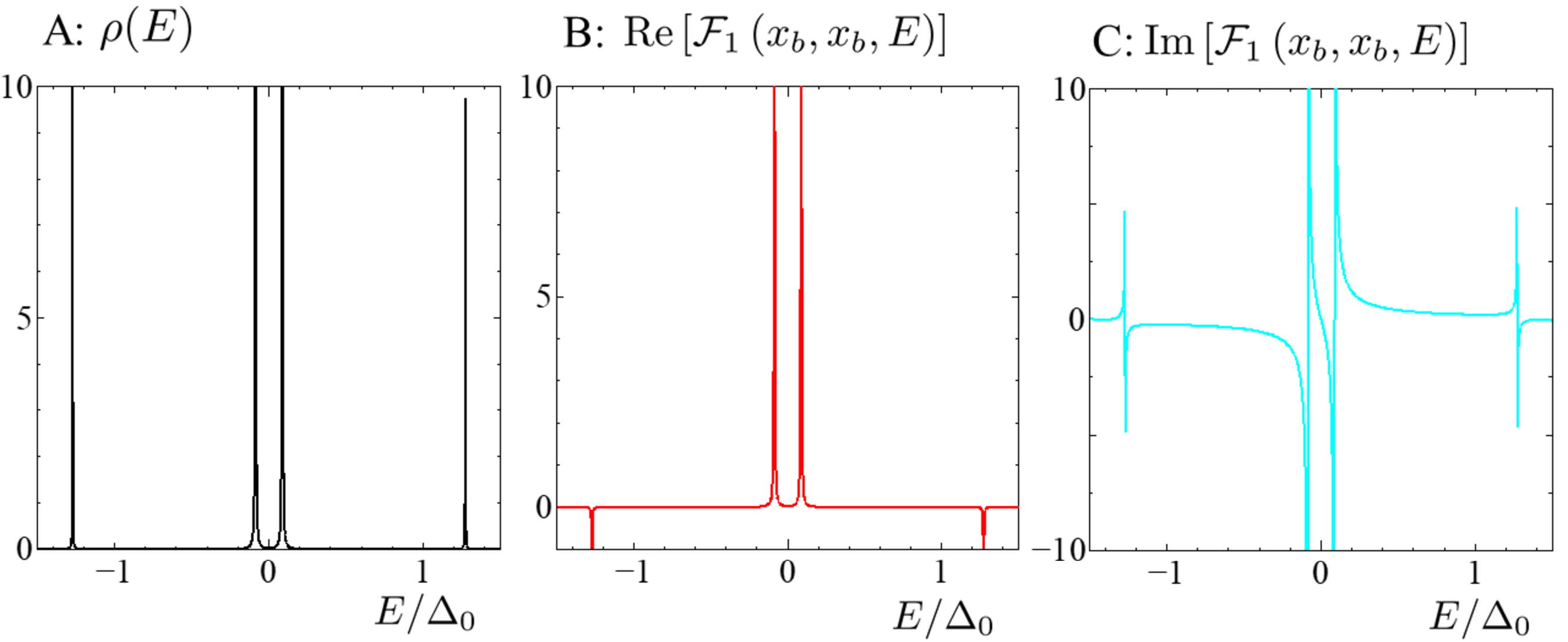}
\end{center}
\caption{Energy dependence of the LDOS and the local OTE pair amplitude in a finite 1D spin-polarized $p$-wave superconductor at $x_{b}=L/1113$ and  $Lk_{F}=1000.5\pi$: A: $\rho(E)=-{\rm Im}[\mathcal{G}(x_{b},x_{b},E)]$,  B: ${\rm Re}[\mathcal{F}_{1}(x_{b},x_{b},E)]$, and C: ${\rm Im}[\mathcal{F}_{1}(x_{b},x_{b},E)]$. Parameters:
$\Delta_{0}=0.002E_{F}$, and  $\delta=0.001\Delta_{0}$.}
\label{FiniteKitaev5}
\end{figure}

Then, by using the incoming and outgoing states given by Eqs.\,(\ref{finiteoutgoingincomingwave}) and (\ref{finiteoutgoingincomingwavetilde}) in Eqs.\,(\ref{McMillan}) and (\ref{boundaryConditions}), we find the retarded Green's function, which, after some manipulations, reads
\begin{equation}
\label{GFFinite}
\begin{split}
 G^{r}\left( x,x', E \right)
&= \left( \frac{m}{\hbar^{2}k_{F}} \right)
\left(1 - \Gamma^{2} \right)^{-1}
\left[
\cos \left( 2\kappa^{-} L \right) 
- \left( a^{2} + b^{2} \cos\left( 2 \kappa^{+}L \right) \right)
\right]^{-1} \\
&\times 
\left\{ 
C_{1}\left(x,x' \right) -2 b C_{2}\left(x,x' \right)
+ a\sin\left( \kappa^{-}L \right) C_{3}\left(x,x' \right)
-2ab \sin\left(\kappa^{+}L \right) C_{4}\left(x,x' \right)
\right\}\,,
\end{split}
\end{equation}
where  
\begin{eqnarray}
&&C_{1}(x,x') 
\nonumber
\\
&=&
\left( 1 + \Gamma^{2} \right)
\left[ 
\cos\left(\kappa^{+} \mid x -x' \mid \right) 
\sin \left[\kappa^{-} \left( \mid x -x' \mid - 2L \right) 
\right]
- a^{2} \cos\left( \kappa^{+} \mid x -x' \mid \right) 
\sin\left( \kappa^{-} \mid x -x' \mid \right)
\right.
\nonumber
\\
&-& 
\left.
b^{2}
\cos\left[ \kappa^{+} 
\left( \mid x -x' \mid -2L \right) \right]
\sin\left( \kappa^{-} \mid x-x' \mid  \right)
\right]
X_{1} 
\nonumber
\\
&+&
\left( 1 - \Gamma^{2}\right)
\left[ \sin \left(\kappa^{+} \mid x - x' \mid \right)
\cos\left( \kappa^{-} \left( \mid x -x' \mid - 2L \right) \right)
-a^{2} \sin\left( \kappa^{+} \mid x -x' \mid \right) 
\cos \left( \kappa^{-} \mid x - x' \mid \right) 
\right.
\nonumber
\\
&-&
\left.
b^{2} \sin \left[ \kappa^{+} 
\left( \mid x - x' \mid - 2L \right) \right]
\cos\left( \kappa^{-} \mid x -x' \mid \right)
\right]
X_{2} 
\nonumber
\\
&+&
2i\Gamma
\left[ \sin\left( \kappa^{+} \left(x -x' \right) \right)
\left[ \sin\left( \kappa^{-}\left( \mid x-x' \mid -2L \right) 
\right)
-a^{2}\sin\left( \kappa^{-} \mid x -x' \mid \right) \right]
\right. 
\nonumber
\\
&-&
\left.
b^{2} \sin\left( \kappa^{-} \left( x -x' \right) \right)
\sin\left(\kappa^{+}\left( \mid x-x' \mid -2L \right) \right)
\right] Y_{1}
\end{eqnarray}
\begin{eqnarray}
&&C_{2}\left( x,x' \right) 
\nonumber
\\
&=&-\left(1 - \Gamma^{2} \right)
\left[ \sin \left( \kappa^{+}L \right)
\cos \left( \kappa^{-}L \right)
\sin\left(\kappa^{+}\left(x +x'-L\right) \right)
\sin\left(\kappa^{-}\left(x +x'-L\right) \right)
\right.
\nonumber
\\
&+& 
\left.
\sin \left( \kappa^{-}L \right)\cos \left( \kappa^{+}L \right)
\cos\left(\kappa^{+}\left(x +x'-L\right) \right)
\cos\left(\kappa^{-}\left(x +x'-L\right) \right)
\right] X_{1}
\nonumber
\\
&+&
\left( 1 + \Gamma^{2} \right)
\left[ \sin \left( \kappa^{+}L \right)\cos \left( \kappa^{-}L \right)
\cos\left(\kappa^{+}\left(x +x'-L\right) \right)
\cos\left(\kappa^{-}\left(x +x'-L\right) \right)
\right.
\nonumber
\\
&+& 
\left.
\sin \left( \kappa^{-}L \right)\cos \left( \kappa^{+}L \right)
\sin\left(\kappa^{+}\left(x +x'-L\right) \right)
\sin\left(\kappa^{-}\left(x +x'-L\right) \right)
\right] X_{2}
\nonumber
\\
&-&
2i\Gamma
\left[
\sin\left(\kappa^{+}L \right)
\cos\left(\kappa^{-}L \right)
\sin\left(\kappa^{+} \left(x+x'-L\right) \right)
\cos\left(\kappa^{-} \left(x+x'-L\right) \right)
\right.
\nonumber
\\
&-&
\left.
\sin\left(\kappa^{-}L \right)
\cos\left(\kappa^{+}L \right)
\sin\left(\kappa^{-} \left(x+x'-L\right) \right)
\cos\left(\kappa^{+} \left(x+x'-L\right) \right)
\right]
Y_{2}
\end{eqnarray}
\begin{eqnarray}
C_{3}\left( x,x' \right)
&=& 4 \Gamma \cos\left[ \kappa^{-} \left( x +x' -L \right) \right]
\cos \left[ \kappa^{+} \left( x - x' \right) \right] X_{1} 
\nonumber
\\
&+& 2 i \left( 1 + \Gamma^{2} \right)
\cos\left[ \kappa^{-} \left(x +x' -L\right) \right]
\sin\left[ \kappa^{+} \left(x -x' \right) \right] Y_{1} 
\nonumber
\\
&+& 2 i \left( 1 - \Gamma^{2} \right)
\sin\left[ \kappa^{-} \left(x +x' -L\right) \right]
\cos\left[ \kappa^{+} \left(x -x' \right) \right] Y_{2} 
\end{eqnarray}
\begin{eqnarray}
C_{4}\left( x,x' \right)
&=& -2 \Gamma \cos\left[ \kappa^{+} \left( x +x' -L \right) \right]
\cos \left[ \kappa^{-} \left( x - x' \right) \right] X_{2} 
\nonumber
\\
&+&  i \left( 1 + \Gamma^{2} \right)
\sin\left[ \kappa^{+} \left(x +x' -L\right) \right]
\cos\left[ \kappa^{-} \left(x -x' \right) \right] Y_{2}
\nonumber
\\
&+&  i \left( 1 - \Gamma^{2} \right)
\cos\left[ \kappa^{+} \left(x +x' -L\right) \right]
\sin\left[ \kappa^{-} \left(x -x' \right) \right] Y_{1}
\end{eqnarray}
and $X_{1}=\hat{\tau}_{0}$, $X_{2}=\hat{\tau}_{3}$, $Y_{1}=\hat{\tau}_{1}$ and $Y_{2}=i\hat{\tau}_{2}$. 
In the above $\kappa_{\pm}$, $\gamma$ and $\Gamma$ are given by 
\begin{equation}
\kappa^{\pm}= \left( k^{+} \pm k^{-} \right)/2 \sim k_{F}
\label{kappa}
\end{equation}
\begin{equation}
\gamma=\Omega/(\hbar v_{F}),  \ 
\Gamma=\frac{\Delta_{0}}{E + \Omega}
\label{gammaGamma}
\end{equation}
with $\Omega$   given by Eq.\,(\ref{Omegapm}).  
At this moment, it is worth highlighting some properties of the retarded Green's function given by Eq.\,(\ref{GFFinite}).  Even though it has an evident complicated structure, it is possible to note that it depends on the length of the superconductor $L$ in an oscillatory fashion via sines and cosines in the square brackets of the first and second lines. Moreover, the Green's function depends on the spatial coordinates $x$ and $x'$ via the coefficients $C_{i}$. Lastly, there is an overall energy dependence, for instance, via $\Gamma=\Delta_{0}/(E+\Omega)$ and $\Omega$   defined in Eq.\,(\ref{Omegapm}). It is a $2\times2$ matrix as expected, with its diagonals and off-diagonals representing the normal and anomalous Green's functions, see Eqs.\,(\ref{GF}). For visualization purposes, it is useful to decompose the total Green's function into its normal and anomalous parts as
\begin{equation}
\begin{split}
 \frac{\hbar^{2}k_{F}}{m} G^{r}\left(x,x',E \right)
&= \frac{1}{2}
\left[ \mathcal{G}\left(x,x',E \right) 
+ \bar{\mathcal{G}}\left(x,x',E \right) \right]X_{1}
+ \frac{1}{2}
\left[ \mathcal{G}\left(x,x',E \right) 
- \bar{\mathcal{G}}\left(x,x',E \right) \right]X_{2}\\
&+
\mathcal{F}_{2}\left(x,x',E \right) Y_{1} + 
+ 
\mathcal{F}_{1}\left(x,x',E\right) Y_{2}
\end{split}
\label{mathcalGandF}
\end{equation}
where $\mathcal{G}(x,x',E)$ corresponds to the normal Green's function, while $\mathcal{F}_{1}(x,x',E)$  and $\mathcal{F}_{2}(x,x',E)$ to the OTE and ETO  pair amplitudes, respectively. 
We now  explore  the correspondence between MZMs and OTE pairing and, for this reason, 
we look at $\mathcal{G}(x,x,E)$ and $\mathcal{F}_{1}(x,x,E)$  
defined in eq. (\ref{mathcalGandF}). 
They are given by 
\begin{equation}
\begin{split}
\mathcal{G}\left( x,x, E \right)
&= 
\left(1 - \Gamma^{2} \right)^{-1}
\left[
\cos \left( 2\kappa^{-} L \right) 
- \left[ a^{2} + b^{2} \cos\left( 2 \kappa^{+}L \right) \right]
\right]^{-1} 
\\
&\times
\left\{
-\left( 1 + \Gamma^{2} \right) 
\sin\left(2 \kappa^{-} L \right)
+ \left( 1 - \Gamma^{2} \right) b^{2} \sin \left( 2 \kappa^{+} L \right) 
\right.
\\
&+ 4 \Gamma a 
\left[ \sin\left(\kappa^{-} L \right) 
\cos \left( \kappa^{-}\left(2x - L \right) \right) 
+ b \sin\left(\kappa^{+} L \right) 
\cos \left( \kappa^{+}\left(2x - L \right) \right)
\right] \\
&-
2b 
\cos\left[ \left(\kappa^{+} + \kappa^{-} \right)
\left(2 x -L \right) \right]
\sin\left[ \left(\kappa^{+} - \kappa^{-} \right) L \right]
\\
&-
\left.
2b \Gamma^{2}
\cos\left[ \left( \kappa^{+} - \kappa^{-} \right)
\left(2 x -L \right) \right]
\sin\left[ \left(\kappa^{+} + \kappa^{-}\right) L \right]
\right\}\,,
\end{split}
\end{equation}

\begin{equation}
\begin{split}
\mathcal{F}_{1}\left( x,x, E \right)
&= 
\left(1 - \Gamma^{2} \right)^{-1}
\left[
\cos \left( 2\kappa^{-} L \right) 
- \left[ a^{2} + b^{2} \cos\left( 2 \kappa^{+}L \right) \right]
\right]^{-1} 
\\
&\times
4bi\Gamma
\left\{
\sin\left(\kappa^{+} L \right) 
\sin \left( \kappa^{+}\left(2x - L \right) \right) 
\left[
\cos\left(\kappa^{-} L \right) 
\cos \left( \kappa^{-}\left(2x - L \right) \right) 
-1 
\right]
\right. 
\\
&- 
\left.
\sin\left(\kappa^{-} L \right) 
\sin \left( \kappa^{-}\left(2x - L \right) \right) 
\left[
\cos\left(\kappa^{+} L \right) 
\cos \left( \kappa^{+}\left(2x - L \right) \right) 
-1 
\right]
\right\}\,.
\end{split}
\end{equation}
By plugging the expressions for $\Gamma$ and $\kappa^{\pm}$ from Eqs.\,(\ref{kappa}) and (\ref{gammaGamma}), we obtain  more compact expressions given by  
\begin{equation}
\begin{split}
\mathcal{G}\left( x,x, E \right)
&= 
\left[
E^{2} \left( \sin^{2} \left(\gamma L \right) 
-\sin^{2} \left(k_{F}L \right) \right)
- \Delta^{2} \sin^{2}\left(k_{F} L \right)
\right]^{-1}
\\
&\times
\left\{
\left( E / \Omega \right)
\left[E^{2}\cos\left(\gamma L \right) 
- \Delta^{2} 
\cos \left[ \gamma 
\left(2x - L \right) 
\right]
\right] \sin \left(\gamma L \right)
\right.
\\
&-
\sin\left(k_{F}L\right)
\left[ \Omega^{2}\cos\left( k_{F}L \right) 
+ \Delta^{2}\cos\left[ k_{F}\left(2x -L \right)
\right]
\right]
\\
&+ E 
\cos\left[k_{F}\left(2x -L \right) \right]
\cos\left[ \gamma \left(2x -L \right)\right]
\left[E \sin\left(k_{F}L\right) \cos\left( \gamma L \right) 
-\Omega \cos \left( k_{F} L \right)\sin \left( \gamma L \right) 
\right]
\\
&+
\left.
E 
\sin\left[k_{F}\left(2x -L \right) \right]
\sin\left[ \gamma \left(2x -L \right)\right]
\left[E \cos\left(k_{F}L\right) \sin\left( \gamma L \right) 
-\Omega \sin \left( k_{F} L \right)\cos \left( \gamma L \right) \right]
\right\}
\end{split}
\end{equation}

\begin{equation}
\begin{split}
\mathcal{F}_{1}\left( x,x, E \right)
&= 
E\Delta i
\left[
E^{2} \left( \sin^{2} \left(\gamma L \right) 
-\sin^{2} \left(k_{F}L \right) \right)
- \Delta^{2} \sin^{2}\left(k_{F} L \right)
\right]^{-1}
\\
&\times
\left\{
\sin\left(k_{F}L\right)
\sin\left[k_{F}\left(2x -L \right) \right]
\left[\cos\left(\gamma L \right)
\cos\left[ \gamma \left(2x -L \right)\right]
-1 \right]
\right.
\\
&-
\left.
\sin\left(\gamma L\right)
\sin\left[\gamma \left(2x -L \right) \right]
\left[\cos\left(k_{F} L \right)
\cos\left[k_{F} \left(2x -L \right)\right]
-1 \right]
\right\}\,.
\end{split}
\label{oddfrequencyanalytical1}
\end{equation}
In the following, we take these two previous expressions and analyze the corresponding LDOS and pair amplitude from the normal and anomalous parts, respectively. Since the spin symmetry is spin-triplet and we are evaluating locally in space, the pair amplitude has an OTE symmetry. In Figs. \ref{FiniteKitaev4} and \ref{FiniteKitaev5}, we plot 
${\rm Im}[\mathcal{G}(x_{b},x_{b},E)]$, 
${\rm Re}[\mathcal{F}_{1}(x_{b},x_{b},E)]$ 
and 
${\rm Im}[\mathcal{F}_{1}(x_{b},x_{b},E)]$
as a function of $E$ at a fixed position close to the edge.  In this case, $\rho(E)=-{\rm Im}[\mathcal{G}(x_{b},x_{b},E)]$ represents the   LDOS normalized by its value in the normal bulk state value.  For $Lk_{F}=1000\pi$, the LDOS  ${\rm Im}[\mathcal{G}(x_{b},x_{b},E)]$  has a sharp peak at zero energy and, interestingly, a similar line shape appears for the real part of the OTE pair amplitude ${\rm Re}[\mathcal{F}_{1}(x_{b},x_{b},E)]$, as shown in  Fig.\,\ref{FiniteKitaev4}(A,B).  
The  imaginary part 
${\rm Im}[\mathcal{F}_{1}(x_{b},x_{b},E)]$ is also enhanced around $E=0$ but it is an odd function of $E$ in  contrast to the ${\rm Re}[\mathcal{F}_{1}(x_{b},x_{b},E)]$, see  Fig.\,\ref{FiniteKitaev4}(C). 
Away from the resonant condition given by $Lk_{F}=1000.5\pi$,  the LDOS and the OTE pairing develop a small splitting, which is reflected in the formation of two peaks around zero energy, see  Fig.\,\ref{FiniteKitaev5}(A,B). This zero energy splitting is a consequence of the finite spatial overlap of the two MZMs emerging at either side of the system, an effect that can be understood by a simple inspection of   Eq.\,(\ref{boundstateapproximation}). 
At the position of the bound states around zero energy,  the amplitudes of both 
${\rm Real}[\mathcal{F}_{1}(x_{b},x_{b},E)]$ and 
${\rm Im}[\mathcal{F}_{1}(x_{b},x_{b},E)](x_{b},x_{b},E)]$ are enhanced, as seen in  Figs.\,\ref{FiniteKitaev5}(B) and Figs. \ref{FiniteKitaev5}(C)

\begin{figure}[tb]
\begin{center}
\includegraphics[width=0.8\columnwidth]{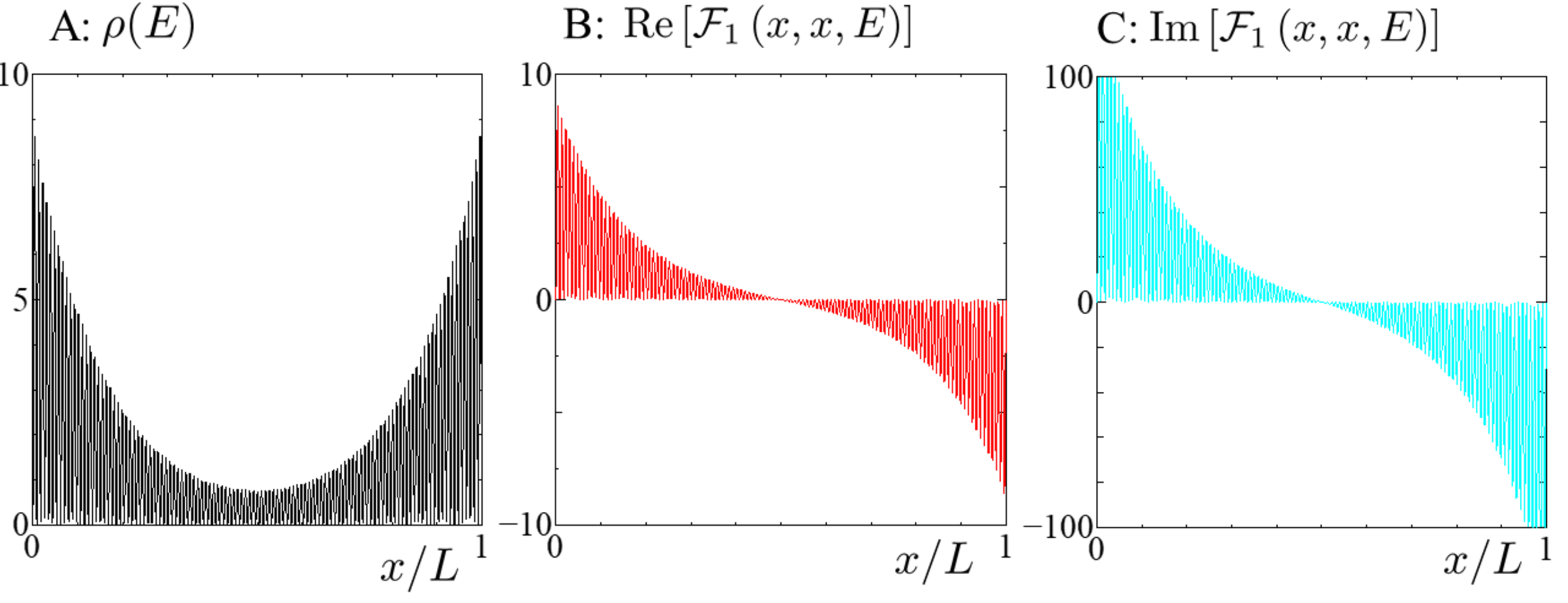}
\end{center}
\caption{
Spatial dependence of the LDOS and the local OTE pair amplitude in a finite 1D spin-polarized $p$-wave superconductor at $E=0.015\Delta_{0}$ and  $Lk_{F}=1000\pi$: A: $\rho(E)=-{\rm Im}[\mathcal{G}(x,x,E)]$,  B: ${\rm Re}[\mathcal{F}_{1}(x,x,E)]$, 
and C: ${\rm Im}[\mathcal{F}_{1}(x,x,E)]$. Parameters: $\Delta_{0}=0.002E_{F}$, and  $\delta=0.001\Delta_{0}$.
}
\label{FiniteKitaevL1}
\end{figure}

\begin{figure}[tb]
\begin{center}
\includegraphics[width=0.8\columnwidth]{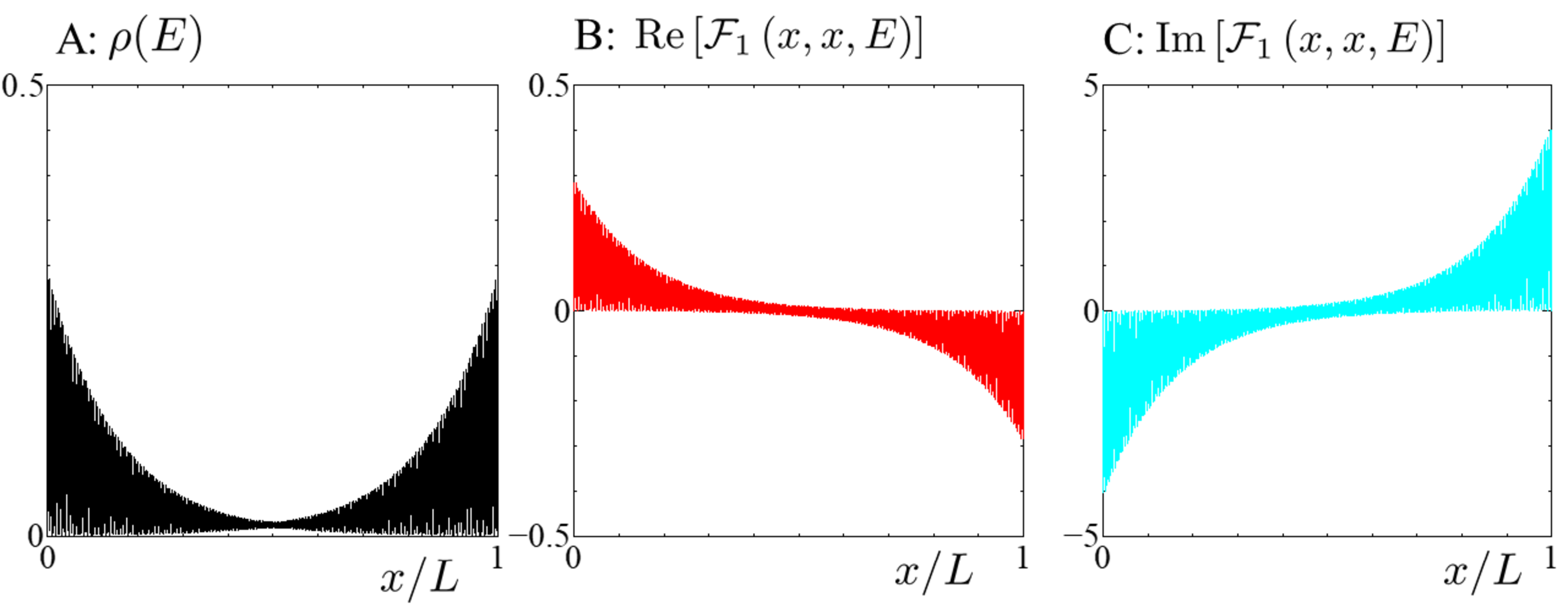}
\end{center}
\caption{
Spatial dependence of the LDOS and the local OTE pair amplitude in a finite 1D spin-polarized $p$-wave superconductor at $E=0.015\Delta_{0}$ and  $Lk_{F}=1000.5\pi$: A: $\rho(E)=-{\rm Im}[\mathcal{G}(x,x,E)]$,  B: ${\rm Re}[\mathcal{F}_{1}(x,x,E)]$, 
and C: ${\rm Im}[\mathcal{F}_{1}(x,x,E)]$. Parameters: $\Delta_{0}=0.002E_{F}$, and  $\delta=0.001\Delta_{0}$.}
\label{FiniteKitaevL2}
\end{figure}

\begin{figure}
\begin{center}
\includegraphics[width=0.8\columnwidth]{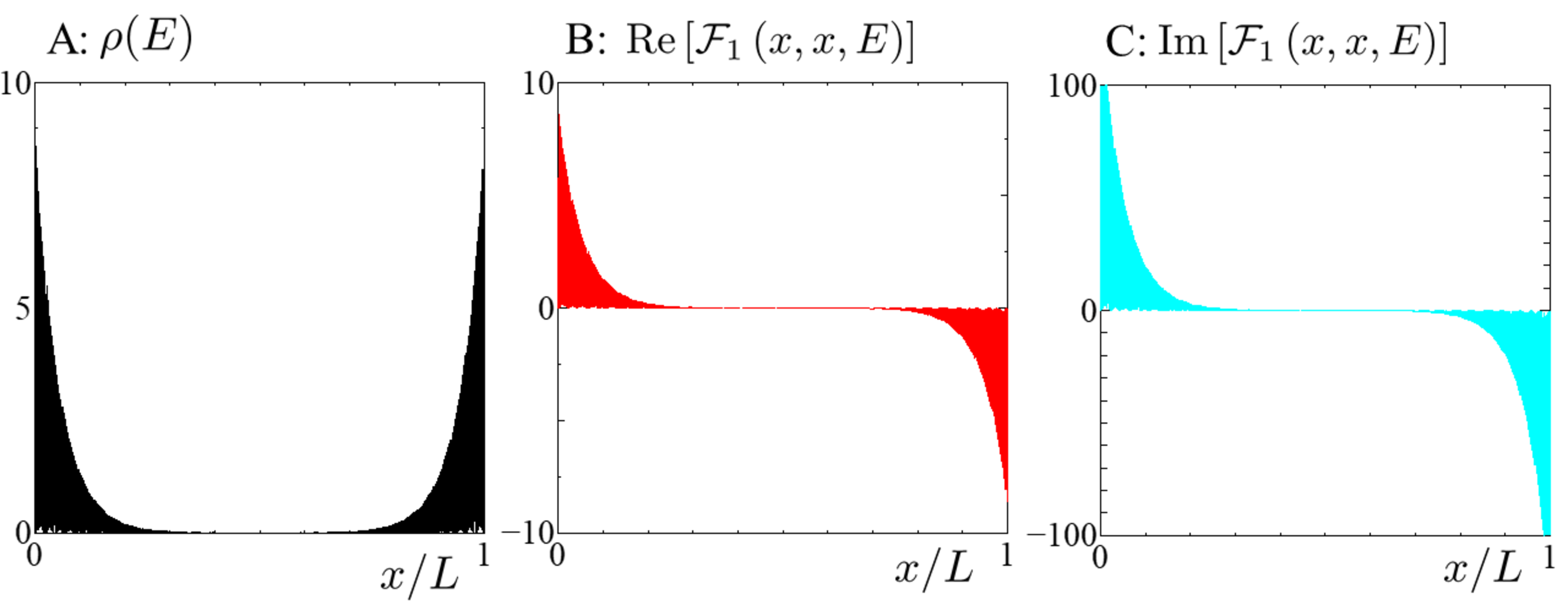}
\end{center}
\caption{
Spatial dependence of the LDOS and the local OTE pair amplitude in a finite 1D spin-polarized $p$-wave superconductor at $E=0.015\Delta_{0}$ and  $Lk_{F}=3000.5\pi$: A: $\rho(E)=-{\rm Im}[\mathcal{G}(x,x,E)]$,  B: ${\rm Re}[\mathcal{F}_{1}(x,x,E)]$, 
and C: ${\rm Im}[\mathcal{F}_{1}(x,x,E)]$. Parameters: $\Delta_{0}=0.002E_{F}$, and  $\delta=0.001\Delta_{0}$.}
\label{FiniteKitaevL3}
\end{figure}

Having seen the relation between local OTE pairing and LDOS as a function of energy, now we explore their spatial dependence.
For this purpose, in Figs.\,\ref{FiniteKitaevL1}-\ref{FiniteKitaevL3}, we plot $\rho(E)$, ${\rm Re}[\mathcal{F}_{1}(x,x,E)]$, and ${\rm Im}[\mathcal{F}_{1}(x,x,E)]$ 
as a function of $x$ for distinct values of the system length.
The first observation is that all these quantities have a 
rapidly oscillatory behavior with a period proportional to the inverse of $k_{F}$ and an exponential decay from the edges with the decay length given by $\hbar v_{F}/\Delta_{0}.$ The exponential decay reflects the localization of MZMs at the edges of the system, where only very long systems host truly MZMs, as indeed seen in Figs.\,\ref{FiniteKitaevL1}, \ref{FiniteKitaevL2}, and \ref{FiniteKitaevL3}. Another feature is that the spatial dependence of  $\rho(E)$, ${\rm Re}[\mathcal{F}_{1}(x,x,E)]$, 
and ${\rm Im}[\mathcal{F}_{1}(x,x,E)]$ is very similar in all cases, albeit the Re and Im parts of the pair amplitudes exhibit distinct signs at the edges. Furthermore, we note that the fact that the pair amplitudes exponentially decay from the edges reveals that this type of pairing is a unique property tied to the edge where MZMs form. As a result of this, both ${\rm Re}[\mathcal{F}_{1}(x,x,E)]$, and ${\rm Im}[\mathcal{F}_{1}(x,x,E)]$ become zero
at $x=L/2$, a common feature of the OTE pair amplitude that is 
independent of $L$.
As seen from Fig. \ref{FiniteKitaevL1}, 
${\rm Re}[\mathcal{F}_{1}(x,x,E)]$ and ${\rm Im}[\mathcal{F}_{1}(x,x,E)]$
become positive (negative) with $L/2>x>0$ $(L/2<x<L)$
for $k_{F}L=1000$. 
This property is a general profile for 
$k_{F}L=n\pi$ where $\sin k_{F}L=0$ is satisfied 
as seen from eq. (\ref{oddfrequencyanalytical1}).  

On the other hand, for $\sin(k_{F}L)=1000.5\pi$, 
as shown in Fig. \ref{FiniteKitaevL2}, 
the magnitudes of 
$\rho(E)$, ${\rm Re}[\mathcal{F}_{1}(x,x,E)]$ and 
${\rm Im}[\mathcal{F}_{1}(x,x,E)]$ are suppressed as compared to 
the case with $\sin(k_{F}L)=1000\pi$ (Fig.\ref{FiniteKitaevL1}). 
It is due to the destructive interference of MZMs localized  left and right edge 
of the system. 
For $k_{F}L=3000.5\pi$, 
$\rho(E)$, ${\rm Re}[\mathcal{F}_{1}(x,x,E)]$ and 
${\rm Im}[\mathcal{F}_{1}(x,x,E)]$ are strongly localized at $x=0$ and $L$, 
and these values are significantly suppressed around $x=L/2$ 
(Fig. \ref{FiniteKitaevL3}). 

\begin{figure}[tb]
\begin{center}
\includegraphics[width=0.95\columnwidth]{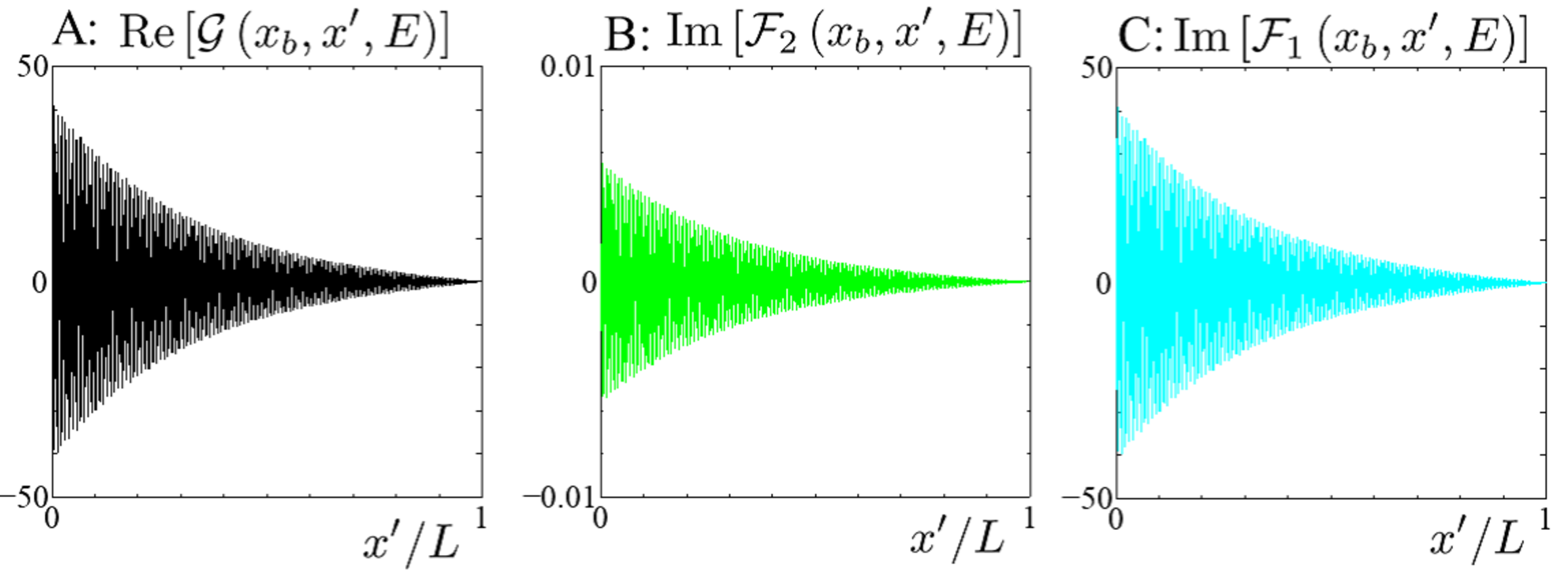}
\end{center}
\caption{Normal and anomalous Green's functions as  functions of $x'$ 
in a finite 1D spin-polarized $p$-wave superconductor at 
$Lk_{F}=1000\pi$.  Distinct panels correspond to:  A: ${\rm Re}[\mathcal{G}(x_{b},x',E)]$,  B: ${\rm Im}[\mathcal{F}_{2}(x_{b},x',E)]$, and 
C: ${\rm Re}[\mathcal{F}_{1}(x_{b},x',E)]$.
Parameters:   $x=x_{b}=L/1113$, $E/\Delta_{0}=0.015\Delta_{0}$, $\Delta_{0}=0.002\mu$, and $\delta=0.001\Delta_{0}$.}
\label{FiniteKitaev1}
\end{figure}

\begin{figure}[tb]
\begin{center}
\includegraphics[width=0.95\columnwidth]{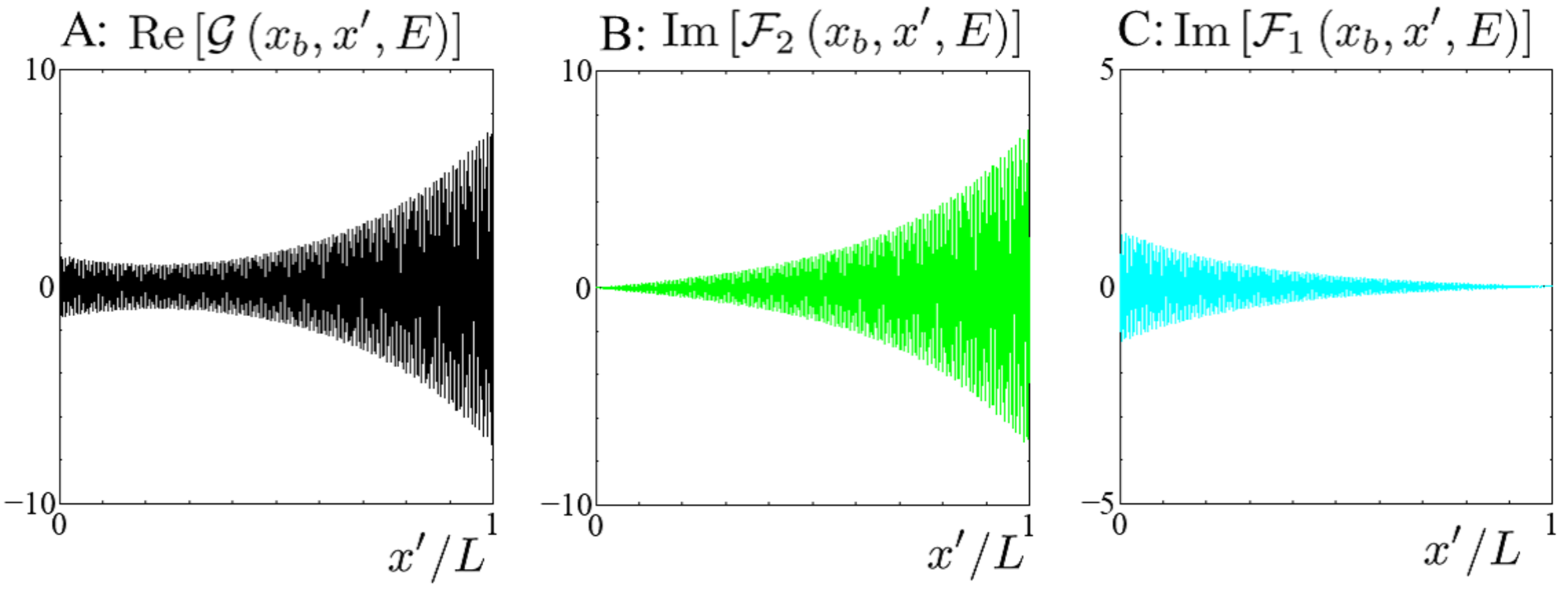}
\end{center}
\caption{Normal and anomalous Green's functions as  functions of $x'$ 
in a finite 1D spin-polarized $p$-wave superconductor 
at $Lk_{F}=1000.5\pi$.  Distinct panels correspond to:  A: ${\rm Re}[\mathcal{G}(x_{b},x',E)]$,  B: ${\rm Im}[\mathcal{F}_{2}(x_{b},x',E)]$, and 
C: ${\rm Re}[\mathcal{F}_{1}(x_{b},x',E)]$.
Parameters:   $x=x_{b}=L/1113$, $E/\Delta_{0}=0.015\Delta_{0}$, $\Delta_{0}=0.002\mu$, and $\delta=0.001\Delta_{0}$.}
\label{FiniteKitaev2}
\end{figure}

\begin{figure}[tb]
\begin{center}
\includegraphics[width=0.95\columnwidth]{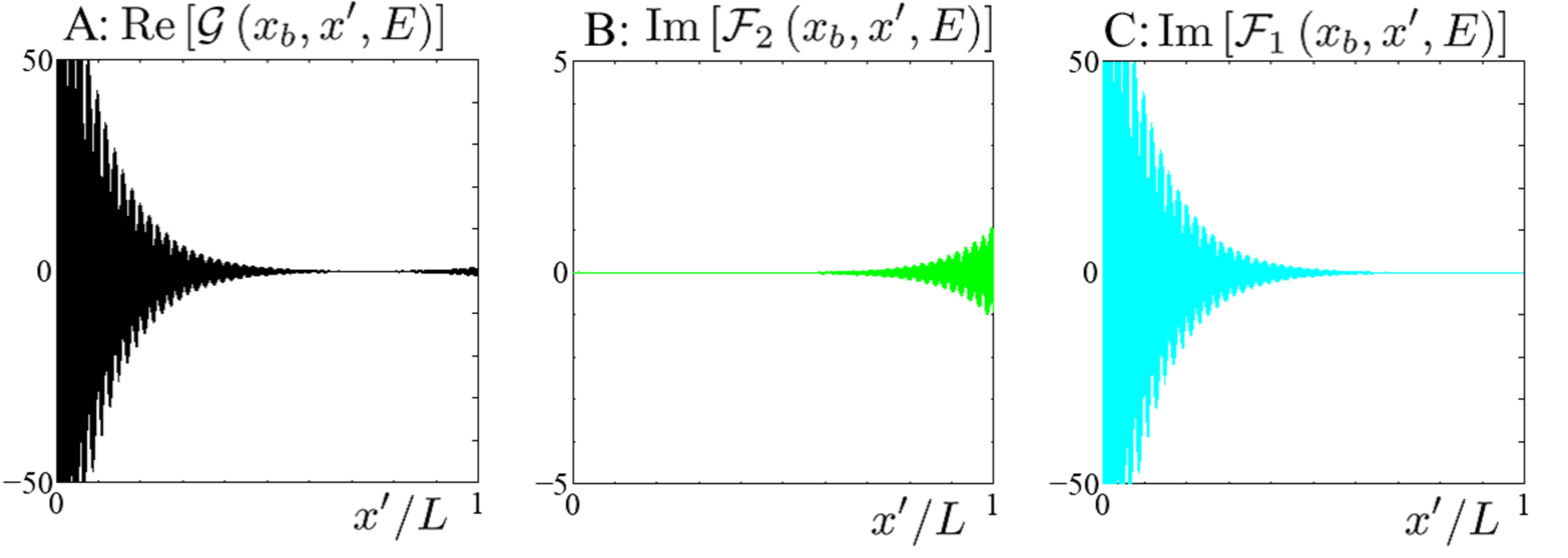}
\end{center}
\caption{Normal and anomalous Green's functions as  functions of $x'$ 
in a finite 1D spin-polarized $p$-wave superconductor 
at $Lk_{F}=3000.5\pi$.  Distinct panels correspond to:  A: ${\rm Re}[\mathcal{G}(x_{b},x',E)]$,  B: ${\rm Im}[\mathcal{F}_{2}(x_{b},x',E)]$, and 
C: ${\rm Re}[\mathcal{F}_{1}(x_{b},x',E)]$.
Parameters:   $x=x_{b}=L/1113$, $E/\Delta_{0}=0.015\Delta_{0}$, $\Delta_{0}=0.002\mu$, and $\delta=0.001\Delta_{0}$.}
\label{FiniteKitaev3}
\end{figure}

We next explore the Green's function components by fixing one spatial coordinate close to the left edge and evolving with the other coordinate as we move towards the bulk of the system. This is particularly useful for obtaining information about nonlocal correlations, which can provide further understanding of MZMs because they exhibit intrinsic spatial nonlocality. In Figs.\, \ref{FiniteKitaev1}, \ref{FiniteKitaev2}, and  \ref{FiniteKitaev3} we plot the spatial dependence of ${\rm Re}[\mathcal{G}(x_{b}, x',E)]$,  ${\rm Im}[\mathcal{F}_{2}(x_{b}, x',E)]$, 
and  ${\rm Im}[\mathcal{F}_{1}(x_{b}, x',E)]$ at $x_{b}=L/1113$ at $E=0.015\Delta_{0}+i\delta$, with $\delta=0.001\Delta_{0}$. Thus, sweeping the values of $x'$ provides information about the nonlocal properties of these Green's functions. Before describing the features of the Green's functions, we note that, at this energy, the real part of $\mathcal{G}(x_{b}, x',E)$  is larger than the imaginary term.   On the other hand, for $\mathcal{F}_{1}(x_{b}, x',E)$ 
and $\mathcal{F}_{2}(x_{b}, x',E)$ the situation is that their imaginary components are larger than their real counterparts.  In Fig.\,\ref{FiniteKitaev1}, we choose $k_{F}L=1000\pi$ where 
the wavefunction of an electron in the normal state 
vanishes at both $x=0$ and $x=L$.  Since we set $\Delta_{0}/\mu=500$, then we get  $L/\xi=\pi$ 
 with $\xi=\hbar v_{F}/\Delta_{0}$.   The first feature we identify is that  ${\rm Re}[\mathcal{G}(x_{b}, x',E)]$ and 
${\rm Im}[\mathcal{F}_{1}(x_{b}, x',E)]$ acquire equally large values,  which exceed those of ${\rm Im}[\mathcal{F}_{2}(x_{b}, x',E)]$. Moreover, we  observe that 
${\rm Re}[\mathcal{G}(x_{b}, x',E)]$, 
${\rm Im}[\mathcal{F}_{1}(x_{b}, x',E)]$, 
and ${\rm Im}[\mathcal{F}_{2}(x_{b}, x',E)]$ 
have an oscillatory behavior, with gradually decaying envelope functions as we move further from the left interface at $x=0$. The order of the period of the rapid oscillations is the inverse of $k_{F}$ and their amplitude decays with a decay length determined by $\xi$. 
When we move to $Lk_{F}=1000.5\pi$, which is slightly different from the condition for $k_{F}L$ discussed in the previous paragraph, the situation changes dramatically, as clearly seen in Fig.\,\ref{FiniteKitaev2}. We first note that the magnitudes of ${\rm Re}[\mathcal{G}(x_{b},x',E)]$ and 
${\rm Im}[\mathcal{F}_{1}(x_{b},x',E)]$ are  suppressed around $x' \sim 0$ as compared to the 
case for $Lk_{F}=1000$ in Fig.\,\ref{FiniteKitaev1}. Interestingly, the amplitudes of the rapid oscillations  is enhanced  with the increase of $x'$  for ${\rm Re}[\mathcal{G}(x_{b},x',E)]$ and 
${\rm Im}[\mathcal{F}_{2}(x_{b},x',E)]$.  On the other hand, for the OTE pair amplitude 
${\rm Re}[\mathcal{F}_{1}(x_{b},x',E)]$ we see that the amplitude of the rapid oscillations 
is suppressed.  
The sensitivity of the $L$ dependence of the Green's functions 
shown in Figs. \ref{FiniteKitaev1} and \ref{FiniteKitaev2} is a consequence of 
the  hybridization of the wave functions of  MZMs localized at the left and  right edges. 
We also note that the suppression of   ${\rm Re}[\mathcal{G}(x_{b},x',E)]$ and 
${\rm Im}[\mathcal{F}_{2}(x_{b},x',E)]$ can be seen at other values of $k_{F}L$, as we show in Fig.\,\ref{FiniteKitaev3} for   $k_{F}L=3000.5\pi$, which is again a result of the reduction of the spatial overlap between MZMs.  
The enhancement of the amplitudes of the rapid oscillations 
shown in Fig. \ref{FiniteKitaev2} is specific to 
1d $p$-wave model and never appears in corresponding 
$s$-wave model. \par

\subsubsection{At resonance}
By focusing on the regime without splitting in the LDOS, namely, at resonance, several simplifications are possible. In particular, this regime implies that we consider 
\begin{equation}
\label{specialcondition1}
\sin\left( \kappa^{+}L \right)=0\,,
\end{equation}
a condition that reflects having a zero-energy bound state $E_{b}=0$ and also reveals the formation of a resonant state. In this case, the following relations are also satisfied
\begin{equation}
\begin{split}
\cos\left( 2\kappa^{+}L \right)&=1\,,\\
\sin\left( 2\kappa^{+}L \right)&=0\,,\\
\sin\left( \kappa^{+} \left( \mid x - x' \mid -2L \right) \right)&=
\sin \left( \kappa^{+} \mid x - x' \mid \right)\,,\\
\cos\left( \kappa^{+} \left( \mid x - x' \mid -2L \right) \right)&=
\cos \left( \kappa^{+} \mid x - x' \mid \right)\,.
\end{split}
\end{equation}
Then, the retarded Greens function $G^{r}(x,x',E)$ is given by a compact expression that reads
\begin{equation}
\label{simplifiedGreens}
\begin{split}
 G^{r}\left( x,x',E \right) 
&=\left( \frac{m}{i \hbar^{2} k_{F}} \right) 
\left( 1 - \exp \left( 2 i \gamma L \right) \right)^{-1} \\
&\times  
\left\{
\frac{E}{\Omega}\cos\left[k_{F} \mid x-x' \mid \right]
\left( \exp\left[ i\gamma \mid x -x' \mid \right]
+ \exp\left[ -i\gamma \left( \mid x -x' \mid -2L \right) \right] 
\right) X_{1}
\right.
\\
&-
\frac{E}{\Omega}\cos\left[k_{F} \left( x + x'  \right) \right]
\left( \exp\left[ i\gamma \left( x + x' \right) \right]
+ \exp\left[ -i\gamma \left(  x + x' -2L \right) \right] 
\right) X_{1}
\\
&-
\frac{2\Delta^{2}}{E \Omega}
\sin\left(k_{F}x \right)\sin\left(k_{F} x' \right)
\left( \exp\left[ i \gamma \left( x +x' \right)\right]
+ \exp\left[ -i \gamma \left( x +x' -2L \right)\right]
\right) X_{1}
\\
&+ i \sin\left[k_{F} \mid x - x' \mid \right]
\left( \exp\left( i\gamma \mid x-x' \mid \right)
-\exp\left( -i\gamma \left[
\mid x -x' \mid -2L \right] \right) \right) X_{2}
\\
&-i \sin \left[k_{F} \left(x +x ' \right) \right]
\left( \exp\left( i\gamma \left( x + x' \right) \right)
-\exp\left( -i\gamma \left(
 x  + x' -2L \right) \right) \right) X_{2}
\\
&+
\frac{i \Delta}{\Omega}
\sin\left[k_{F} \left(x - x' \right) \right]
\left( \exp\left( i\gamma \mid x-x' \mid \right)
+\exp\left( -i\gamma \left[
\mid x -x' \mid -2L \right] \right) 
\right.
\\
&-
\left. 
\exp\left( i\gamma \left( x + x' \right) \right)
-\exp\left( -i\gamma \left(
 x  + x' -2L \right) \right) 
\right)
Y_{1}
\\
&-
\left.
\frac{2\Delta}{E}
\sin\left(k_{F}x \right)\sin\left(k_{F} x' \right)
\left( \exp\left[ i \gamma \left( x +x' \right)\right]
-\exp\left[ -i \gamma \left( x +x' -2L \right)\right]
\right) Y_{2}
\right\}\,.
\end{split}
\end{equation}
Thus, for $L \rightarrow \infty$, we can see that Eq.\,(\ref{simplifiedGreens})   reduces to the expression for a semi-infinite $p$-wave superconductor presented in  Eq.\,(\ref{semi-infiniteKitaev}). Furthermore, it is important to remark that, at  special lengths   $L$ satisfying Eq.\,(\ref{specialcondition1}), the Green's function for a finite $p$-wave superconductor can be   expressed by  the semi-infinite Green's functions defined for $x>0$  and that for $x<L$. 

\subsection{Summary}
We have presented a detailed discussion about the properties of retarded Green's functions in 1D spin-polarized $p$-wave superconductors. In particular,  we have focused on bulk, semi-infinite, and finite length systems, and demonstrated that their respective retarded Green's functions reveal interesting functionalities due to the formation of MZMs. In the case of the bulk system, we have shown that the symmetry of the superconducting pair correlations exhibit the expected even-frequency spin-triplet $p$-wave symmetry which corresponds to the ETO pair symmetry class, which reflects the nature of the pair potential. Interestingly, for a semi-infinite system, we have found that the LDOS and OTE pairing at the boundary exhibit a large peak at zero energy due to the emergence of a MZM, exhibiting an exponential and oscillatory decay, and can be used as detection probes. We have also highlighted that having a largely dominant OTE pairing means that the superconducting pair symmetries in the presence of MZMs are distinct from those of the bulk, which reveals yet another signature that can be further exploited for detecting the still elusive Majorana quasiparticles in condensed matter systems. In the case of a finite system, we have further shown that having a finite length produces MZMs decaying from both ends which can then spatially overlap and produce finite oscillations around zero energy in both the OTE and LDOS.  Even though very large systems are desirable for achieving MZMs and for their potential applications, having finite systems, although unavoidable, seems to be the key for probing the spatial nonlocality of MZMs and thus for distinguishing them from other topologically trivial zero-energy states.

\section{Detecting Majorana zero modes via tunneling spectroscopy}
\label{section5}
This part is devoted to analyze how MZMs and generic ZESABSs manifest when electrons tunnel across a normal-superconductor (NS) junction, where N represents a normal metal lead while S is a superconductor with a given pair potential.  In particular, we will focus on conductance, which is perhaps the simplest transport observable and represents the central concept in tunneling spectroscopy \cite{tanaka1991theory,kashiwaya1995symmetry,kashiwaya1995tunneling,TK95,YTK97,PhysRevLett.98.077001}.  For completeness, we first address tunneling spectroscopy when S represents a conventional superconductor with spin-singlet $s$-wave pair potential. Then, we inspect tunneling spectroscopy for superconductors with spin-singlet $d$-wave and spin-triplet $p$-wave pair potentials. Before we go further, we note that to discuss tunneling effects, we must take into account elementary scattering processes occurring at the  NS interface due to the incidence of quasiparticles.  To treat these scattering processes, we will employ the BdG equations given by Eq.\,(\ref{generalizedBdG}), where superconducting systems with a  spatially non-uniform pair potential, 
as junctions, can be investigated.

\subsection{Tunneling spectroscopy in conventional superconductors}
\label{section51}
Conventional superconductors exhibit a pair potential that is spin-singlet 
and isotropic in space ($s$-wave). Thus, considering the 1D case for simplicity but without loss of generality, the BdG equations given by Eq.\,(\ref{generalizedBdG}) can be written as,
\begin{equation}
\begin{split}
E u(x) &= 
\left(-\frac{\hbar^{2}}{2m}\frac{d^{2}}{dx^{2}} + U(x)-\mu\right)u(x) 
+ \Delta(x)v(x)
\\
E v(x)& = 
-\left(-\frac{\hbar^{2}}{2m}  \frac{d^{2}}{dx^{2}} + U(x)-\mu\right)v(x) 
+ \Delta^{*}(x)u(x)
\end{split}
\end{equation}
where $u (v)$ are electron (hole) components of the wavefunction,  $\mu$ is the chemical potential,  $U(x)$ represents a one-body potential, and $\Delta(x)$ is the superconducting spin-singlet $s$-wave pair potential. Since we are interested in inspecting tunneling across an NS junction, the pair potential is chosen to be 
\begin{equation}
\Delta\left(x \right)=
\left \{
\begin{array}{ll}
0 & x \leq 0 \\
\Delta_{0} & x>0 
\end{array}
\right.
\end{equation}
which shows that the NS interface is located at $x=0$ and that superconductivity in N is absent. Moreover, the nature of the interface at $x=0$ is determined by the profile of the one-body potential which we here consider it to be just a delta barrier, namely,  $U\left(x \right)=H_{b}\delta\left(x \right)$, with $H_{b}$ the strength of the barrier. 

\begin{figure}[b]
\begin{center}
   \includegraphics[width=0.7\columnwidth]{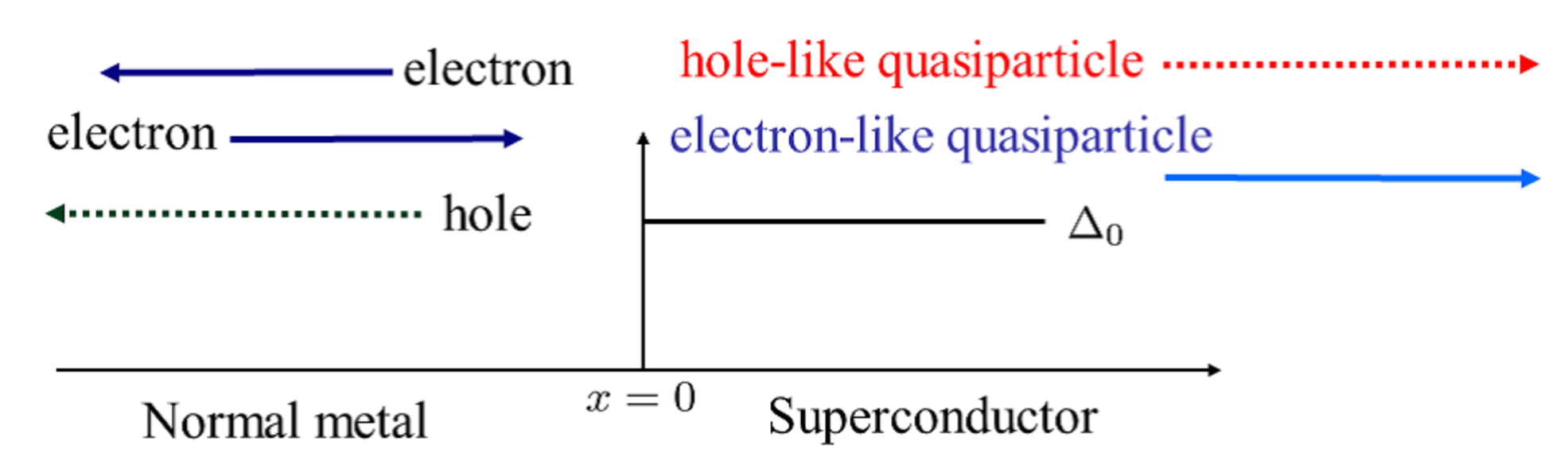} 
\end{center}
\caption{Scattering processes at the interface of a 1D NS junction  along $x$, 
where the direction of the arrow denotes 
the direction of the group velocity and the interface is located at $x=0$: 
an incident electron (solid dark blue arrow) from the normal metal 
can be reflected back into the normal metal as an electron (solid dark blue arrow) and also as a hole (dotted black arrow),  known as normal and Andreev reflections, respectively. 
Moreover, the incident electron can be transmitted into the superconductor as electron-like (solid light blue arrow) and hole-like (dotted red arrow) quasiparticles, processes known as normal transmission and crossed Andreev reflection, respectively.}
\label{fig:1dscattering}
\end{figure}

\begin{figure}[tb]
\begin{center}
\includegraphics[width=0.8\columnwidth]{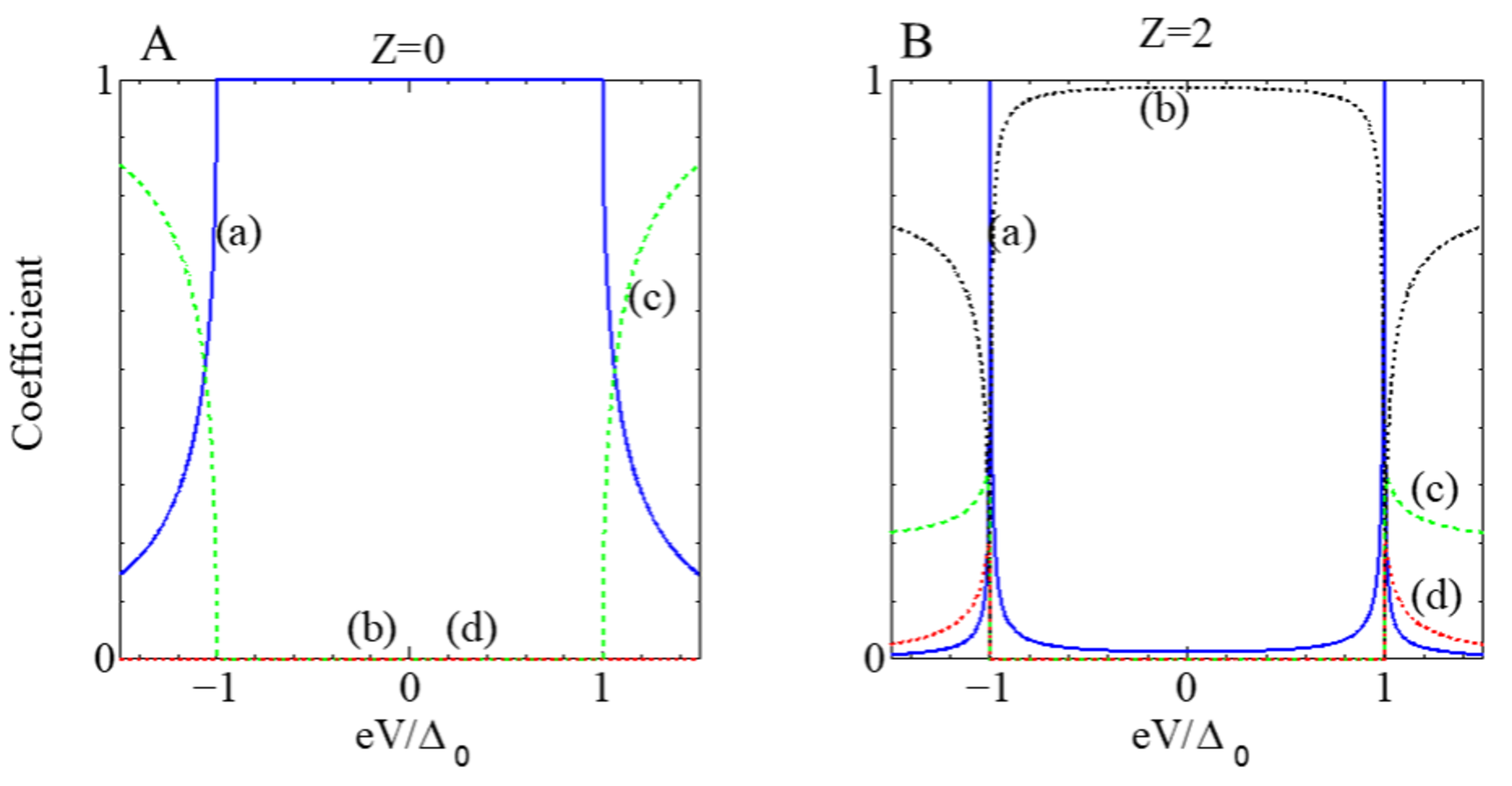}
\end{center}
\caption{Scattering probabilities across a NS interface with S being a conventional spin-singlet $s$-wave superconductor. (a): Andreev reflection probability, 
(b): Normal reflection probability, (c): Transmission probability of 
electron-like quasiparticle, 
(d): Transmission probability of hole-like quasiparticle. 
(A)$Z=0$, (B)$Z=2$. }
\label{1dswaveABCD}
\end{figure}

We are now in a position to discuss transport across the NS junction, where multiple scattering processes occur and are determined by incident particles at the interface \cite{BTK82}. For instance, an incident electron from N  can be reflected back to N as an electron or hole, processes known as normal and Andreev reflections. The Andreev reflection is the most fundamental transport phenomenon in superconducting junctions  \cite{andreev1964thermal,andreev1965thermal} and the key concept to understand the superconducting proximity effect \cite{pannetier2000andreev,klapwijk2004proximity}, see also Ref.\,\cite{asano2021andreev}. Moreover, the incident electron can be also transmitted to S into electron-like and hole-like quasiparticles. These processes  are schematically shown in 
Fig.\,\ref{fig:1dscattering}; similar processes happen for incident holes from N or quasi-electrons (quasi-holes) from S. Thus, for an incident electron from N, the corresponding wave function is given by 
\begin{equation}
\label{wavefunctionNSjunction}
\Psi\left(x \right)=
\left \{
\begin{array}{ll}
\exp \left(ip^{+}x \right)
\begin{pmatrix}
1 \\
0
\end{pmatrix}
+ 
a\left(E \right) \exp \left(ip^{-}x\right)
\begin{pmatrix}
0 \\
1
\end{pmatrix}
+ 
b\left(E \right) \exp \left(-ip^{+}x\right)
\begin{pmatrix}
1 \\
0
\end{pmatrix}\,,
& 
x \leq 0\,, \\
c\left( E \right)
\exp \left(ik^{+}x\right)
\begin{pmatrix}
u_{0} \\
v_{0}
\end{pmatrix}
+
d\left( E \right)
\exp \left(-ik^{-}x \right)
\begin{pmatrix}
v_{0} \\
u_{0}
\end{pmatrix}\,,
&
x>0\,,
\end{array}
\right.
\end{equation}
where the coefficients $a(E)$, $b(E)$, $c(E)$, and $d(E)$ represent to the amplitudes of  Andreev reflection,   normal reflection,   electron transmission, and hole transmission, respectively. These coefficients are found from the boundary conditions  by integrating the BdG equations around $x=0$, which give 
\begin{equation}
\label{boundarycondition1d}
\begin{split}
[\Psi\left(x\right)]_{x=0_{-}}&=[\Psi\left(x\right)]_{x=0_{+}}\,,\\
\Big[\frac{d}{dx}\Psi\left(x \right)\Big]_{x=0+} 
- \Big[\frac{d}{dx}\Psi\left(x \right)\Big]_{x=0_{-}}
&=\frac{2mH_{b}}{\hbar^{2}}[\Psi\left(x \right)]\Big|_{x=0}\,,
\end{split}
\end{equation}
where $k^{\pm}=\sqrt{2m(\mu \pm \Omega)/\hbar^{2}}$,  $p^{\pm}=\sqrt{2m(\mu \pm E)/\hbar^{2}}$, 
$u_{0}=\sqrt{(1 + \Omega/E)/2}$,  $v_{0}=\sqrt{(1 - \Omega/E)/2}$, and
\begin{equation}
\Omega 
\equiv \lim_{\delta \rightarrow 0} 
\sqrt{\left(E + i\delta \right)^{2} - \Delta_{0}^{2}}
= 
\displaystyle
\left \{
\begin{array}{ll}
\sqrt{E^{2} - \Delta_{0}^{2}} & \quad E  \geq \Delta_{0} \\
i \sqrt{\Delta_{0}^{2} - E^{2}} &  \quad -\Delta_{0} 
\leq E \leq \Delta_{0} \\ 
-\sqrt{E^{2} - \Delta_{0}^{2}} & \quad E  \leq -\Delta_{0} 
\end{array}
\right.
\label{Omega}
\end{equation}
with $\delta>0$ being an infinitesimal positive number. For simplicity and also for pedagogical purposes, we now focus on the Andreev approximation,  where the chemical potential is the largest energy scale, namely, $\mu \gg |E|$ and  $\mu \gg |\Omega|$. This approximation then implies that 
$p^{+} \sim p^{-} \sim k^{+} \sim k^{-} \sim k_{F}$, which then allow us to obtain simple and compact expressions for the coefficients of the wavefunction given by Eq.\,(\ref{wavefunctionNSjunction}), which are given by \cite{BTK82}
\begin{equation}
\label{reflectioncoefficientswave}
\begin{split}
a\left( E \right)&=\frac{\Gamma}{(1+Z^{2})-Z^{2}\Gamma^{2}}\,,\\
b\left( E \right)&=\frac{iZ(1-iZ)(\Gamma^{2}-1)}{(1 + Z^{2}) - Z^{2}\Gamma^{2}}\,,\\
c\left(E \right)&=\frac{1}{u_{0}}\frac{1-iZ}{(1+Z^{2})-Z^{2}\Gamma^{2}}\,,\\
d\left( E \right)&=\frac{1}{u_{0}}\frac{iZ\Gamma}{(1 + Z^{2}) - Z^{2}\Gamma^{2}}
\end{split}
\end{equation}
where $\Gamma=v_{0}/u_{0}$ and $Z=m H_{b}/\hbar^{2} k_{F}$. We can now obtain the probabilities of the scattering processes, which are defined by 
\begin{equation}
\label{reflectioncoefficientswavex}
\begin{split}
A& \equiv |a(E)|^{2}\,, \\  
B& \equiv |b(E)|^{2}\,, \\ 
C& \equiv (|u_{0}|^{2} - |v_{0}|^{2})
|c(E)|^{2}\,, \\ 
D& \equiv(|u_{0}|^{2} - |v_{0}|^{2})
|d(E)|^{2}\,,
\end{split}
\end{equation}
where  $A$, $B$, $C$, and $D$ characterize the probabilities of Andreev reflection,  of normal reflection, of transmission as an electron-like quasiparticle, and that of transmission as a hole-like quasiparticle (or crossed Andreev reflection). The conservation of current then implies that these probabilities satisfy the following condition
\begin{equation}
A + B + C + D =1\,,
\end{equation}
which also reflects the unitarity of the so-called scattering matrix \cite{datta1997electronic}. We can now write the scattering probabilities as
\begin{eqnarray}
A&=&\frac{\sigma^{2}_{N}|\Gamma|^{2}}
{|1 - (1-\sigma_{N})\Gamma^{2}|^{2}}\,, 
\label{probabilitysA}
\\ 
B&=& \frac{(1 -\sigma_{N})|1 - \Gamma^{2}|^{2}}
{|1 - (1-\sigma_{N})\Gamma^{2}|^{2}}\,, 
\label{probabilitysB}
\\
C&=&\frac{\sigma_{N}(1 - |\Gamma|^{2})}{|1 - (1-\sigma_{N})\Gamma^{2}|^{2}}\,, 
\label{probabilitysC}
\\ 
D&=& \frac{\sigma_{N}(1 -\sigma_{N})|\Gamma|^{2}(1 - |\Gamma|^{2})}
{|1 - (1-\sigma_{N})\Gamma^{2}|^{2}}\,,
\label{probabilitysD}
\end{eqnarray}
where we have used the transmissivity in the normal state $\sigma_{N}=1/(1+Z^{2})$ with $Z$ characterizing the strength of the barrier at the NS interface. To visualize the energy dependence of the scattering probabilities given by Eqs.\,(\ref{probabilitysA}) to (\ref{probabilitysD}), in Fig.\,\ref{1dswaveABCD} we plot them as a function of $E$ at $Z=0$ and $Z=2$. For $Z=0$, we have $\sigma_{N}=1$ which represents full transmissivity and the interface is fully transparent. In this full transparent regime, the normal reflection and hole transmission probabilities vanish for all energies, namely, $B=D=0$, as seen in dotted black and dotted red curves denoted by (b,d) in Fig.\,\ref{1dswaveABCD}(A). We also see that perfect Andreev reflection occurs for subgap energies, where $A=1$ and $C=0$  for  $|E| \leq \Delta_{0}$. In the case of low transparency, $Z\neq0$ induces reduced values of $\sigma_{N}$, which has serious implications on transport across the NS interface. For instance, the Andreev reflection probability  $A$, which is proportional to $\sigma_{N}^{2}$, is seriously suppressed for $Z=2$ except at the gap edges  $E=\pm\Delta_{0}$, see blue curve in Fig.\,\ref{1dswaveABCD}(B).  In contrast, the normal reflection probability $B$ is dominant in almost all energies except for $E=\pm \Delta_{0}$, see black dotted curve in Fig.\,\ref{1dswaveABCD}(B). We finally point out that in both cases of $Z=0$ and $Z\neq0$, the electron and hole transmission probabilities vanish, where  $C=0$ and $D=0$ for $|E|<\Delta_{0}$, an effect that occurs because quasiparticles can not enter into superconductor region as a traveling wave, see red and green curves in Fig.\,\ref{1dswaveABCD}(A,B).

In what follows, we inspect the tunneling conductance which can be obtained using the probabilities given by Eqs.\, (\ref{probabilitysA}) and (\ref{probabilitysB})
using the so-called BTK formula \cite{BTK82}. The essence of this formula is that the conductance of the junction $\sigma_{S}$  is determined by the interface resistance. Thus, if the number of channels through the interface is fixed, the resulting conductance is determined by the effective transparency at the interface. At sufficiently low temperatures,  
the differential conductance $dI_{NS}/dV$ is   determined by the tunneling conductance 
$E=eV$ is given by 
\begin{equation}
\frac{dI_{NS}}{dV}=\frac{2e^{2}}{h}\sigma_{S}(E=eV),
\label{dIdV}
\end{equation} 
where the conductance is obtained as $\sigma_{S}(E)= 1 + A -B$. Then, by using $A$ and $B$ from Eqs.\,(\ref{probabilitysA}) and (\ref{probabilitysB}), we get
\begin{figure}[tb]
\begin{center}
\includegraphics[width=0.5\columnwidth]{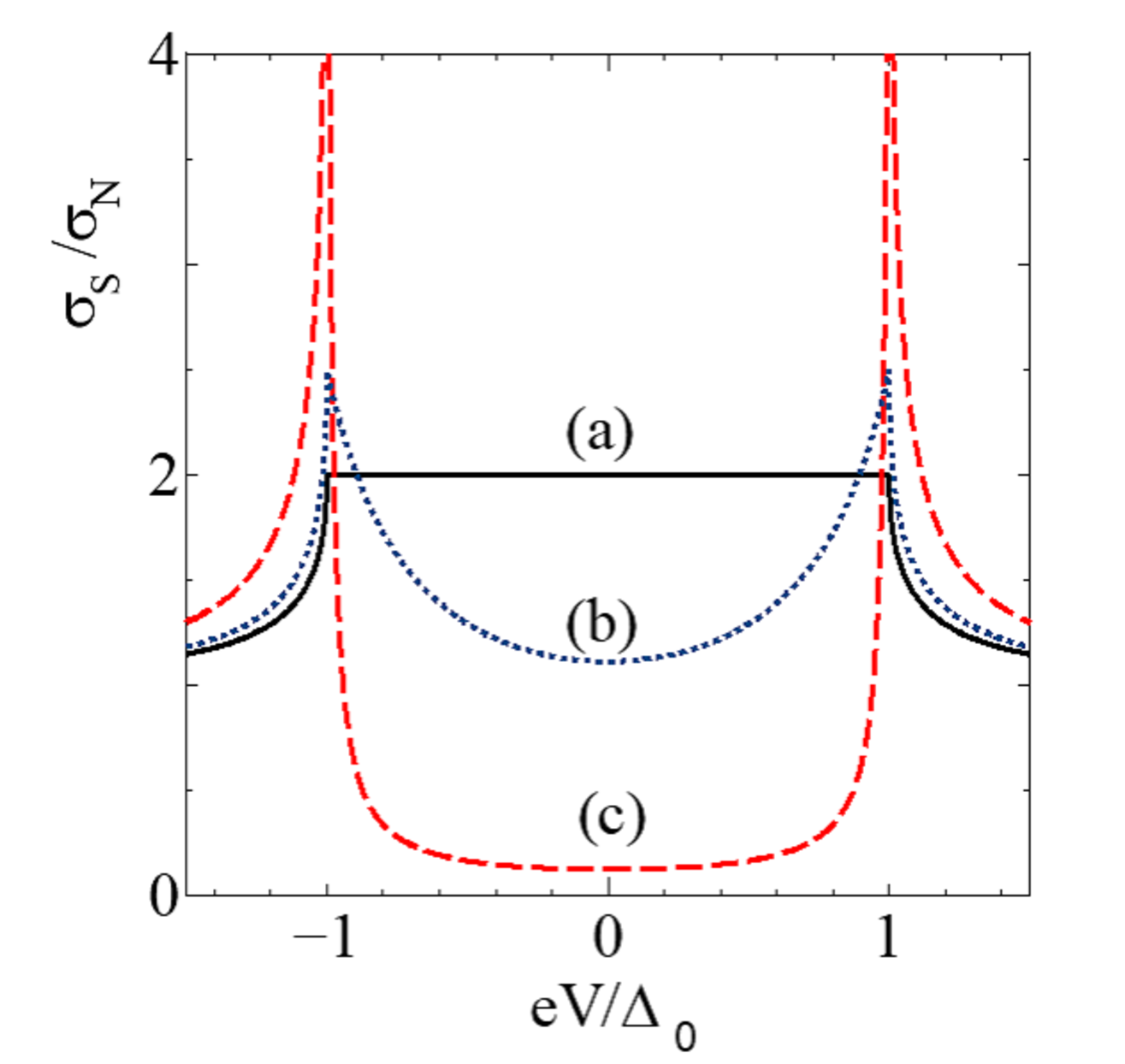}
\end{center}
\caption{Normalized conductance across an NS junction with S being a conventional spin-singlet $s$-wave superconductor. Different curves correspond to (a) $Z=0$, (b) $Z=0.5$, and (c) $Z=2$.}
\label{1dswave}
\end{figure}
\begin{equation}
\sigma_{S}(E) = \sigma_{N} 
\displaystyle\frac{1 + \sigma_{N} \mid \Gamma \mid^{2} - \left(1 - \sigma_{N} \right) \mid \Gamma \mid^{4}}
{ \mid 1 - \left(1 - \sigma_{N} \right)\Gamma^{2} \mid^{2}}
=
\begin{dcases}
\displaystyle\frac{2 \sigma_{N}^{2}\Delta_{0}^{2} }
{  (\sigma_{N} - 2)^{2}\Delta_{0}^{2} - 4(1-\sigma_{N})E^{2}}\,, 
& \mid E \mid \leq \Delta_{0}\,, \\
\displaystyle\frac{2 \sigma_{N} E }
{  (2 - \sigma_{N})\Omega  + \sigma_{N} E}\,, 
& \mid E \mid > \Delta_{0}\,,
\end{dcases}
\label{BTKtunnelconductance}
\end{equation}
where $\Omega$ is given by Eq.\,(\ref{Omega}). In Fig.\,\ref{1dswave}, we present the normalized conductance obtained here as a function of $eV$ for different values of $Z$ that describe full, intermediate, and tunnel transparencies of the interface.   For $\sigma_{N}=1 (Z=0)$, the conductance is quantized for subgap energies such that   $\sigma_{S}(E)=2$ is satisfied for $|E| \leq \Delta_{0}$, see solid black curve in Fig.\,\ref{1dswave}. By taking account of the spin degeneracy, having $\sigma_{S}(E)=2$ means that the differential conductance  in Eq.\,(\ref{dIdV})  reaches $4e^{2}/h$.  As $Z$ takes finite values, $\sigma_{N}$ reduces, which produces lower conductance values, as depicted by blue dotted and red dashed curves in Fig.\,\ref{1dswave}. For sufficiently large $Z$, the normal transmissivity $\sigma_{N}$ is very small inside the superconducting gap, as a result of $\sigma_{S}(E)$ being proportional to $\sigma_{N}^{2}$. Thus, the differential conductance at subgap energies can acquire values within the range of $0$ up to $4e^{2}/h$. Note also that for energies outside the superconducting gap, $|E| > \Delta_{0}$,  the conductance $\sigma_{S}(E)$ approaches   the 
normal state value $\sigma_{N}$.  To summarize our discussion for very large $Z$, and thus low normal transmissivity  ($\sigma_{N} \ll 1$), it is possible to obtain a simple expression for $\sigma_{S}(E)$, which reads
\begin{equation}
\sigma_{S}(E)/ \sigma_{N} \approx  {\rm Re}[E/\Omega]\,.
\end{equation}
This expression means that tunneling conductance normalized by its value in 
the normal state expresses the DOS of a bulk superconductor in the low transparent junction and is insensitive to the phase of the pair potential. In what follows, we will see that this situation is dramatically changed for NS junctions with unconventional superconductors.  

\subsection{Tunneling spectroscopy in unconventional superconductors}
\label{section52}
\begin{figure}[b]
\begin{center}
\includegraphics[width=9.5cm,clip]{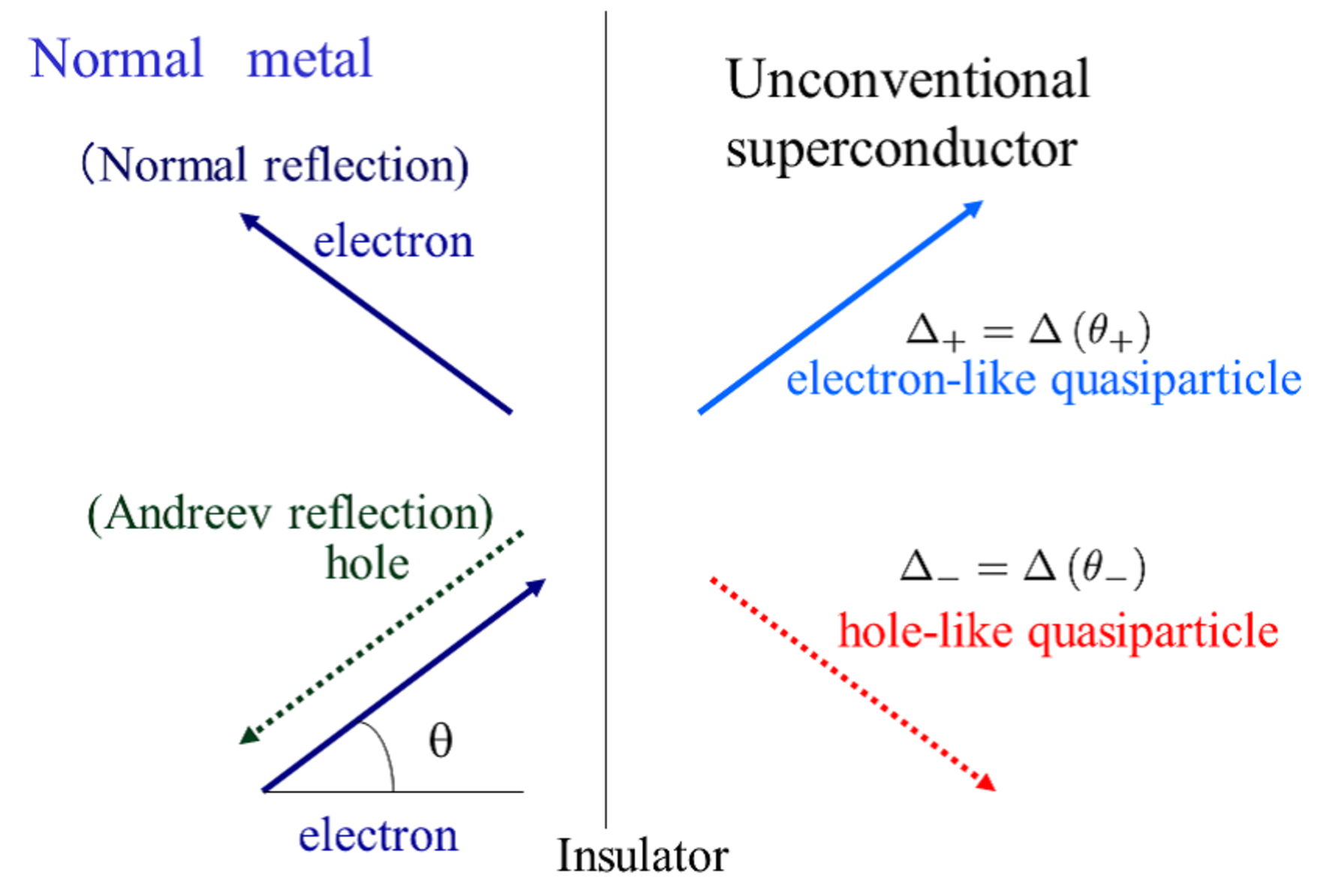}
\end{center}
\caption{Scattering processes at a two-dimensional normal metal/unconventional 
superconductor junction along $x$, where the direction of the arrow denotes the direction of the group velocity and the interface is located at $x=0$: an incident electron with incident angle $\theta$ from the normal metal can be reflected back into the normal metal as an electron and also as a hole, processes 
known as normal and Andreev reflections. Moreover, the incident electron can be transmitted into the superconductor as electron-like and hole-like quasiparticles, processes known as normal transmission and crossed Andreev reflection, respectively.  The pair potential of the unconventional superconductor is parametrized by the incident angle as $\theta_{+}=\theta$, $\theta_{-}=\pi-\theta$, which represent pair potentials for electron-like and hole-like quasiparticles.}
\label{fig:81}
\end{figure}

In this part we analyze conductance in NS junctions with  N being a semi-infinite normal metal and S  being a semi-infinite unconventional superconductor. In general, we proceed as in the previous subsection taking into account that unconventional superconductors have anisotropic pair potentials. 
We consider a 2D NS junctions along $x$, with a flat interface  located at $x=0$ and N (S) defined for $x<0$ ($x>0$), see Fig.\,\ref{fig:81}. Moreover, we assume that the interface is determined by an insulating barrier, which here we model by a delta barrier potential $H_{b}\delta(x)$; the perpendicular direction along $y$ is translationally invariant. With this information, we are now in a position to employ the methodology presented in Section \ref{section2} to obtain the scattering states whose coefficients will determine the charge transport in the same spirit as that discussed in the previous subsection.

Since the NS junction now is a 2D system, there is an incident angle with respect to $x$ axis, denoted by $\theta$, at which quasiparticles hit the NS interface, see Fig.\,\ref{fig:81}. Now, we use the BdG equations in the quasiclassical limit to obtain the scattering states and inspect conductance across the NS interface, see Eqs.\,(\ref{envelopefunction}) and the discussion at the beginning of subsection \ref{subsection2d}. In this limit, for an incident electron, the total wavefunction becomes,

\begin{equation}
\label{NSjunctionUnconvNS}
\Psi(\theta,x)=
\left \{
\begin{array}{ll}
\Bigg[
\begin{pmatrix}
1\\
0
\end{pmatrix}
\exp\left(i\gamma_{0} x\right)+
a\left(E,\theta\right)
\begin{pmatrix}
0\\
1
\end{pmatrix}
\exp\left(-i\gamma_{0} x\right)
\Bigg]\exp\left(ik_{F}{\rm cos}\theta x\right)&\\
+
b\left(E,\theta\right)
\begin{pmatrix}
1\\
0
\end{pmatrix}
\exp\left(-i\gamma_{0} x\right)\exp\left(-ik_{F}{\rm cos}\theta x\right)&\,,\quad x\leq0\\
c\left(E,\theta\right)
\begin{pmatrix}
u_{+}\\
v_{+}
\end{pmatrix}
\exp\left(i\gamma_{+} x\right)\exp\left(ik_{F}{\rm cos}\theta x\right)&\\
+
d\left(E,\theta\right)
\begin{pmatrix}
v_{-}\\
u_{-}
\end{pmatrix}
\exp\left(i\gamma_{-} x\right)\exp\left(-ik_{F}{\rm cos}\theta x\right)&\,,\quad x>0
\end{array}
\right.
\end{equation}
where $u_{\pm}$, $v_{\pm}$, $\gamma_{0}$, and $\gamma_{\pm}$ are given by Eqs.\,(\ref{coherenceuv}), (\ref{gammapm}), and (\ref{Omegapm}). The coefficients in Eqs.\,(\ref{NSjunctionUnconvNS})   represent the following processes: $a$, $b$, $c$, and $d$ correspond to  Andreev reflection, normal reflection, normal transmission, and   Andreev transmission (or crossed Andreev reflection), respectively. To fully determine the wavefunction $\Psi(\theta,x)$, we still need to find the coefficients $a$, $b$, $c$, and $d$, which is done by employing the corresponding boundary condition at the NS interface. This condition is given by 
\begin{equation}
\begin{split}
{\Psi}(\theta,x) \Big|_{x=0_{-}}
&={\Psi}(\theta,x) \Big|_{x=0_{+}}\,,\\
\frac{d}{dx} {\Psi}\left(\theta,x\right)\Big|_{x=0_{+}}
-\frac{d}{dx} {\Psi}\left(\theta,x \right)\Big|_{x=0_{-}}
&=\frac{2mH_{b}}{\hbar^{2}}{\Psi}\left(\theta,x \right)\Big|_{x=0_{+}}\,.
\end{split}
\end{equation}

Then, by taking $k_{Fx}=k_{F}\cos\theta \gg |\gamma_{0}|$, 
$k_{Fx} \gg |\gamma_{\pm}|$, the coefficients 
$a(E,\theta)$, $b(E,\theta)$, 
$c(E,\theta)$, and $d(E,\theta)$ 
acquire simpler expressions that read,
\begin{equation}
\label{reflection_coefficient_general}
\begin{split}
a\left(E,\theta \right)&=\frac{\sigma_{N}\left(\theta \right) \Gamma_{+}}
{1 - \left[ 1 - \sigma_{N}\left(\theta \right) \right] \Gamma_{+} \Gamma_{-}}\,, \\ 
b\left(E,\theta \right)&=
\frac{i Z_{\theta}}{1 + iZ_{\theta}} 
\frac{\Gamma_{+}\Gamma_{-} -1}
{1 - \left[1 - \sigma_{N}\left(\theta \right) \right]
\Gamma_{+} \Gamma_{-}}\,,\\
c\left(E,\theta \right)&=
\frac{1}{\left(1 + iZ_{\theta}\right)u_{+}} 
\frac{\Gamma_{+}\Gamma_{-} -1}
{1 - \left[1 - \sigma_{N}\left(\theta \right) 
\right] \Gamma_{+} \Gamma_{-}}\,,\\
d\left(E,\theta \right)&=
\frac{i Z_{\theta}\Gamma_{+}}
{\left( 1 + Z^{2}_{\theta} \right) u_{-}} 
\frac{\Gamma_{+}\Gamma_{-} -1}
{1 - \left[1 - \sigma_{N}\left(\theta \right) \right] \Gamma_{+} \Gamma_{-}}\,,
\end{split}
\end{equation}
where $Z_{\theta}=Z/\cos\theta$, with $Z=mH_{b}/(\hbar^{2}k_{F})$  
and
\begin{equation}
\label{TKformulaGamma}
\begin{split}
\Gamma_{+}&=\frac{v_{+}}{u_{+}}=\frac{\Delta^{*}(\theta_{+})}{E + \Omega_{+}}\,,\\
\Gamma_{-}&=\frac{v_{-}}{u_{-}}=\frac{\Delta(\theta_{-})}
{E + \Omega_{-}}, \\
\sigma_{N}(\theta)&=\frac{\cos^{2}\theta}{Z^{2}+\cos^{2}\theta}\,.
\end{split}
\end{equation}

After finding the coefficients of the total wavefunction in Eq.\,(\ref{NSjunctionUnconvNS}), we are now in a position to obtain the tunneling conductance, which now needs to be averaged over the angles $\theta$. Thus, tunneling conductance, normalized by its value in the normal state, is obtained as
\begin{equation}
\sigma_{T}\left(E \right)=
\frac{\int^{\pi/2}_{-\pi/2}d\theta 
\sigma_{S}(E,\theta)
\cos\theta}
{\int^{\pi/2}_{-\pi/2} d\theta \sigma_{N}\left(\theta \right) \cos\theta}
\label{TKformula0}
\end{equation}
where $\sigma_{S}$ and $\sigma_{R}$ are obtained as
\begin{equation}
\sigma_{S}(E,\theta)
=\sigma_{N}\left(\theta \right)\sigma_{R}\left(E,\theta \right)
=1 + \mid a\left(E,\theta \right)\mid^{2} - 
\mid b\left(E,\theta \right)\mid^{2}
\label{Andreevcondunctanceformula}
\end{equation}
and 
\begin{equation}
\sigma_{S}\left(E,\theta \right) 
=\frac{ 1 + \sigma_{N}\left(\theta \right) \mid \Gamma_{+}\mid^{2}
- \left[1 - \sigma_{N}\left(\theta \right) \right]\mid\Gamma_{+}\Gamma_{-}\mid^{2} }
{\mid 1 - 
\left[1 - \sigma_{N}\left(\theta \right)\right]
\Gamma_{+}\Gamma_{-} \mid^{2}},
\label{TKformula}
\end{equation}
with $E=eV$  \cite{TK95}.  In this formula, 
$\sigma_{S}(E,\theta)$  expresses the effective transmissivity in the  superconducting state. An interesting fact to note in Eq.\,(\ref{TKformula}) is that it reveals the formation of the SABSs we studied in subsection \ref{subsection2d}.  In fact,  the condition for the formation of SABSs is given by the  zeroes of the conductance denominator in Eq. \,
(\ref{TKformula}) when the N and S regions are separated  $\sigma_{N}(\theta) \rightarrow 0$ \cite{TK95,KT96,kashiwaya00}. This permits us to find the bound state condition equal to $1=\Gamma_{+}\Gamma_{-}$ which is exactly the same as the one given by Eq.\,(\ref{SABSrelation}) found in subsection \ref{subsection2d}. Thus, the bound state energy found from the zeroes in the denominator of the conductance is $E_{b}=\Delta \,{\rm cos}(\delta_{\phi}/2)\,{\rm sgn}[{\rm sin}(\delta_{\phi}/2)] $, which is the same as the expression given by  Eq.\,(\ref{ABSUNSC}) and we assumed   $|\Delta_{\pm}|=\Delta_{0}$, with $\Delta_{\pm}=\Delta(\theta_{\pm})$; here also $\delta_{\phi}=\phi_{-}-\phi_{+}$ and ${\rm exp}(i\phi_{\pm})=\Delta_{\pm}/|\Delta_{\pm}|$. See subsection \ref{subsection2d} for a detailed discussion on the SABSs.

It is useful now to express the conductance $\sigma_{S}$ given by Eq.\,(\ref{TKformula}) in terms of the energy of the SABS given by Eq.\,(\ref{ABSUNSC}). For simplicity, we denote $|\Delta_{\pm}|=\Delta_{0}$ and consider the regime with $|E|\leq\Delta_{0}$. Hence, we obtain
\begin{equation}
\label{conductanceEB}
\sigma_{S}(E,\theta)=\frac{2\gamma^{2}\Delta_{0}^{4}}{\gamma^{2}\Delta_{0}^{4}+\big[ E_{b}\sqrt{\Delta_{0}^{2}-E^{2}}-E\sqrt{\Delta^{2}_{0}-E_{b}^{2}}\big]^{2}}
\end{equation}
where $\gamma=\sigma_{N}(\theta)/(2\sqrt{1-\sigma_{N}(\theta)})$ 
\cite{Tanaka2021}. Then, for a SABS with zero energy, namely, $E_{b}=0$, the conductance given by Eq.\,(\ref{conductanceEB}) becomes
\begin{equation}
\label{TKformularesonant}
\sigma_{S}(E,\theta)=
\frac{2\gamma^{2} \Delta_{0}^{2}}
{E^{2} + \gamma^{2} \Delta_{0}^{2}}\,.
\end{equation}
Therefore, when the energy $E\equiv eV$ is tuned to the energy of the zero-energy SABS, namely, when $E=0$, we see that Eq.\,(\ref{TKformularesonant}) becomes $\sigma_{S}(E=0,\theta)=2$, irrespective of the value of $\sigma_{N}(\theta)$. Moreover, the width of resonance at zero energy decreases with the  decrease of  $\sigma_{N}(\theta)$ 
\cite{TK95,KT96,kashiwaya00}.  Thus, the conductance expression given by Eq.\,(\ref{TKformularesonant}) reveals a remarkable property of tunneling spectroscopy of unconventional superconductors in the presence of ZESABSs. We also anticipate that this expression also captures the essence of tunneling spectroscopy in the 
presence of MZMs 
\cite{Haim2015,Haim2019}, as we will discuss later.  

\subsubsection{For spin-polarized  $p$-wave superconductors}
In the case of spin-polarized $p$-wave superconductors, we now know that they host MZMs at their edges in the 1D case \cite{kitaev}. Thus, in an NS junction formed by these types of superconductors and a normal metal, tunneling occurs into an MZM at the interface. To the impact of the MZM on conductance, we only consider the contribution from 
$\theta=0$, which simplifies Eqs.\,(\ref{reflection_coefficient_general}) and leads to scattering probabilities given by
\begin{equation}
\begin{split}
A&=\frac{\sigma^{2}_{N}\mid \Gamma \mid^{2}}
{\mid 1 + (1-\sigma_{N})\Gamma^{2} \mid^{2}}, \\ 
B&=\frac{(1 -\sigma_{N})\mid 1 + \Gamma^{2} \mid^{2}}
{\mid 1 + (1-\sigma_{N})\Gamma^{2} \mid^{2}}, \\
C&=\frac{\sigma_{N}(1 - \mid \Gamma \mid^{2})}
{\mid 1 + (1-\sigma_{N})\Gamma^{2}\mid^{2}}, \\ 
D&=
\frac{\sigma_{N}(1 -\sigma_{N})\mid \Gamma \mid^{2}(1 - \mid \Gamma \mid^{2})}
{\mid 1 + (1-\sigma_{N})\Gamma^{2} \mid^{2}}. 
\label{probabilitiespwave}
\end{split}
\end{equation}

\begin{figure}[tb]
\begin{center}
\includegraphics[width=0.8\columnwidth]{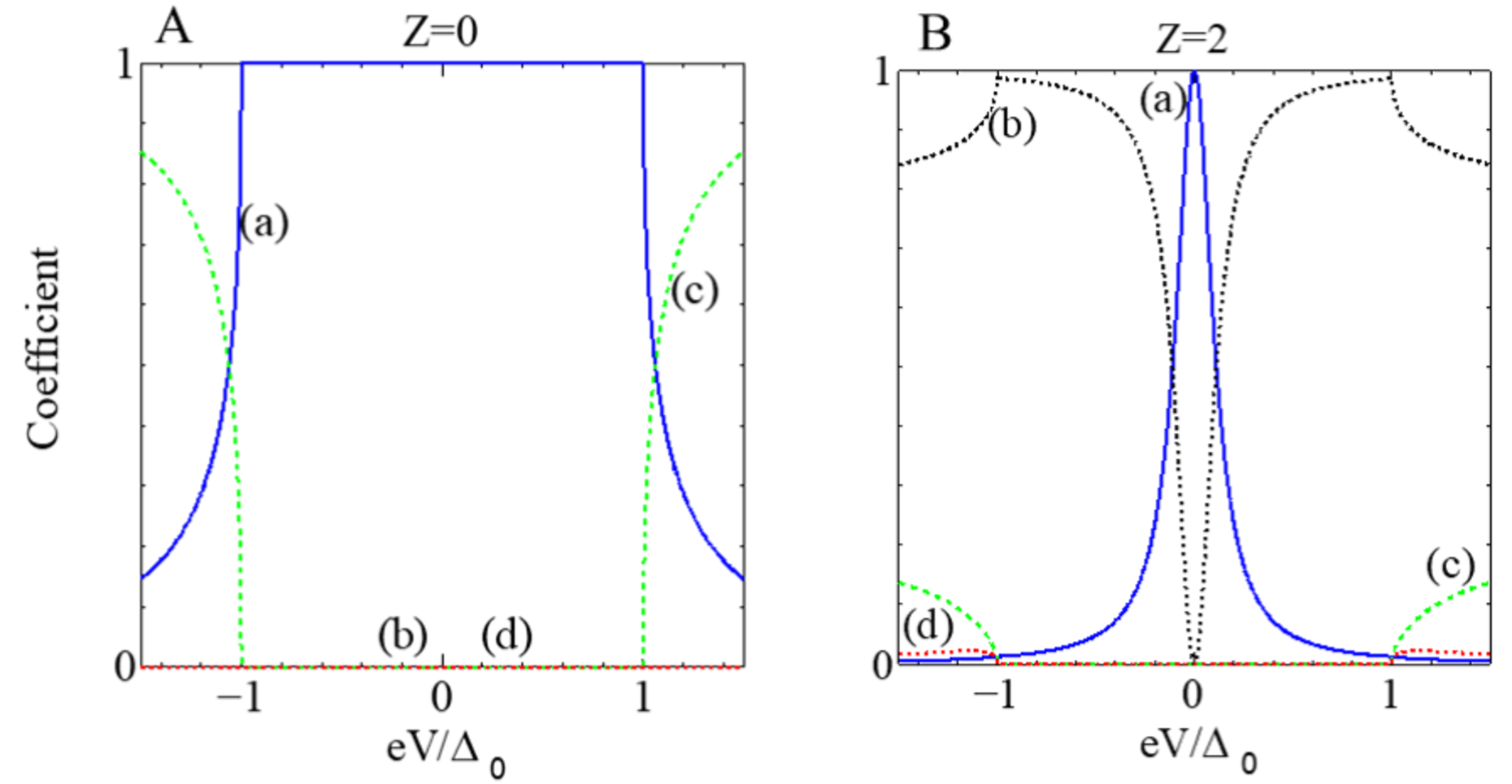}
\end{center}
\caption{Scattering probabilities for a NS junction with a $p$-wave superconductor, Eqs.\,(\ref{probabilitiespwave}) for  $Z=0$ (A) and  $Z=2$ (B). Different curves represent the following processes: (a): Andreev reflection probability, 
(b): Normal reflection probability, (c): Transmission probability of 
electron-like quasiparticle, 
(d): Transmission probability of hole-like quasiparticle for 
spin-triplet $p$-wave case. For $Z=0$, $B=D=0$ independent of  $E$. }
\label{1dpwaveABCD}
\end{figure}

In  contrast to the obtained probabilities for NS junctions with conventional spin-singlet 
$s$-wave pair potential and given by Eqs.\,(\ref{probabilitysA}) to (\ref{probabilitysD}), 
for $p$-wave superconductors in Eqs.\,(\ref{probabilitiespwave}) there is   an additional $-$ in front of the factor  $(1 - \sigma_{N})$ in the denominator.  Moreover,  the sign in front of  $\mid \Gamma \mid^{2}$ in the numerator of $B$ is changed from $-$ to $+$.  As we will see below, the impact of the MZM located at the NS interface has its origin in these changes of signs. In Fig.\,\ref{1dpwaveABCD}, we plot coefficients $A$,$B$,$C$,and $D$ as a function of energy for distinct values of $Z$. The first observation is that the probabilities for $Z=0$ in  Fig.\,\ref{1dpwaveABCD} behave exactly the same as in NS junctions with $s$-wave superconductors shown in Fig.\,\ref{1dswaveABCD}. Interestingly,  at $Z=2$, the Andreev and normal reflection probabilities exhibit a sharp peak and dip around zero energy $eV=0$, respectively. This behavior is in stark contrast to what occurs in NS junctions with $s$-wave superconductors. We highlight that irrespective of the value of $Z$, there is a perfect Andreev probability with $A=1$ at $eV=0$, while zero normal reflection probability  $B=0$.

\begin{figure}[tb]
\begin{center}
\includegraphics[width=0.5\columnwidth]{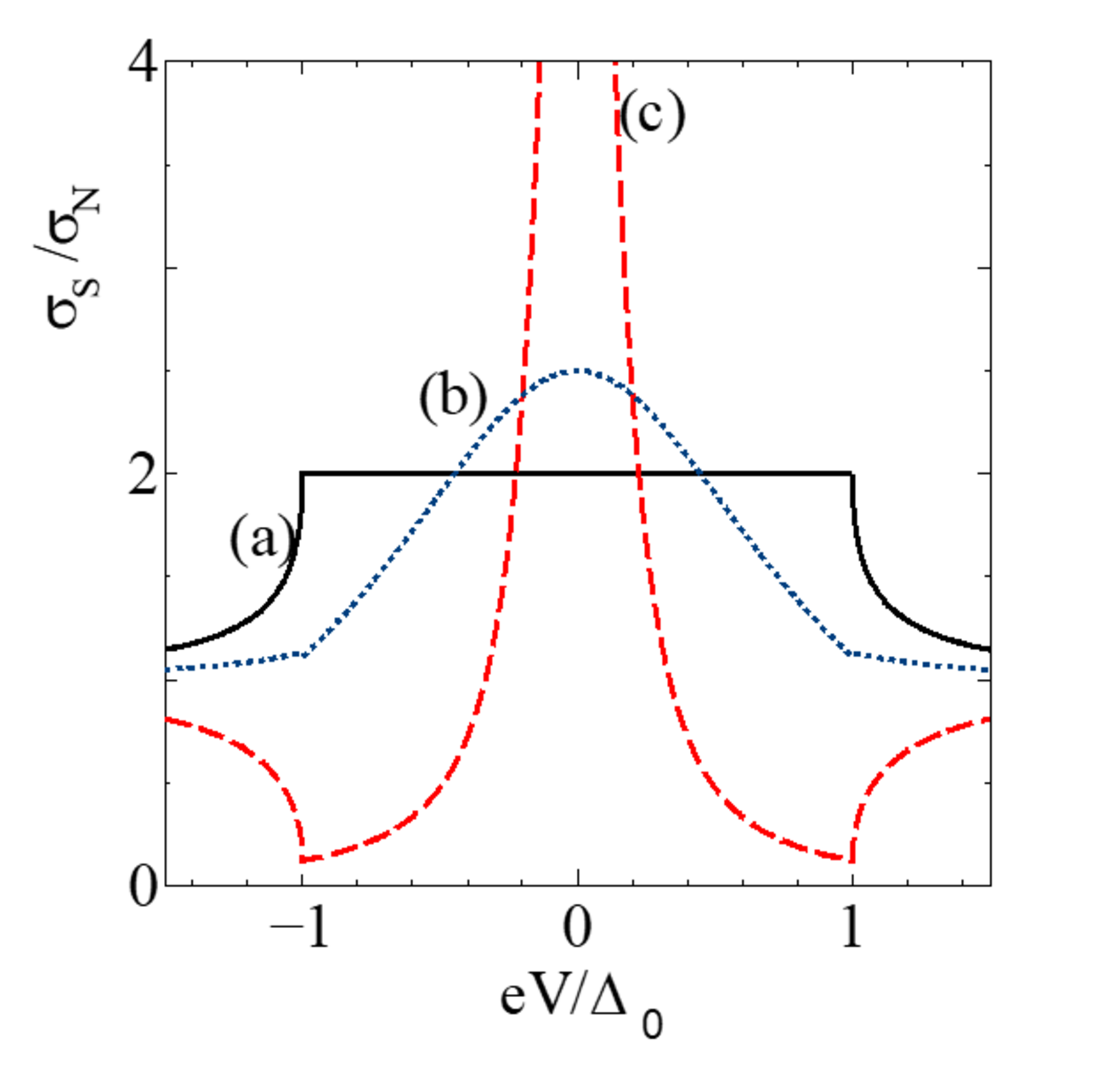}
\end{center}
\caption{Normalized conductance across an NS junction with a $p$-wave superconductor by its value in the normal state. (a)$Z=0$, (b)$Z=0.5$, (c)$Z=2$.}
\label{1dpwave}
\end{figure}

Next, we calculate the  conductance $\sigma_{S}(eV)=1+A-B$ which, by making use of the probabilities $A$ and $B$ given by Eqs.\,(\ref{probabilitiespwave}), we obtain  
\begin{equation}
\label{BTKtunnelconductancepwave}
\begin{split}
\sigma_{S}(E) &= \sigma_{N} 
\displaystyle\frac{1 + \sigma_{N} \mid \Gamma \mid^{2} - \left(1 - \sigma_{N} \right) \mid \Gamma \mid^{4}}
{ \mid 1 + \left(1 - \sigma_{N} \right)\Gamma^{2} \mid^{2}} \\
&=
\begin{dcases}
\displaystyle\frac{2 \sigma_{N}^{2}\Delta_{0}^{2} }
{  \sigma_{N}^{2}\Delta_{0}^{2} + 4\left(1 -\sigma_{N} \right)E^{2}}\,, 
& \mid E \mid \leq \Delta_{0}\,, \\
\displaystyle
\frac{2 \sigma_{N} E  
	\left( \left(2 - \sigma_{N} \right)\Omega  + \sigma_{N} E \right)}
{\left( \left(2 - \sigma_{N} \right)\Omega  + \sigma_{N} E \right)^{2} 
	+ 4 \left( 1 - \sigma_{N} \right) \Delta^{2}_{0}}\,, 
& \mid E \mid > \Delta_{0}\,.
\end{dcases}
\end{split}
\end{equation}
This expression represents the conductance across NS junctions formed by a metallic lead N and a semi-infinite spin-polarized $p$-wave superconductor \cite{Yakovenko2001}, which was initially predicted in the tunneling conductance of unconventional pairing with ZESABSs in Refs.\,\cite{TK95,kashiwaya00}. See also Refs.\,\cite{PhysRevLett.98.237002, PhysRevLett.103.237001, PhysRevB.82.180516}. To see the properties Eq.\,(\ref{BTKtunnelconductancepwave}) reveal, it is instructive to compare Eq.\,(\ref{BTKtunnelconductancepwave}) with the conductance expression for    NS junctions with $s$-wave superconductors given by Eq.\,(\ref{BTKtunnelconductance}). For   $\sigma_{N}=1$, we find no difference between the two expressions given by Eq.\,(\ref{BTKtunnelconductance}) and Eq.\,(\ref{BTKtunnelconductancepwave}). Remarkably,  the difference appears with the decrease of $\sigma_{N}$. In the case of NS junction with $p$-wave superconductors,  the conductance $\sigma_{S}(E)$ develops a peak around $E=eV=0$ for $\sigma_{N}<1$, giving rise to a quantized value  of $\sigma_{S}(E=0)=2$, see Fig.\,\ref{1dpwave}. The fact that there is a factor of $2$ is a consequence coming from the spin-polarized nature of the junction. This zero-energy peak (ZEP), sometimes referred to as zero-bias peak (ZBP), is a manifestation of the emergence of an MZM at the interface and, therefore, it represents a strong Majorana signature.  It is worth pointing out that this zero-bias conductance peak (ZBCP) stems from having a perfect Andreev reflection probability, as shown by the blue curve in Fig.\,\ref{1dpwaveABCD}(B). Moreover, the height of the ZBCP of 
normalized conductance $\sigma_{T}(eV)=\sigma_{S}/\sigma_{N}$ increases with the increase of $Z$ 
(decrease of $\sigma_{N}$), as depicted by the red curve in Fig.\,\ref{1dpwave}.  
We can therefore conclude that MZMs at the edge of topological superconductors produce ZBCPs and these peaks can be seen as robust evidence of Majorana physics. A similar ZBCP is expected at the right end of the superconductor, which should emerge as well due to the spatial nonlocality of MZMs. The detection of ZBCPs at both ends is thus necessary and it must comply with the nonlocality effects shown by MZMs.   These features are of course complementary to the peaks at zero energy shown in the LDOS and odd-frequency pairing in Section \ref{section3}. We further point out that ZBCPs can be also taken as evidence of the having dominant OTE pairing  \cite{Asano2013}, see also Refs.\,\cite{tanaka12,Cayao_2020,PhysRevB.106.L100502}.

It is important to mention that, after conceiving the physical realization of spin-polarized $p$-wave superconductivity in superconductor-semiconductor hybrid systems, there has been an impressive number of theoretical works predicting ZBCPs as signatures of MZMs and topological superconductivity, see e.g., Refs.\,\cite{PhysRevLett.109.227006,PhysRevB.86.180503,PhysRevB.86.224511,PhysRevB.87.024515,PhysRevResearch.3.023221,cao2023recent,lutchyn2018majorana,zhang2019next,prada2019andreev,frolov2020topological,flensberg2021engineered},  non-sinusoidal current-phase curves \cite{San-Jose:11a,PhysRevLett.112.137001,PhysRevB.94.085409,PhysRevB.96.165415,PhysRevB.96.205425,Ilan_2014,cayao2018andreev,cayao2018finite,luethi2023majorana}. Along these lines, experimental advances have also reported interesting results \cite{Mourik:S12,Chang2015Hard, Higginbotham2015, deng2016majorana, Albrecht16, PhysRevLett.119.176805, PhysRevLett.119.136803, Gul2018,zhang2019next,zhang2021large}, although the unambiguous identification of MZMs still remains a challenging task \cite{prada2019andreev}. The main difficulty in this case is that  ZBCPs similar to those expected due to MZMs in Eq.\,(\ref{BTKtunnelconductancepwave}) also appear in the presence of topologically trivial ZESABSs \cite{PhysRevB.86.100503, PhysRevB.86.180503,PhysRevB.91.024514,San-Jose2016, PhysRevB.96.075161, PhysRevB.96.195430, PhysRevB.97.155425, PhysRevB.98.245407, PhysRevLett.123.107703, PhysRevB.100.155429, PhysRevLett.123.217003, 10.21468/SciPostPhys.7.5.061, Avila2019, PhysRevResearch.2.013377, PhysRevLett.125.017701, PhysRevLett.125.116803, PhysRevB.102.245431, Yu2021, prada2019andreev, Pal2018, doi:10.1126/science.abf1513, PhysRevB.101.195303, PhysRevB.98.155314, PhysRevB.97.161401, PhysRevB.101.014512, PhysRevB.104.134507, PhysRevB.104.L020501, Marra_2022x, PhysRevB.105.035148, Schuray2020, PhysRevB.102.045111, Grabsch2020, PhysRevB.102.245403, PhysRevB.103.144502,chen2022topologically,PhysRevB.105.144509,PhysRevB.106.014522,PhysRevB.107.184519,PhysRevLett.130.207001}. These trivial ZEABSs and their associated ZBCPs, however, do not share all the properties of MZMs. Thus,  exploiting intrinsic Majorana properties, such as their spatial nonlocality, could open a way to distinguish them from trivial zero-energy states \cite{PhysRevB.96.085418,PhysRevB.97.161401,PhysRevB.98.085125,PhysRevLett.123.117001,PhysRevB.102.045111,PhysRevB.104.L020501,baldo2023zero,PhysRevB.105.035148,PhysRevB.109.045132,PhysRevLett.130.116202,sugeta2023enhanced}. Despite the evident open questions, nevertheless, the experimental progress made in recent years, as well as the ongoing efforts following a bottom-up approach to fabricate minimal Kitaev chains \cite{dvir2023realization,zatelli2023robust,bordin2023crossed}, is very likely to have a positive impact on the realization, detection, and application of MZMs in the near future. 
 
\subsubsection{For spin-singlet $d$-wave superconductors}
\label{subsecdwavejunction}
\begin{figure}[tb]
\begin{center}
\includegraphics[width=15cm,clip]{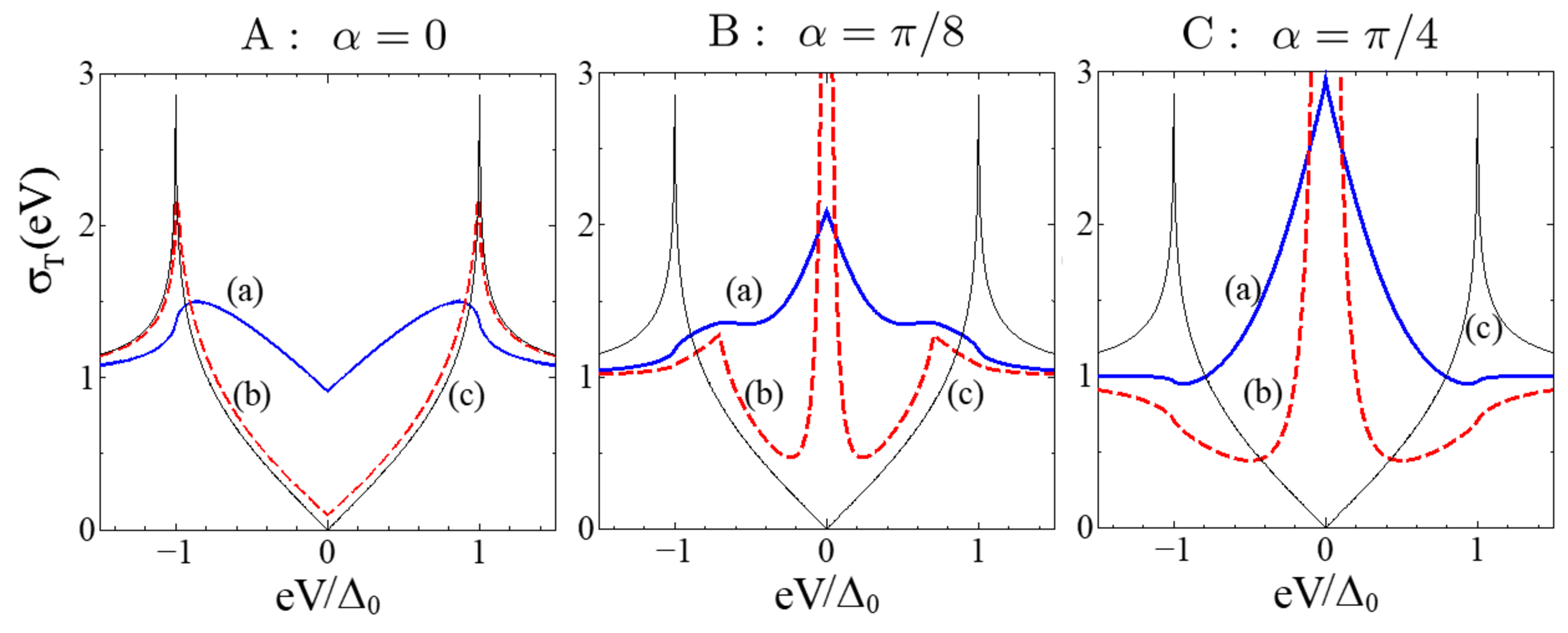}
\end{center}
\caption{Normalized conductance across a junction formed with a $d$-wave superconductor. Different panels correspond to different misorientation angles such that A: $\alpha=0$, B: $\alpha=\pi/8$, C: $\alpha=\pi/4$. Moreover, the solid blue, dashed red, and solid black correspond to distinct values of $Z$, where
(a) $Z=0.5$, (b) $Z=2$, and (c) Bulk DOS.}
\label{2ddwave}
\end{figure}

For NS junctions with a superconductor having spin-singlet $d$-wave pair potential, we consider that the $d$-wave pair potential is expressed by the angle measured from the $a$-axis of the basal plane since the one of the lobe direction coincides with that of $a$-axis. 
We also  consider the junction with misorientation angle $\alpha$, 
which denotes the angle between the normal to the interface and the $a$-axis of a $d$-wave superconductor. The schematic of the surface of $\alpha=0$ and $\alpha=\pi/4$ is shown in Fig.\,\ref{SABSdwave}. In this case, the pair potential $\Delta_{\pm}$  felt by quasiparticles with an injection angle $\theta$   is given by  
\begin{equation}
\Delta_{\pm}=\Delta_{0}\cos\left(2\theta \mp 2\alpha \right),
\label{dwavemisorientation2}
\end{equation}
Having defined the pair potential, we plot the normalized tunneling conductance $\sigma_{T}(eV)$ 
in Eq. (\ref{TKformula0})
as a function of $eV$ in Fig.\,\ref{2ddwave} for distinct misorientation angles $\alpha$ and finite  $Z$ (small  $\sigma_{N}(\theta)$) \cite{TK95}
.  For $\alpha=0$, the normalized conductance exhibits a finite value having a $V$ shape and reaching its minimum at $eV=0$, see Fig. \ref{2ddwave}(A). While intermediate values of $Z$ induce finite minimum in the conductance at zero energy, large values of $Z$ produce almost vanishing conductance at zero energy; in this case,  the line shape of $\sigma_{T}(eV)$  approaches  almost similar to that by LDOS of bulk $d$-wave superconductor with the increase of $Z$, as can be seen by comparing the different curves in Fig.\,\ref{2ddwave}(A). Thus, no ZBCP appears which is consistent with the absence of ZESABS at $\alpha=0$, see subsection \ref{dwaveSABS}.   Interestingly, for nonzero $\alpha$, a ZBCP in $\sigma_{T}(eV)$ 
always appears and it becomes prominent with the increase of $Z$. Its peak height and width becomes maximum for $\alpha=\pi/4$, as can be really seen in Fig. \ref{2ddwave}(B,C), see Refs.\, 
\cite{TK95,KT96,kashiwaya00}. The ZBCP $\alpha=\pi/4$ is a consequence of the formation of a ZESABS, which, for this misorientation angle, forms a zero-energy flat band for any  $\theta$.  It is worth noting that a similar zero-energy flat band   appears for spin-triplet $p_{x}$-wave pairing  both for $S_{z}=0$ and $S_{z}=1$ cases where 
$\Delta(\theta_{\pm})$ is given by $\Delta(\theta_{\pm})=\pm \Delta_{0}
\cos \theta$, see Refs.\,\cite{YTK98,Yakovenko2004,odd3b}.  Before closing this part, we would like to briefly mention that the ZBCP discuss in this part has been observed in many experiments of tunneling spectroscopy of high $T_{c}$ cuprates, as reported in Refs.\, 
\cite{Experiment1,Experiment2,Experiment3,Experiment4,Experiment5,Sharoni_2001,Experiment6,Experiment7,Millo2018,Bouscher_2020,10.1119/1.3549729}, and also in other strongly correlated superconductors like CeCoIn$_{5}$ \cite{Rourke} and PuCoIn$_{5}$ \cite{Daghero2012}.  \par 
 
\subsection{Summary}
This section has focused on analyzing tunneling conductance across NS junctions based on unconventional superconductors. In particular, we discussed that tunneling into a spin-polarized $p$-wave superconductor produces a quantized ZBCP  with high $2e^{2}/h$ as a strong signature of the emergence of an MZM at the end of the system in the topological phase. We have shown that this conductance quantization at zero energy is robust against interface imperfections and results from perfect Andreev reflections,  an effect that does not occur in spinful junctions with conventional spin-singlet superconductors where conductance can take values between $0$ and $4e^{2}/h$. Then, we have addressed tunneling into the surface of a spin-singlet $d$-wave superconductor can also give rise to a pronounced peak at zero-energy due to the formation of a ZESABS which, however, depends on the misorientation angle. We have also highlighted the intense theoretical and experimental efforts in superconductor-semiconductor hybrids aiming to detect ZBCPs, which, despite the challenges, remain one of the most promising scenarios to realize and detect   Majorana physics and topological superconductivity.
In this section, we study tunneling conductance for spin-triplet and 
spin-singlet pairing, separately.  The existence of  ZBCP by ZESABS has been predicted in non-centrosymmetric superconductor junctions with mixed parity pairing state \cite{Iniotakis2007,
TanakaNagaosaBalatsky,TanakaYada2010,Yada2011}. 
ZBCP has been reported in a recent experiment 
of AuSn$_{4}$ where mixed parity surface 
superconductivity is promising \cite{Zhu2023}.  
\section{Detecting Majorana zero modes via the anomalous proximity effect}
\label{section6}
In this section, we consider NS junctions formed by a finite diffusive N region, denoted as DN, and a semi-infinite unconventional superconductor, and then investigate signatures of MZMs and ZESABSs in the anomalous proximity effect. Before going any further, it is important to clarify what we mean by anomalous proximity effect. Under general grounds, the superconducting proximity effect occurs when a superconductor, placed in proximity to another material such as a normal metal, induces superconducting correlations or Cooper pairs which can have completely different symmetries associated with the nature of the normal metal and interface. Thus, non-superconducting materials can acquire superconducting properties, with the coupling between the two systems determining the amount of proximity-induced Cooper pairs. Thus, the proximity effect is present in the NS junctions discussed in Section \ref{section5}, where proximity-induced Cooper pairs in the ballistic N region are naturally characterized by Andreev reflections    \cite{nagato1993theory,pannetier2000andreev,klapwijk2004proximity,klapwijk2014direct}.  To detect the proximity effect, conductance and LDOS in N are commonly carried out, which are then able to reveal the properties of the induced superconductivity.  At this point, we note that, when the superconductor is coupled to diffusive normal metal DN,  the resistance of the DN gets influenced by the penetration of Cooper pairs due to the proximity effect. Thus, the LDOS and conductance exhibit a behavior that is different from junctions with ballistic N metals. For these reasons, the changes in the penetration of Cooper pairs, LDOS, and conductance in DNS junctions are referred to as anomalous proximity effect \cite{Proximityp,Proximityp2,Proximityp3}, which is the focus of this part for unconventional superconductors \cite{TextTanaka2021}. For pedagogical purposes, we first  address the proximity effect in conventional spin-singlet $s$-wave superconductors and then inspect the anomalous proximity effect in unconventional superconductors.

\subsection{Proximity effect in conventional superconductors}
We consider a DNS junction along $x$ formed by a DN metal of length $L$ and a semi-infinite conventional superconductor with uniform spin-singlet $s$-wave pair potential in S and then explore charge conductance with an N electrode attached to DN. The electrode, the DN metal, and the superconductor are located at  $x<-L$, $-L<x<0$, and $x>0$, respectively, with an insulating interface barrier at $x=0$  given by a $\delta$-function. Here, DN is assumed to be in the dirty limit where the mean free path $\ell$ is much smaller than the thermal diffusion length $\sqrt{\hbar D/(2\pi k_{B} T)}$ with Boltzmann constant $k_{B}$, where $D$ is the diffusion constant and $T$ is the temperature. Within this energy scale, the coherence between electrons and holes is good and the retro-reflectivity of the Andreev process is satisfied. Moreover, $D$ introduces an energy scale known as the Thouless energy $E_{Th}=\hbar D/L^{2}$. Moreover, we consider $L\gg \ell$, and its resistance in the normal state is denoted as $R_{d}$. Taking all this information into account, the calculation of conductance is carried out using the quasiclassical Green's function framework \cite{TextTanaka2021}. 

Under these conditions, the conductance is obtained from the charge current    \cite{Volkov1993,Yip1995}. In our case,  probed at the left side of DN,  it is given \cite{TextTanaka2021}
\begin{equation}
I_{el}=I_{el}(V)=
\displaystyle\frac{1}{e}\displaystyle\int^{\infty}_{0}
dE 
\displaystyle\frac{f_{t}\left(x=-L \right)}
{R}
\label{currentjunctionKLfinal}
\end{equation}
with $R$ being the resistance of the junction at any temperature and $f_{t}$ is 
given by 
\[
f_{t}\left(x=-L \right)= \frac{1}{2} \left\{
\tanh \left[\left(E +eV \right)/\left(2 k_{B} T \right)\right]
- \tanh \left[\left(E -eV \right)/\left(2k_{B} T \right) \right]
\right\},
\]
where $V$ is the bias voltage. 
Here, 
$f_{t}(x=-L)$ is the Fermi distribution function at the left interface where the electrode is attached without any barrier. 
At low temperatures, 
the resistance of the junction
$R$ is given by 
\begin{equation}
\label{Resis}
R=\displaystyle\frac{R_{b}}{I_{K}} 
+ \displaystyle\frac{R_{d}}{L} \displaystyle\int^{0}_{-L}
\displaystyle\frac{dx}{\cosh^{2} \zeta_{\rm Im}\left(x \right)}\,.
\end{equation}
where the first (second) term in the expression for $R$ describes the resistance 
contributions of the interface (diffusive normal metal DN). 
Here, $R_{b}$ and $R_{d}$  represent the normal state resistances at the interface and in DN, respectively, while $\zeta_{\rm Im}$ is the imaginary part of $\zeta(x)$ which characterizes the proximity effect, namely, $\zeta_{\rm Im}(x)={\rm Im}[\zeta(x)]$.  
It is noted that $\cos\zeta(x)$ and $\sin\zeta(x)$ express the quasiclassical Green's function.  
The current  $I_{K}$ and $\zeta(x)$ are found from
\begin{equation} 
\label{EqsUsadel}
 \begin{split}
\hbar D
 \frac{\partial^2}{\partial x^2}
\zeta \left(x \right) + 2iE 
\sin \left[\zeta \left(x \right)\right] &= 0\,,\\
D\frac{\partial}{\partial x}
\left[
\frac{\partial f_{t}\left(x \right)}{\partial x}
\mathrm{cosh^{2}} \zeta_{\rm Im}\left(x \right)
\right]
&=0\,,   \\
 I_{K}&=
\rm{Re}\left(\cos \zeta_{N}\right)\rm{Re}\left(\cos \zeta_{S} \right)
+ \rm{Re}\left(\sin \zeta_{N}\right)\rm{Re}\left(\sin \zeta_{S} \right). 
 \end{split}
 \end{equation}
where $\zeta_{S(N)}$ are solutions $\zeta(x)$   in S(N); note that $\zeta(x)$ is a function characterizing the proximity effect and is obtained by solving the Usadel equation described above Eq.\,(\ref{EqsUsadel}), see also Ref.\,\cite{Golubov2003,TextTanaka2021} for more details. Then, using the knowledge about resistance in Eqs.\,(\ref{Resis}), conductance in the junction can be found as its inverse. In particular,  the conductance, normalized by its value in the normal state, is given by
\begin{equation} 
\label{conductanceDN}
\sigma_{T}(eV)=\frac{R_{N}}{R_{S}}
\end{equation}
where $R_{N(S)}$ represents the resistance of the junction in the normal (superconducting) state at sufficiently low temperatures obtained from Eqs.\,(\ref{Resis}). 

\begin{figure}[tb]
	\begin{center}
\includegraphics[width=0.8\columnwidth]{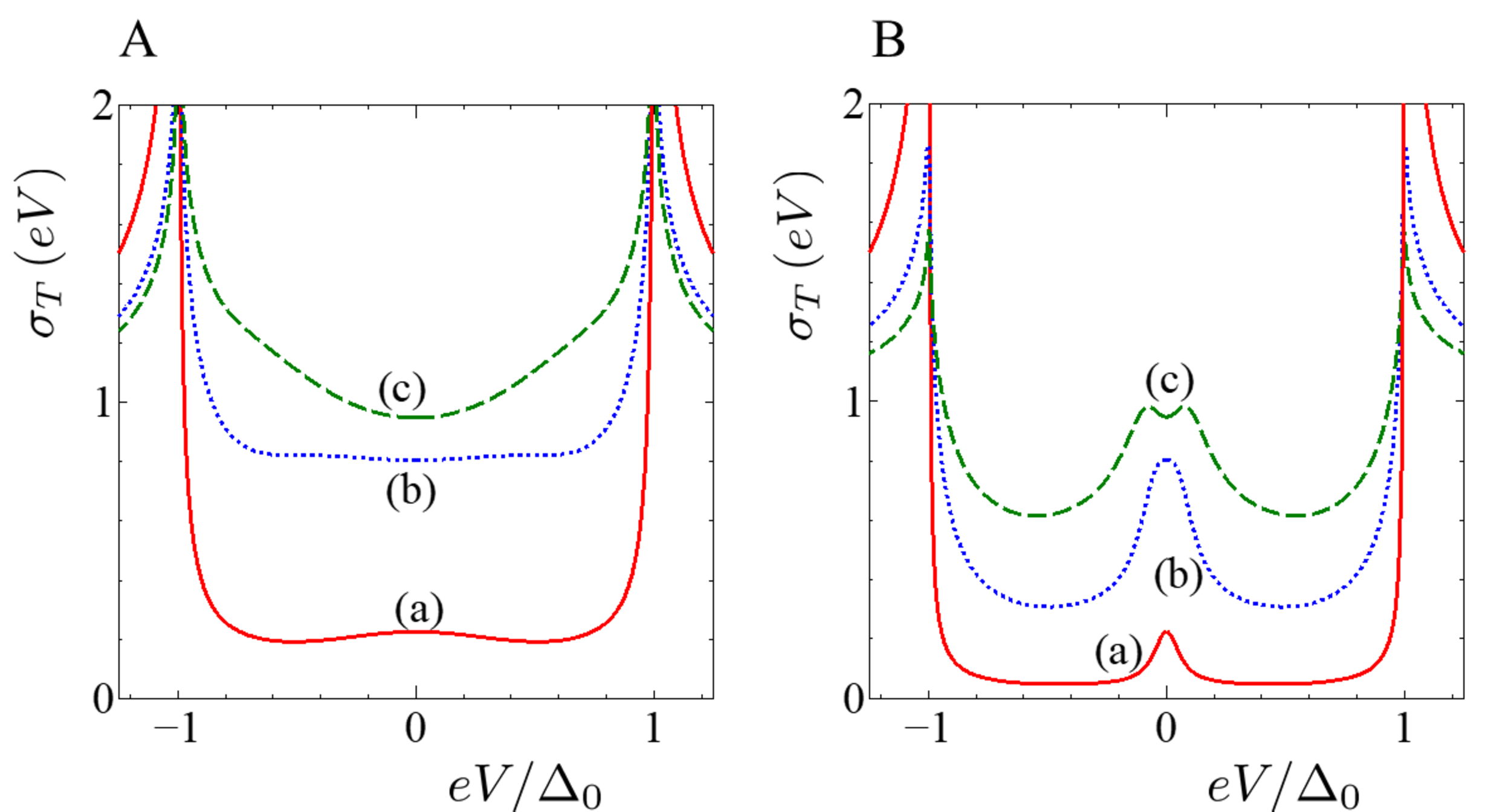}
\end{center}
\caption{Normalized conductance in DNS junctions where S is a spin-singlet $s$-wave superconductor and DN is a diffusive normal metal.  Left and right panels correspond to distinct values of the Thouless energy: A: $E_{Th}=0.5\Delta_{0}$. B: $E_{Th}=0.05\Delta_{0}$. Different curves correspond to distinct values of $R_{d}/R_{b}$: (a) $ R_{d}/R_{b}=0.2$, (b) $R_{d}/R_{b}=1$, (c) $R_{d}/R_{b}=2$.}
\label{swaveKLboundaryconductance}
\end{figure}

\begin{figure}[tb]
	\begin{center}
\includegraphics[width=0.8\columnwidth]{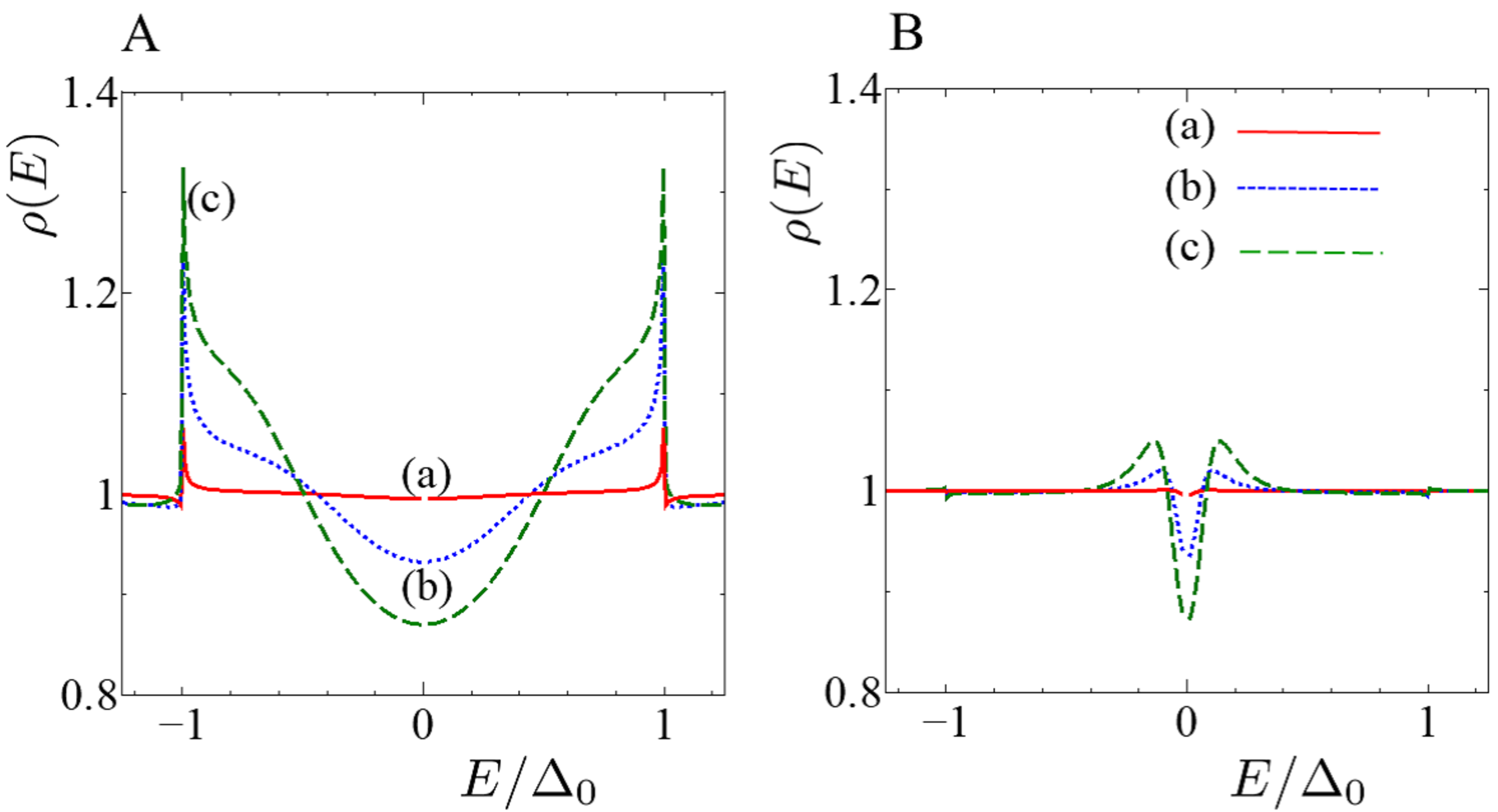}
	\end{center}
\caption{LDOS  of the diffusive normal metal DN at $x=-L/2$ for a DNS junction, where  S is a spin-singlet $s$-wave superconductor. Left and right panels correspond to distinct values of the Thouless energy: A: $E_{Th}=0.5\Delta_{0}$. B: $E_{Th}=0.05\Delta_{0}$. The LDOS is normalized by its value in the normal state. Different curves correspond to distinct values of $R_{d}/R_{b}$: (a) $R_{d}/R_{b}=0.2$, (b) $R_{d}/R_{b}=1$, (c) $R_{d}/R_{b}=2$.}
\label{swaveKLboundaryLDOS}
\end{figure}

To visualize the behavior of the conductance given by Eq.\,(\ref{conductanceDN}), in  Fig.\,\ref{swaveKLboundaryconductance} we plot it as a function of $eV$ for distinct values of  $R_{d}/R_{b}$ at two representative large and low values of $E_{Th}$.  For completeness, in Fig.\,\ref{swaveKLboundaryLDOS} we also show the normalized LDOS $\rho$ normalized by its value in the normal state 
in the middle of  DN at $x=-L/2$ as a function of $E=eV$ for the same cases of Fig.\,\ref{swaveKLboundaryconductance}.  We note that $\rho$ 
is obtained by $\rho={\rm Real}[\cos \zeta]$ 
using $\zeta(x)$ from the same quasiclassical Green's functions employed to find the charge current in  Eq.\,(\ref{currentjunctionKLfinal}), see Ref.\,\cite{TextTanaka2021} for more details. At large Thouless energies with $E_{Th}=0.5\Delta_{0}$,  the conductance $\sigma_{T}(eV)$ has an almost U-shaped structure, acquiring lower values when the ratio $R_{d}/R_{b}$ is small, see    Fig. \ref{swaveKLboundaryconductance}(A). Note that the conductance at $R_{d}/R_{b}=0.2$ in  Fig. \ref{swaveKLboundaryconductance}(A) has a similar behavior as that in ballistic junctions at low-transparencies presented in Fig.\,\ref{1dswave}. As the ratio $R_{d}/R_{b}$ increases, the subgap conductance values also acquire higher values but still maintain a U shape, with a minimum near $eV=0$ that is only developed for very large $R_{d}/R_{b}$, see blue and green curves in Fig. \ref{swaveKLboundaryconductance}(A). When it comes to the LDOS in this regime with $E_{Th}=0.5\Delta_{0}$, it has a clear U-shaped profile at small ratio $R_{d}/R_{b}$ but, as such ratio increases, the LDOS gets suppressed at low voltages and develops a clear minimum around $eV=0$, see Fig.\,\ref{swaveKLboundaryLDOS}(A). 
We can thus conclude that in this regime, the penetration of Cooper pairs into DN increases with the increase of $R_{d}/R_{b}$ which then enhance conductance which is, however, not visible in the LDOS. 

In contrast to large Thouless energies $E_{Th}=0.5\Delta_{0}$ discussed in previous paragraph,  the conductance  $\sigma_{T}(eV)$ at low Thouless energies $E_{Th}=0.05\Delta_{0}$ develops a very different behavior, as can be seen in Fig.\,\ref{swaveKLboundaryconductance}(B). In fact, $\sigma_{T}(eV)$ exhibits a peak at zero voltage bias for $R_{d}/R_{b}=0.2$ and $1$, as depicted by red and blue curves in Fig.\,\ref{swaveKLboundaryconductance}(B). It is interesting to note that this ZBCP becomes prominent at $R_{d}/R_{b}=1$, although it does not exhibit any robustness as its high strongly depends on the ratio $R_{d}/R_{b}$ \cite{Volkov1993,Yip1995}. The ZBCP can be understood as follows: When the energy of an injected electron in DN is near the Fermi surface, the reflected hole by Andreev reflection interferes with the electron due to retroreflectivity, forming a standing wave that helps electrical conduction \cite{vanWees,Kastaltsky}. 
As a result, the conductance increases within the range in which the energy of the incident electron is smaller than the Thouless energy as measured from the Fermi surface.  On the contrary, when the energy of the incident electron exceeds the Thoules energy, the conductance rapidly decreases because standing waves are no longer formed. 
Moreover, the height of the  ZBCP has been shown to never exceed unity  \cite{Volkov1993,Yip1995} and it is very sensitive to the applied magnetic field \cite{Volkov1993}. We can thus see that the ZBCP in Fig. \,\ref{swaveKLboundaryconductance}(B) has a distinct origin to those ZBCPs due to MZMs and ZESABSs of unconventional superconductor junctions  discussed in  
 Section \ref{section5}.   Moreover,  the corresponding LDOS $\rho$ intriguingly shows a dip-like (gap-like) structure around $eV=0$ for all cases of the ratio $R_{d}/R_{b}$, reaching a minimum at $eV=0$ independent of the value of $R_{d}/R_{b}$. We can thus conclude that strong Thouless energies promote an enhanced penetration of Cooper pairs around zero energy, which is reflected in the large 
 value of conductance at zero voltage. This enhancement is, however, not seen in the LDOS, which we interpret 
to happen 
 because the LDOS does not capture the interference effects among quasiparticle paths in DN, in contrast to what conductance does. 

\subsection{Anomalous proximity effect in unconventional superconductors}
In the case of DNS junctions with unconventional superconductors, special care needs to be taken due to the injection angles $\theta$. As before, conductance is obtained from the resistance $R$, which here reads  \cite{Proximityp2}
\begin{equation}
R=\frac{R_{b}}{\left<I_{K} \right>}
+\frac{R_{d}}{L}\int_{-L}^{0}\frac{dx}{\cosh
^{2}\zeta_{\rm{Im}}\left(x \right)}\,,
\label{resistance}
\end{equation}
where $\zeta_{\rm{Im}}(x)$ is a function characterizing the proximity effect and obtained using Eq.\,(\ref{EqsUsadel}) with appropriate boundary conditions, while $\langle I_{K} \rangle$ represents the angular average over many channels of $I_{K}$. For    spin-singlet     superconductors, $I_{K}$ is given by \cite{Kokkeler2022}
\begin{equation}
I_{K} =\frac{(\sigma_{N}/2)  C_{1}}
{\left| 1 - (1 - \sigma_{N}) \Gamma_{+}\Gamma_{-}
+ \sigma_{N} \sin \left(\frac{\zeta_{N}}{2}\right)
\left[-(1 +\Gamma_{+}\Gamma_{-})
\sin\left( \frac{\zeta_{N}}{2} \right)
+ i ( \Gamma_{+} + \Gamma_{-})
\cos\left( \frac{\zeta_{N}}{2} \right)
\right]
\right|^{2}
}
\end{equation}
with
\begin{equation}
\label{Ib0spinsinglet}
\begin{split}
C_{1} &= 2\left[ 1 + \sigma_{N} \mid \Gamma_{+}\mid^{2} + 
\left(\sigma_{N}-1 \right)\mid \Gamma_{+} \Gamma_{-}\mid^{2} \right]  + \sigma_{N} \sinh^{2}\left(\zeta_{Ni}\right)
\left(1 + \mid \Gamma_{+} \mid^{2} \right)
\left(1 + \mid \Gamma_{-} \mid^{2} \right)  \\
& + 
\left(2 - \sigma_{N} \right)\left(1 - \mid \Gamma_{+} \Gamma_{-}\mid^{2} \right)
\left[ \cos \zeta_{Nr} \cosh \zeta_{Ni} - 1 \right]   \\
&- 
\left(2 - \sigma_{N} \right) \cosh \zeta_{Ni} \sin \zeta_{Nr} 
{\rm Im}\left[ \left(\Gamma_{+} + \Gamma_{-} \right)\left(1 - \Gamma^{*}_{+}\Gamma^{*}_{-}\right) 
\right]  \\
& - 
\sigma_{N}\cosh \zeta_{Ni} \sinh \zeta_{Ni} {\rm Re}
\left[ \left(\Gamma_{+} + \Gamma_{-} \right)
\left(1 + \Gamma^{*}_{+}\Gamma^{*}_{-}\right)  
\right]\,.
\end{split}
\end{equation}

On the other hand,  for spin-triplet superconductors is given by  \cite{Proximityp,Proximityp2,Kokkeler2022}
\begin{equation}
  I_{K} =
\frac{(\sigma_{N}/2) C_{1}}
{\left| 1 - (1 - \sigma_{N}) \Gamma_{+}\Gamma_{-}
	+ \sigma_{N} \sin \left(\frac{\zeta_{N}}{2}\right)
	\left[-(1 +\Gamma_{+}\Gamma_{-})
	\sin\left( \frac{\zeta_{N}}{2} \right)
	-(\Gamma_{+} - \Gamma_{-})
	\cos\left( \frac{\zeta_{N}}{2} \right)
	\right]
	\right|^{2}
}
\end{equation}
with $C_{1}$ given by
\begin{equation}
\label{Ib0spintriplet}
\begin{split}
C_{1} &= 2\left[ 1 + \sigma_{N} \mid \Gamma_{+}\mid^{2} + 
\left(\sigma_{N}-1 \right)\mid \Gamma_{+} \Gamma_{-}\mid^{2} \right] +  
\sigma_{N} \sinh^{2}\left(\zeta_{Ni}\right)
\left(1 + \mid \Gamma_{+} \mid^{2} \right)
\left(1 + \mid \Gamma_{-} \mid^{2} \right) \\
& + 
\left(2 - \sigma_{N} \right)\left(1 - \mid \Gamma_{+} \Gamma_{-}\mid^{2} \right)
\left[ \cos \zeta_{Nr} \cosh \zeta_{Ni} - 1 \right]\\
&- 
\left(2 - \sigma_{N} \right) \cosh \zeta_{Ni} \sin \zeta_{Nr} 
{\rm Re}\left[ \left(\Gamma_{+} - \Gamma_{-} \right)\left(1 - \Gamma^{*}_{+}\Gamma^{*}_{-}\right) 
\right] \\
& -  
\sigma_{N}\cosh \zeta_{Ni} \sinh \zeta_{Ni} {\rm Im}
\left[ \left(\Gamma^{*}_{+} - \Gamma^{*}_{-} \right)
\left(1 + \Gamma_{+}\Gamma_{-}\right) \
\right]\,.
\end{split}
\end{equation}
Here, $ \sigma_{N}=1/(1 + Z^{2})$, $\Gamma_{\pm}=\Delta_{\pm}/(E + \Omega_{\pm})$, $\zeta_{Nr}$, $\zeta_{Ni}$ denotes the real and imaginary part of $\zeta_{N}$, which characterizes the proximity induced superconducting correlations in N, see Eq.\,(\ref{EqsUsadel}) and Refs.\,\cite{TextTanaka2021,Kokkeler2022} for details. We remember that having $\zeta_{N}=0$  indicates the absence of proximity effect in DN. It can be verified that for a DN region of zero length,  $L=0$, then $R_{d}=0$, $\zeta_{N}=0$, $\zeta_{Nr}=0$, and $\zeta_{Ni}=0$ are satisfied. Then, the resistance of the junction is determined only at the interface and we reproduce the results in ballistic junctions in Eqs. \,(\ref{TKformula0})-(\ref{TKformula}). Therefore, in general, Eq.\,(\ref{resistance}), combined with Eq.\,(\ref{Ib0spinsinglet}) and Eq.\,(\ref{Ib0spintriplet}), including also Eqs.\,(\ref{EqsUsadel}), enable the calculation of conductance for DNS junctions with spin-singlet and spin-triplet unconventional superconductors as given by Eq.\,(\ref{conductanceDN}) but with the resistance here obtained from Eqs.\,(\ref{resistance}).  

\subsubsection{Anomalous proximity effect in 1D spin-polarized $p$-wave superconductors}
\begin{figure}
\centering
\includegraphics[width=0.85\columnwidth]{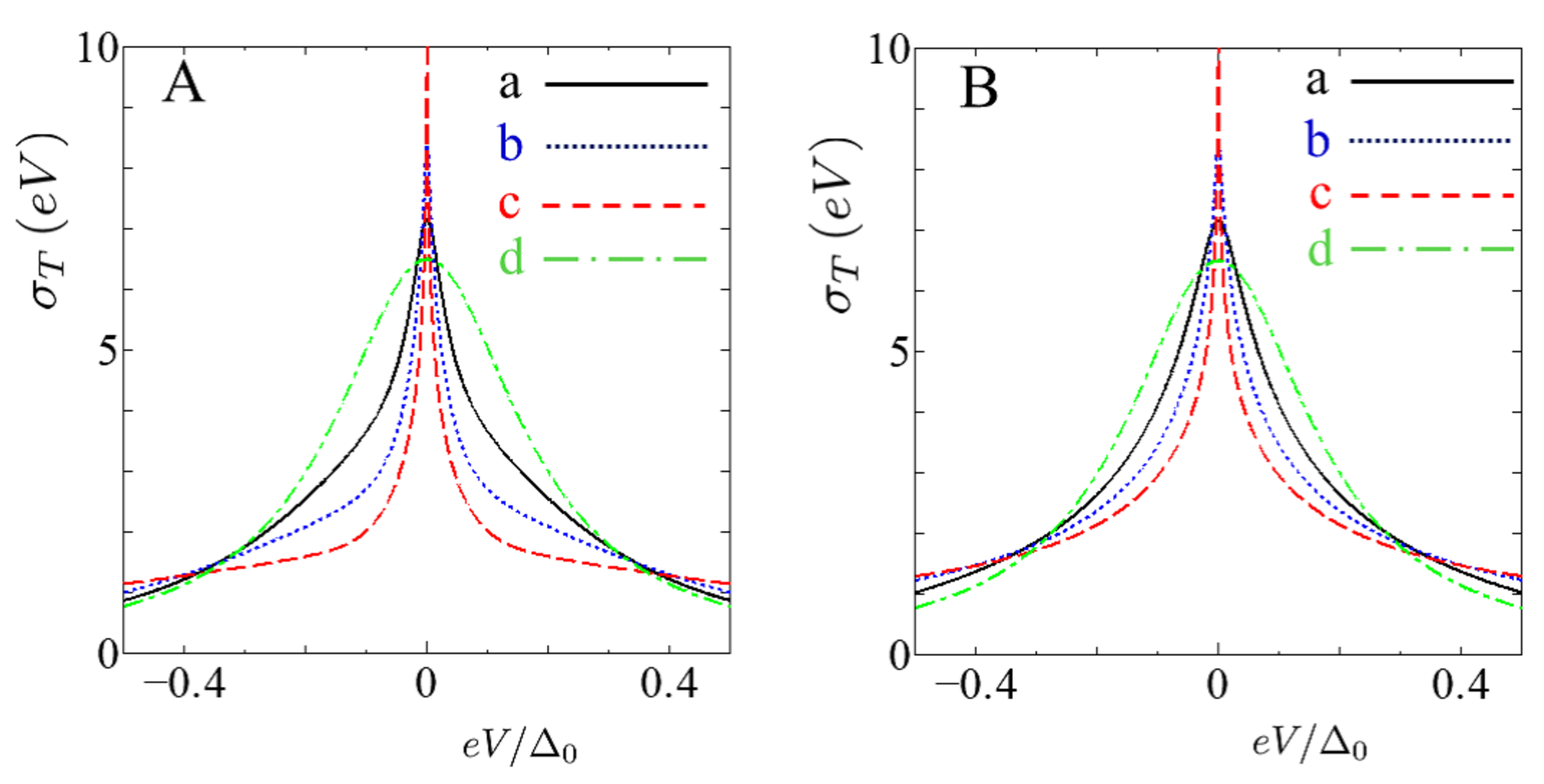} 
\caption{Normalized conductance   $\sigma_{T}(eV)$ in a DNS junction where S is a 1D spin-polarized $p$-wave superconductor and DN is a diffusive normal metal. Here, $\sigma_{T}(eV)$ is normalized by its value in a normal state for $Z=1.5$. Left and right panels correspond to distinct values of the Thouless energy: A: $E_{Th}=0.05\Delta_{0}$, B: $E_{Th}=0.5\Delta_{0}$. Different curves in each panel correspond to distinct values of $R_{d}/R_{b}$: (a) $R_{d}/R_{b}=0.1$, (b) $R_{d}/R_{b}=0.3$, (c) $R_{d}/R_{b}=1$, and 
(d) $R_{d}/R_{b}=0$ (ballistic). }
\label{pxwaveNazarovconductance1}
\end{figure}

In this subsection, we consider the 1D limit of $p$-wave 
superconductor junctions \cite{Kokkeler2022}.  It corresponds to the Kitaev chain in the continuous limit of the topological phase, where the pair potential is expressed by $\Delta_{+}=-\Delta_{-}=\Delta_{0}$. Since $\Gamma_{+}=-\Gamma_{-}=\Gamma$ with $\Gamma=\Delta_{0}/(E + \Omega)$,  the expression for  $I_{K}$  from Eq.\,(\ref{Ib0spintriplet}) acquires a simpler form given by 
\cite{Kokkeler2022}
\begin{equation}
I_{K}=\frac{\sigma_{N}}{2} 
\frac{C_{1}}
{\left| 1 + \left(1 - \sigma_{N} \right) \Gamma^{2}
	+ \sigma_{N} \sin \left(\frac{\zeta_{N}}{2}\right)
	\left[-\left(1 - \Gamma^{2} \right)
	\sin\left( \frac{\zeta_{N}}{2} \right)
	-2 \Gamma
	\cos\left( \frac{\zeta_{N}}{2} \right)
	\right]
	\right|^{2}
}
\end{equation}
with 
\begin{equation}
\label{Ib0spintripletKitaevchain}
\begin{split}
C_{1} &= 2\left[ 1 + \sigma_{N} \mid \Gamma \mid^{2} + 
\left(\sigma_{N}-1 \right)\mid \Gamma \mid^{4} \right]  + 
\sigma_{N} \sinh^{2}\left(\zeta_{Ni}\right)
\left(1 + \mid \Gamma \mid^{2} \right)^{2}\\
& + 
\left(2 - \sigma_{N} \right)\left(1 - \mid \Gamma \mid^{4} \right)
\left[ \cos \zeta_{Nr} \cosh \zeta_{Ni} - 1 \right]  - 
\left(2 - \sigma_{N} \right) \cosh \zeta_{Ni} \sin \zeta_{Nr} 
{\rm Re}\left[ 2\Gamma \left(1 + \left( \Gamma^{*}_{+} \right)^{2} \right) 
\right] \\
&-
2 \sigma_{N}\cosh \zeta_{Ni} \sinh \zeta_{Ni} {\rm Im}
\left[ \Gamma^{*}_{+} \left(1 - \Gamma^{2} \right) \
\right]\,.
\end{split}
\end{equation}
Then,   $\langle I_{K}\rangle$ is obtained as $\langle I_{K}\rangle= I_{K}/\sigma_{N}$ and used in Eq.\,(\ref{resistance}) to finally find the normalized conductance  $\sigma_{T}(eV)$  following Eq.\,(\ref{conductanceDN}).  In Fig.\,\ref{pxwaveNazarovconductance1} we present    $\sigma_{T}(eV)$   at $Z=1.5$ for distinct values of the ratio $R_{d}/R_{b}$ but fixed $R_{b}$. As a reference, we plot $\sigma_{T}(eV)$  with $R_{d}=0$  corresponding to the ballistic junction case discussed in Fig. \ref{1dpwave}.  First, at low Thousless energies such as  $E_{Th}=0.05\Delta_{0}$ in Fig.\,\ref{pxwaveNazarovconductance1}(A), we identify three clear voltage regions with distinct conductance profiles. At $eV \sim 0$, the normalized conductance $\sigma_{T}(eV)$ exhibits a peak that increases as $R_{d}/R_{b}$ increases, reaching a maximum value  at $R_{d}/R_{b}=1$. For voltages in the interval $0.05\Delta_{0} < |eV| \lesssim  0.4\Delta_{0}$, the conductance develops a maximum value for $R_{d}/R_{b}=0$ while for $R_{d}/R_{b}=1$ it is seriously suppressed as compared to 
the ballistic case with $R_{d}/R_{b}=0$.  For $|eV| > 0.4\Delta_{0}$, the conductance  becomes maximum for $R_{d}/R_{b}=1$ again, as depicted by red curve in Fig.\,\ref{pxwaveNazarovconductance1}(A). In the case of larger Thouless energies shown in Fig.\,\ref{pxwaveNazarovconductance1}(B) for 
 $E_{Th}=0.5\Delta_{0}$, we observe an overall similar behaviour at lower  $E_{Th}$. For instance, by a direct inspection, we observe that in the ballistic regime $R_{d}/R_{b}=0$, there is no change between the conductance at $E_{Th}=0.05\Delta_{0}$ and $E_{Th}=0.5\Delta_{0}$ shown by green curves in Fig.\,\ref{pxwaveNazarovconductance1}(A) and (B) respectively. For $R_{d}/R_{b}\neq0$, the ZBCP at $E_{Th}=0.5\Delta_{0}$  undergoes a slight reduction in its sharpness in contrast to what is seen at $E_{Th}=0.05\Delta_{0}$ but the overall profile is more or less preserved, see curves (a-c) in Fig.\,\ref{pxwaveNazarovconductance1}(A,B).

\begin{figure}
\centering
\includegraphics[width=0.85\columnwidth]{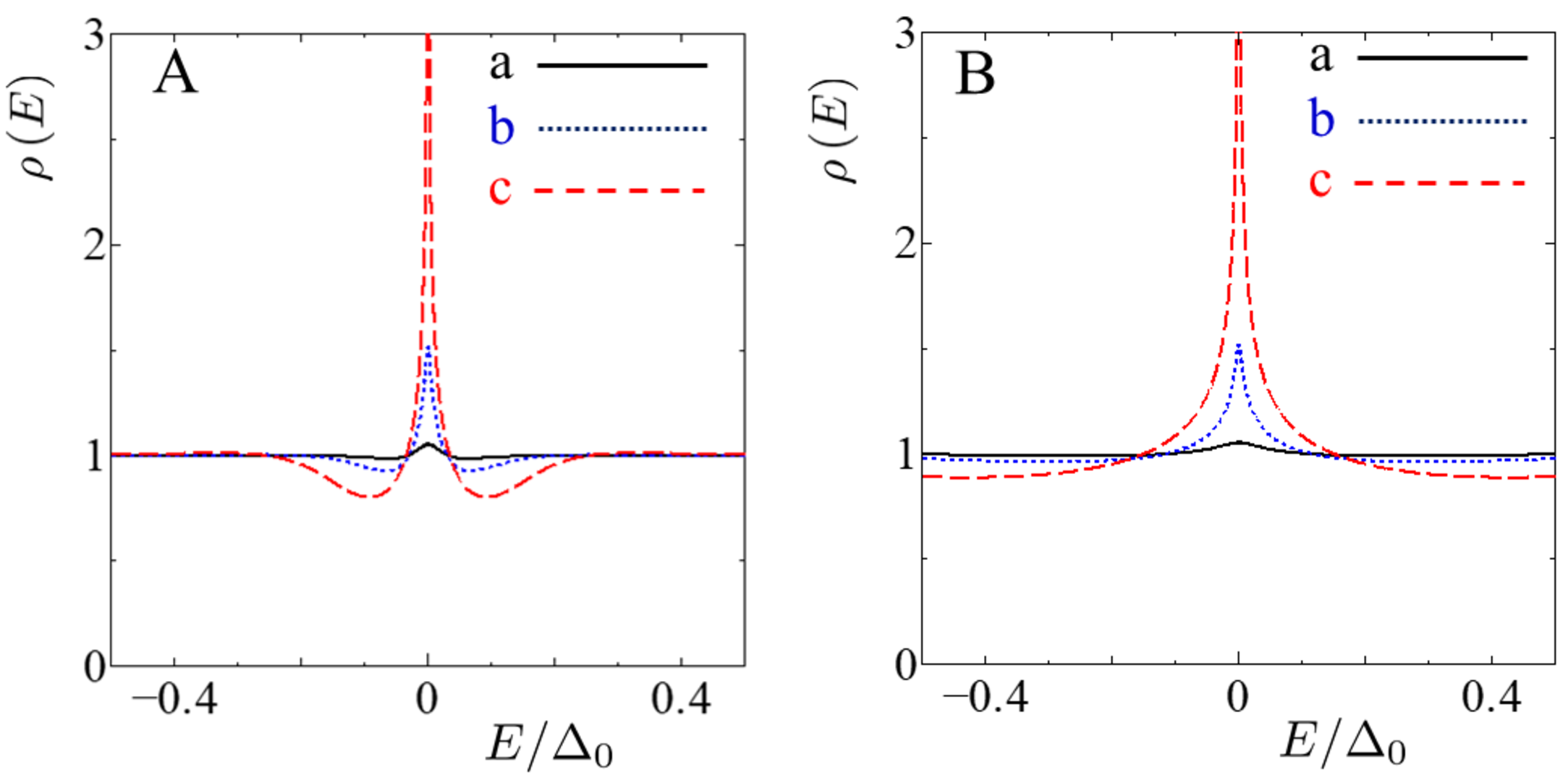}
\caption{LDOS $\rho(E)$  at $x=-L/2$ in a DNS junction where S is a 1D spin-polarized $p$-wave superconductor and DN is a diffusive normal metal. Here, $\rho(E)$  is normalized by its value in the normal state for $Z=1.5$. Left and right panels correspond to distinct values of the Thouless energy: A: $E_{Th}=0.05\Delta_{0}$, B: $E_{Th}=0.5\Delta_{0}$. Different curves in each panel correspond to distinct values of $R_{d}/R_{b}$: (a) $R_{d}/R_{b}=0.1$, (b) $R_{d}/R_{b}=0.3$, (c) $R_{d}/R_{b}=1$.}
\label{proximitypxwaveldos1}
\end{figure}

In order to further understand the proximity effect in the diffusive region DN,  we plot the LDOS 
$\rho(E)$ normalized by its value in the normal state 
at middle of DN in $x= -L/2$ in Fig.\,\ref{proximitypxwaveldos1}. 
For all cases, the resulting $\rho(E)$ has a sharp ZEP  \cite{Proximityp,Proximityp2,Kokkeler2022} and 
completely different from the standard proximity effect in conventional superconductor junction which commonly has a zero energy dip (gap) as seen in Fig.\,\ref{swaveKLboundaryLDOS}.  For the ballistic regime with $R_{d}=0$, the normalized LDOS $\rho(E)=1$ is satisfied due to $\zeta(x)=0$, which indicates the absence of the proximity effect.  The height of the ZEP becomes larger with the increase of $R_{d}/R_{b}$, as seen in curves (c) of Figs.\,\ref{proximitypxwaveldos1}(A) and (B). The functionalities of the ZEP can be understood by solving the Usadel equation Eqs.\,(\ref{EqsUsadel}) for $\zeta(x)$, with appropriate boundary conditions, which then gives the normalized LDOS $\rho\left(E \right) ={\rm Re}\left[ \cos\zeta\left(x \right) \right]$ 
\cite{Proximityp,Proximityp2}. Thus, at $E=0$ gives $\rho\left( 0 \right)
={\rm cosh}\left[{2R_{d}}/({R_{b}\sigma_{N}})\right]$, which clearly demonstrates that the ZEP in the LDOS depends on $R_{d}$, $R_{b}$, and $\sigma_{N}$. It is therefore evident that $\rho(0)$ gets enhanced with the increase of $R_{d}/R_{b}$ at fixed values of $\sigma_{N}$ as it happens in Fig.\,\ref{proximitypxwaveldos1}. Furthermore, it is worth pointing out that the LDOS for $E_{Th}=0.05\Delta_{0}$ exhibits a dip-like structure at $E= \pm E_{d}$  with   $0.05\Delta_{0}<E_{d} < 0.1\Delta_{0}$, see Fig.\,\ref{proximitypxwaveldos1}(A), with a prominent size for $R_{d}/R_{b}=1$, but, however, tends to disappear for $E_{Th}=0.5\Delta_{0}$. The fact that the LDOS develops a   ZEP  in the DN region as a result of the proximity effect due to the $p$-wave superconductor makes it different from the dip-like structure seen in conventional junctions and that is the reason why this effect is called anomalous proximity effect \cite{Proximityp,Proximityp2,Proximityp3,Tshape}. 

\begin{figure}
\centering
\includegraphics[width=12cm,clip]{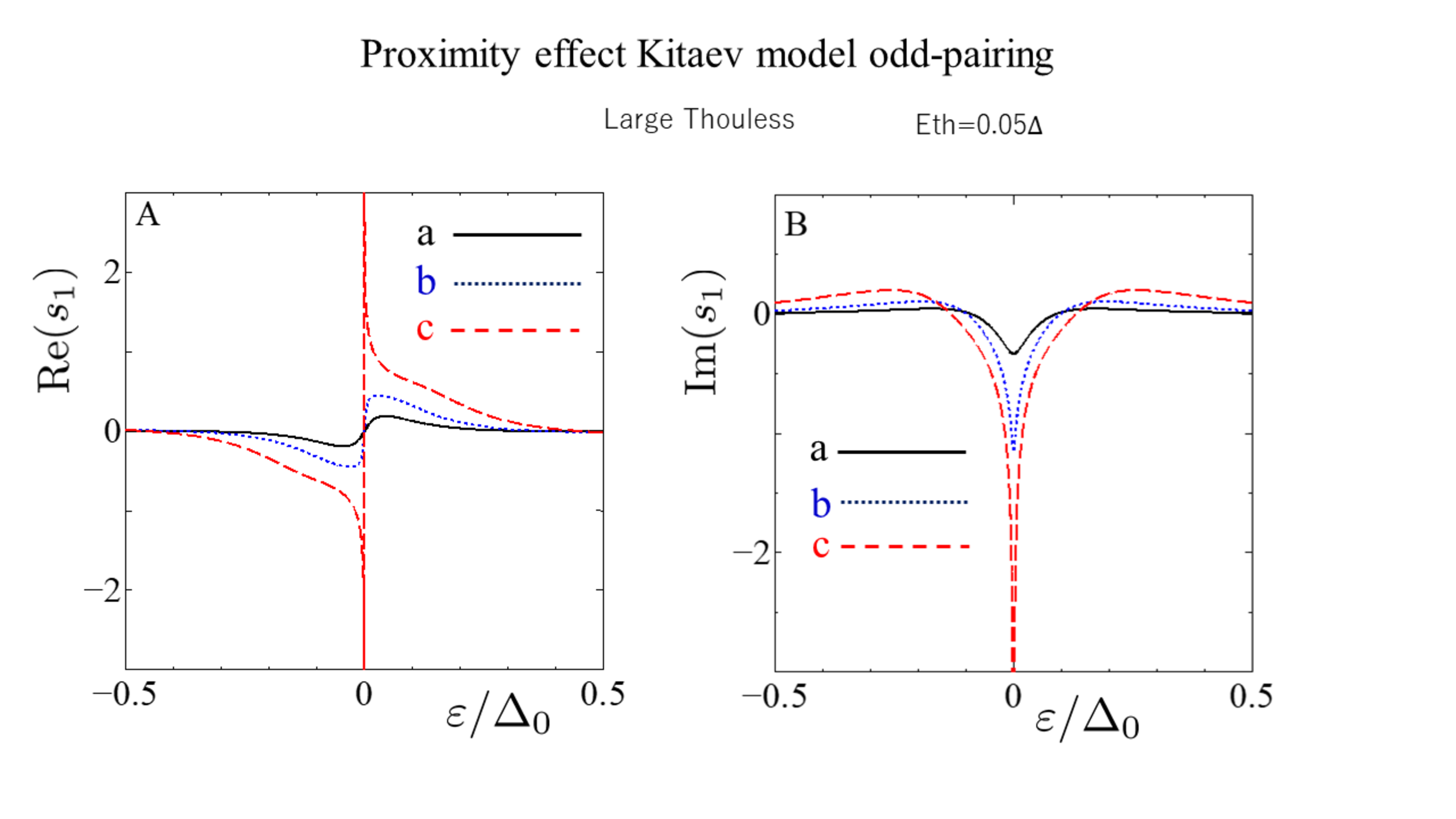}
\caption{Energy dependence of the OTE pair amplitude  at $E_{Th}=0.05\Delta_{0}$ in a DNS junction where S is a spin-polarized $p$-wave superconductor and DN a diffusive normal metal. Panels A and B show the real and imaginary parts, respectively. Different curves correspond to different values of  $R_{d}/R_{b}$: (a) $R_{d}/R_{b}=0.1$, (b) $R_{d}/R_{b}=0.3$,  (c) $R_{d}/R_{b}=1$.}
\label{proximityKitaevodd1}
\end{figure}

\begin{figure}
\centering
\includegraphics[width=0.9\columnwidth]{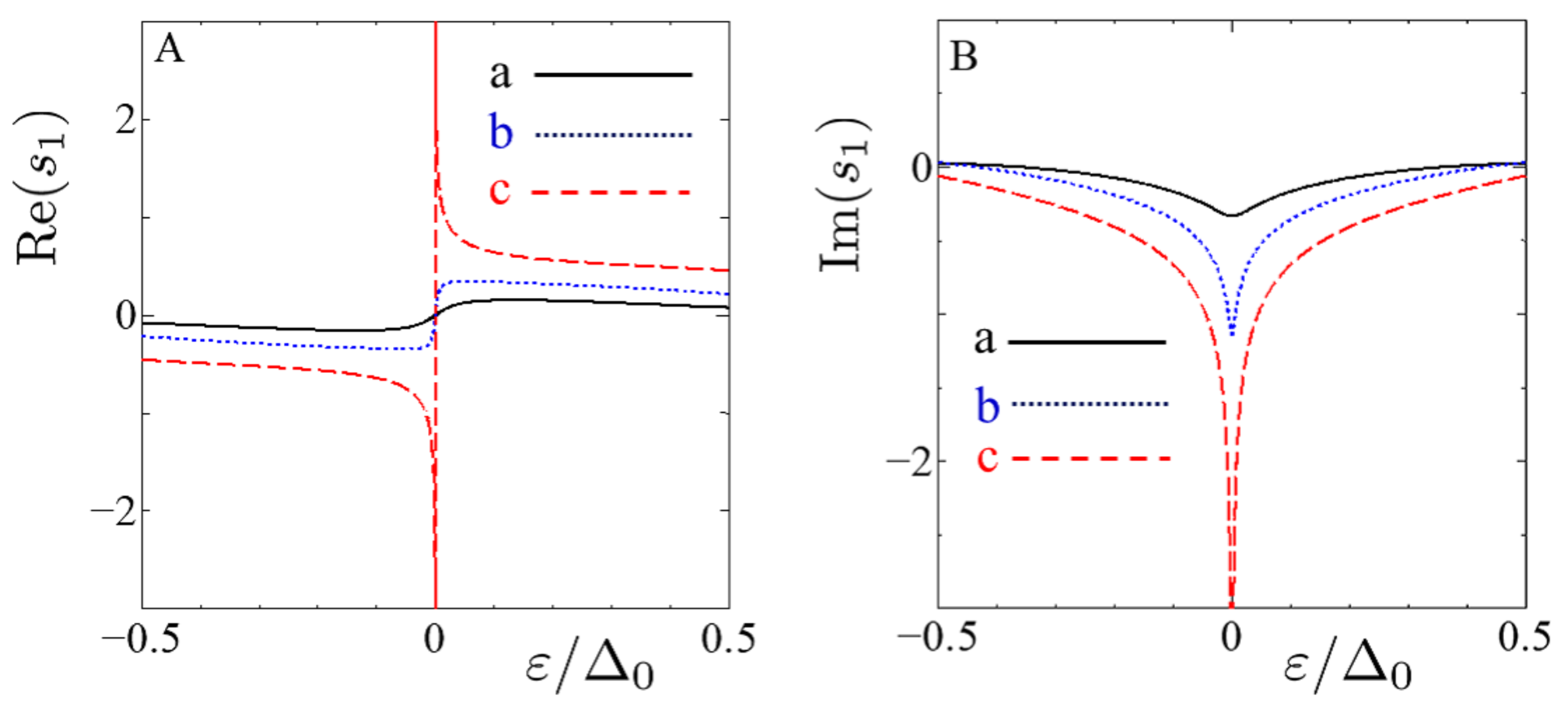}
\caption{Energy dependence of the OTE pair amplitude  at $E_{Th}=0.5\Delta_{0}$ in a DNS junction where S is a spin-polarized $p$-wave superconductor and DN a diffusive normal metal. Panels A and B show the real and imaginary parts, respectively. Different curves correspond to different values of  $R_{d}/R_{b}$: (a) $R_{d}/R_{b}=0.1$, (b) $R_{d}/R_{b}=0.3$, 
(c) $R_{d}/R_{b}=1$.}
\label{proximityKitaevodd1b}
\end{figure}

The anomalous proximity effect in this case is also characterized by having an induced pair amplitude in the DN region with odd-frequency spin-triplet $s$-wave symmetry \cite{odd1,odd3,tanaka12}. Since the edge of the $p$-wave superconductor hosts an MZM, the anomalous proximity effect can be seen as another Majorana effect which, in turn, can help detecting MZMs  \cite{tanaka12,Kokkeler2022}. The odd-frequency pair amplitude $s_{1}$ is plotted as a function of $E$ in Figs.\,\ref{proximityKitaevodd1} 
and \ref{proximityKitaevodd1b} for $E_{Th}=0,05\Delta_{0}$ and $E_{Th}=0.5\Delta_{0}$, respectively. In this case, the real part Re($s_{1}$) becomes an odd function of $E$ while the imaginary part Im($s_{1}$) becomes an even function of $E$.  The magnitudes of both Re($s_{1}$) and 
Im($s_{1}$) are enhanced around $E = 0$ \cite{odd1}.  This means that the ZEP and MZM always accompany the formation of odd-frequency pairing 
\cite{odd1,odd3,tanaka12,Asano2013,Spectralbulk}. 
 
\begin{figure}
\centering
\includegraphics[width=0.8\columnwidth]{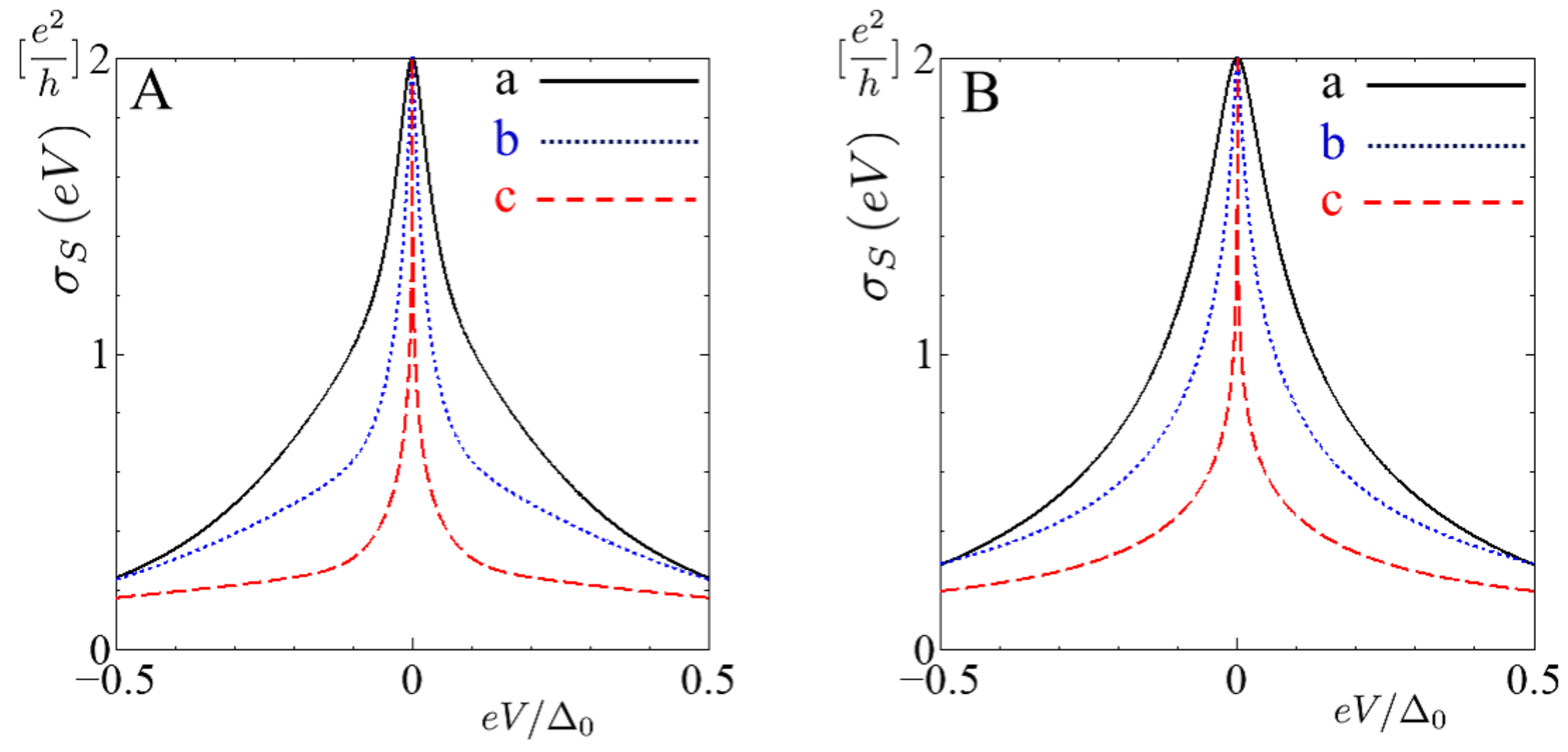}
\caption{Charge  conductance $\sigma_{S}(eV)$ in DN/$p$-wave superconductor junctions. 
(a)$R_{d}/R_{b}=0.1$, (b)$R_{d}/R_{b}=0.3$, 
(c)$R_{d}/R_{b}=1$. 
A: $E_{Th}=0.05\Delta_{0}$, B: $E_{Th}=0.5\Delta_{0}$ 
}
\label{quantizedpxwave}
\end{figure}

Finally, we address   the perfect resonance of charge conductance 
at zero voltage. For this purpose, we look at $\sigma_{S}(eV)=1/R$ 
which is not a normalized conductance, given by
\begin{equation}
\sigma_{S}\left( eV \right) 
= \frac{\sigma_{T} \left(eV \right)}{R_{d}+ R_{b}}
\end{equation}
As shown in Fig. \ref{quantizedpxwave}, the zero bias conductance $\sigma_{S}(0)$ is quantized and its value is $2e^{2}/h$ independent of any value of $R_{d}/R_{b}$ and $E_{Th}$. This is a very interesting and robust result that can be shown analytically  \cite{Proximityp,Proximityp2,Asano2013}. 

In fact, for a spin-triplet $p$-wave superconductor, it is possible to show that
\begin{equation}
\begin{split}
\int^{0}_{-L}dx \frac{1}{{\rm \cosh}^{2} 
\left[ \zeta_{Ni} \left(x + L \right)/L \right] } 
&= \frac{\sigma_{N}R_{b}}{2}
{\rm tanh}\zeta_{Ni}\,,\\
I_{K}&= 1+ \exp\left(2\zeta_{Ni}\right).
\end{split}
\end{equation}
where $
\zeta_{Ni}=\frac{2R_{d}}{\sigma_{N}R_{b}}$. Then, the total resistance of the junction $R$ at $eV=0$ becomes
\begin{equation}
R=\frac{1}{2}R_{0}\tanh(\zeta_{Ni})+R_{0}\frac{1}{1 + \exp(2\zeta_{Ni})}
\label{totalresistance}
\end{equation}
with $R_{0}=R_{b}\sigma_{N}$.  In the present case, $R_{0}$ becomes 
$h/e^{2}$ since the number of channel is only one. The total resistance of the junction then acquires the value  
$R=R_{0}/2$ and 
we obtain 
\begin{equation}
\sigma_{S}(eV=0)=\frac{2}{R_{0}}=
\frac{2}{R_{b}\sigma_{N}}=\frac{2e^{2}}{h}.
\label{quantization}
\end{equation}
In Eq.\,(\ref{totalresistance}),  the first term comes from the resistance in DN and 
it reduces to be $R_{0}/2$ for $R_{d} \rightarrow \infty$. 
On the other hand, the second term is from the resistance 
at the interface and it reduces to zero at large $R_{d}$ limit. 
It is remarkable, that the obtained conductance (resistance) at zero voltage 
is completely independent of $R_{d}$ and $R_{b}$. 
It is a strong feature of the proximity effect of MZM 
into DN \cite{Proximityp,tanaka12,Asano2013,Takagi2020}. 
The quantization of $\sigma_{S}(eV)$ can be  also explained 
from the viewpoint of the index theorem \cite{Ikegaya2016}.

Finally, we would like to address the question that the observed quantization is robust against the boundary condition at the left interface between the electrode and DN. In the present analysis, the resistance between the electrode and the 
DN is set to be zero. It is possible to extend in the case where the 
resistance at the interface at $x=-L$ 
is $R_{b'}$ and the trasmissibity at the interface is $\sigma_{N2}$ 
given by 
\[
\sigma_{N2}=\frac{1}{Z'^{2}+ 1}
\]
with $R_{b'}=R_{0}/\sigma_{N2}$.  

The total resistance of the junction is then given by 
\begin{equation}
R= \frac{R_{b'}}{\langle I'_{K}\rangle} 
+\frac{R_{d}}{L}\int_{-L}^{0}\frac{dx}{\cosh
^{2}\zeta_{\rm{Im}}\left(x \right)}
+ 
\frac{R_{b}}{\left<I_{K} \right>}
\label{totalextension}
\end{equation}
and after a cumbersome calculation, it is possible to obtain  at $E=0$, 
\begin{equation}
\label{xxxxxxx}
\begin{split}
\langle I'_{K} \rangle
&=-\frac{2}{\sigma_{N2}}
\frac{\left( 2 - \sigma_{N2} \right)^{2} + \sigma_{N2}^{2} }
{\left( 2 - \sigma_{N2} \right)^{2} - \sigma_{N2}^{2}}\,,\\
\frac{R_{b'}}{\langle I'_{K} \rangle}&
= -\frac{R_{0}}{2}
\frac{
\left(2 - \sigma_{N2} \right)^{2} - \sigma_{N2}^{2} 
}
{\left(2 - \sigma_{N2} \right)^{2} + \sigma_{N2}^{2}}
=\frac{R_{0}}{2}{\rm tanh} \zeta_{\rm Im}\left(x=-L \right)\,,\\
\frac{R_{d}}{L}\int_{-L}^{0}\frac{dx}{\cosh
^{2}\zeta_{\rm{\rm Im}}\left(x \right)}
&=\frac{R_{0}}{2}
\left( {\rm tanh}\zeta_{\rm Im}\left(x=0_{-} \right)
-{\rm tanh}\zeta_{\rm Im}\left(x=-L \right)
\right)\,,\\
\frac{R_{b}}{\langle I_{K} \rangle}
&= \frac{R_{0}}{1 + \exp\left[2\zeta_{\rm Im}\left(x=0_{-}\right)\right]}\,.
\end{split}
\end{equation}
Then, by plugging Eqs.\,(\ref{xxxxxxx}) into Eq.\,(\ref{totalextension}), we obtain 
$R=R_{0}/2$ and from Eq.\,(\ref{quantization}) it is not difficult to see that the perfect conductance quantization due to MZMs is verified. We note that the anomalous proximity effect is absent in DN-S when S is a $d$-wave superconductor \cite{Proximityd,Proximityd2}; in this case, the resulting proximity effect becomes the conventional proximity effect realized in junctions between DN and spin-singlet $s$-wave superconductors. It is worth noting that the anomalous proximity effect has been also studied in non-centrosymmetric superconductor junctions with coexistent spin-singlet $s$-wave and spin-triplet $p$-wave pair potentials \cite{Kokkeler2022,Kokkeler2023}, which showed that the anomalous proximity effect remains when the $p$-wave component is dominant. It is thus possible to establish a direct relationship between the robustness of the anomalous proximity effect and $p$-wave nature of superconductivity which gives rise to MZMs.

Before closing this section, we would like to briefly mention about 
the anomalous proximity effect appearing in the zero-voltage charge conductance in 1D $p$-wave superconductors, where an MZM appears at the edge. Although ZBCP appears in the non-topological superconductor, the present ZBCP by the anomalous proximity effect shows remarkable features. 
The height of ZBCP can become extremely  larger as compared to the 
background value of high voltage value above energy gap. 
Here, we denote the conductance as a function of $eV$ and $T$ as $\sigma_{S}(eV,T)=\sigma_{S}(eV)$. 
First, we focus on the voltage dependence for $T=0$. 
As seen from the perfect resonance independent of the 
resistance in diffusive normal metal $R_{D}$ and that at the insulating barrier $R_{b}$ \cite{Proximityp,Proximityp2,Asano2013}
\begin{equation}
\sigma_{S}(0,0)/\sigma_{S}(eV_{c},0) =2e^{2}(R_{B} + R_{D})/h
\label{ratiovoltage}
\end{equation}
is satisfied. Here, $eV_{c}$ is sufficiently large as compared to $\Delta_{0}$. With the increase of $R_{B} + R_{D}$, the 
ratio in Eq. (\ref{ratiovoltage}) 
increases monotonically which is completely different from 
conventional junctions and can be taken as a signature of MZMs. 
Interestingly, if we look at the temperature dependence of $\sigma_{S}(eV=0,T)$,
\begin{equation}
\sigma_{S}(0,0)/\sigma_{S}(0,T) =2e^{2}(R_{B} + R_{D})/h 
\end{equation}
for $T>T_{c}$, we obtain the same expression 
as Eq. (\ref{ratiovoltage}).  
Here, $T_{c}$ represents the 
critical temperature of the superconductor
where the resistance of the junction is 
$R_{D} +R_{B}$. In this regard, temperature dependence is also a good indicator of MZMs. We emphasize that the ZBCP cannot be taken as the only indicator of MZMs; the temperature and voltage dependencies of the normalized conductance discussed here also provide complementary evidence on MZMs. The ZBCP and the enhancement of conductance at zero voltage are  results of topologically protected MZMs accompanying odd-frequency spin-triplet $s$-wave pairing \cite{Proximityp,odd1,tanaka12}

\subsection{Summary}
In this part we have discussed charge transport and LDOS in NS junctions formed by a finite diffusive normal metal N and a superconductor S. We have shown that when the superconductor has a spin-singlet $s$-wave pair potential,   the resulting conductance as a function of energy can exhibit a U-shaped profile but it can also have a peak at zero energy due to impurity scattering, while the LDOS in this case can produce a dip-like feature. The ZBCP, however, is always below perfect quantization and it is not robust under the variation of parameters.   
Interestingly, for spin-polarized $p$-wave superconductors, the conductance develops a   ZBCP that is robust against impurity scattering. Moreover, this quantized ZBCP is also accompanied by a zero energy peak in the LDOS and dominant odd-frequency spin-triplet pair correlations, which, together with the ZBCP, define the anomalous proximity effect. Thus, the anomalous proximity effect could be taken as an alternative signature to further explore the detection of MZMs and topological superconductivity \cite{tanaka12,Spectralbulk}.

\section{Detecting Majorana states via the Josephson effect}
\label{section7}
The Josephson effect is one of the most fundamental 
phenomena in superconductivity and was predicted by Brian Josephson more than 60 years ago 
~\cite{Josephson}.  The  Josephson effect reveals the flow of Cooper pairs, known as supercurrent, between two superconductors forming a Josephson junction simply due to a finite phase difference between their pair potentials and without any applied voltage bias \cite{Kulik1970,ishii1970josephson,ishii1972thermodynamical,bardeen1972josephson,svidzinsky1973concerning,kulik1975,RevModPhys.51.101}. In 1991, Akira Furusaki demonstrated that the d.c. Josephson effect can be understood as a result of Andreev reflection processes happening on both interfaces, which then naturally account for the transfer of Cooper pairs between superconductors \cite{Furusaki91}; see also Ref.\,\cite{furusaki1990unified,furusaki1991current,PhysRevB.45.10563}. This study has paved the way to understand the Josephson effect as a result of phase-dependent ABSs \cite{Furusaki91,furusaki1990unified,furusaki1991current,PhysRevB.45.10563,Beenakker91,PhysRevB.45.10563,Beenakker:92,Furusaki_1999}, which are formed between the superconductors and exhibit unique profiles for superconductors with unconventional pair potentials \cite{kashiwaya00,PhysRevB.64.224515,PhysRevLett.96.097007,RevModPhys.76.411,sauls2018andreev,mizushima2018multifaceted}. Thus, since Cooper pairs are involved in the transport across the junction \cite{RevModPhys.51.101,Tinkham}, the Josephson effect has the potential to reveal the type of emergent superconductivity. It is worth mentioning that the Josephson effect is also at the core of interesting quantum applications, see Refs.\,\cite{devoret2005implementing,acin2018quantum,krantz2019quantum,aguado2020perspective,benito2020hybrid,aguado2020majorana,doi:10.1146/annurev-conmatphys-031119-050605,PRXQuantum.2.040204,siddiqi2021engineering} for more details. In this part, we are interested in exploring the d.c. Josephson effect in junctions based on unconventional superconductors and how  ZESABSs and MZMs can be identified using phase-biased Josephson transport.

\subsection{Josephson current in 1D spin-singlet  $s$-wave superconductor junctions}
To set a pedagogical ground, we first focus on the Josephson effect in a 1D Josephson junction formed by a superconductor-insulator-superconductor (SIS) structure. To describe the Josephson effect, we will first obtain the ABSs and then the supercurrent, see Refs.\,\cite{Beenakker:92,cayao2018andreev}. Thus, to find the ABSs we need to employ the BdG equations discussed in Section \ref{subsection11} with spin-singlet $s$-wave pair potential. The insulating region is modeled by a delta function barrier potential without superconductivity, while the S regions have a finite and constant spin-singlet $s$-wave pair potential.  Thus, the delta potential and the pair potential are given by  
\begin{equation}
\label{Smodeldeltafunction}
\Delta(x)=
\left \{
\begin{array}{ll}
\Delta_{L}\,, & x < 0\,, \\
\Delta_{R}\,, & x>0\,, 
\end{array}
\right.\quad
U(x)=H_{b}\delta(x)\,, 
\end{equation}
where   $H_{b}$ is the strength of the delta barrier, $\Delta_{L(R)}=\Delta_{0}\exp(i\varphi_{L(R)})$ the pair potential of the left (right) superconductor with $\varphi_{L(R)}$ being its associated superconducting phase and $\Delta_{0}$   the isotropic spin-singlet $s$-wave pair potential amplitude. Without loss of generality, we take  $\varphi_{L}=\varphi$ and $\varphi_{R}=0$, such that there is a finite phase difference $\varphi$ across the junction. We focus on subgap ABS energies, which means that wavefunctions have to exponentially decay as $x$ moves away from either side of the junction. In this case, the wavefunction of the BdG equations is given by 
\begin{equation}
\Psi(x)=
\left \{
\begin{array}{ll}
A\exp (-ik^{+}x)
\begin{pmatrix}
1 \\
\Gamma \exp(-i\varphi)
\end{pmatrix}
+ 
B \exp (ik^{-}x)
\begin{pmatrix}
\Gamma \exp(i\varphi) \\
1
\end{pmatrix}\,,
&
x \leq 0\,, \\
C \exp (ik^{+}x)
\begin{pmatrix}
1 \\
\Gamma
\end{pmatrix}
+
D \exp (-ik^{-}x)
\begin{pmatrix}
\Gamma \\
1
\end{pmatrix}\,,
&
x>0\,,
\end{array}
\right.
\label{wavefunctionSIS}
\end{equation}
where $k^{\pm}=\sqrt{\frac{2m}{\hbar^{2}}(\mu \pm \Omega)}$, 
and $\Gamma=E/(E + \Omega)$. The coefficients $A$, $B$, $C$, and $D$ are then found from the boundary conditions for $\Psi(x)$  given by
\begin{equation}
\begin{split}
\Psi\left( x=0_{-} \right) &= \Psi\left( x=0_{+} \right),\,,\\
\frac{d}{dx}\Psi\left( x=0_{+} \right)
- \frac{d}{dx}\Psi\left( x=0_{-} \right)
&=\frac{2m}{\hbar^{2}}\Psi \left( x=0_{+} \right)
\label{boundaryconditionj}
\end{split}
\end{equation}
The ABS energies are then obtained from the nontrivial solutions of Eq.\,(\ref{boundaryconditionj}), in the same way as we have done to obtain Eq.\,(\ref{ABSUNSC}) in subsection \ref{subsection2d}. Moreover, we consider the limit of  $\mu \gg \Delta_{0}$ where wavevectors as simply approximated as the Fermi wavevector $k^{\pm}\sim k_{F}$. Thus, by taking these conditions into account, we obtain the ABS energies located at $x=0$ to be given by \cite{furusaki1990unified}
\begin{equation}
E_{b}(\varphi)= \pm \Delta_{0}\sqrt{ 1 - \sigma_{N} \sin^{2}\left(\frac{\varphi}{2}\right)}\,, 
\label{Andreevboundstate1dswave}
\end{equation}
where $\sigma_{N}$ is the normal transparency at the interface 
given by $\sigma_{N}=1/(1 + Z^{2})$, and $Z=mH_{b}/(\hbar^{2}k_{F})$. These ABS energies are degenerate in spin due to the considered model. We note that there are two ABSs,  with positive and negative energies.  When $Z\ll1$, which corresponds to $\sigma_{N}=1$ and defines the fully transparent regime, the ABSs energies read $E_{b}(\varphi)=\pm\Delta_{0}{\rm cos}(\varphi/2)$: this means that the ABSs reach zero energy at $\varphi=\pi$. In contrast, for non-zero barrier strengths $Z\neq0$, the ABSs energies $\varphi=\pi$ become  $E_{b}(\varphi)=\pm\Delta_{0}\sqrt{1-\sigma_{N}}$, leading to a finite energy gap.

The supercurrent across the SIS Josephson junction can be obtained from the thermodynamic relation \cite{zagoskin}
\begin{equation}
I\left(\varphi \right)
=\frac{1}{\Phi} \frac{\partial F\left(\varphi \right)}{\partial \varphi} \,,
\end{equation}
where $\Phi=\hbar/2e$ is the reduced superconducting  flux 
quantum, while $F(\varphi)$ is the free energy that in general depends on other parameters of the system. Assuming spin degeneracy, the free energy can be calculated as
\begin{equation}
F\left(\varphi \right)
=-2 k_{B}T 
\log \left[2 \cosh\left(\frac{E_{b}}{2k_{B}T} \right) \right]\,,
\end{equation}
with ${\rm log}$ represents the natural logarithm and $T$ temperature. Thus,   by calculating the derivative of $F(\varphi)$ with respect to $\varphi$, we arrive at \cite{Furusaki91}
\begin{equation}
I\left(\varphi \right)
=\frac{1}{R_{N}}\frac{\pi \Delta_{0}}{2e}
\frac{\sin(\varphi)}{\sqrt{1 - \sigma_{N} \sin^{2}\frac{\varphi}{2}}}
{\rm tanh}\left(
\frac{\Delta_{0}}{2k_{B}T}
\sqrt{ 1 - \sigma_{N}\sin^{2}\frac{\varphi}{2}} \right) \,, 
\label{currentphase1}
\end{equation}
where $R_{N}={\pi \hbar}/{(e^{2}\sigma_{N})}$  is the resistance of the junction in the normal state. For sufficiently low transparent junction with large $Z$, 
$R_{N}I(\varphi)$ becomes 
\begin{equation}
R_{N}I\left(\varphi \right)=\frac{\pi \Delta_{0}}{2e} \sin(\varphi) {\rm tanh}
\left(\frac{\Delta_{0}}{2k_{B}T} \right)\,,
\end{equation}
where current phase relation becomes simple sinusoidal
relation given by Ambegaokar and Baratoff ~\cite{Ambegaokar}. 
On the other hand, for fully transparent limit with $Z=0$ and $\sigma_{N}=1$, 
$R_{N}I(\varphi)$ becomes 
\begin{equation}
R_{N}I\left(\varphi \right)
=\frac{\pi \Delta_{0}}{e} \sin\left(\frac{\varphi}{2}\right)\,
{\rm tanh}\left(\frac{\Delta_{0}\cos\left(\frac{\varphi}{2}\right)}{2k_{B}T} \right)\,,
\label{JosephsonKulik}
\end{equation}
where the current phase relation is proportional to 
$\sin(\varphi/2)$ for $-\pi< \varphi < \pi$ at sufficiently low temperature~\cite{Kulik}. Before going further, it is useful to note that $\Delta_{0}$ depends on temperature $T$  obeying
\begin{equation}
\Delta_{0}=\Delta_{0}\left(T=0 \right)
{\rm tanh}
\left( 1.74 \sqrt{\frac{T_{C}-T}{T}} \right)
\end{equation}
where $\Delta_{0}\left(T=0 \right)$ is the zero temperature pair potential, while $T_{C}$ the critical temperature. 
Now, to visualize the behaviour of the Josephson current, in Fig.\,\ref{1dJosephsonphase}(A) we show its phase dependence in the tunnel ($Z=3$) and transparent ($Z=0$) regimes. We clearly observe that $I(\varphi)$ has a simple sinusoidal behaviour at low transparencies, while it develops a strong skewness in the transparent regime, alike a sawtooth profile, see red and blue curves in  Fig \ref{1dJosephsonphase}(A). Moreover, the maximum supercurrents $I_{C}$, known as critical currents, are plotted in Fig.\,\ref{1dJosephsontemp}(A) as a function of temperature for the tunnel and transparent regimes. Here we notice that $R_{N}I_{C}$ decreases as a temperature increases, acquiring a similar decreasing behaviour near the critical temperature; at low temperatures however, the transparent regime exhibits larger values. 
 
\subsection{Josephson current in 1D  spin-polarized $p$-wave superconductor junctions}
In this part, we explore the Josephson effect in Josephson junctions between 1D superconductors having spin-triplet $p$-wave  pair potentials. The structure of the junction is the same as previous subsection described by Eqs.\,(\ref{Smodeldeltafunction}) but now with the pair potentials for a 1D $p$-wave superconductor \cite{Yakovenko2004}, namely, $\Delta_{L(R)}=\hat{k}\Delta_{0} \exp(i\varphi_{L(R)})$  with $\hat{k}=k/k_{\rm F}$.  We then obtain the ABS energies and supercurrent following the same steps as described in previous subsection but using the BdG equations within the quasiclassical approximation pointed out in Subsection \ref{subsection13}. Then,   the corresponding wavefunction can be written as
\begin{equation}
\Psi(x)=
\left \{
\begin{array}{ll}
A\exp (-ik^{+}x)
\begin{pmatrix}
1 \\
-\Gamma \exp(-i\varphi)
\end{pmatrix}
+ 
B \exp (ik^{-}x)
\begin{pmatrix}
\Gamma \exp(i\varphi) \\
1
\end{pmatrix}\,,
&
x \leq 0\,, \\
C \exp (ik^{+}x)
\begin{pmatrix}
1 \\
\Gamma
\end{pmatrix}
+
D \exp (-ik^{-}x)
\begin{pmatrix}
-\Gamma \\
1
\end{pmatrix}\,,
&
x>0\,.
\end{array}
\right.
\label{wavefunctionPIP}
\end{equation}
\begin{figure}[t]
\begin{center}
\includegraphics[width=0.8\columnwidth]{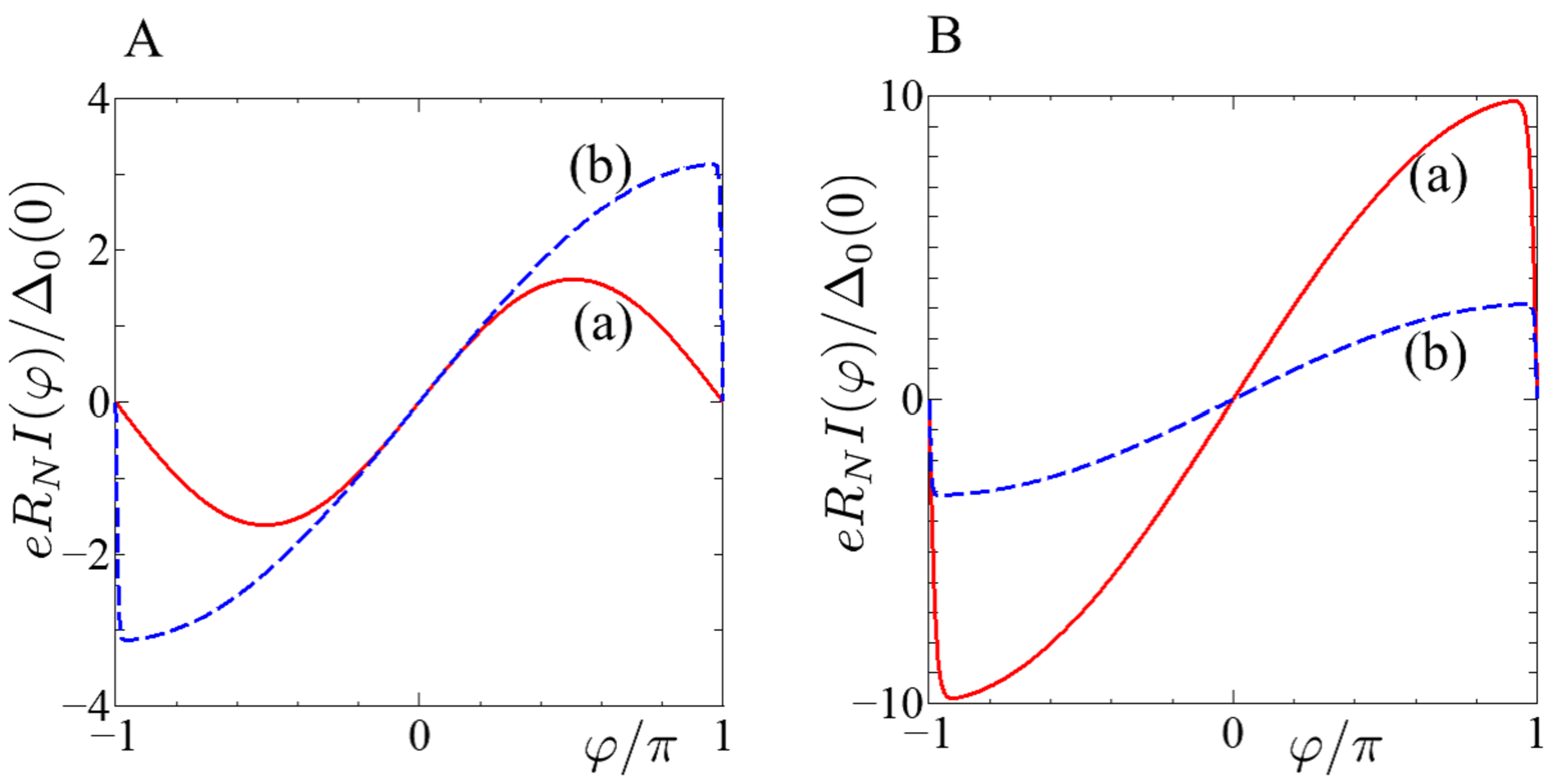}
\end{center}
\caption{Phase dependent supercurrents $I(\varphi)$ in 1D Josephson junctions
  for A: $s$-wave superconductors and B: $p$-wave superconductors.  The curves in each panel correspond to (a)$Z=3$, (b)$Z=0$.  $\Delta_{0}(0)$ is the magnitude of the pair potential $\Delta_{0}$ at 
zero temperature.}
\label{1dJosephsonphase}
\end{figure} 
Hence, by imposing the boundary conditions given by Eqs.\,(\ref{boundaryconditionj}) and for subgap energies with $\mu\gg\Delta_{0}$,  the energy  of the ABS is given by \cite{Yakovenko2004}
\begin{equation}
E_{b}=\pm \Delta_{0} \sqrt{\sigma_{N}} 
\cos\left(\frac{\varphi}{2}\right)\,,
\label{Majoranaboundstate}
\end{equation}
where $\sigma_{N}$ is the normal transparency. Interestingly, the junction in the regime under investigation is already topological, see Subsection \ref{section3}, which implies that the ABS energies found in Eq.\,(\ref{Majoranaboundstate}) correspond to topological ABSs.  These ABSs are gapless no matter what value the normal transparency $\sigma_{N}$ acquires, thus developing a robust zero energy crossing at $\varphi=\pi$ 
unlike the ABSs in conventional Josephson junctions of the previous subsection; the only effect of $\sigma_{N}$  here is to reduce the ABSs from the gap edges. The zero-energy protected crossing signals the emergence of a pair of MZMs at the inner sides of the junction. These topological  ABSs $E_{b}$  given by  Eq. \,(\ref{Majoranaboundstate}) have also been  
explored due to the hybridization of ZESABSs in 
$d$-wave superconductor junctions \cite{TK96a,TK97,kashiwaya00}. 
A consequence of the zero-energy crossing is that these topological ABSs exhibit a $4\pi$-periodicity  with $\varphi$, a remarkable property that was initially derived in the context of $d$-wave superconductor junctions in 1996 \cite{TK96a}.  Then, the  supercurrent due to the topological ABSs is given by  
\begin{equation}
R_{N}I\left( \varphi \right)
= \frac{\pi\Delta_{0}}{e\sqrt{\sigma_{N}}}\,\sin\left(\frac{\varphi}{2}\right)\,
{\rm tanh} \left(\frac{\Delta_{0}\sqrt{\sigma_{N}} \cos\left(\frac{\varphi}{2}\right)}
{2k_{B}T} \right)\,,
\label{currentphase1p}
\end{equation}
\begin{figure}[t]
\begin{center}
\includegraphics[width=0.8\columnwidth]{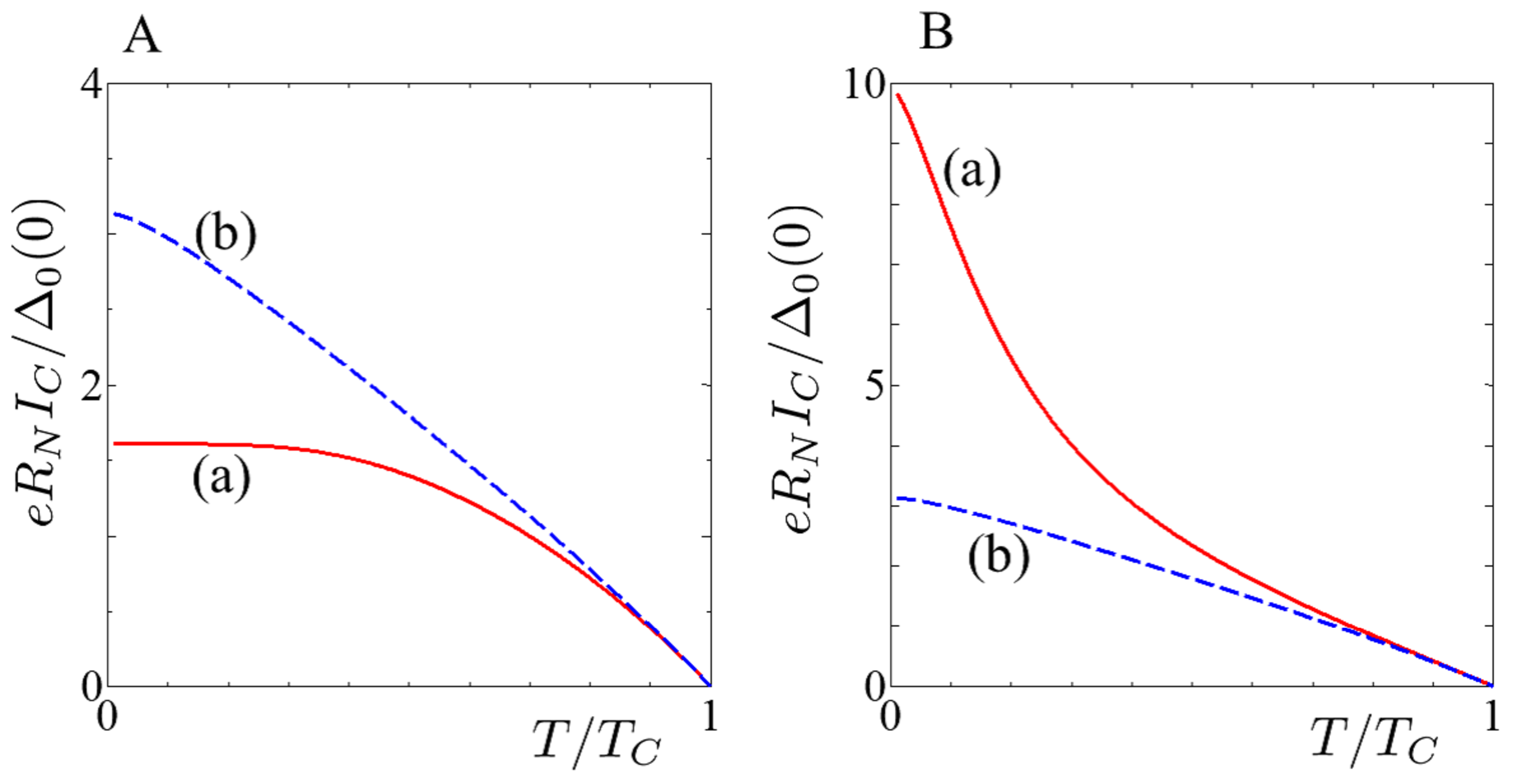}
\end{center}
\caption{Temperature dependence of the critical currents $I_{C}$ in 1D Josephson junctions
  for A: $s$-wave pair superconductors and B: $p$-wave superconductors. The curves in each panel correspond to (a)$Z=3$, and (b)$Z=0$. $\Delta_{0}(0)$ is the magnitude of the pair potential $\Delta_{0}$ at  zero temperature.}
\label{1dJosephsontemp}
\end{figure} 
where $R_{N}=\pi\hbar/(e^{2}\sigma_{N})$ is the junction's resistance in the normal state. For $\sigma_{N}=1$, Eq.\,(\ref{currentphase1p}) 
coincides with Eq.\,(\ref{JosephsonKulik})
for SIS junctions with $s$-wave superconductors.  The phase-dependent supercurrents are presented in Figs.\,\ref{1dJosephsonphase}(B) in the tunnel and transparent regimes. In Fig.\,\ref{1dJosephsontemp}(B) we also show the temperature dependence of their associated critical currents. The first observation is that the supercurrent exhibits a strong skewness near $\varphi=\pm\pi$ as a result of the robust zero-energy crossing of their ABSs and does not depend on the junction transparency, unlike what occurs in conventional Josephson junctions. Thus, the current phase relation seriously deviates from a simple sinusoidal one and can be approximated by   $\sin \varphi/2$ within $-\pi \leq \varphi \leq \pi$. 
Nevertheless, the supercurrent remains $2\pi$-periodic as a function of $\varphi$; under fermion parity conservation, however, the zero-energy  crossing at $\varphi=\pi$ is parity protected and the supercurrent becomes $4\pi$-periodic with $\varphi$ \cite{Yakovenko,kitaev}.
In relation to the temperature dependence, the supercurrent  $R_{N}I(\varphi)$  at sufficiently low $\sigma_{N}$ is proportional to $1/T$, see red curve in Fig.\,\ref{1dJosephsontemp})(B). At     low temperatures, however, the supercurrent  $R_{N}I(\varphi)$ is proportional to $1/\sqrt{\sigma_{N}}$ \cite{TK97,Yakovenko2004}. 
 Thus, the magnitude of Josephson curren for a topological Josephson junction made of 1D spin-polarized $p$-wave superconductors is seriously enhanced as compared to conventional Josephson junctions.  This remarkable feature was initially reported during 1996-1997 in the context of    $d$-wave superconductor junctions, see Refs. \,\cite{TK96a,TKJosephson,TK97,YBJosephson}. The critical current enhancement at low temperatures seen in $p$-wave Josephson junctions, opposite to what occurs in conventional Josephson junctions, is due to the presence of MZMs; due to ZESABSs in $d$-wave Josephson junctions. 
\begin{figure}[t]
\begin{center}
\includegraphics[width=0.99\columnwidth]{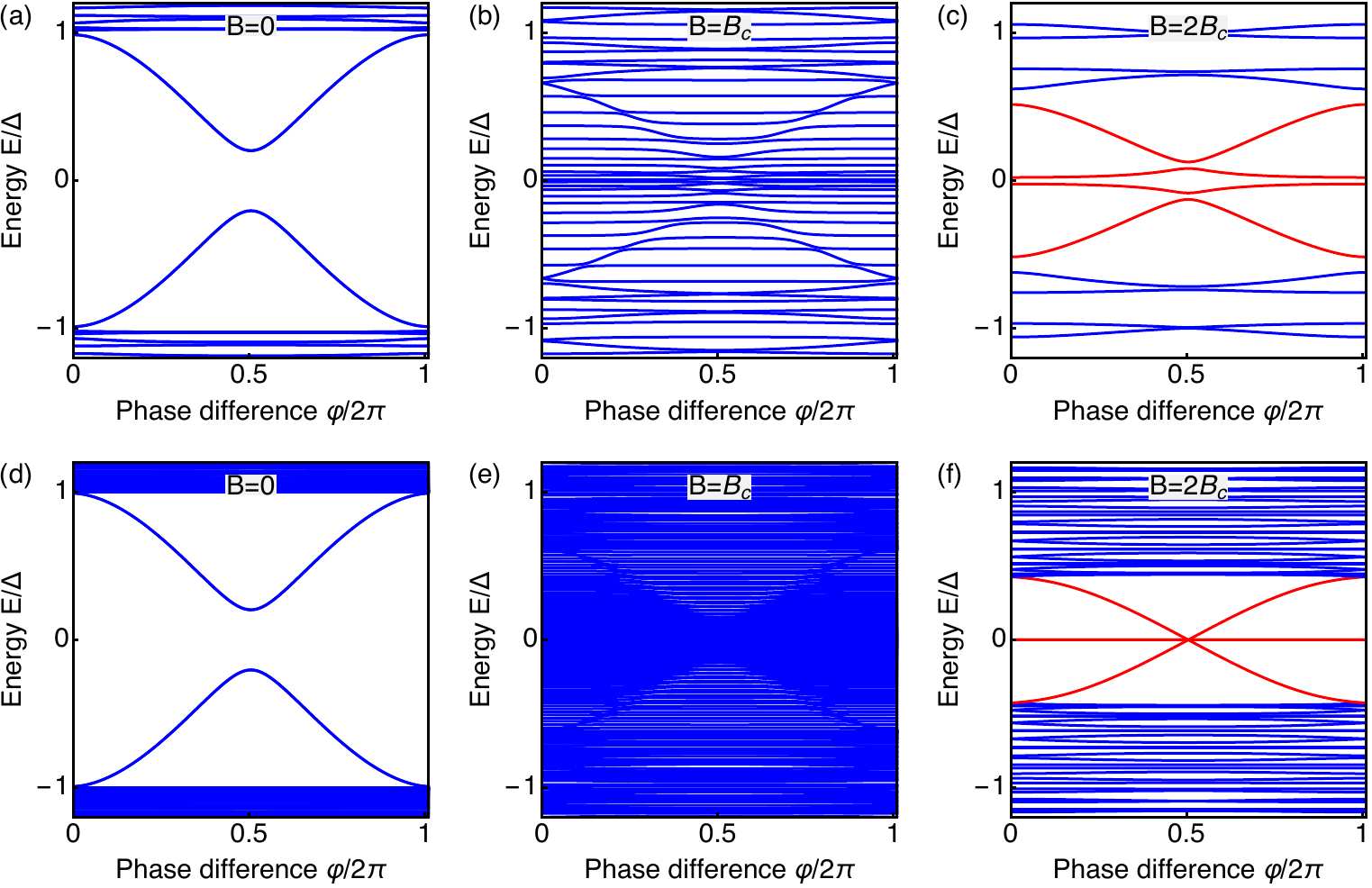}
\end{center}
\caption{Phase dependent low-energy Andreev spectrum in a  SNS Josephson junction based on nanowires with SOC for S region length with $L_{S}=2\mu$m (a-c) and $L_{S}=10\mu$m (d-f). Different panels show the evolution with the Zeeman field $B$. The red curves in (c,f) indicate the four lowest ABSs giving rise to four MZMs at $\varphi=\pi$ for long S regions. Parameters: $\alpha_{R}=20$meVnm, $\Delta=0.25$meV, $\mu_{N(S)}=0.5$meV, $L_{N}=20$nm, $L_{S}=2\mu$m. }
\label{SNSRashbaABS}
\end{figure} 

To close this part, we briefly discuss 1D SNS phase-biased Josephson junctions with the ingredients used to engineer $p$-wave superconductivity, namely,  nanowires SOC, $s$-wave superconductivity, and under the presence of Zeeman field, see subsection \ref{subsection3b}. These junctions are commonly modeled by using the tight-binding description discussed in subsection \ref{subsection3b}, where the 1D lattice is separated into a middle N region without pair potential and two S regions on the left and right with a finite pair potential. Moreover, a finite phase difference is considered across the junction. For details on the modeling, we refer the reader to Refs.\,\cite{cayao2018andreev,PhysRevB.96.205425}. We first analyze the evolution of the low-energy phase-dependent spectrum as the Zeeman field increases, which is presented in Fig.\,\ref{SNSRashbaABS} for short and long S regions, both with a very short N region. At zero Zeeman $B=0$, a pair of ABSs within the gap $\Delta$ emerge developing a finite gap $\varphi=\pi$, as expected since the considered parameters place the junction away from the Andreev approximation. As $B$ increases, the ABSs split, and one of them lower in energy as $B$ approaches the critical field $B_{c}$ which now is given as $B_{c}=\sqrt{\mu_{S}^{2}+\Delta^{2}}$, where $\mu_{S}$ is the chemical potential in S and $\Delta$ is the pair potential in S. For $B<B_{c}$ the ABSs do no depend on $L_{S}$, as seen by comparing Fig.\,\ref{SNSRashbaABS}(a,d). At $B=B_{c}$, the gap closes and the S regions undergo a topological phase transition into a topological phase for $B>B_{c}$. We see that the gap closing is more noticeable for longer S regions. Interestingly, as the S regions enter into the topological phase, the spectrum shows four  ABSs, which are now topological, within the lowest gap which is different from $\Delta$, see Fig.\,\ref{SNSRashbaABS}(c). There appear two levels around zero energy with a weak dependence for all $\varphi$, while the other two levels approach zero energy $\varphi=\pi$. Quite remarkably, for longer S regions, the energy levels closest to zero become flat at zero energy, while the other subgap pair of levels really reach zero energy at $\varphi=0$,  see Fig.\,\ref{SNSRashbaABS}(c). It has been shown that the dispersionless energy levels define two MZMs at $\varphi=0$ located at the outer ends of the S regions,  while the four topological ABSs define four MZMs at $\varphi=\pi$, with two additional MZMs located at the inner sides of the S regions, see Refs.\,\cite{San-Jose:11a,PhysRevB.96.205425,cayao2018andreev}. We stress that the dependence of the energy splitting at around zero energy on  $L_{S}$ reflects its intrinsic nature, that it comes from MZMs located at the ends of the S regions. By increasing $L_{S}$, the wavefunctions of the MZMs  overlap less and become more closer to zero energy. It is thus possible to see the length dependence of the zero-energy splitting as a consequence of the Majorana nonlocality.

\begin{figure}[t]
\begin{center}
\includegraphics[width=0.8\columnwidth]{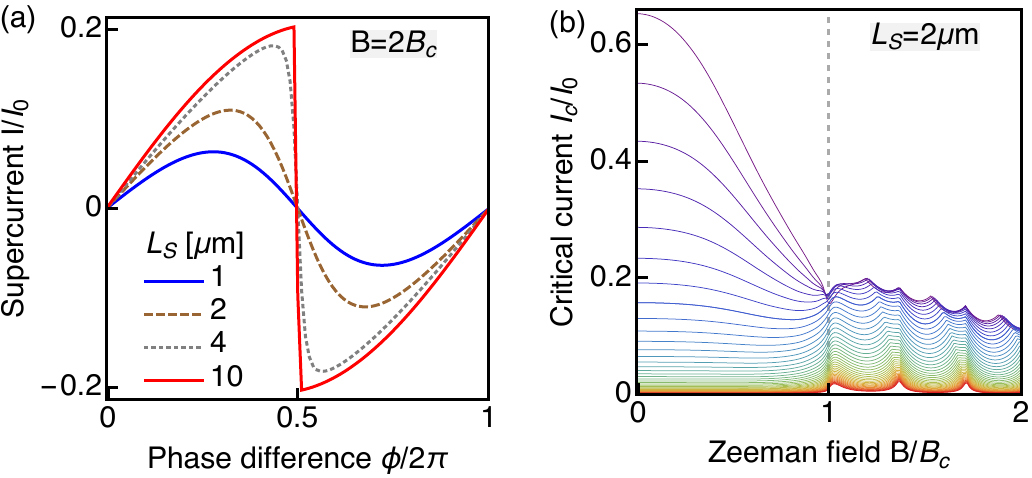}
\end{center}
\caption{(a) Current-phase curves in the topological phase ($B>B_{c}$) of a  SNS Josephson junction based on nanowires with SOC for S regions with distinct lengths. (b) Zeeman dependent on the critical currents at $L_{S}=2\mu$m  for different values of the normal transmission, ranging from the tunnel regime (bottom red curve) to the full transparent regime (top violet curve).  Parameters: $\alpha_{R}=20$meVnm, $\Delta=0.25$meV, $\mu_{N(S)}=0.5$meV, $L_{N}=20$nm, $I_{0}=e\Delta/\hbar$. Data adapted from Ref.\,\cite{PhysRevB.96.205425}.}
\label{IcSNSRashba}
\end{figure}

In relation to the supercurrents, in Fig.\,\ref{IcSNSRashba}(a) we present the current-phase curves $I(\varphi)$ in the topological phase for different $L_{S}$. In Fig.\,\ref{IcSNSRashba}(b) we show the critical current $I_{C}$ (maximum supercurrents) as a function of the Zeeman field for different values of the normal transmission, from the tunnel up to the full transparent regime. The first feature to notice is that the current-phase curves in the topological phase $B>B_{c}$ are $2\pi$-periodic. Moreover, they exhibit a strong dependence on the length of the S regions, developing a sawtooth profile near $\varphi=\pi$ for very long S regions. This sawtooth profile has been shown to be robust against variations of the normal transmission, temperature, and also against scalar disorder \cite{PhysRevB.96.205425,cayao2018andreev}, revealing that measuring the length dependence of $I(\varphi)$ in realistic nanowire-based Josephson junctions is an unambiguous indicator of Majorana physics. When it comes to the critical currents, at full transparencies, the critical current reduces at $B$ increases, remains finite and forms a kink at $B=B_{c}$, and then develops finite oscillations in the topological phase $B>B_{c}$, see top violet curve in Fig.\,\ref{IcSNSRashba}(b). The oscillations above $B_{c}$ were shown to occur due to the zero-energy splitting at $\varphi=\pi$ seen in Fig.\,\ref{SNSRashbaABS}(c), and therefore intimately related to the formation of MZMs \cite{PhysRevB.96.205425}. Thus, $I_{C}$ traces the gap closing and reveals the emergence of MZMs \cite{PhysRevB.96.205425,san2013multiple}, thus offering a way to unambiguously detect them even when the junction hosts trivial ZESs \cite{PhysRevB.104.L020501,baldo2023zero} and even use them for enhancing the efficiency of Josephson diodes \cite{PhysRevB.109.L081405}. When the normal transparency of the junction is reduced, the critical currents reduce, with a faster rate in the trivial regime $B<B_{c}$. In the tunneling regime, bottom red curve in Fig.\,\ref{IcSNSRashba}(b), the critical current in the trivial regime is vanishing small, while, interestingly, it exhibits a reentrant behavior at $B=B_{c}$ and still capturing the Majorana oscillations for $B>B_{c}$. This intriguing effect has been interpreted to be due to the fact that the contribution to the critical current from MZMs is proportional to $\sim\sqrt{\sigma_{N}}$, while from ABS the critical current is proportional to $\sim\sigma_{N}$, where $\sigma_{N}$ is the normal transmission \cite{PhysRevLett.112.137001,PhysRevB.96.205425}. Thus, for $\sigma_{N}\ll1$, the critical current from MZMs becomes a larger quantity. Another feature of Fig.\,\ref{IcSNSRashba}(b) is the change in periodicity of the oscillations for $B>B_{c}$,  where the period of the oscillations is doubled as the transparency of the junction is reduced. In this case, the tunnel regime promotes the presence of two independent pairs of Majorana state which are not coupled via the N region and oscillate with the same periodic as a function of $B$. Again, this period doubling effect is only attributed to a topological Josephson junction hosting four MZMs and its detection could help distinguish MZMs from trivial ZESs \cite{baldo2023zero}. 
Signatures of MZMs in supercurrents have also been explored in the non-equilibrium regime, with interesting studies due to multiple Andreev reflections \cite{san2013multiple,PhysRevLett.107.177002} and intriguing Shapiro steps \cite{badiane2013ac,PhysRevLett.111.046401}, but experiments, despite the efforts \cite{rokhinson2012fractional, wiedenmann20164,PhysRevX.7.021011,laroche2019observation}, so far seem not conclusive \cite{dartiailh2021missing,zhang2022missing}.
Before ending this part,  we highlight that the Zeeman dependent of the critical currents discussed here has been also reported experimentally in Refs.\,\cite{Tiira2017,PhysRevLett.126.036802}, which is compatible with the topological nature of the Josephson junction with MZMs. It is therefore encouraging to further explore phase-biased Josephson transport for the detection of MZMs and topological superconductivity.

\begin{figure}[tb]
\begin{center}
\includegraphics[width=0.7\columnwidth]{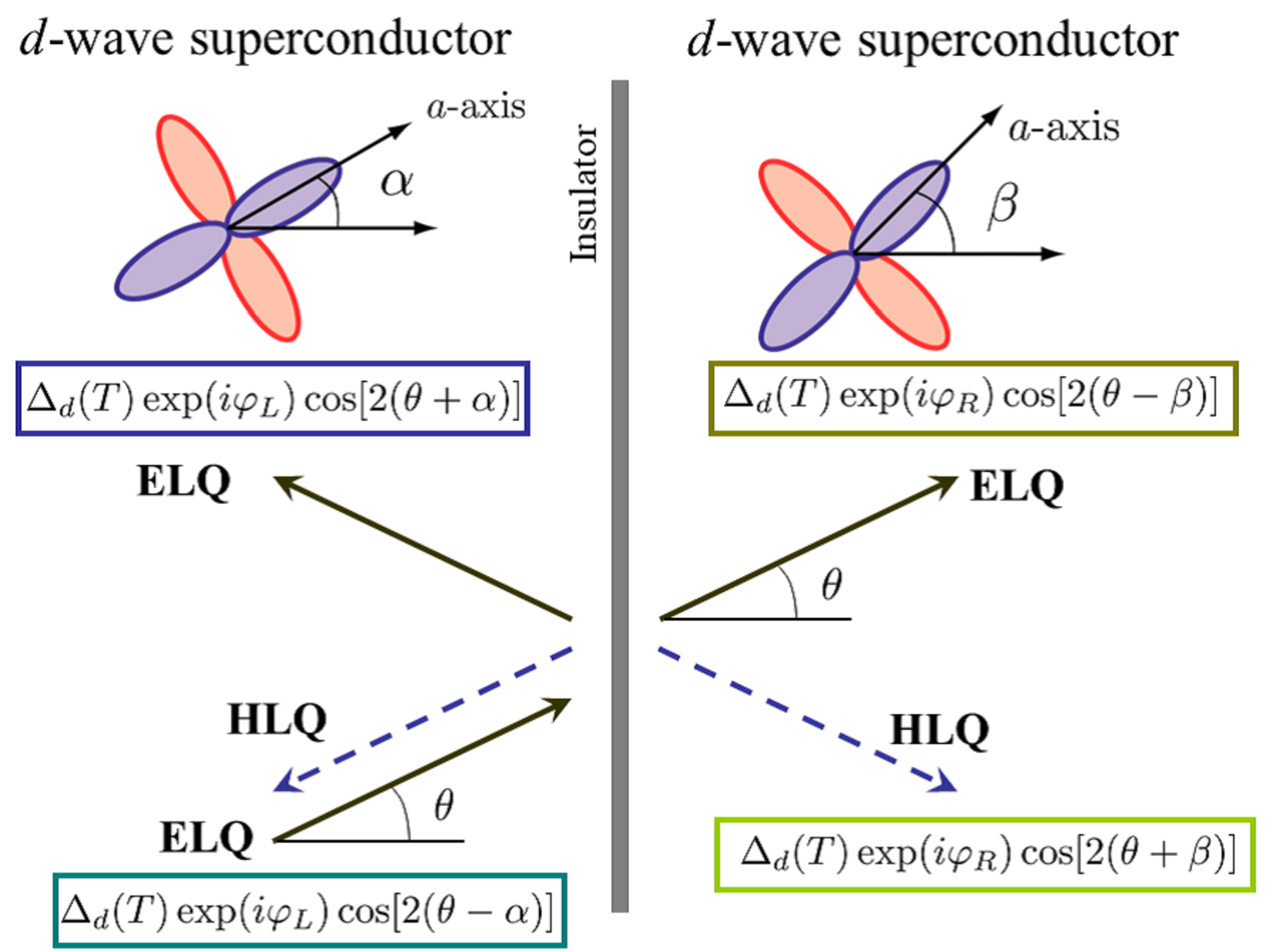}
\end{center}
\caption{Sketch of a Josephson junction formed by $d$-wave superconductors.  ELQ and HLQ denote electron-like quasiparticle and  hole-like one, respectively. 
Here, $\alpha$ ($\beta$) are angles between the crystal axis of 
$d$-wave superconductor with the normal to the interface, while $\theta$ is an incidence angle of electron-like quasiparticle from the left side of the superconductor.}
\label{dwavedwavejunction}
\end{figure}

\subsection{Josephson current in 2D spin-singlet  $d$-wave superconductor junctions} 
We here turn our attention to the d.c. Josephson effect in Josephson junctions formed between superconductors with spin-singlet  $d$-wave pair potentials, as schematically shown in Fig.\,\ref{dwavedwavejunction}. As in previous subsections, here we consider a SIS Josephson junction with the insulating region located at $x=0$ and determined by $U(x)=H_{b}\delta(x)$, while the S regions having  a $d$-wave  pair potential with a finite phase difference, which, within quasiclassical approximation, can be expressed as 
\begin{equation}
\label{2.3}
\Delta(\theta,x)
=
\left\{\begin{array} {ll}
\Delta_{L}\left(\theta \right)\exp\left(i\varphi_{L} \right), & x<0 \\
\Delta_{R}\left(\theta \right)\exp\left(i\varphi_{R} \right), & x>0
\end{array}
\right.
\end{equation}
where $\Delta_{L(R)}(\theta)=\Delta_{0}(\hat{k}^{2}_{x} - \hat{k}^{2}_{y})$ is a pair potential felt by quasiparticle where the direction of its motion is determined by the angle $\theta$ such that $\exp(i\theta)=\hat{k}_{x} + i\hat{k}_{y}$, with  $\hat{k}_{x}=k_{Fx}/k_{F}={\rm cos}(\theta)$, $\hat{k}_{y}=k_{Fy}/k_{F}={\rm sin}(\theta)$, and  $k_{F}=|\bm{k}_{\rm F}|$. We consider the situation where the 
lobe direction of the $d$-wave pair potential (crystal axis) is tilted from the 
normal to the interface by an angle $\alpha (\beta)$ as shown in Fig.\,\ref{dwavedwavejunction}. Moreover, $\varphi_{L(R)}$ is the superconducting phase of the left (right) superconductor. We are interested in the energies of the ABS and in the supercurrent carried by them. It has been shown that alternatively to the Free energy calculation discussed in the previous subsection, the Josephson current can be calculated using Andreev reflection coefficients as, 
 \begin{equation}
\label{5.1.7}
I(\varphi)=
\frac{\pi \bar{R}_{N} k_{B}T}{R_{N} e }  \sum_{\omega_{n}}
\int^{\pi/2}_{-\pi/2}
\left[ \frac{a_{e}\left(\theta,i\omega_{n},\varphi \right)}
{\tilde{\Omega}_{n,L,+}} \mid {\Delta}_{L}\left(\theta_{+} \right) \mid
- \frac{a_{h}\left(\theta, i\omega_{n},\varphi \right)}
{\tilde{\Omega}_{n,L,-}} \mid {\Delta}_{L}\left(\theta_{-} \right) \mid\right] \cos\theta
d\theta\,, 
\end{equation}
where $a_e (\theta,{\rm i}\omega_n, \varphi)$ [$a_h (\theta, {\rm i}\omega_n, \varphi)$] 
represents the Andreev reflection coefficient of   an incident electron (hole)   from the  left   superconductor~\cite{TKJosephson,TK97}, 
$\tilde{\Omega}_{n,L,\pm}={\rm sgn}
\left(\omega_{n}\right) 
\sqrt{\mid \Delta_{L}\left(\theta_{\pm}\right) \mid^{2}+ \omega_{n}^{2}}$, with $\omega_{n}$ being Matsubara frequencies,  $\theta_{+}=\theta$,  $\theta_{-}=\pi-\theta$, and  $R_{N}$ is the normal state  resistance   of the 
2D junction  and $\bar{R}_{N}$ is obtained as
\begin{equation}
\label{5.1.8}
\begin{split}
\bar{R}_{N}^{-1} = \int^{\pi/2}_{-\pi/2} \sigma_{N}(\theta) 
\cos\theta d\theta\,,
\end{split}
\end{equation}
where $\sigma_{N}(\theta)=
1/(1 + Z_{\theta}^{2})$, $Z_{\theta}={Z}/{\cos \theta}$, $Z= {mH_{b}}/{(\hbar^{2}k_{F})}$. In what follows, we set  $\sigma_{N}(\theta)\equiv\sigma_{N}$ 
and discuss  results for $d$-wave Josephson junctions. Before doing that, however, it is important to mention that, generally speaking, for the considered  $d$-wave Josephson junctions, it is possible to have distinct types of pair potentials. For instance, for  scattering  from an electron-like quasiparticle 
injected from the left  superconductor,  four types of pair potentials appear, which can be summarized as
\begin{equation}
\begin{split}
\Delta_{L}(\theta_\pm)&=
\Delta_{0}\cos[2(\theta \mp \alpha)]\exp(i\varphi_{L})\,,\\
\Delta_{R}(\theta_\pm)&
=\Delta_{0}\cos[2(\theta \mp \beta)]\exp(i\varphi_{R})\,.
\end{split}
\end{equation}
where $\Delta_{0}$ is the pair potential amplitude, $\theta$ is the angle of incidence with respect to the $x$-axis, while   $\alpha$ and $\beta$ represent 
the angles between the lobe directions of the $d$-wave pair potentials 
in the left and right superconductors and the normal to the interface 
of the junction, see Fig.\,\ref{dwavedwavejunction} for details.

\begin{figure}[tb]
\begin{center}
\includegraphics[width=0.8\columnwidth]{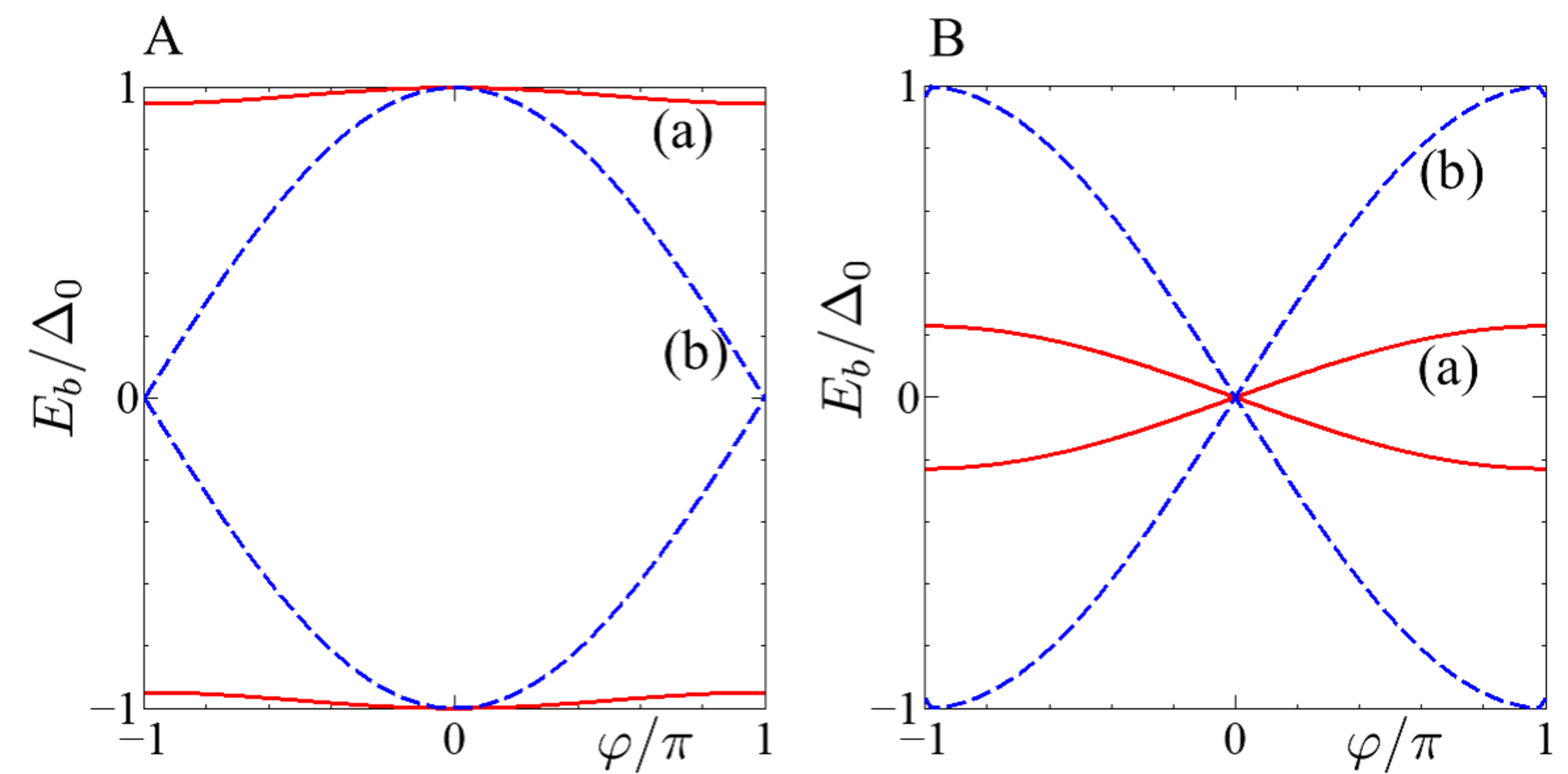}
\end{center}
\caption{Energies of the ABSs in Josephson junctions with $d$-wave superconductors as a function of $\varphi$ at fixed $\theta$. Different panels correspond to: (A) $\theta=0$ and $\alpha=0$ and  (B) $\theta=\pi/4$ and $\alpha=\pi/4$. Red and blue curves correspond to (a) $Z=3$ and (b) $Z=0$.}
\label{dwavedwaveboundstate1}
\end{figure}

\begin{figure}[tb]
\begin{center}
\includegraphics[width=0.9\columnwidth]{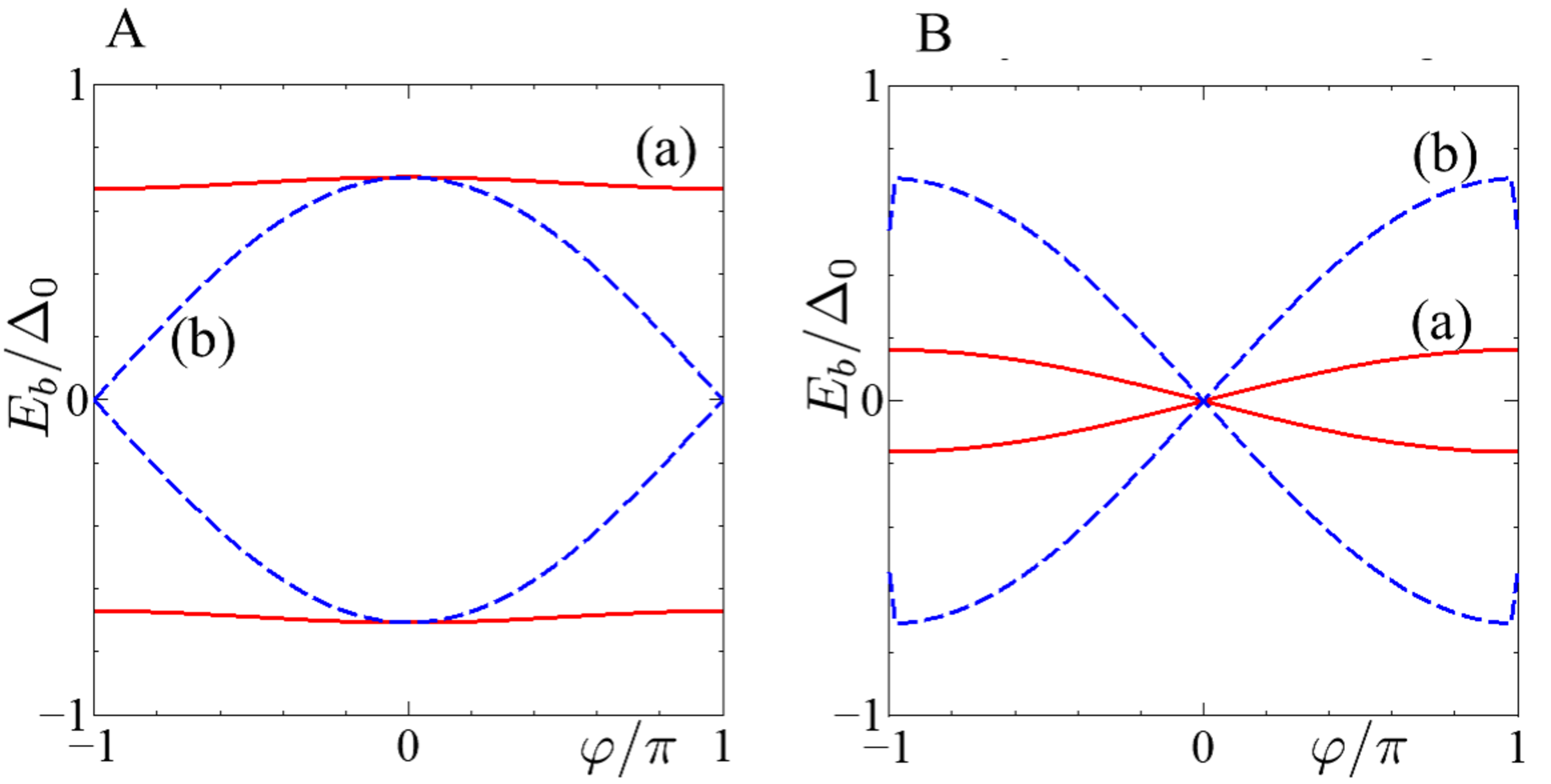}
\end{center}
\caption{Energies of the ABSs in Josephson junctions with $d$-wave superconductors as a function of $\varphi$ at fixed $\theta$. Different panels correspond to: (A) $\theta$=0 and $\alpha=\pi/8$ and 
(B) $\theta=\pi/4$ and $\alpha=\pi/8$. Red and blue curves correspond to (a) $Z=3$ and (b) $Z=0$.}
\label{dwavedwaveboundstate2}
\end{figure}

\begin{figure}[tb]
\begin{center}
\includegraphics[width=0.9\columnwidth]{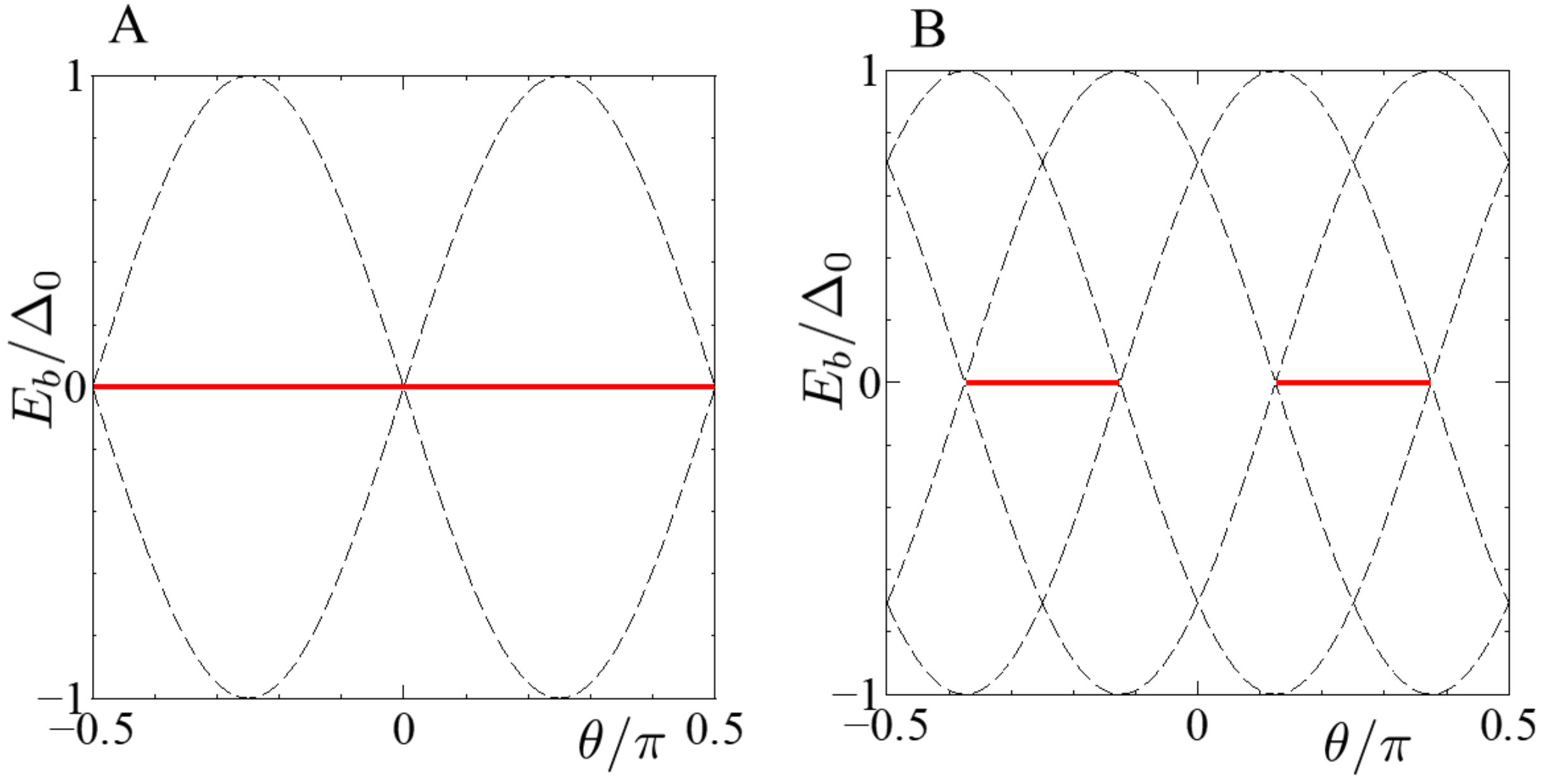}
\end{center}
\caption{Energies $E_{b}$ of the ABSs in a Josephson junction with $d$-wave superconductors as a function of $\theta$ at $Z=3$ and distinct values of $\varphi$: (A) $\varphi=0$ and $\alpha=\pi/4$, 
(B) $\varphi=0$ and $\alpha=\pi/8$. The solid line is $E_{b}$
and dashed curves correspond to    $\pm \Delta_{L+}(\theta)$ and $\pm \Delta_{L-}(\theta)$.}
\label{dwavedwaveboundstatetheta}
\end{figure}

Now we consider a mirror-type Josephson junction, where   $\alpha=-\beta$,
where the energies of the ABSs can be analytically obtained. In particular, for these types of junctions, the Josephson current has been found to be equal to \cite{TKJosephson,TK97}
\begin{equation}
\label{eq2.44}
R_{N}I\left(\varphi\right)= \frac{\pi \bar{R}_{N} k_{B}T}{e }
\sum_{\omega_{n}}
\int^{\pi/2}_{-\pi/2} F\left(\theta,i\omega_{n},\varphi\right) \sin(\varphi) \,
\sigma_{N}\cos(\theta) d\theta\,,
\end{equation}
where $\varphi=\varphi_{L}-\varphi_{R}$, $\bar{R}_{N}$ is given by Eq.\,(\ref{5.1.8}), and 
\begin{equation}
F\left(\theta,i\omega_{n},\varphi\right)
=\frac{2 \Delta_{L+}\left(\theta\right) \Delta_{L-}\left(\theta\right) }
{ \left[\tilde{\Omega}_{n,L,+} \tilde{\Omega}_{n,L,-} + \omega_{n}^{2}
+ \left(1-2\sigma_{N}\sin^{2} \varphi/2\right)
\Delta_{L+}\left(\theta\right)
\Delta_{L-}\left(\theta\right)\right]}\,.
\label{mirrorjunction1}
\end{equation}
with  $\Delta_{L\pm}(\theta)=\Delta_{R\mp}(\theta)$.
To find the ABS, we can search for the zeroes of the denominator of Eq.\,(\ref{mirrorjunction1}) in real energies, $i\omega_{n}\rightarrow E$, whose expression then reads,
\begin{equation}
\label{ABSdwaveEq}
-\Omega_{L+}\Omega_{L-} -E^{2}   
+ \left(1-2\sigma_{N}\sin^{2} \varphi/2\right)
\Delta_{L+}\left(\theta\right)
\Delta_{L-}\left(\theta\right)\,,
\end{equation}
where $\Omega_{L\pm}$ are obtained as
\begin{equation}
\Omega_{L\pm}
\equiv \lim_{\delta \rightarrow 0} 
\sqrt{\left(E + i\delta \right)^{2} - \Delta_{L\pm}^{2}\left(\theta\right)}
= \displaystyle
\left \{
\begin{array}{ll}
\sqrt{E^{2} - \Delta_{L\pm}^{2}\left(\theta\right)}\,, & 
\quad E  \geq |\Delta_{L\pm}\left(\theta\right)|\,, \\
i \sqrt{\Delta_{L\pm}^{2}\left(\theta\right) - E^{2}}\,, &  
\quad -|\Delta_{L\pm}\left(\theta\right)| \leq E \leq 
|\Delta_{L\pm}\left(\theta\right)|\,, \\ 
-\sqrt{E^{2} - \mid \Delta_{L\pm}\left(\theta\right)^{2}}\,, & \quad E  \leq -
|\Delta_{L\pm}\left(\theta\right)|\,.
\end{array}
\right.
\label{Omegapmxx}
\end{equation}
Then, by solving Eq.\,(\ref{ABSdwaveEq}) for $E$, we find the energies of the ABSs given by \cite{kashiwaya00,Tanaka2021}
\begin{equation}
\begin{split}
E_{b}(\varphi)&=\pm 2\sqrt{\sigma_{N}}
\sin\left(\frac{\varphi}{2}\right) |\Delta_{L+}(\theta)| 
|\Delta_{L-}(\theta)|\\
&\times
\sqrt{\frac{1 - \sigma_{N} \sin^{2}\left(\frac{\varphi}{2}\right)}
{[\Delta_{L+}(\theta)-\Delta_{L-}(\theta)]^{2} + 
4 \sigma_{N} \sin^{2}\left(\frac{\varphi}{2}\right) \Delta_{L+}(\theta)
\Delta_{L-}(\theta)}}\,,
\end{split}
\end{equation}
with $\mid E_{b}(\varphi) \mid \leq {\rm min}{\rm}
(|\Delta_{L+}(\theta)|, |\Delta_{L-}(\theta)|)$. The ABSs energies acquire simpler forms at fixed angles. For instance,  for $\alpha=0$ and $\alpha=\pi/4$, 
$E_{b}$ can be simply expressed as \cite{kashiwaya00,TK96a}
\begin{equation}
E_{b}(\varphi)=
\left \{
\begin{array}{lll}
\pm \Delta_{eff} \sqrt{\cos^{2}\left(\frac{\varphi}{2}\right) + \left(1 - \sigma_{N}\right)\sin^{2}\left(\frac{\varphi}{2}\right)}\,, &\quad  \Delta_{eff}=\Delta_{0}\cos 2\theta\,, &\quad \alpha=\beta=0\,, \\ 
\pm \Delta_{eff} \sqrt{\sigma_{N}}\sin\left(\frac{\varphi}{2}\right)\,, & \quad 
\Delta_{eff}=\Delta_{0}\sin 2\theta\,, &\quad \alpha=-\beta=\pm\pi/4\,. 
\end{array}
\right.
\end{equation}
From these two expressions, we can already draw some important conclusions. For instance, for $\alpha=0$,  the obtained phase-dependent ABS $E_{b}(\varphi)$  is essentially equivalent to that of spin-singlet $s$-wave superconductor junction discussed in Eq.\,(\ref{Andreevboundstate1dswave}).
On the other hand, for $\alpha=\pm \pi/4$, the ABS energy $E_{b}(\varphi)$ always exhibits a $\sin(\varphi/2)$ dependence which is independent of the interface transparency and forms a $4\pi$ periodicity with $\varphi$ \cite{TK96a,TKJosephson}. This $4\pi$ periodicity appears in spin-triplet $p_{x}$-wave superconductor junctions discussed in Eq.\,(\ref{Majoranaboundstate})  and has been understood to be the key for detecting MZMs in the ac Josephson current~\cite{Yakovenko2004}. To understand this discussion, in  Figs.\,\ref{dwavedwaveboundstate1} and
\ref{dwavedwaveboundstate2}, we present the ABS energies as a function of phase $E_{b}(\varphi)$ for distinct values of $\theta$, $\alpha$, and $Z$. For completeness, in Fig.\,\ref{dwavedwaveboundstatetheta} we also show the ABS energies as a function of $\theta$ at fixed $\varphi=0$. For $\alpha=\pi/4$, the ZESABS with $E_{b}=0$ appears for  $-\pi/2< \theta <\pi/2$ except for  the nodal direction with $\theta=0$, as indicated by red color in Fig.\,\ref{dwavedwaveboundstatetheta}(A).   For $\alpha=\pi/8$, the ZESABS appears for  $-\pi/8<\mid \theta \mid <3\pi/8$, as depicted by red color in Fig.\,\ref{dwavedwaveboundstatetheta}(B). It is also mentioned that the superconducting diode effect can appear for 
$d$-wave superconductor junction on the surface of the topological insulator 
where ZESABS plays an important role \cite{Tanakadiode2022}.

\begin{figure}[tb]
\begin{center}
\includegraphics[width=0.45\columnwidth]{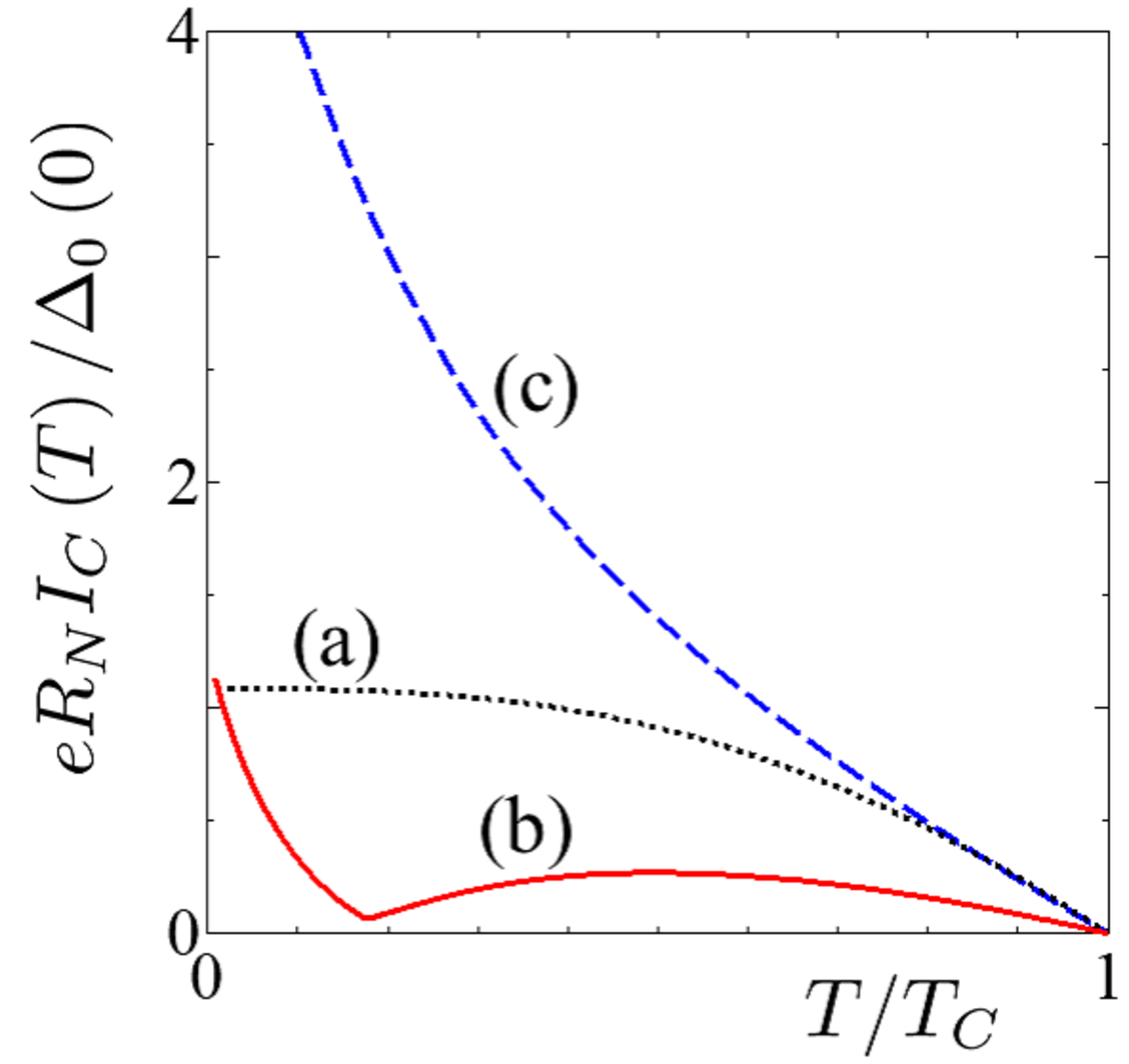}
\end{center}
\caption{Temperature dependence of the critical current $I_{C}(T)$ of mirror type 
$d$-wave Josephson junction where $\alpha=-\beta$ at $Z=2$. Different curves correspond to: (a) $\alpha=0$, (b) $\alpha=0.1\pi$, (c) $\alpha=0.25\pi$.  Here, $\Delta_{0}(0)$ is the magnitude of the pair potential $\Delta_{0}$ at zero temperature.}
\label{dwaveJosephson}
\end{figure}

Having discussed the ABSs,  we now calculate the Josephson current using Eq.\,(\ref{eq2.44}).  It is found that for $\alpha =0$ and $\alpha=\pm \pi/4$,  
it is possible to sum over Matsubara frequencies analytically in Eq.\, (\ref{mirrorjunction1}).  In the case of $\alpha=0$, the phase-dependent supercurrent $I(\varphi)$ is given by  
\begin{equation}
\label{eq2.48}
R_{N}I(\varphi)
=\frac{\pi \bar{R}_{N}}{e}
\int^{\pi/2}_{-\pi/2}
\frac{\Delta_{0}\cos2\theta \sigma_{N}\cos\theta \sin\varphi}
{2\sqrt{1 -\sigma_{N}\sin^{2}\left(\varphi/2\right)}}
{\rm tanh} \left[\frac{\Delta_{0}\cos2\theta
\sqrt{1 -\sigma_{N}\sin^{2}\left(\varphi/2\right)}}
{2k_{B}T}\right]
d\theta\,.
\end{equation}
The temperature dependence of the critical current $I_{C}(T)$ is 
shown in Fig.~\ref{dwaveJosephson}, which essentially becomes  equivalent to the case with spin-singlet $s$-wave case found by Ambegaokar-Baratoff \cite{Ambegaokar} and shown in Fig.~\ref{1dJosephsontemp}(A). 

On the other hand, for $\alpha=-\beta=\pi/4$, the phase-dependent supercurrent 
$R_{N}I(\varphi)$ becomes
\begin{equation}
\label{5.3.2}
R_{N}I\left(\varphi\right)
=-\frac{\pi \bar{R}_{N}}{e}
\int^{\pi/2}_{-\pi/2}
\frac{\Delta_{eff}\left(T,\theta\right) \sin\varphi}
{2\sqrt{\sigma_{N}}\cos\left(\varphi/2\right)}
{\rm tanh} \left[ \frac{\Delta_{eff}
\left(T,\theta\right)
\cos\left(\varphi/2\right)\sqrt{\sigma_{N}}}
{2k_{B}T} \right] \sigma_{N}\cos\theta d\theta,
\end{equation}
with $\Delta_{eff}(T,\theta)= \Delta_{0}\sin(2\theta)$.  In the limit of 
$\sqrt{\sigma_{N}} |\Delta_{eff}(T,\theta)|\ll 2k_{B}T$, the critical current $I_{C}(T)$ is approximated to be 
\begin{equation}
\label{5.3.3}
R_{N}I_{C}\left(T\right)
= \frac{\pi \bar{R}_{N}}{4ek_{B}T}
\int ^{\pi/2}_{-\pi/2} \Delta^{2}_{eff}\left(T,\theta\right)\sigma_{N}
\cos\theta d\theta.
\end{equation}
which increases with the decrease of the temperature \cite{TKJosephson,TK97,YBJosephson}, as can be seen in   
Fig.~\ref{dwaveJosephson}(c). This temperature behaviour is also seen in the case of $p$-wave Josephson junction discussed in the last subsection.

For sufficiently low temperatures, with $k_{B}T\ll \sqrt{\sigma_{N}} |\Delta_{eff}(T,\theta)|$, the critical current acquires the following form 
\begin{equation}
\label{5.3.4}
R_{N}I_{C}\left(T \right)
\sim \frac{\pi \bar{R}_{N}}{e}
\int ^{\pi/2}_{-\pi/2}
\mid \Delta^{2}_{eff}\left(T,\theta \right) \mid \sqrt{\sigma_{N}}
\cos \theta d\theta.
\end{equation}
$R_{N}I_{C}(T)$  is proportional to $\sqrt{R_{N}}$ 
(inverse of $\sqrt{\sigma_{N}}$) 
similar to the 1D $p$-wave superconductor junction discussed in the previous subsection.  Furthermore, we also note that the critical currents exhibit a non-monotonic temperature dependence for $\alpha$ within $0 \leq \alpha \leq \pi/4$, which is shown in Fig.~\ref{dwaveJosephson}(b), see Refs.\,\cite{TKJosephson,TK97,YBJosephson,Tanaka2021}

In order to understand this exotic behavior of the critical currents discussed in the previous paragraph, we decompose  $R_{N}I(\varphi)$ into negative and positive parts, denoted as $G_{n}(\varphi)$   and  
$G_{p}(\varphi)$, respectively, for     $0<\varphi<\pi$
\cite{TK97,Tanaka2021}. Then, these positive and negative parts are given as
\begin{equation}
\label{5.3.9}
\begin{split}
G_{n}\left(\varphi \right)&=  \frac{\bar{R}_{N} \pi k_{B}T}{e }
\Big\{ \sum_{\omega_{n}}
\int^{-\pi/4+ |\alpha|}_{-\pi/4-|\alpha|}
F\left(\theta,i\omega_{n},\varphi\right)
\sigma_{N}\cos\theta d\theta \\
&+\int^{\pi/4 + |\alpha| }_{\pi/4- |\alpha|}
F\left(\theta,i\omega_{n},\varphi\right)
\sigma_{N}\cos\theta d\theta \Big\} \sin \varphi\,,\\
G_{p}(\varphi)&=R_{N}I(\varphi) - G_{n}(\varphi)\,,
\end{split}
\end{equation}
where $F(\theta,i\omega_{n},\varphi)$ is given by Eq.~(\ref{mirrorjunction1}). Here, the term $G_{p}$ denotes the contribution  to $R_{N}I(\varphi)$ coming from    the domain of $\theta$ in  $\pi/4 + |\alpha| < |\theta|<\pi/2$ and $-\pi/4 + |\alpha|< \theta < \pi/4 -|\alpha|$. By taking Eq.~(\ref{mirrorjunction1})   into account, it is not difficult to see that, when    $\sigma_{N}<1/2$, the  denominator of $F(\theta,i\omega_{n},\varphi)$ becomes small $\pm \pi/4 - |\alpha| < \theta < \pm \pi/4 + |\alpha|$ because
  the sign of   $\tilde{\Omega}_{n,L,+} \tilde{\Omega}_{n,L,-}$ and   
 $\Delta_{L+}\left(\theta\right)\Delta_{L-}\left(\theta\right)$  are different. 
In particular, in the low transparent limit, when $\sqrt{\sigma_{N}}|\Delta_{eff}(T,\theta)|\ll |\omega_{n}| \ll |\Delta_{eff}(T,\theta)|$, the negative part
$G_{n}(\varphi)$ is given as \cite{TK97}
\begin{equation}
G_{n}\left(\varphi\right)
= \frac{-8\pi \bar{R}_{N} k_{B}T}{e}
\int^{\pi/4 +|\alpha|}_{\pi/4- |\alpha|}
\sum_{\omega_{n}}
\frac{|{\Delta}_{L}\left(\theta_{+}\right)|^{2}
|{\Delta}_{L}\left(\theta_{-}\right)|^{2} }
{\omega_{n}^{2}
[|{\Delta}_{L}\left(\theta_{+}\right)|
+ |{\Delta}_{L}\left(\theta_{-}\right)| ]^{2}}
\sigma_{N}\cos \theta d\theta
\propto \frac{1}{T}\,.
\end{equation}
This expression reveals that the magnitude of $G_{n}(\varphi)$ is enhanced with the decrease of temperature \cite{TK97,Tanaka2021}.  On the other hand,  $G_{p}(\varphi)$ becomes almost constant at low temperatures since the denominator of $G_{p}(\varphi)$ is not suppressed with the decrease in temperature due to the absence of ZESABS since $\Delta_{L}(\theta_{+})\Delta_{L}(\theta_{-})\geq0$ is satisfied.  At high temperatures,
the phase-dependent supercurrent $I(\varphi)$ becomes positive for $0<\varphi<\pi$ because  the magnitude of $G_{p}(\varphi)$ exceeds   that of $G_{n}(\varphi)$. On the other hand, at low temperatures, 
$|G_{n}(\varphi)| > G_{p}(\varphi)$ is satisfied and the resulting phase-dependent supercurrent  $I(\varphi)$ becomes negative for $0<\varphi<\pi$. 
Another feature to note is that, with the decrease of the temperature, a crossover from $0$-to-$\pi$-junction occurs \cite{TK97}.  To close, we point out that the non-monotonic temperature dependence of the Josephson current has been experimentally observed in Josephson junctions of high $T_{c}$ cuprate ~\cite{Ilichev,Testa} and it is qualitatively consistent with curve (b) in Fig.\,\ref{dwaveJosephson}.

\subsection{Summary}
We have discussed the Josephson effect in phase-biased Josephson junctions and showed that they can be very useful in detecting MZMs. We first showed the ABSs and current-phase curves in Josephson junctions with semi-infinite spin-singlet $s$-wave superconductors. Then, we addressed the ABSs, current-phase curves in the topological phase of Josephson junctions with semi-infinite $p$-wave superconductors, showing that the topological ABSs develop a robust zero-energy crossing at $\varphi=\pi$ that leads to current-phase curves with a skewed profile that is robust to barrier imperfections. Moreover, in this part, we also showed that the ABSs and supercurrents in finite  Josephson junctions based on nanowires with SOC provide crucial information to detect MZMs in realistic systems. Specially, these systems host four MZMs whose spatial nonlocality permits to detect them in an unambiguous manner via current-phase curves at $\varphi=\pi$  and critical currents, even when topologically trivial ZESs are also present. Lastly, we showed that Josephson junctions made of $d$-wave superconductors host intriguing ABSs, which can even become completely flat at zero energy and promote large critical currents at low temperatures. 

\section{Summary and outlook}
In this review article, we have summarized some theoretical aspects of MZMs from the viewpoint of SABSs in unconventional superconductors. We have started by discussing that MZMs are a special type of ZESABS in spin-polarized $p$-wave superconductors emerging at zero-energy located at the edges and are a self-conjugate quasiparticle. We have then clarified that the regime where MZMs appear corresponds to a topological phase which can be also realized by using semiconductors with SOC, conventional spin-singlet $s$-wave superconductivity and a large Zeeman field. We have later shown that, by using Green's functions of $p$-wave superconductors, the LDOS can reveal the emergence of MZMs. Moreover, it also helps to understand that MZMs always accompany the formation of dominant odd-frequency spin-triplet $s$-wave pair correlations, which represent yet another intriguing but less explored property of MZMs. We have also discussed that MZMs produce unique signatures in charge conductance, anomalous proximity effect, and skewed phase-biased Josephson currents. While independently these effects can help to identify MZMs, it is necessary to explore all  Majorana properties together, such as their zero-energy nature, their spatial nonlocality, and their odd-frequency pairing.  

When exploring conductance, it is necessary to perform simultaneous measurements of ZBCPs at both ends of the system, which should exhibit a strong dependence on the length of the system as a signature of spatial nonlocality. Short systems should exhibit an oscillatory conductance around zero energy, which becomes pinned at zero energy only for long wires and originate the ZBCPs. When trying to detect MZMs in supercurrents, a length-dependent profile of the current-phase curves should be inspected to probe Majorana nonlocality. Along these lines, MZMs and the topological phase can be also revealed by exploiting the Zeeman and system's length dependencies of the critical currents. With these ideas, it might be possible to access the zero-energy nature of MZMs and also their spatial nonlocality. Of course that apart from the predicted Majorana signatures, it is necessary to also test them against effects where it is expected that they do not survive. For instance, the Majorana signatures can be tested under the presence of a Zeeman field parallel to the SOC, which is expected to reduce the induced gap where MZMs appear and thus destroy them. To test the odd-frequency pairing of MZMs, it could be possible to further explore the anomalous proximity effect, perhaps in the same way as the long-range proximity effect in superconductor-ferromagnet junctions \cite{Efetov2} because in both cases the spin symmetry of the pair amplitude is spin-triplet. Another less explored signature is the topological phase transition which is expected to occur in the bulk, and should be carried out along with the measurements of the MZMs at the edges. 

While the ideas discussed here might not solve all the difficulties, they could certainly help overcome some of the problems in relation to the unambiguous identification of MZMs when topologically trivial zero-energy states also appear.  Despite the challenges,  all the theoretical and experimental studies are clearly advancing our understanding of the phenomena present in Majorana devices. 

\section*{Acknowledgments}
We thank  R. Aguado, Y. Asano, O. Awoga, M. Benito, S. Bergeret, A. Black-Schaffer, M. Cuoco, P. Burset, D. Chakraborty, Y. Fukaya, A. Furusaki, P. Gentile, A.A. Golubov, S. Hoshino, S. Ikegaya, T. Kokkeler, S. Kashiwaya, Y. Kawaguchi, S. Kobayashi, B. Lu, S. Matsuo, T. Mizushima, N. Nagaosa, R. Nakai, K. Nomura, S. Nakosai, M. Leijnse, M. Sato, B. Sothmann, P. Stano, S. Suzuki, B. Trauzettel, K. Yada, and A. Yamakage for insightful discussions.  Y. T. acknowledges financial support from JSPS with Grants-in-Aid for Scientific Research (KAKENHI Grants No. 20H00131, No. 23K17668, and 24K00583.). S. T. thanks for the support of 
the Würzburg-Dresden Cluster of Excellence ct.qmat, EXC2147, project-id 390858490, the DFG (SFB 1170), and the Bavarian Ministry of Economic Affairs, Regional Development and Energy within the High-Tech Agenda Project “Bausteine für das Quanten Computing auf Basis topologischer Materialen”. 
J. C. acknowledges financial support from the Swedish Research Council (Vetenskapsr{\aa}det Grant No. 2021-04121)  and the Carl Trygger’s Foundation (Grant No. 22: 2093). 

\vspace{0.2cm}
\noindent

\let\doi\relax
\bibliographystyle{ptephy}
\bibliography{BiblioMajorana}
%



\end{document}